\documentclass{aa} 
\usepackage[T1]{fontenc} 
\usepackage{siunitx}
\usepackage{amsmath}
\usepackage{amssymb}
\usepackage{graphicx}
\usepackage{longtable} 
\usepackage{lscape}
\usepackage[export]{adjustbox}
\usepackage{natbib}
\usepackage{subcaption}
\usepackage{calrsfs}
\usepackage{array}
\usepackage{hyperref}
\usepackage{multirow}
\usepackage{gensymb}
\usepackage[capitalise]{cleveref} 

\bibpunct{(}{)}{;}{a}{}{,}

\newcommand{\Mjup}{M$_{\rm Jup}$\,}
\newcommand{\Mjupv}{M$_{\rm Jup}$}

\newcommand{\Msun}{M$_{\sun}$}

\newcommand{\msini}{$m_{\rm p}\sin{i}$\,}
\newcommand{\rhk}{log$R'_{\rm HK}$}

\begin{document} 

 \title{Multi techniques approach to identify and/or constrain radial velocity sub-stellar companions}


 \author{F. Philipot\inst{1}\fnmsep\thanks{Please send any request to florian.philipot@obspm.fr}
 \and
 A.-M. Lagrange\inst{1,} \inst{2} 
 \and
 F. Kiefer \inst{1} 
 \and
 P. Rubini\inst{3} 
 \and
 P. Delorme\inst{2}
 \and
 A. Chomez \inst{1,2}
 }

 \institute{
LESIA, Observatoire de Paris, Universit\'{e} PSL, CNRS, 5 Place Jules Janssen, 92190 Meudon, France
\and
Univ. Grenoble Alpes, CNRS, IPAG, 38000 Grenoble, France
\and
Pixyl S.A. La Tronche, France
 }
 \date{Received 07 April 2023 / Accepted 8 August 2023}

 
 \abstract
 {Although more than one thousand sub-stellar companions have already been detected with the radial velocity (RV) method, many new companions remain to be detected in the public RV archives.}
 {We wish to use the archival data obtained with the ESO/HARPS spectrograph to search for sub-stellar companions.}
 {We use the astronomic acceleration measurements of stars obtained with the Hipparcos and Gaia satellites to identify anomalies that could be explained by the presence of a companion. Once hints for a companion are found, we combine the RV data with absolute astrometry and, when available, relative astrometry data, using a Markov Chain Monte Carlo (MCMC) algorithm to determine the orbital parameters and mass of the companion.}
 {We find and characterize three new brown dwarfs (GJ660.1 C, HD73256 B, and HD165131 B) and six new planets (HD75302 b, HD108202 b, HD135625 b, HD185283 b, HIP10337 b, and HIP54597 b) with separations between 1 and 6 au and masses between 0.6 and 100 \Mjupv. We also constrain the orbital inclination of ten known sub-stellar companions and determine their true mass. Finally, we identify twelve new stellar companions. This shows that the analysis of proper motion anomalies allows for optimizing the RV search for sub-stellar companions and their characterization.}
 {}


 \maketitle

%

\section{Introduction}

Since the first discovery of a planet orbiting a solar-type star \citep{1995Natur.378..355M}, various radial velocity (RV) surveys have detected hundreds of extra-solar planets and brown dwarfs (\cite{2011arXiv1109.2497M}, \cite{2016A&A...588A.145H}, \cite{2020MNRAS.492..377W}, \cite{2021ApJS..255....8R}). In the long term, these surveys aim at estimating the minimum mass and radial distributions of these companions and comparing these values with results obtained by different population synthesis models (\cite{2018haex.bookE.143M}, \cite{2021A&A...656A..69E}). 

Despite the many detections, numerous RV data available in the public archives have never been published and could potentially contain the signals of unknown companions. The proper motion acceleration measurements published in Gaia data release 3 for more than 1.4 billion stars can be used to select the targets with the highest probability of having a companion. Indeed, the study of the star's proper motion accelerations has recently allowed the detection in high contrast imaging (hereafter HCI) of a few sub-stellar companions (e.g., \cite{2022MNRAS.513.5588B}, \cite{2022AJ....164..152S}, \cite{2023AJ....165...39F}, \cite{2022ApJ...934L..18K}).

Absolute astrometry based on position and proper motion measurements obtained with Gaia \citep{2020yCat.1350....0G} and Hipparcos (\cite{1997A&A...323L..49P}, \cite{2007A&A...474..653V}) can furthermore remove the indetermination on the companion orbital inclination that is left with the RV method. For instance, using only the RV data, \cite{2019AJ....157..252K} reported a planet candidate with a minimum mass of about 10 \Mjup while, combining the RV data with absolute astrometry, \cite{2021AJ....162..266L} found a highly inclined orbit and a true mass of about 20 \Mjup. Since the first Gaia data release, the combination of these two techniques has allowed us to improve the characterization of tens of sub-stellar companions (\cite{2019A&A...629C...1G},\cite{2019AJ....158..140B}, \cite{2019A&A...631A.125K}, \cite{2020A&A...642A..31D}, \cite{2021AJ....162...12V}, \cite{2021AJ....161..179B}, \cite{2021AJ....162..301B}, \cite{2021A&A...645A...7K}, \cite{2021AJ....162..266L}, \cite{2022ApJS..262...21F}, \cite{2023A&A...670A..65P}).

However, the precise mass of a companion can only be determined if the orbital parameters are properly constrained, which remains difficult when the orbital period (P) is much larger than the RV baseline \cite{2023arXiv230500047L}. In the case of poorly constrained periods, the addition of relative astrometry from HCI or interferometry allows us to significantly improve the characterization of the companion. This is, for instance, the case of HD 211847 B which was first reported using only RV data as a low-mass brown dwarf candidate with very poorly constrained orbital parameters and minimum mass \citep{2011A&A...525A..95S}, then reported as a massive brown dwarf adding absolute astrometry \citep{2022ApJS..262...21F} and finally identified as a low-mass star by adding relative astrometry data (\cite{2023A&A...670A..65P}).

In this paper, we study the proper motion anomalies of the stars observed with the HARPS/ESO spectrograph, with the aim of detecting new sub-stellar companions. We also combine the available RV data and absolute astrometry with Hipparcos and Gaia measurements plus, when available, relative astrometry to further constrain the orbital and mass properties of the newly detected and the already known sub-stellar and stellar companions identified. 

In Section 2, we describe the selection of our sample. In Section 3, we present the RV, astrometric, and HCI data and the algorithm used to perform the orbital fits. We then provide the orbital parameters and masses of the new companions, as well as the update of these parameters for the known companions, in Sections 4 and 5, respectively. In Section 6, we summarize the results obtained for stellar mass companions. Finally, we summarize the results and conclude in Section 6.

\section{Sample}

\subsection{Proper motion anomaly for stars observed by HARPS-South}

We first select the 734 stars observed with the High Accuracy Radial Velocity Planet Searcher (HARPS) spectrograph mounted on the ESO 3.6 m telescope at La Silla \citep{2003Msngr.114...20M} that are listed in the HARPS-RVBANK archive \citep{2020A&A...636A..74T} with more than 30 RV data points spread on a temporal baseline greater than 1000 days. Secondly, we use the astrometric acceleration of stars (so-called proper motion anomaly, \cite{2019A&A...623A..72K}) determined by the difference in proper motion (hereafter PM) between the average Hipparcos-Gaia PM (24.5 yr time baseline) and the "instantaneous" PM published in the Gaia data release 3 (hereafter DR3) to select systems with the highest probability of hosting a sub-stellar companion. To do so, we select the stars with a proper motion anomaly (hereafter PMa) by 2$\sigma$ larger than the noise. We simulate many samples of Gaia DR3 measurements for single stars of similar $G$-magnitudes and $Gb-Gr$ colors, out of which we fit a PMa solely based on pure noise. For any of the 734 stars, we obtain PMa distributions of those stars as if they were single. The percentile of the PMa, measured in \cite{2022A&A...657A...7K}, within these distributions determines by how many $\sigma$ they exceed the noise. To focus our study on the detection of sub-stellar companions, we then exclude systems for which the predicted mass of a companion orbiting at 10 au that could account for the measured PMa value is larger than 200 \Mjupv. We obtain a sample of 116 stars.

\subsection{Selection of systems with a substellar companion}

For the selected sample of stars, we then retrieve all the RV measurements available in the public archives of twelve different spectrographs (see below) and analyze them to identify the presence of companion(s). For this step, we use the DPASS genetic algorithm\footnote{DPASS uses an evolutionary algorithm to fit RV data with one or more Keplerian orbits. This algorithm converges to the solution corresponding to the smallest root-mean-square (hereafter rms) of the RV residuals.} \citep{{2019NatAs...3.1135L}} to detect periodicity in the data and make a rapid estimate of the orbital properties and the minimum mass of the companion(s). Thus, we find 20 multiple systems (two or more companions) and 80 single systems (one companion).

Then, we select the systems for which the orbital parameters of the companion are properly determined and the mass is lower than about 80 \Mjupv, corresponding to sub-stellar companions. To do so, we combine the RV data with astrometry to determine the orbital inclination of the companion and obtain its true mass. We use an MCMC algorithm to estimate the uncertainties of the different parameters. In the first step, we use the orvara algorithm \citep{Brandt_2021_orvara} which allows combining the RV data with the absolute astrometry and the available HCI data. orvara converges faster than most other algorithms \citep{2021AJ....162..186B} and allows us to quickly estimate the orbital parameters and mass of the companion and exclude stellar companions. However, \cite{2023arXiv230500047L} have shown that, when the orbital period is not well covered by the RV data points, MCMC algorithms have difficulties exploring very eccentric solutions. Therefore, we then use our MCMC algorithm (described in Section 3.4) based on the MCMC sampling tool used by \cite{2023arXiv230500047L} to allow a better exploration of the possible solutions. 

In 27 cases, the coupling of the different methods does not properly constrain the orbit of companions, mainly due to the poor coverage of their periods by RV data. We also report 25 stellar companions, for which the orbital parameters and the mass can be retrieved in Table \ref{table_summary_stellar}. Finally, we obtain a sample of 27 sub-stellar companions for which it is possible to estimate all orbital parameters and true masses.

\section{Data available on the sample systems}

\subsection{Radial velocity data}

We use the archival data obtained with the HARPS spectrograph available in the European Southern Observatory (ESO) archives\footnote{http://archive.eso.org}. To characterize as well as possible the different systems, we combine the HARPS data with all other publicly available RV data points. We retrieve from the literature the data obtained with the University College London Echelle Spectrograph (UCLES, \cite{1990SPIE.1235..562D}), the Lick Observatory Hamilton Echelle Spectrometer (Lick, \cite{1987PASP...99.1214V}), the Automated Planet Finder (APF, \cite{2014PASP..126..359V}), the Coudé Echelle Spectrometer (CES, \cite{1982SPIE..331..232E}) long camera (LC) and very long camera (VLC), the Magellan Inamori Kyocera Echelle spectrograph (MIKE, \cite{2003SPIE.4841.1694B}), the Carnegie Planet Finder Spectrograph (PFS, \cite{2010SPIE.7735E..53C}), the High-Resolution Spectrograph (HRS, \cite{1998SPIE.3355..387T}), and the High-Resolution Echelle Spectrometer (HIRES, \cite{1994SPIE.2198..362V}). We retrieve from the DACE archive\footnote{https://dace.unige.ch} the data obtained with the CORALIE echelle spectrograph \citep{1999astro.ph.10223Q} and from the Observatoire de Haute-Provence (OHP) archive\footnote{http://atlas.obs-hp.fr} the data obtained with the ELODIE \citep{1996A&AS..119..373B} and SOPHIE \citep{2008SPIE.7014E..0JP} spectrographs. For the 53 systems with a characterized companion, the considered RV datasets are listed in Appendix B (Table \ref{table_RV_full}). 

Due to a major optical fiber upgrade in 2015 \citep{2015Msngr.162....9L}, an RV offset may be present between HARPS data obtained before and after this upgrade. We, therefore, separate these data into two different sets as if they were obtained from two different instruments. We label "H03" the data sets obtained before May 19, 2015 and "H15" the data sets obtained after June 3, 2015. As for HARPS, a major upgrade has been performed on the HIRES \citep{2019MNRAS.484L...8T} and SOPHIE \citep{2013A&A...549A..49B} spectrographs and two major upgrades have been performed on the CORALIE spectrograph \citep{2010A&A...511A..45S}. We label "Hir94" and "Hir04" the HIRES data sets obtained before and after August 7, 2004, respectively, "SOPHIE" and "SOPHIE+" the SOPHIE data sets obtained before and after June 17, 2011, respectively, and "C98", "C07", and "C14" the CORALIE data sets obtained before June 2007, between June 2007 and November 2014, and after November 2014, respectively.

\subsection{Relative astrometry data}

Whenever possible, we use high-contrast imaging data available in the literature as well as new relative astrometry measurements obtained from observations in the NAOS+CONICA (NACO; \cite{2003SPIE.4839..140R}, \cite{2003SPIE.4841..944L}) and the Spectro-Polarimetric High-contrast Exoplanet REsearch (SPHERE; \cite{2019A&A...631A.155B}) archives. The observing logs for the analyzed observations are summarized in Table \ref{obs_log}.

\begin{figure*}[]
 \centering
\includegraphics[width=1.0\textwidth]{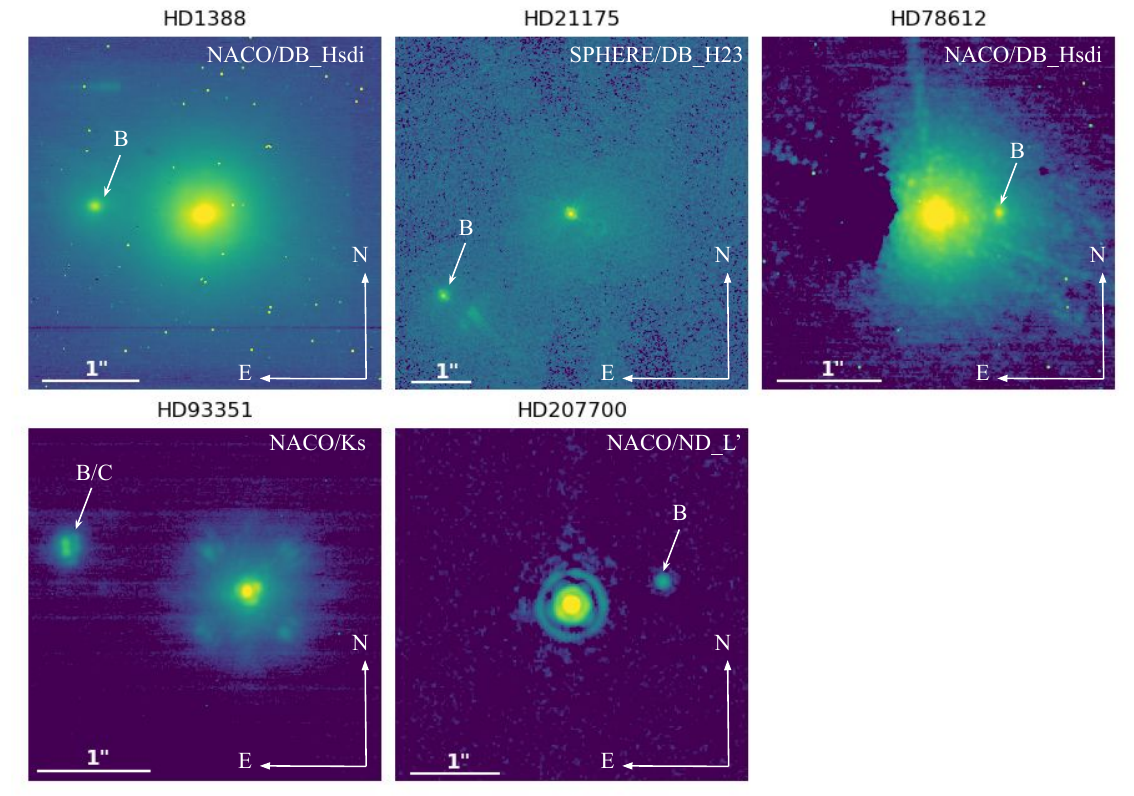}
\caption{Images of new stellar companions detected in HCI. The observations considered come from the NACO and SPHERE ESO archives.
\label{Image_HCI}} 
\end{figure*}

For the NACO observations, even when they are quite shallow, these HCI data are precious to identify large separation stellar companions that could mimic the astrometric signature of closer-in sub-stellar objects. Given the low contrast and wide separation of such stellar companions, we carried out a very simple reduction, by just stacking and subtracting one to another raw data cubes acquired at two different offset positions within the same observation, which subtracts the dark and thermal background. In spite of this simple reduction, we detect five high S/N stellar companions (Fig. \ref{Image_HCI}) in the systems HD1388, HD78612, HD93351, and HD207700. Since we choose to use only non-coronographic observation even when deeper coronographic data did exist, we are also able to measure very accurately the position of the host star and of the high S/N stellar companions, using moffatian fit of their diffraction-limited core. 

For the SPHERE observations, we detect a massive stellar companion in the system HD21175. This companion is outside the field of view of the integral field spectrograph (IFS), hence only the data from the near-infrared imager and spectrograph (IRDIS) are used. All the pre-reduction steps (dark and background subtraction, flat-field correction, bad pixel correction, and centering) are performed on the SPHERE data center \citep{delorme17sphere} using tools provided in the ESO data reduction and handling pipeline \citep{Pavlolv_drh}. Due to the stellar nature of the companion, its point spread function (PSF) on the residual image provided by angular differential imaging (ADI) based algorithms using the science sequence is completely saturated and prevents any fitting of it. Thus, we choose to extract the astrometry of the companion using the off-axis unsaturated PSF frames of the primary star taken before and after the coronographic sequence. We created a 4D dataset containing only the PSF frames. We used a simple No-ADI algorithm embedded into the SPHERE speckle calibration tool (SPECAL) software \citep{galicher2018specal} installed in the SPHERE data center. The astrometry and photometry are extracted by SPECAL from the residual stack produced by the No-ADI reduction, taking into account the neutral density filter used during the off-axis calibration frames (ND\_3.5).

The separations and position angles (hereafter SEP and PA) found for the new detections can be retrieved in Table \ref{table_relative_astrometry}. For GJ680, HD101198, HD111031, HD195564, HD221146, and HIP113201, we use the separations and position angles published in the literature. All considered relative astrometry values are summarized in Table \ref{table_relative_astrometry_full}.

\begin{table*}[]
\centering
\caption{Characteristics of the reduced HCI observations.}
\begin{adjustbox}{width=\textwidth}
\begin{tabular}{cccccccccc}
\hline 
Star & Date Obs. & Instrument & Filter & NDIT$\times$DIT(s)$\times$Nframe$^{*}$ & Seeing (") & Airmass & $\tau_0$ (ms)$^{a}$ & Program ID \\ 
\hline 
HD1388 & 2006-07-24 & NACO & DB\_Hsdi & 40x4x1 & 1.74 & 1.03 & 0.0008 & 077.C-0293(A) \\ [0.1cm]
HD78612 & 2006-01-04 & NACO & DB\_Hsdi & 25x7x1 & 1.00 & 1.034 & 0.0037 & 076.C-0762(A) \\ [0.1cm]
HD93351 & 2012-06-15 & NACO & Ks & 200x0.15x1 & 0.81 & 1.377 & 0.0018 & 089.C-0641(A) \\ [0.1cm]
HD207700 & 2012-06-14 & NACO & ND\_$L'$ & 200x0.2x5 & -1.00 & 1.517 & 0.0019 & 089.C-0641(A) \\ [0.1cm]
HD21175 & 2016-10-04 & SPHERE & DB\_H23 & 16x8x2 & 0.75 & 1.090 & 0.0019 & 098.C-0739(A) \\ 
\hline 
\vspace{0.05cm}
\end{tabular}
\end{adjustbox}
\textbf{Notes:} $^*$: NDIT is the number of sub integration within a DIT, DIT is the detector integration time per frame. $\tau_0$ corresponds to the coherence time. 
\label{obs_log}
\end{table*}

\begin{table}[]
\centering
\caption{New relative astrometry.}
\begin{adjustbox}{width=0.45\textwidth}
\begin{tabular}[h!]{cccccc}
\hline
Companion & JD - 2400000 & SEP (mas) & PA ($\deg$) \\ 
\hline
HD1388 B & 53940.39 & $1490 \pm 10$ & $85.6 \pm 1$ \\ [0.1cm]
HD21175 B & 57665.35 & $2628.0 \pm 2.5$ & $122.75 \pm 0.08$ \\ [0.1cm]
HD78612 B & 53739.29 & $613 \pm 5$ & $272.3 \pm 1$ \\ [0.1cm]
HD93351 B/C & 56093.02 & $1684 \pm 30$ & $76.3 \pm 1$ \\ [0.1cm]
HD207700 B & 56092.36 & $1090 \pm 20$ & $284.2 \pm 1$ \\
\hline
\end{tabular}
\end{adjustbox}
\textbf{Notes:} Since the companion of HD93351 is binary, we consider the average separation in RA and DEC of the two companions.
\label{table_relative_astrometry}
\end{table}

\subsection{Absolute astrometry data }

We use the position and proper motion values from the Hipparcos-Gaia Catalog of Accelerations (HGCA) published by \cite{2021ApJS..254...42B} from Hipparcos and Gaia early DR3 (\cite{2016A&A...595A...1G}, \cite{2021A&A...649A...1G}) measurements to determined the stellar acceleration of each selected target. For each star, the difference between the position measurements obtained by Hipparcos and Gaia divided by the mean interval between the two measurements ($\sim$25 years), gives a more accurate tangential proper motion. The corresponding values are provided in Table \ref{PMvalues}.

\subsection{Orbit fitting}

We use our MCMC algorithm presented in \cite{2023A&A...670A..65P} to fit the RV and/or relative astrometry and/or absolute astrometry data assuming Keplerian orbits. Our tool is based on the emcee 3.0 python package \citep{Foreman_Mackey_2013}, and the HTOF package \citep{Brandt_2021_htof} to fit the Hipparcos/Gaia intermediate astrometric data. It follows large sections of the orvara (Orbits Using Radial Velocity, Absolute, and/or Relative Astrometry) code \citep{Brandt_2021_orvara} for the likelihood computation. In particular:

\begin{equation}\begin{split}LLH_{sep,pa}^{2} = -0.5 *[&\frac{\Delta_{sep}^2}{\sigma_{sep}^2(1-\rho_{sep,pa}^2)}\\
& + \frac{\Delta_{pa}^2}{\sigma_{pa}^2(1-\rho_{sep,pa}^2)}\\
& + \frac{2\rho_{sep,pa} * \Delta_{sep} * \Delta_{pa}}{\sigma_{sep}\ \sigma_{pa}\ (1-\rho_{sep,pa}^2)}]\end{split}\end{equation}

\begin{equation}LLH_{RV}^2 = -0.5*\left[\frac{\Delta_{RV}^2}{\sigma_{RV}^2+ jitter^2} + \log(\sigma_{err}^2 + jitter^2)\right]\end{equation}

\begin{equation}\begin{split}LLH_{HG}^{2} =  (&\mu_{H,o}-\bar{\mu}-p\mu_{H})^{T}C^{-1}_{H}(\mu_{H,o}-\bar{\mu}-p\mu_{H})\\
& + (\mu_{HG,o}-\bar{\mu}-p\mu_{HG})^{T}C^{-1}_{HG}(\mu_{HG,o}-\bar{\mu}-p\mu_{HG})\\
& + (\mu_{G,o}-\bar{\mu}-p\mu_{G})^{T}C^{-1}_{G}(\mu_{G,o}-\bar{\mu}-p\mu_{G})\end{split}\end{equation}
where $\Delta_{sep}$, $\Delta_{pa}$, and $\Delta_{RV}$ are the difference between the measured and predicted separations, position angle, and RV. $\rho_{sep,pa}$ is the correlation coefficient between the separation and the position angle measurement. $\sigma_{sep}$, $\sigma_{pa}$, and $\sigma_{err}$ correspond to the uncertainty on the measured separation, the PA, and the RV, respectively. $\sigma_{RV}$ is the variance of the RV, and jitter is the stellar jitter measured. For the likelihood of the proper motions measurements, $\mu_{x,o}$ and $\mu_{x}$ correspond to the predicted and observed proper motion, p is the parallax, $C^{-1}_{x}$ is the inverse of the covariance matrix, and $\bar{\mu}$ corresponds to the center-of-mass motion of the system's barycenter.

Unlike orvara or other similar packages, our code uses modular arithmetic computations for angles, wherever appropriate, to alleviate artifacts at the range boundaries. To minimize multi-modality effects, the MCMC code uses a mix of move functions, in particular the Differential Evolution (DE) move function suggested in the Emcee package documentation for these situations \citep{Foreman_Mackey_2013}.

We considered ten free parameters for each system: the host star mass, the parallax\footnote{The host star mass and the parallax considered are those reported by \cite{2022A&A...657A...7K}.}, the semi-major axis (\textit{a}), the eccentricity (ecc), the orbital inclination (\textit{I}), the longitude of ascending node ($\Omega$), the argument of periastron ($\omega$), the periastron passage time ($P_{time}$)\footnote{The periastron passage time is calculated considering an initial time at BJD-2454000}, the companion mass, and a stellar jitter. It is important to note that in the fitting process, we consider $\sqrt{ecc} \cos\omega$ and $\sqrt{ecc} \sin\omega$ instead of ecc and $\omega$. In addition, to combine data from different instruments, we added an instrumental offset for each instrument as a free parameter of the model (see above). We considered uniform priors for all fitting parameters, except for the host star mass and the distance of the system for which we considered Gaussian priors and for the orbital inclination, for which we considered a sin(\textit{I}) prior.

When the orbital period of a companion is much shorter than Gaia DR3 observing window (1038 days), the PMa does not allow us to constrain the orbital inclination. In such cases, we fit Keplerian orbits with our MCMC, considering only the RV data points. We then use GASTON (Gaia Astrometric Noise Simulation to derive Orbit incliNation; \citep{2019A&A...631A.125K,2021A&A...645A...7K}), updated to the improved Gaia DR3 accuracy, to model the DR3 astrometric excess noise (hereafter AEN) with a motion of the system's photocenter induced by the orbiting RV companion at an optimal orbital inclination. This leads to a new estimate of the true mass of the companion. The principle of GASTON is to assume the presence of this companion and i) simulate Gaia's observations of the system with many possible orbits of the photocenter, ii) fit out the 5-parameters linear model, as performed by Gaia, that includes position, proper motion, and parallax, iii) calculate the AEN from the residuals, and iv) compare the DR3 AEN to the outcomes of the simulations and perform an optimization of the orbital inclination. The GASTON method has been improved since the EDR3 release. It now accounts for the dependence of the noise intrinsic to Gaia's acquisitions and data reductions (instrumental jitter, spacecraft's attitude, and modeling) on magnitude ($G$-band) and color ($Gb-Gr$)~\citep{Lindegren2021}. It also accounts for a parallax offset due to the orbital motion and produces a corrected parallax. 

\section{New companions}

\subsection{Brown dwarfs}

\subsubsection{GJ660.1}

GJ660.1 is a binary system composed of a M1V star (GJ660.1 A) and another low mass star of type M$9.5 \pm 0.3$ (GJ660.1 B, \cite{2016AJ....151...46A}). The projected separation between the two M dwarfs is estimated at $122 \pm 9$ au \citep{2011ApJ...743..109S}. For this system, 59 RV data points were obtained with the HARPS spectrograph between 2009 and 2019 for GJ660.1 A. A periodicity is clearly present in the periodogram of the data with a peak at about 615 days. Using DPASS, we find the best solution, corresponding to an additional brown dwarf candidate with a minimum mass of 16 \Mjup and a period of 620 days (Fig. \ref{RV_DPASS_GJ660_1}). 

\begin{figure}[]
 \centering
\includegraphics[width=0.45\textwidth]{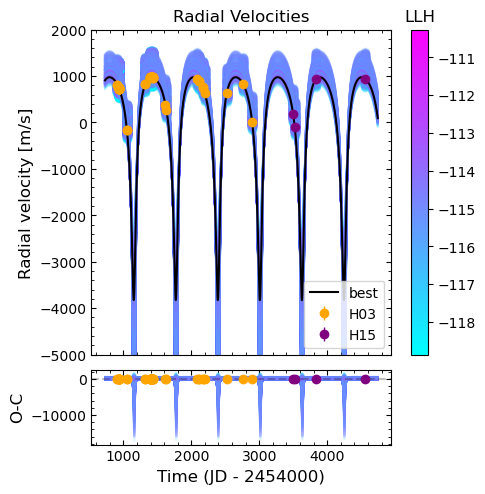}
\caption{Fit of the GJ660.1 A RV measurements. The black curve corresponds to the best fit. The color bar indicates the log-likelihood corresponding to the different fits plotted.
\label{RV_GAS_GJ660_1}} 
\end{figure}

The significant AEN in the Gaia DR3 data for GJ660.1 of 2.633\,mas (83$\sigma$ larger than noise) is dominated by orbital signals with a period shorter than 100\,yr (\textit{a} < 16 au). The effect of GJ660.1 B is thus negligible, and we assume that the AEN is only due to GJ660.1 C.
We used Gaston to fit the AEN considering the RV solution found with our MCMC algorithm (Fig. \ref{RV_GAS_GJ660_1}), considering 300000 iterations (burn-in = 50000) and 1000 walkers, and varying the orbital inclination and the longitude of ascending node. We find a semi-major axis of $1.12 \pm 0.02$ au (P = $618.6 \pm 0.1$ d), an eccentricity of $0.65_{-0.08}^{+0.11}$, a minimum mass of $59_{-15}^{+34}$ \Mjupv, and an inclination of $87_{-43}^{+39}$°. This leads to a true mass within $78_{-16}^{+38}$ \Mjupv. The posteriors obtained for each free parameter of the MCMC are reported in Table \ref{table_summary_BD} and the corner plot is available in Appendix D (Fig. \ref{Corner_GJ660_1}). Note that, even if the orbital period is properly covered by the RV data, additional measurements near the periastron are necessary to better constrain the mass and eccentricity of GJ660.1 C and define the precise nature of the companion.

Looking at the periodogram of the RV data, corrected for the signal of GJ660.1 C, we observe a significant peak at about 200 days. No correlation between RV variations and bisector velocity span (BVS) variations that may explain this peak is observed. This signal could therefore correspond to the presence of an additional companion located at 0.5 au (P = 206 d) with a minimum mass of 0.3 \Mjupv (Fig. \ref{RV_DPASS_GJ660_1}). It should be noted that, given the mass and semi-major axis of this candidate companion, the AEN induced by it would be negligible compared with that induced by GJ 660.1 C.

\subsubsection{HD73256}

82 RV data points were obtained for the G8IV star HD73256, 26 RV HARPS measurements were obtained between 2003 and 2012, 38 RV CORALIE measurements were obtained between 2001 and 2003, and 18 RV HIRES measurements were obtained between 2004 and 2016. \cite{2003A&A...407..679U} reported a Hot Jupiter with a minimum mass of $1.87 \pm 0.49$ \Mjup and a period of $2.54858 \pm 0.00016$ days, considering the CORALIE data points. Considering all the RV data points available, a periodicity appears in the periodogram of the data with a strong peak around 2.5 days corresponding to the companion identified by \cite{2003A&A...407..679U}. Once the planet signal is removed, a periodicity appears in the periodogram with a peak around 3000 d. Using DPASS, we find the best two-planet solution with inner planet orbital parameters and minimum mass close to those reported by \cite{2003A&A...407..679U} and a new, outer giant planet candidate with a minimum mass of 10 \Mjup and a period of 2777 days (Fig. \ref{RV_DPASS_HD73256}). After removing both planets' signals, a periodicity is again present around 14 days, which corresponds to the rotation period of the host star \citep{2003A&A...407..679U}.

\begin{figure}[]
 \centering
\includegraphics[width=0.45\textwidth]{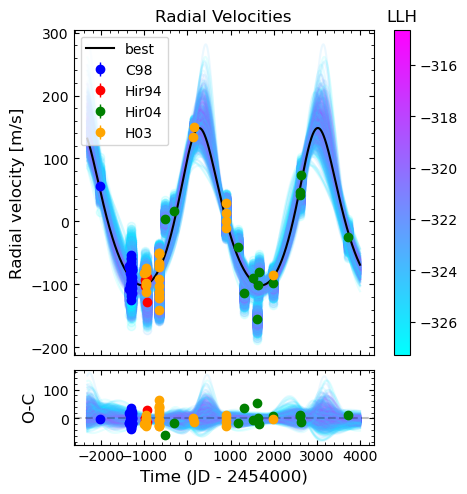}
\includegraphics[width=0.5\textwidth]{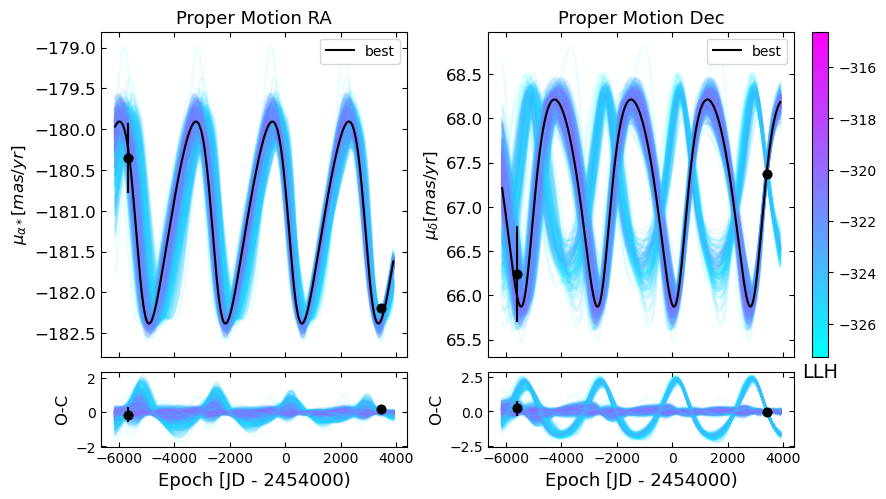}
\caption{Fit of the orbit of HD73256. \textit{Top}: Fit of the HD73256 RV measurements after subtracted the signal of the inner planet. \textit{Bottom}: Fit of the HD73256 astrometric acceleration in right ascension (left) and declination (right). The black points correspond to the Hipparcos and Gaia EDR3 proper motion measurements. In each plot, the black curve corresponds to the best fit. The color bar indicates the log-likelihood corresponding to the different fits plotted. 
\label{RV_AA_HD73256}} 
\end{figure}

As the orbital period of HD73256 b is much smaller than the duration of Gaia DR3, the proper motion induced by the inner planet can be neglected. Thus, we fit simultaneously all the RV data points available, corrected for the signal of the inner planet, with the absolute astrometry (Fig. \ref{RV_AA_HD73256}) with our MCMC, considering 300000 iterations (burn-in = 50000) and 1000 walkers. We find a semi-major axis of $3.8 \pm 0.1$ au (P = $2690_{-102}^{+60}$ d) and an eccentricity of $0.16 \pm 0.07$ and constrain the orbital inclination of HD73256 c to either $29_{-3}^{+5}$° or $152_{-7}^{+8}$°. We find a true mass of $16 \pm 1$ \Mjupv, corresponding to a low-mass brown dwarf. The posteriors obtained for each free parameter of the MCMC are reported in Table \ref{table_summary_BD} and the corner plots are available in Appendix (Fig. \ref{Corner_HD73256} and \ref{Corner_HD73256_sup}).

\subsubsection{HD165131}

69 RV data points were obtained with the HARPS spectrograph between 2006 and 2021 for the G3/5V star HD165131. A periodicity is clearly present in the periodogram of the data, with a peak at about 2350 days. Using DPASS, we find the best solution corresponding to a brown dwarf candidate with a minimum mass of 18 \Mjup and a period of 2344 days (Fig. \ref{RV_DPASS_HD165131}).

\begin{figure}[]
 \centering
\includegraphics[width=0.45\textwidth]{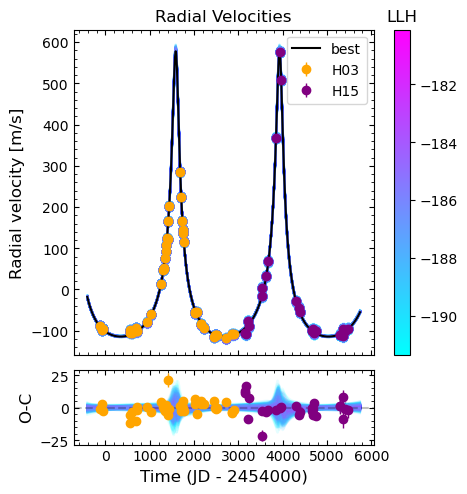}
\includegraphics[width=0.5\textwidth]{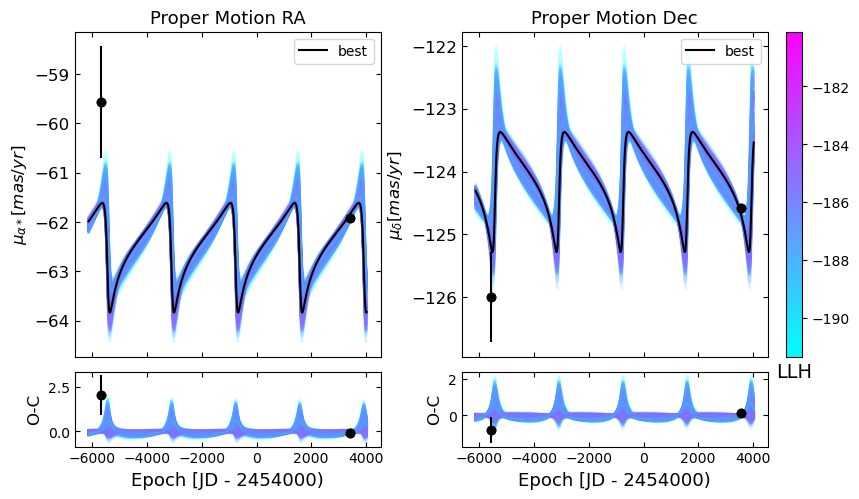}
\caption{Fit of the orbit of HD165131. \textit{Top}: Fit of the HD165131 RV measurements. \textit{Bottom}: Fit of the HD165131 astrometric acceleration in right ascension (left) and declination (right). The black points correspond to the Hipparcos and Gaia EDR3 proper motion measurements. In each plot, the black curve corresponds to the best fit. The color bar indicates the log-likelihood corresponding to the different fits plotted. 
\label{RV_AA_HD165131}} 
\end{figure}

By simultaneously fitting RV HARPS measurements and absolute astrometry with our MCMC (fig. \ref{RV_AA_HD165131}), considering 300000 iterations after 50000 iterations, 1000 walkers, we characterize the orbital parameters of the companion and constrain the orbital inclination to either $70_{-9}^{+11}$° or $109_{-12}^{+10}$°. We find a semi-major axis of $3.59_{-0.07}^{+0.06}$ au (P = $2343 \pm 2$ d), an eccentricity of $0.672 \pm 0.003$ and a true mass of $19_{-1}^{+2}$ \Mjupv. The posteriors obtained for each free parameter of the MCMC are reported in Table \ref{table_summary_BD} and the corner plot is available in Appendix D (Fig. \ref{Corner_HD165131}).

\begin{table}[h!]
\centering
\caption{Summary of posteriors for new brown dwarfs.}
\begin{adjustbox}{width=0.5\textwidth}
\begin{tabular}[h!]{cccc}
\hline
 Parameter & GJ660.1 C & HD73256 C & HD165131 B \\ 
\hline 
 \textit{a} (au) & $1.12 \pm 0.02$ & $3.8 \pm 0.1$ & $3.59_{-0.07}^{+0.06}$ \\ [0.1cm]
 P (days) & $618.6 \pm 0.1$ & $2690_{-102}^{+60}$ & $2343 \pm 2$ \\ [0.1cm]
 Ecc & $0.65_{-0.08}^{+0.11}$ & $0.16 \pm 0.07$ & $0.672 \pm 0.003$ \\ [0.1cm]
 \textit{I} (°) & $87_{-43}^{+39}$$^{*}$ & $29_{-3}^{+5}$ or $152_{-7}^{+8}$ & $93_{-43}^{+39}$ \\ [0.1cm]
 Mass (\Mjupv) & $78_{-16}^{+38}$$^{*}$ & $16 \pm 1$ & $19_{-1}^{+2}$ \\ [0.1cm]
 $\Omega$ (°) & $112_{-123}^{+122}$$^{*}$ & $303_{-6}^{+5}$ or $269_{-7}^{+8}$ & $141_{-19}^{+20}$ \\ [0.1cm]
 $\omega$ (°) & $188_{-2}^{+1}$ & $300_{-53}^{+37}$ & $3.1 \pm 0.4$ \\ [0.1cm]
 $P_{time}$ (days) & $540_{-3}^{+2}$ & $244_{-179}^{+134}$ & $1588 \pm 2$ \\ [0.1cm]
 $\sqrt{ecc}cos(\omega)$ & $-0.80_{-0.07}^{+0.06}$ & $0.42_{-0.12}^{+0.08}$ & $0.818_{-0.001}^{+0.002}$ \\ [0.1cm]
 $\sqrt{ecc}sin(\omega)$ & $-0.12_{-0.01}^{+0.02}$ & $-0.07_{-0.14}^{+0.13}$ & $0.044 \pm 0.006$ \\ [0.1cm]
 Jitter (m/s) & $3.4_{-0.5}^{+0.6}$ & $23_{-2}^{+3}$ & $5.6 \pm 0.6$ \\ [0.1cm]
 rms (m/s) & $4.6_{-0.2}^{+1.6}$ & $24.2_{-1.6}^{+3.8}$ & $6.3_{-0.1}^{+0.2}$ \\ [0.1cm]
\hline
 & H03 = $-50500_{-224}^{+141}$ & H03 = $59 \pm 6$ & H03 = $46204 \pm 1$ \\ [0.1cm]
 Instrumental & H15 = $-50505_{-224}^{+142}$ & C98 = $74_{-16}^{+13}$ & H15 = $46218_{-2}^{+1}$ \\ [0.1cm]
 offset (m/s) & & Hir94 = $101 \pm 13$ & \\ [0.1cm]
 & & Hir04 = $34 \pm 9$ & \\ [0.1cm]
\hline
\end{tabular}
\end{adjustbox}
\textbf{Notes:} The median and $1 \sigma$ confidence interval are given for each parameter. $^{*}$Values obtained with Gaston.
\label{table_summary_BD}
\end{table}

\subsection{New planets}

\subsubsection{HD75302}

82 RV data points were obtained for this G5V star HD75302, 35 RV HARPS measurements were obtained between 2012 and 2018, 26 RV ELODIE measurements were obtained between 1997 and 2005, and 21 RV SOPHIE measurements were obtained between 2008 and 2018. A periodicity appears in the periodogram of the data, with a weak peak around 2000 days (Fig. \ref{Perio_DPASS_HD75302}). Using DPASS, we find the best solution corresponding to an eccentric planet candidate with a period of 4300 days, a minimum mass of 5.3 \Mjup, and an eccentricity of 0.88. However, we observe a significant correlation between RV variations and bisector velocity span (BVS) variations (Fig. \ref{RV_BVS_HD75302}) indicating a possible stellar activity signal. Correcting the RV data with the linear regression between RVs and the bisector velocity span \citep{2007A&A...467..721M}, we obtain a new RV dataset for which a strong peak is present in the periodogram around 3000 days. Considering these data corrected for the stellar activity, we find the best solution for the companion corresponding to a less massive and eccentric planet (\msini = 2.7 \Mjup and e = 0.44) but with a similar period (fig. \ref{RV_DPASS_HD75302}).

\begin{figure}[]
 \centering
\includegraphics[width=0.45\textwidth]{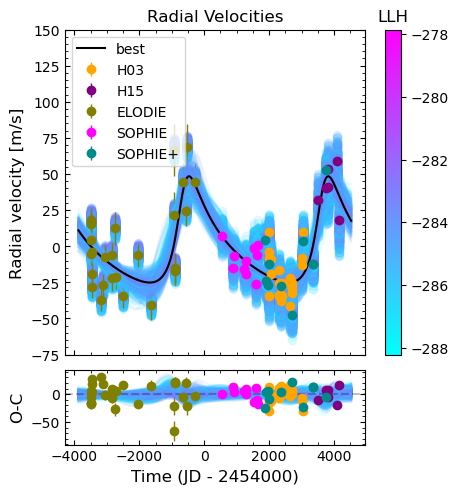}
\includegraphics[width=0.5\textwidth]{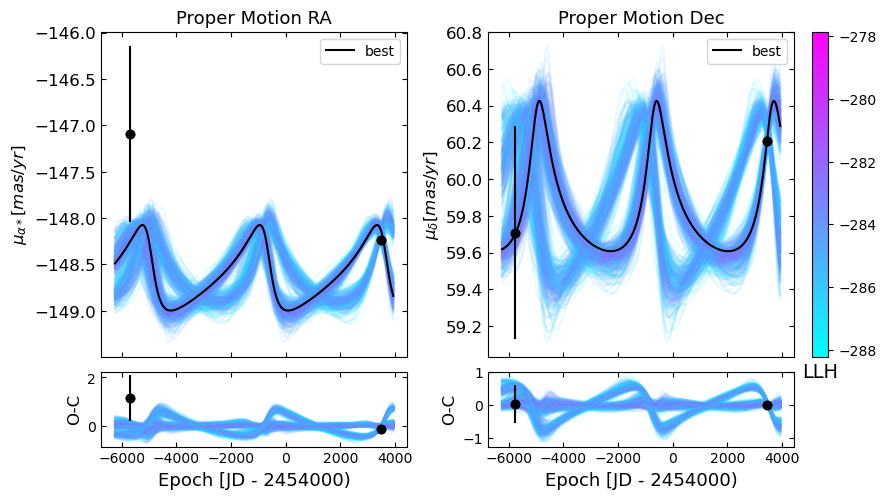}
\caption{Fit of the orbit of HD75302. \textit{Top}: Fit of the HD75302 RV measurements. \textit{Bottom}: Fit of the HD75302 astrometric acceleration in right ascension (left) and declination (right). The black points correspond to the Hipparcos and Gaia EDR3 proper motion measurements. In each plot, the black curve corresponds to the best fit. The color bar indicates the log-likelihood corresponding to the different fits plotted. 
\label{RV_AA_HD75302}} 
\end{figure}

Combining the corrected RV data with absolute astrometry with our MCMC (fig. \ref{RV_AA_HD75302}), considering 300000 iterations (burn-in = 50000) and 1000 walkers, we find a well-constrained orbital solution for HD75302 b and we constrain its orbital inclination to either $29 \pm 5$° or $150_{-6}^{+5}$°. We find a semi-major axis of $5.3_{-0.1}^{+0.2}$ au (P = $4356_{-112}^{+173}$ d), an eccentricity of $0.39_{-0.11}^{+0.10}$ and a true mass of $5.4_{-0.4}^{+0.5}$ \Mjupv. The posteriors obtained for each free parameter of the MCMC are reported in Table \ref{table_summary_planets} and the corner plot is available in Appendix D (Fig. \ref{Corner_HD75302}).

\subsubsection{HD108202}

42 RV data points were obtained with the HARPS spectrograph between 2005 and 2022 for the K4/5V star HD108202. A periodicity is present in the periodogram of the data, with a peak of around 3000 days. Using DPASS, we find the best solution corresponding to a giant planet candidate with a minimum mass of 2.4 \Mjup and a period of 2990 days (Fig. \ref{RV_DPASS_HD108202}).

\begin{figure}[]
 \centering
\includegraphics[width=0.45\textwidth]{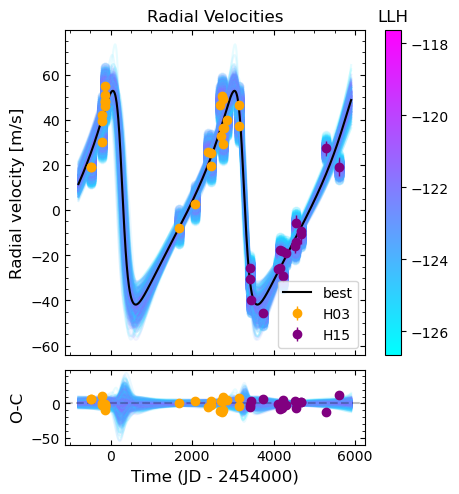}
\includegraphics[width=0.5\textwidth]{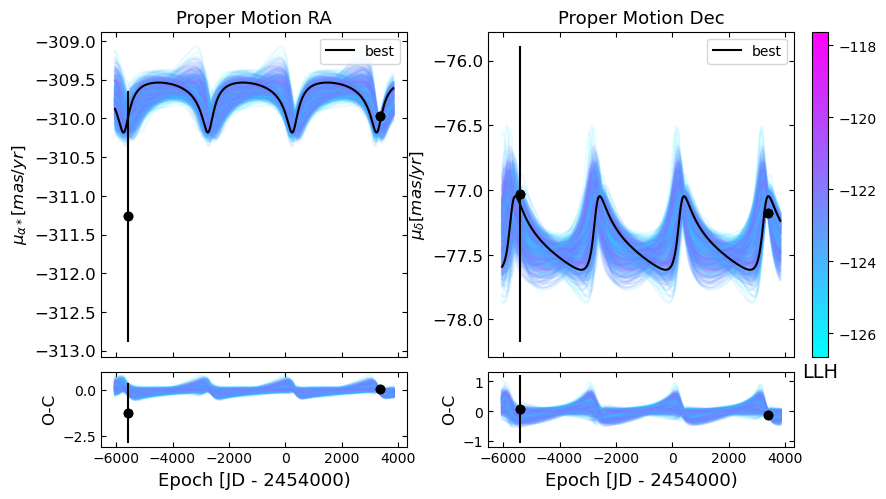}
\caption{Fit of the orbit of HD108202. \textit{Top}: Fit of the HD108202 RV measurements. \textit{Bottom}: Fit of the HD108202 astrometric acceleration in right ascension (left) and declination (right). The black points correspond to the Hipparcos and Gaia EDR3 proper motion measurements. In each plot, the black curve corresponds to the best fit. The color bar indicates the log-likelihood corresponding to the different fits plotted. 
\label{RV_AA_HD108202}} 
\end{figure}

By simultaneously fitting RV data and absolute astrometry with our MCMC (fig. \ref{RV_AA_HD108202}), considering 200000 iterations (burn-in = 50000) and 1000 walkers, we characterize the orbital parameters of the companion. We find an orbital inclination of either $56_{-13}^{+19}$° or $130_{-24}^{+14}$°, a semi-major axis of $3.7 \pm 0.1$ au (P = $2990_{-51}^{+54}$ d), an eccentricity of $0.52_{-0.03}^{+0.04}$ and a true mass of $3.0_{-0.5}^{+0.7}$ \Mjupv. The posteriors obtained for each free parameter of the MCMC are reported in Table \ref{table_summary_planets} and the corner plot is available in Appendix D (Fig. \ref{Corner_HD108202}).

\subsubsection{HD135625}

99 RV data points were obtained with the HARPS spectrograph between 2004 and 2021 for the G3IV/V star HD135625. A periodicity is clearly present in the periodogram of the data, with a strong peak at about 4500 days. Using DPASS, we find the best solution corresponding to a giant planet candidate with a minimum mass of 2 \Mjup and a period of 4060 days (Fig. \ref{RV_DPASS_HD135625}).

\begin{figure}[]
 \centering
\includegraphics[width=0.45\textwidth]{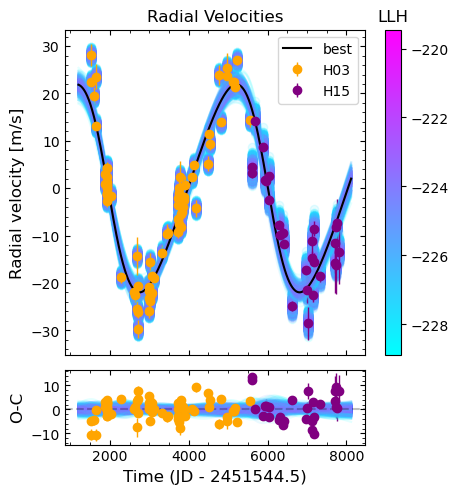}
\includegraphics[width=0.5\textwidth]{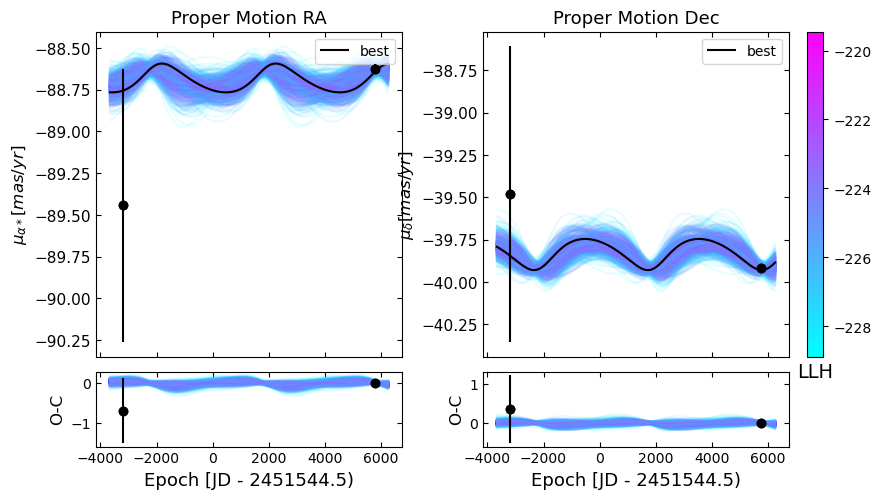}
\caption{Fit of the orbit of HD135625. \textit{Top}: Fit of the HD135625 RV measurements. \textit{Bottom}: Fit of the HD135625 astrometric acceleration in right ascension (left) and declination (right). The black points correspond to the Hipparcos and Gaia EDR3 proper motion measurements. In each plot, the black curve corresponds to the best fit. The color bar indicates the log-likelihood corresponding to the different fits plotted. 
\label{RV_AA_HD135625}} 
\end{figure}

By simultaneously fitting RV data and absolute astrometry with our MCMC (fig. \ref{RV_AA_HD135625}), considering 200000 iterations (burn-in = 30000) and 1000 walkers, we characterize the orbital parameters of the companion. Absolute astrometry does not allow us to properly constrain the orbital inclination. We can only exclude the solutions corresponding to a high inclined orbit and we find an orbital inclination of $93_{-43}^{+39}$°. We find a semi-major axis of $5.4 \pm 0.1$ au (P = $4055_{-44}^{+45}$ d), an eccentricity of $0.16_{-0.03}^{+0.04}$, and a true mass of $2.3_{-0.3}^{+0.8}$ \Mjupv. The posteriors obtained for each free parameter of the MCMC are reported in Table \ref{table_summary_planets} and the corner plot is available in Appendix D (Fig. \ref{Corner_HD135625}).

\subsubsection{HD185283}

78 RV data points were obtained with the HARPS spectrograph between 2004 and 2021 for the K3V star HD185283. A periodicity is clearly present in the periodogram of the data, with a strong peak at about 4500 days. Using DPASS, we find the best solution corresponding to a giant planet candidate with a minimum mass of 1.5 \Mjup and a period of 4000 days (Fig. \ref{RV_DPASS_HD185283}).

\begin{figure}[]
 \centering
\includegraphics[width=0.45\textwidth]{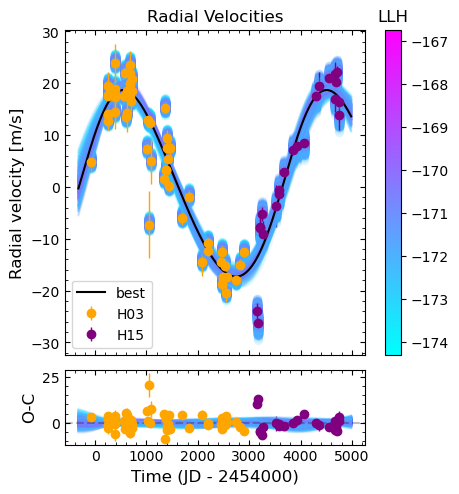}
\includegraphics[width=0.5\textwidth]{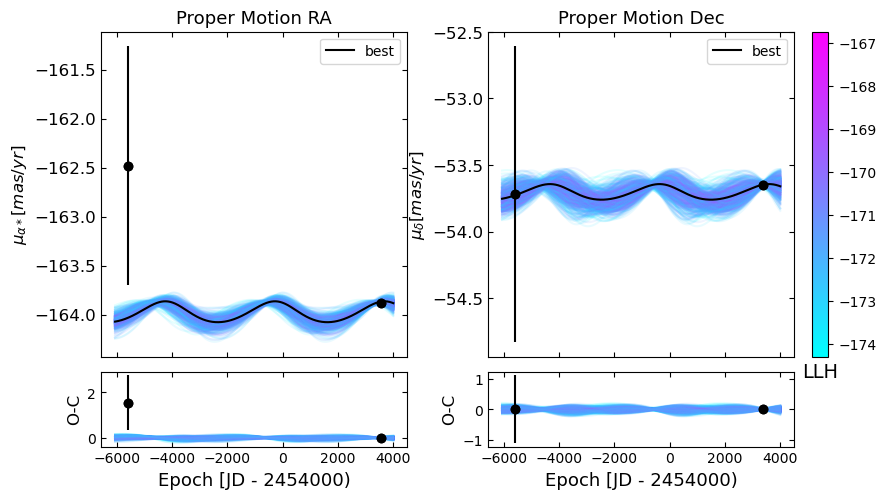}
\caption{Fit of the orbit of HD185283. \textit{Top}: Fit of the HD185283 RV measurements. \textit{Bottom}: Fit of the HD185283 astrometric acceleration in right ascension (left) and declination (right). The black points correspond to the Hipparcos and Gaia EDR3 proper motion measurements. In each plot, the black curve corresponds to the best fit. The color bar indicates the log-likelihood corresponding to the different fits plotted. 
\label{RV_AA_HD185283}} 
\end{figure}

By simultaneously fitting RV data and absolute astrometry with our MCMC (fig. \ref{RV_AA_HD185283}), considering 300000 iterations (burn-in = 30000) and 1000 walkers, we characterize the orbital parameters of the companion and we constrain the orbital inclination to be of $91_{-31}^{+30}$°. We find a semi-major axis of $4.6_{-0.1}^{+0.2}$ au (P = $4060_{-112}^{+148}$ d), an eccentricity of $0.07_{-0.04}^{+0.05}$, and a true mass of $1.3_{-0.1}^{+0.3}$ \Mjupv. The posteriors obtained for each free parameter of the MCMC are reported in Table \ref{table_summary_planets} and the corner plot is available in Appendix D (Fig. \ref{Corner_HD185283}).

\subsubsection{HIP10337}

108 RV data points were obtained for the K7V star HIP10337, 90 RV HARPS measurements were obtained between 2005 and 2022, and 18 RV HIRES measurements were obtained between 2008 and 2014. A periodicity is present in the periodogram of the data, with a large peak at about 8000 days. Using DPASS, we find the best solution corresponding to a giant planet candidate with a minimum mass of 2.9 \Mjup and a period of 6400 days (Fig. \ref{RV_DPASS_HIP10337}).

\begin{figure}[]
 \centering
\includegraphics[width=0.45\textwidth]{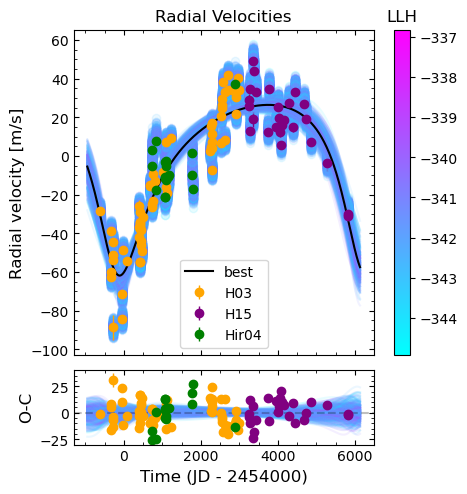}
\includegraphics[width=0.5\textwidth]{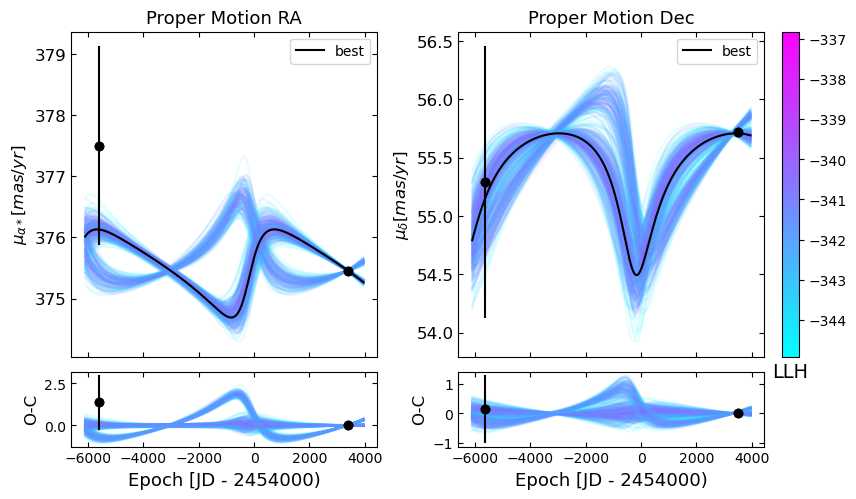}
\caption{Fit of the orbit of HIP10337. \textit{Top}: Fit of the HIP10337 RV measurements. \textit{Bottom}: Fit of the HIP10337 astrometric acceleration in right ascension (left) and declination (right). The black points correspond to the Hipparcos and Gaia EDR3 proper motion measurements. In each plot, the black curve corresponds to the best fit. The color bar indicates the log-likelihood corresponding to the different fits plotted. 
\label{RV_AA_HIP10337}} 
\end{figure}

By simultaneously fitting RV data and absolute astrometry with our MCMC (fig. \ref{RV_AA_HIP10337}), considering 300000 iterations (burn-in = 30000)and 1000 walkers, we characterize the orbital parameters of the companion and constraining the orbital inclination to either $33_{-4}^{+5}$° or $146_{-5}^{+4}$°. We find a semi-major axis of $5.9 \pm 0.1$ au (P = $6240_{-156}^{+151}$ d), an eccentricity of $0.43 \pm 0.05$, and a true mass of $4.8 \pm 0.6$ \Mjupv. The posteriors obtained for each free parameter of the MCMC are reported in Table \ref{table_summary_planets} and the corner plots are available in Appendix (Fig. \ref{Corner_HIP10337} and \ref{Corner_HIP10337_sup}).

\subsubsection{HIP54597}

69 RV data points were obtained with HARPS between 2006 and 2022 for the K5V star HIP54597. A periodicity is clearly present in the periodogram of the data, with a strong peak at about 3300 days. Using DPASS, we find the best solution corresponding to a giant planet candidate with a minimum mass of 2.3 \Mjup and a period of 3270 days (Fig. \ref{RV_DPASS_HIP54597}).

\begin{figure}[]
 \centering
\includegraphics[width=0.45\textwidth]{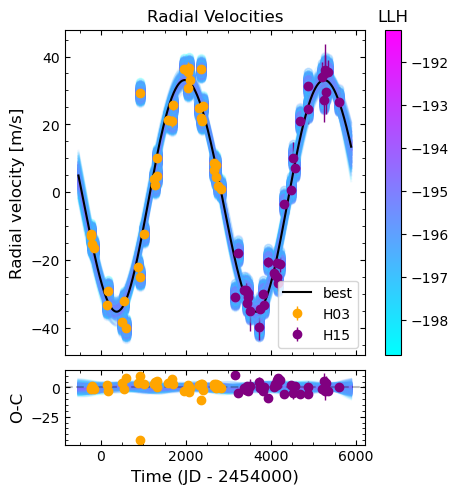}
\includegraphics[width=0.5\textwidth]{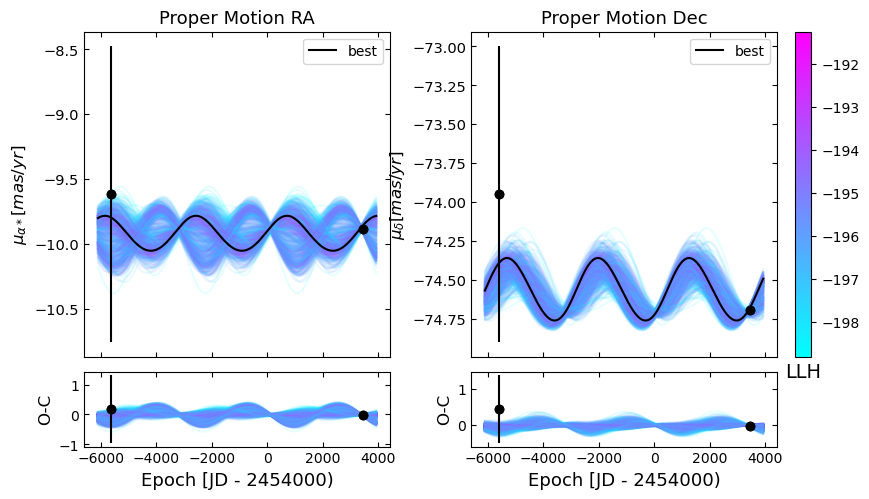}
\caption{Fit of the orbit of HIP54597. \textit{Top}: Fit of the HIP54597 RV measurements. \textit{Bottom}: Fit of the HIP54597 astrometric acceleration in right ascension (left) and declination (right). The black points correspond to the Hipparcos and Gaia EDR3 proper motion measurements. In each plot, the black curve corresponds to the best fit. The color bar indicates the log-likelihood corresponding to the different fits plotted. 
\label{RV_AA_HIP54597}} 
\end{figure}

By simultaneously fitting RV data and absolute astrometry with our MCMC (fig. \ref{RV_AA_HIP54597}), considering 300000 iterations (burn-in = 30000) and 1000 walkers, we characterize the orbital parameters of the companion and constraining the orbital inclination to either $62_{-13}^{+16}$° or $116_{-14}^{+13}$°. We find a semi-major axis of $4.0 \pm 0.1$ au (P = $3274_{-46}^{+47}$ d), an eccentricity of $0.03_{-0.02}^{+0.03}$, and a true mass of $2.4_{-0.2}^{+0.4}$ \Mjupv. The posteriors obtained for each free parameter of the MCMC are reported in Table \ref{table_summary_planets} and the corner plot is available in Appendix D (Fig. \ref{Corner_HIP54597}).

\begin{table*}[h!]
\centering
\caption{Summary of posteriors for new planets.}
\resizebox{1.\textwidth}{!}{
\begin{tabular}[h!]{ccccccccccccc}
\hline
 Parameter & HD75302 b & HD108202 b & HD135625 b & HD185283 b & HIP10337 b & HIP54597 b \\ 
\hline
 \emph{a} (au) & $5.3_{-0.1}^{+0.2}$ & $3.7 \pm 0.1$ & $5.4 \pm 0.1$ & $4.6_{-0.1}^{+0.2}$ & $5.9 \pm 0.1$ & $4.0 \pm 0.1$ \\ [0.1cm]
 P (days) & $4356_{-112}^{+173}$ & $2990_{-51}^{+54}$ & $4055_{-44}^{+45}$ & $4060_{-112}^{+148}$ & $6240_{-156}^{+151}$ & $3274_{-46}^{+47}$ \\ [0.1cm]
 Ecc & $0.39_{-0.11}^{+0.10}$ & $0.52_{-0.03}^{+0.04}$ & $0.16_{-0.03}^{+0.04}$ & $0.07_{-0.04}^{+0.05}$ & $0.43 \pm 0.05$ & $0.03_{-0.02}^{+0.03}$ \\ [0.1cm]
 \textit{I} (°) & $29 \pm 5$ or $150_{-6}^{+5}$ & $56_{-13}^{+19}$ or $130_{-24}^{+14}$ & $93_{-43}^{+39}$ & $91_{-31}^{+30}$ & $33_{-4}^{+5}$ or $146_{-5}^{+4}$ & $62_{-13}^{+16}$ or $116_{-14}^{+13}$ \\ [0.1cm]
 Mass (\Mjupv) & $5.4_{-0.4}^{+0.5}$ & $3.0_{-0.5}^{+0.7}$ & $2.3_{-0.3}^{+0.8}$ & $1.3_{-0.1}^{+0.3}$ & $4.8 \pm 0.6$ & $2.4_{-0.2}^{+0.4}$ \\ [0.1cm]
 $\Omega$ (°) & $211 \pm 11$ or $246_{-10}^{+11}$ & $308_{-9}^{+10}$ or $320 \pm 8$ & $141_{-19}^{+20}$ & $238_{-21}^{+20}$ & $49 \pm 14$ or $261 \pm 14$ & $118_{-21}^{+31}$ or $204_{-23}^{+20}$\\ [0.1cm]
 $\omega$ (°) & $307_{-37}^{+22}$ & $78_{-8}^{+9}$ & $89 \pm 12$ & $281 \pm 38$ & $174 \pm 9$ & $164_{-103}^{+123}$ \\ [0.1cm]
 $P_{time}$ (days) & $3607_{-245}^{+160}$ & $266_{-57}^{+67}$ & $1946_{-126}^{+125}$ & $136_{-399}^{+378}$ & $6193_{-156}^{+195}$ & $482_{-712}^{+672}$ \\ [0.1cm]
 $\sqrt{ecc}cos(\omega)$ & $0.38_{-0.23}^{+0.19}$ & $0.14 \pm 0.11$ & $0.01_{-0.09}^{+0.08}$ & $0.16_{-0.14}^{+0.09}$ & $-0.61_{-0.05}^{+0.08}$ & $-0.08_{-0.28}^{+0.13}$ \\ [0.1cm]
 $\sqrt{ecc}sin(\omega)$ & $-0.46_{-0.10}^{+0.15}$ & $0.70 \pm 0.03$ & $0.39_{-0.05}^{+0.04}$ & $-0.18_{-0.10}^{0.16}$ & $0.20_{-0.13}^{0.10}$ & $-0.04_{-0.12}^{+0.15}$ \\ [0.1cm]
 Jitter (m/s) & $13.9_{-1.3}^{+1.5}$ & $6.5_{-0.8}^{+1.0}$ & $4.2_{-0.3}^{+0.4}$ & $3.7 \pm 0.4$ & $12 \pm 1$ & $7.3_{-0.6}^{+0.7}$ \\ [0.1cm]
 rms (m/s) & $17.0_{-0.8}^{+1.4}$ & $7.1_{-0.5}^{+1.3}$ & $4.7_{-0.1}^{+0.2}$ & $4.5_{-0.1}^{+0.2}$ & $12.1_{-0.4}^{+0.9}$ & $7.2 \pm 0.2$ \\ [0.1cm]
\hline
 & H03 = $20614 \pm 6$ & H03 = $-5776_{-3}^{+4}$ & H03 = $-14961 \pm 1$ & H03 = $51782 \pm 1$ & H03 = $3203 \pm 2$ & H03 = $51754_{-2}^{+1}$ \\ [0.1cm]
 Instrumental & H15 = $20804_{-9}^{+11}$ & H15 = $-5766_{-3}^{+2}$ & H15 = $-14949 \pm 1$ & H15 = $51792_{-2}^{+1}$ & H15 = $3201 \pm 1$ & H15 = $51762_{-2}^{+1}$ \\ [0.1cm]
 offset (m/s) & ELODIE = $12147 \pm 4$ & & & & Hir04 = $4_{-4}^{+3}$ & \\ [0.1cm]
 & SOPHIE = $20524_{-6}^{+5}$ & & & & & \\ [0.1cm]
 & SOPHIE+ = $20566_{-5}^{+6}$ & & & & & \\ [0.1cm]
\hline
\end{tabular}}
\textbf{Notes:} The median and $1\sigma$ confidence interval are given for each parameter. 
\label{table_summary_planets}
\end{table*}

\subsection{Long-term stellar activity}

The long-term magnetic activity of the host star can lead to variations in the RVs over periods similar to the orbital periods of the companions detected in our study (2-15 years, \cite{2019A&A...627A..56M}). In order to estimate the amplitude of these RV variations, \cite{2019A&A...628A.125M} simulated the impact of the magnetic activity for F-G-K type stars considering a $LogR'_{HK}$ between -5.1 and -4.6. They show that long-term magnetic activity can induce an RV jitter between 1 m/s for quiet stars and up to 10 m/s for the most active stars. These RV variations can thus significantly impact the detection of sub-Jupiter planets with a minimum mass lower than typically 0.6 \Mjup \citep{2023arXiv230500047L}. All the newly detected companions have a minimum mass higher than 1.5 \Mjupv, corresponding to RV amplitudes higher than 40 m/s, well above the expected activity signal. 

Concerning eventual short-term activity, we did not note a significant correlation between BVS and VR for any of these systems (except for HD75302, see above). This result is consistent with the fact that these companions orbit around mature stars (> typ. 1 Gyr) whose stellar jitter is generally smaller than for young systems \citep{2021AJ....161..173T}.

\section{Improved parameters of known RV companions}

\subsection{Long-period companions (P > 1000 d)}

By combining RV and absolute astrometry data, we properly characterize 14 sub-stellar companions that have already been reported in the literature. We significantly improve the characterization and determine the exact nature of ten companions, and we obtain similar results to those previously reported for four companions: HD140901 c, HD169830 c, HD221420 B, and HIP70849 b.

\subsubsection{HD16905}

52 RV data points were obtained with the HARPS spectrograph between 2003 and 2020 for the K3V star HD16905. By coupling HARPS RV measurements, obtained before February 2018 (Julian day: 2458160), with absolute astrometry, \cite{2022ApJS..262...21F} reported an eccentric massive giant planet with a mass of $9.1_{-1.4}^{+1.3}$ \Mjupv, an orbital inclination of $19.7_{-1.9}^{+3.3}$°, a semi-major axis of $6.4_{-0.4}^{+1.6}$ au and an eccentricity of $0.67_{-0.07}^{+0.06}$.

Fitting all the RV data available and the absolute astrometry data with our MCMC (fig. \ref{RV_AA_HD16905} and \ref{RV_AA_HD16905_sup}), considering 300000 iterations (burn-in = 50000) and 1000 walkers, we find significantly different solutions, with a semi-major axis of $8.8_{-0.3}^{+0.4}$ au (P = $10256_{-522}^{+618}$ d), a mass of $11.3_{-0.7}^{+0.6}$ \Mjupv, and an orbital inclination of either $43 \pm 3$° or $136 \pm 3$°.

\subsubsection{HD28254}

72 RV data points were obtained with the HARPS spectrograph between 2003 and 2021 for the G5V star HD28254. \cite{2010A&A...523A..15N} reported a very eccentric giant planet candidate with a minimum mass of $1.16_{-0.06}^{+0.10}$ \Mjupv, a semi-major axis of $2.15_{-0.05}^{+0.04}$ au and an eccentricity of $0.81_{-0.02}^{+0.05}$ based on the RV HARPS data obtained before April 2009 (Julian day: 2454931). 

The baseline of RV data used by \cite{2010A&A...523A..15N} fully covers the period of HD28254 b. Combining all HARPS RV measurements available with absolute astrometry with our MCMC (fig. \ref{RV_AA_HD28254} and \ref{RV_AA_HD28254_sup}), considering 600000 iterations (burn-in = 100000) and 1000 walkers, we find a solution with a semi-major axis very close to that reported by \cite{2010A&A...523A..15N} but more eccentric: \textit{a} = $2.45_{-0.04}^{+0.03}$ au (P = $1333 \pm 4$ d) and ecc = $0.95_{-0.04}^{+0.03}$. We also estimate the orbital inclination at either $21_{-11}^{+38}$° or $162_{-27}^{+7}°$ and the true mass between 1.5 and 6.5 \Mjupv. Hence, HD28254 b is a giant planet.

\subsubsection{HD62364}

87 RV data points were obtained with the HARPS spectrograph between 2004 and 2021 for the F7V star HD62364. By coupling the HARPS RV measurements obtained before February 2018 (Julian day: 2458160) with absolute astrometry, \cite{2022ApJS..262...21F} reported two companions: a first, $17.4_{-1.7}^{+1.6}$ \Mjup brown dwarf at $18.9_{-1.1}^{+1.6}$ au and a second $24.9_{-2.9}^{+2.7}$ \Mjup brown dwarf at $36.98_{-2.8}^{+3.0}$ au. 

By fitting the RV data considering a one companion system with DPASS, using 3 years of additional HARPS RV measurements, we find a significantly different result: a single companion at approximately 6.4 au and no additional signal in the RV residuals. Moreover, when considering a two-companions system, we find an offset between the HARPS RV measurements obtained before and after the fiber upgrade of about 78 m/s while the average offset for a star of this spectral type is smaller than 20 m/s \citep{2015Msngr.162....9L}, which is in agreement with the offset of 17 m/s found for a one-companion system. We conclude that there is no evidence for the presence of a second companion.

Combining all the RV data points available and the absolute astrometry data with our MCMC, considering a one-companion system (fig. \ref{RV_AA_HD62364} and \ref{RV_AA_HD62364_sup}) and 300000 iterations (burn-in = 30000), 1000 walkers, we find an eccentric brown dwarf with a mass of $19_{-1}^{+2}$ \Mjupv, a semi-major axis of $6.4 \pm 0.1$ au (P = $5172 \pm 30$ d), an eccentricity of $0.610_{-0.005}^{+0.004}$ and an orbital inclination of either $42 \pm 2$° or $131 \pm 2$°.

\subsubsection{HD89839}

98 RV data points were obtained with HARPS between 2004 and 2021 for the F7V star HD89839. \cite{2011A&A...527A..63M} reported a giant planet candidate based on the RV HARPS measurements obtained before June 2010 (Julian day: 2455353). However, this dataset does not fully cover the orbit of the companion, which is consequently poorly characterized. The reported orbital parameters are therefore poorly constrained, with a semi-major axis of $6.8_{-2.4}^{+3.3}$ au and an eccentricity of $0.32 \pm 0.20$. The available data allow us to reasonably well constrain the amplitude of RV variations, corresponding to a minimum mass of $3.9 \pm 0.4$ \Mjupv. 

The additional data points obtained between 2010 and 2021 now allow us to properly constrain the orbital parameters of HD89839 b and, by simultaneously fitting all the RV data points available with the absolute astrometry with our MCMC (Fig. \ref{RV_AA_HD89839} and \ref{RV_AA_HD89839_sup}), considering 300000 iterations (burn-in = 30000) and 1000 walkers, we constrain the orbital inclination to either $53_{-9}^{+16}$° or $129_{-15}^{+9}$°. Thus, we find a semi-major axis of $4.9 \pm 0.1$ au (P = $3448 \pm 26$ d), an eccentricity of $0.20 \pm 0.02$, and a true mass of $5.2 \pm 0.8$ \Mjupv. Hence, HD89839 b is a giant planet.

\subsubsection{HD114783}

282 RV data points were obtained for the K1V star HD114783, 91 RV HARPS measurements were obtained between 2004 and 2016, 157 RV HIRES measurements were obtained between 1998 and 2019, and 34 RV HRS measurements were obtained between 2004 and 2007. Firstly, \cite{2002ApJ...568..352V} reported a giant planet candidate with a minimum mass of 1 \Mjup and a period of 501 days. \cite{2016ApJ...821...89B} reported an outer companion with a minimum mass of $0.611_{-0.053}^{+0.056}$ \Mjup and a period of $4319_{-130}^{+151}$ days. Considering the RV HIRES measurements, \cite{2021ApJS..255....8R} reported a two planets system at $1.164 \pm 0.016$ au and $4.97_{-0.11}^{+0.12}$ au (P = $4344_{-14}^{+33}$ d) with a minimum mass of $1.033_{-0.033}^{+0.034}$ \Mjup and $0.660_{-0.047}^{+0.046}$ \Mjup respectively.

As the orbital period of HD114783 b is much smaller than the duration of Gaia DR3, the proper motion induced by the planet can be neglected. Thus, we fit simultaneously all the RV data points available, corrected for the signal of the inner planet, with the absolute astrometry with our MCMC (Fig. \ref{RV_AA_HD114783} and \ref{RV_AA_HD114783_sup}), considering 300000 iterations (burn-in = 100000) and 1000 walkers. We find orbital parameters close to those reported by \cite{2021ApJS..255....8R} with a semi-major axis of $5.0 \pm 0.1$ au (P = $4352_{-76}^{+88}$ d) and an eccentricity of $0.05_{-0.03}^{+0.04}$. We constrain the orbital inclination of HD114783 c to either $21_{-4}^{+7}$° or $159_{-6}^{+4}$°. We find a true mass of $2.0 \pm 0.4$ \Mjupv. Hence, HD114783 c is a giant planet.

\subsubsection{HD140901}

443 RV data points were obtained for the G7IV star HD140901, 52 RV HARPS measurements were obtained between 2004 and 2007, 321 RV AAT measurements were obtained between 1998 and 2014, and 70 RV PFS measurements were obtained between 2011 and 2022. Considering all the RV data points available and the absolute astrometry, \cite{2022ApJS..262...21F} reported a massive planet with a semi-major axis of $7.42_{-0.58}^{+0.32}$ au, an inclination of $169.4_{-55.4}^{+2.2}$°, a mass of $6.3_{-5.0}^{+1.3}$ \Mjup and an eccentricity of $0.61_{-0.04}^{+0.03}$. They also reported the presence of an inner planet with a minimum mass of $0.049_{-0.008}^{+0.002}$ \Mjup located at $0.085_{-0.004}^{+0.003}$ au. However, the study did not detail the analysis of the system. Although a significant peak is present in the periodogram of the data around 10 days, once the signal of the external planet is subtracted, the amplitude of the dispersion of the RV measurements is greater than the signal induced by the potential planet. It is not possible to confirm the presence of the planet.

In any case, as the orbital period of the planet candidate is much smaller than the duration of Gaia DR3, the proper motion induced by this potential planet is negligible. Thus, fitting simultaneously the same RV dataset as \cite{2022ApJS..262...21F} and the absolute astrometry with our MCMC (Fig. \ref{RV_AA_HD140901} and \ref{RV_AA_HD140901_sup}), considering 400000 iterations (burn-in = 100000) and 1000 walkers, we find a semi-major axis of $11.8_{-2.5}^{+4.1}$ au (P = $14386_{-4415}^{+8099}$ d), an inclination of either $40_{-11}^{+18}$° or $138_{-21}^{+12}$°, a mass of $1.8 \pm 0.5$ \Mjup and an eccentricity of $0.77_{-0.07}^{+0.06}$. This solution is within the error bars associated with the values found by \cite{2022ApJS..262...21F} and confirms that HD140901 b is a giant planet. 

\subsubsection{HD143361}

120 RV data points were obtained for the G6V star HD143361, 47 RV HARPS measurements were obtained between 2007 and 2019, 44 RV CORALIE measurements were obtained between 2010 and 2014, and 29 RV MIKE measurements were obtained between 1998 and 2005. \cite{2017MNRAS.466..443J} reported a giant planet candidate with a minimum mass of $3.48 \pm 0.24$ \Mjupv, a semi-major axis of $1.98 \pm 0.07$ au and an eccentricity of $0.193 \pm 0.015$, considering the RV MIKE and CORALIE measurements and the RV HARPS measurements obtained before June 2013 (Julian day: 2456462).

Fitting simultaneously all the RV data points available with the absolute astrometry with our MCMC (Fig. \ref{RV_AA_HD143361} and \ref{RV_AA_HD143361_sup}), considering 300000 iterations (burn-in = 50000) and 1000 walkers, we find orbital parameters close to those reported by \cite{2017MNRAS.466..443J} with a semi-major axis of $2.05 \pm 0.03$ au (P = $1040 \pm 1$) and an eccentricity of $0.198_{0.006}^{+0.007}$. We constrain the orbital inclination of HD143361 b to either $50_{-12}^{+19}$° or $128_{-27}^{+18}$°. Thus, we find a true mass between 3.9 and 6.3 \Mjupv. Hence, HD143361 b is a giant planet. 

\subsubsection{HD167677}

44 RV data points were obtained with HARPS between 2005 and 2019 for this G5V star HD167677. \cite{2011A&A...527A..63M} reported a giant planet candidate with a minimum mass of $1.36 \pm 0.12$ \Mjupv, a semi-major axis of $2.90 \pm 0.12$ au and an eccentricity of $0.17 \pm 0.07$. They considered the RV HARPS measurements obtained before October 2010 (Julian day: 2455495) which properly cover the period of the companion.

Fitting simultaneously all the HARPS RV measurements available with the absolute astrometry with our MCMC (Fig. \ref{RV_AA_HD167677} and \ref{RV_AA_HD167677_sup}), considering 200000 iterations (burn-in = 50000) and 1000 walkers, we find orbital parameters close to those reported by \cite{2011A&A...527A..63M} with a semi-major axis of $2.9 \pm 0.1$ au (P = $1804 \pm 73$ d) and an eccentricity of $0.35_{-0.15}^{+0.25}$. We constrain the orbital inclination of HD167677 b to either $20_{-6}^{+14}$° or $155_{-12}^{+6}$°. Thus, we find a true mass of $3.7_{-1.1}^{+1.6}$ \Mjupv. Hence, HD167677 b is a giant planet.

\subsubsection{HD169830}

277 RV data points were obtained for the F7V star HD169830, 86 RV HARPS measurements were obtained between 2004 and 2009, 106 RV CORALIE measurements were obtained between 1999 and 2003, and 85 RV HIRES measurements were obtained between 2000 and 2019. Firstly, \cite{2001A&A...375..205N} reported a giant planet candidate with a minimum mass of $2.94 \pm 0.12$ \Mjupv, a semi-major axis of 0.82 au, and an eccentricity of $0.35 \pm 0.04$, considering the RV CORALIE measurements obtained before May 2000 (Julian day: 2451668). A second giant planet candidate was then identified by \cite{Mayor_2004} and characterized by \cite{2021ApJS..255....8R}, considering all available RV data points. They found a minimum mass of $3.51 \pm 0.12$ \Mjupv, a semi-major axis of $3.283_{-0.036}^{+0.035}$ au and an eccentricity of $0.257 \pm 0.019$, for HD169830 c. Recently, combining all the RV data points with absolute astrometry, \cite{2022ApJS..262...21F} reported an orbital inclination of $24.47_{-7.21}^{+12.74}$° and a true mass of $7.67_{-2.76}^{+1.94}$ \Mjupv, for HD169830 c\footnote{As the orbital period of HD169830 b is much smaller than the duration of Gaia DR3, the proper motion induced by this planet is negligible.}.

Fitting simultaneously the same RV dataset as \cite{2022ApJS..262...21F}, corrected for the signal of HD169830 b, and the absolute astrometry \textbf{with our MCMC (Fig. \ref{RV_AA_HD169830} and \ref{RV_AA_HD169830_sup}), considering 300000 iterations (burn-in = 50000) and 1000 walkers}, we find a semi-major axis of $3.29_{-0.05}^{+0.04}$ au \textbf{(P = $1832 \pm 7$ d)}, an inclination of either $23_{-6}^{+12}$° or $158_{-9}^{+5}$°, a mass of $8.9_{-2.6}^{+2.8}$ \Mjup and an eccentricity of $0.23_{-0.02}^{+0.03}$. This solution is within the error bars associated with the values found by \cite{2022ApJS..262...21F} and confirms that HD169830 c is a giant planet.

\subsubsection{HD196050}

131 RV data points were obtained for the G3V star HD196050 A, 56 RV HARPS measurements were obtained between 2004 and 2021, 31 RV CORALIE measurements were obtained between 1999 and 2003, and 44 RV UCLES measurements were obtained between 1998 and 2005. Based on NACO high contrast imaging observations, \cite{2007A&A...474..273E} reported a binary companion orbiting at about 500 au from HD196050 A and composed of an M1.5-M4.5 low mass star with a mass of $0.29 \pm 0.02$ \Msun and an M2.5-M5.5 low-mass star with a mass of $0.19 \pm 0.02$ \Msun. Given the separation and mass of the HD196050 B/C pair, their movement will induce negligible variations in RV and proper motion. \cite{2002MNRAS.337.1170J} reported a giant planet candidate with a minimum mass of $2.8 \pm 0.5$ \Mjupv, a semi-major axis of $2.4 \pm 0.5$ au and an eccentricity of $0.19 \pm 0.09$, considering the RV UCLES measurements obtained before June 2002 (Julian day: 2452455). Considering the RV CORALIE measurements, \cite{2004A&A...415..391M} reported orbital parameters and minimum mass for HD196050 Ab close to those reported by \cite{2002MNRAS.337.1170J}.

Fitting simultaneously all the RV data points available with the absolute astrometry with our MCMC (Fig. \ref{RV_AA_HD196050} and \ref{RV_AA_HD196050_sup}), considering 300000 iterations (burn-in = 30000) and 1000 walkers, we improve the orbital parameters with a semi-major axis of $2.65 \pm 0.03$ au (P = $1392 \pm 4$ d) and an eccentricity of $0.20 \pm 0.01$. We constrain the orbital inclination of HD196050 Ab to either $42_{-7}^{+11}$° or $141_{-12}^{+7}$°. Thus, we find a semi-major axis of $2.65 \pm 0.03$ au, an eccentricity of $0.20 \pm 0.01$, and a true mass of $4.7 \pm 0.8$ \Mjupv. Hence, HD196050 Ab is a giant planet.

\subsubsection{HD204961}

238 RV data points were obtained for the M2/3V star HD204961, 183 RV HARPS measurements were obtained between 2003 and 2019, 39 RV AAT measurements were obtained between 1998 and 2013, and 16 RV PFS measurements were obtained between 2011 and 2013. \cite{2009ApJ...690..743B} reported a giant planet candidate with a minimum mass of $0.64 \pm 0.06$ \Mjup and a period of $3416 \pm 131$ days. An inner super-earth was reported by \cite{2014ApJ...791..114W} with a minimum mass of $5.4 \pm 1$ $M_{Earth}$. However, the signal corresponding to this potential inner planet was then identified as corresponding to the rotation period of the host star by \cite{2022A&A...66arithmetics4A..64G}. 

Fitting simultaneously all the RV data points available with the absolute astrometry with our MCMC (Fig. \ref{RV_AA_HD204961} and \ref{RV_AA_HD204961_sup}), considering 300000 iterations (burn-in = 30000) and 1000 walkers. We find orbital parameters close to those reported by \cite{2022A&A...664A..64G} with a semi-major axis of $3.7 \pm 0.1$ au (P = $3853_{-47}^{+51}$ d) and an eccentricity of $0.05 \pm 0.03$. We constrain the orbital inclination of HD204961 b to either $51 \pm 3$° or $134 \pm 3$°. We find a true mass of $0.99_{-0.08}^{+0.09}$ \Mjupv. Hence, HD204961 b is a giant planet.

\subsubsection{HD216437}

160 RV data points were obtained for the G1V star HD216437, 100 RV HARPS measurements were obtained between 2007 and 2021, 31 RV CORALIE measurements were obtained between 1999 and 2003, and 39 RV UCLES measurements were obtained between 1998 and 2005. \cite{2002MNRAS.337.1170J} reported a giant planet candidate with a minimum mass of $2.1 \pm 0.3$ \Mjupv, a semi-major axis of $2.4 \pm 0.5$ au and an eccentricity of $0.33 \pm 0.09$, considering the RV UCLES measurements obtained before June 2002 (Julian day: 2452456). Considering the RV CORALIE measurements, \cite{2004A&A...415..391M} reported orbital parameters and minimum mass for HD216437 b close to those reported by \cite{2002MNRAS.337.1170J}.

Fitting simultaneously all the RV data points available with the absolute astrometry with our MCMC (Fig. \ref{RV_AA_HD216437} and \ref{RV_AA_HD216437_sup}), considering 400000 iterations (burn-in = 100000) and 1000 walkers, we improve the orbital parameters and constrain the orbital inclination of HD216437 b to either $36_{-6}^{+8}$° or $148_{-7}^{+5}$°. Thus, we find a semi-major axis of $2.57 \pm 0.04$ au (P = $1359 \pm 1$ d), an eccentricity of $0.31 \pm 0.01$, and a true mass of $4.2 \pm 0.7$ \Mjupv. Hence, HD216437 b is a giant planet. 

\subsubsection{HD221420}

203 RV data points were obtained for the G1V star HD221420, 77 RV HARPS measurements were obtained between 2003 and 2018, 88 RV AAT measurements were obtained between 1998 and 2015, and 38 RV PFS measurements were obtained between 2012 and 2021. \cite{2019AJ....157..252K} reported a giant planet candidate with a minimum mass of $9.7_{-1.0}^{+1.1}$ \Mjupv, a semi-major axis of $18.5 \pm 2.3$ au, and an eccentricity of $0.42_{-0.07}^{+0.05}$, considering the RV AAT measurements. Recently, combining the RV AAT and HARPS measurements with absolute astrometry, \cite{2021AJ....162..266L} reported an orbital inclination of either $17.8_{-2.8}^{+2.9}$° or $162.2_{-2.9}^{+2.8}$° and a true mass of $20.6_{-1.6}^{+2.0}$ \Mjupv, corresponding to a brown dwarf. Moreover, they obtained a better constraint on the other orbital parameters and found a semi-major axis of $9.99_{-0.70}^{+0.74}$ au and an eccentricity of $0.162_{-0.030}^{+0.035}$.

Fitting simultaneously the all available RV data points with the absolute astrometry with our MCMC (Fig. \ref{RV_AA_HD221420} and \ref{RV_AA_HD221420_sup}), considering 300000 iterations (burn-in = 50000) and 1000 walkers, we find a semi-major axis of $10.1 \pm 0.7$ au (P = $10066_{-1125}^{+929}$ d), an inclination of either $14 \pm 1$° or $162 \pm 2$°, a mass of $22 \pm 2$ \Mjup and an eccentricity of $0.12_{-0.03}^{+0.04}$. This solution is within the error bars associated with the values found by \cite{2021AJ....162..266L} and confirms that HD221420 B is a brown dwarf.

\subsubsection{HIP70849}

59 RV data points were obtained with HARPS between 2006 and 2021 for the K7V star HIP70849. \cite{2011A&A...535A..54S} reported the detection of a very poorly constraint companion with a minimum mass between 3 and 15 \Mjupv, a semi-major axis between 4.5 and 36 au, and an eccentricity between 0.47 and 0.96 ($3\sigma$ confidence interval), considering the RV HARPS measurements obtained before June 2010 (Julian day: 2455374). In \cite{2023A&A...670A..65P}, we already used our MCMC (Fig. \ref{RV_AA_HIP70849}) to combine available HARPS data points with absolute astrometry, considering 300000 iterations (burn-in = 30000) and 1000 walkers. We found a semi-major axis of $3.99_{-0.07}^{+0.06}$ au (P = $3649 \pm 18$ d), an inclination of $96 \pm 16$°, a mass of $4.5_{-0.3}^{+0.4}$ \Mjup and an eccentricity of $0.65_{-0.01}^{+0.02}$, confirming that HIP70849 b is a giant planet. In this study, as we consider the same dataset and the same fitting tools as \cite{2023A&A...670A..65P}, we obtain the same solution.

\subsection{Short-period companions (P < 1000 d)}

Four companions (HD44219 b, HD100777 b, HD111998 b, and HD215497 c) have an orbital period smaller than the duration of Gaia data as delivered in the DR3. It is, therefore, not possible to determine their inclination and, therefore, their true mass precisely from the variations of proper motion. We have therefore fitted only the radial velocities with \textit{I} fixed at 90° and $\omega$ fixed at 0° to obtain the other orbital parameters and the minimum mass of the four companions. We then use GASTON to estimate the maximum orbital inclination and the maximum mass, which can explain the AEN measured by the Gaia DR3. 

\subsubsection{HD44219}

65 RV data points were obtained with HARPS between 2003 and 2021 for the G3V star HD44219. \cite{2010A&A...523A..15N} reported a giant planet candidate with a minimum mass of $0.58_{-0.04}^{+0.06}$ \Mjupv, a semi-major axis of $1.19 \pm 0.02$ au, and an eccentricity of $0.61_{-0.09}^{+0.07}$, considering the HARPS data points obtained before April 2009 (Julian day: 2454932).

Fitting all the HARPS data points available with our MCMC, considering 200000 iterations (burn-in = 50000) and 1000 walkers (Fig. \ref{RV_short_period}), we find orbital parameters close to those reported by \cite{2010A&A...523A..15N} with a semi-major axis of $1.26_{-0.07}^{+0.02}$ au (P = $472_{-4}^{+3}$ d) and an eccentricity of $0.41_{-0.14}^{+0.12}$. Using Gaston, we could estimate that the orbital inclination of HD44219 b is between 4° and 176° (confidence interval at 3$\sigma$) and, thus, that the maximum mass of the planet is 11 \Mjupv. Hence, HD44219 b is a giant planet.

\subsubsection{HD100777}

37 RV data points were obtained with HARPS between 2004 and 2007 for the G8V star HD100777. \cite{2007A&A...470..721N} reported a giant planet candidate with a minimum mass of $1.16 \pm 0.03$ \Mjupv, a semi-major axis of $1.03 \pm 0.03$ au, and an eccentricity of $0.36 \pm 0.02$, considering the all the available HARPS data.

Fitting the same dataset as the one considering by \cite{2007A&A...470..721N} with our MCMC, considering 200000 iterations (burn-in = 50000) and 1000 walkers (Fig. \ref{RV_short_period}), we find orbital parameters close to those reported by \cite{2007A&A...470..721N} with a semi-major axis of $1.06_{-0.01}^{+0.02}$ au (P = $383 \pm 1$ d) and an eccentricity of $0.38 \pm 0.02$. Using Gaston, we could estimate that the orbital inclination of HD100777 b is between 4° and 176° (confidence interval at 3$\sigma$) and, thus, that the maximum mass of the planet is 18 \Mjupv. Hence, the use of the AEN measured by Gaia does not allow us to determine with certainty the nature of the companion. Moreover, it is important to note that the PMa does not seem to agree with that induced by HD100777 b. This difference could be explained by the presence of an external companion. However, there is no significant signal present in the RV data. This can be due to the short RV temporal baseline ($\sim$1200 d).

\subsubsection{HD111998}

150 RV data points were obtained with the HARPS spectrograph between 2006 and 2014 for HD111998. As HD111998 is a fast-rotating F6V star, the HARPS DRS is not able to properly compute the radial velocity of the star. For this reason, \cite{Borgniet_2017} used the SAFIR (Spectroscopic data via Analysis of the Fourier Interspectrum Radial velocities) software \citep{2005A&A...443..337G} to compute the RV. They reported a giant planet candidate with a minimum mass of $4.51 \pm 0.50$ \Mjupv, a semi-major axis of $1.82 \pm 0.07$ au, and an eccentricity of $0.03 \pm 0.04$.

Fitting the same dataset as the one considering by \cite{Borgniet_2017} with our MCMC, considering 300000 iterations (burn-in = 100000) and 1000 walkers (Fig. \ref{RV_short_period}), we find orbital parameters close to those reported by \cite{Borgniet_2017} with a semi-major axis of $1.91_{-0.03}^{+0.02}$ au (P = $810_{-5}^{+6}$ d) and an eccentricity of $0.13_{-0.04}^{+0.05}$. Using Gaston, we could estimate that the orbital inclination of HD111998 b is between 23° and 154° (confidence interval at 3$\sigma$) and, thus, that the maximum mass of the planet is 12.8 \Mjupv. Hence, HD111998 is a giant planet. Morover, it is important to note that as for HD100777, the PMa measured for HD111998 does not seem to agree with that induced by the HD111998 b. This difference could be due to an additional companion, which could explain the long-term trend reported in \cite{Borgniet_2017}.

\subsubsection{HD215497}

115 RV data points were obtained with HARPS between 2004 and 2009 for the K3V star HD215497. \cite{2010A&A...512A..48L} reported a Super-earth candidate with a minimum mass of 6.6 $M_{Earth}$, a semi-major axis of 0.047 au and en eccentricity of $0.16 \pm 0.09$ and a giant planet candidate with a minimum mass of 0.33 \Mjupv, a semi-major axis of 1.28 au, and an eccentricity of $0.49 \pm 0.04$, considering all the available HARPS data points.

HD215467 b is not massive enough to induce a significant PMa or AEN. We have therefore considered only HD215497 c. Fitting the HARPS data points, corrected for the signal of HD215497 b, with our MCMC, considering 400000 iterations (burn-in = 100000) and 1000 walkers (Fig. \ref{RV_short_period}), we find a semi-major axis of $1.31 \pm 0.02$ au \textbf{(P = $566 \pm 4$ d)}, a minimum mass of $0.35 \pm 0.02$ \Mjupv, and an eccentricity of $0.47_{-0.04}^{+0.03}$, the solution found by \cite{2010A&A...512A..48L} is within the error bars associated with our values. Using Gaston, we could estimate that the orbital inclination of HD215467 c is between 4° and 176° (confidence interval at 3$\sigma$) and, thus, that the maximum mass of the planet is 5.0 \Mjupv. Hence, HD215497 c is a giant planet.

\section{Stellar companions}

Among the 116 selected stars, we detect and confirm the nature of 27 stellar companions, including 12 new detections (Table \ref{table_summary_stellar}). These companions orbit between 0.56 and 90 au and have masses between 0.1 and 1.1 \Msun (uncertainties included). 

Even if the addition of absolute astrometry can allow us to lift the sin(\textit{I}) indetermination and determine the true mass of a companion, the fit of the variations of the proper motion of a star does not always allow us to choose between a prograde and a retrograde orbit for the companion. For almost all the systems presented in this study, when the orbital period is properly constrained, we find very close semi-major axes and masses for the prograde and retrograde solutions. One exception is the system HIP39470 for which the 32 RV data obtained with HARPS between 2004 and 2019 do not fully cover the orbital period of the companion. Thanks to the significant PMa of the star, adding the absolute astrometry allows us to significantly improve the constraints on the companion's orbital parameters compared to the RV alone. However, considering a prograde and a retrograde solution (\textit{I} = $22_{-3}^{+2}$° and \textit{I} = $170 \pm 1$° respectively), we find two different solutions with a semi-major axis of $8.8 \pm 0.2$ or $7.6 \pm 0.1$ au and a mass of $0.138_{-0.008}^{+0.007}$ or $0.147 \pm 0.005$ \Msun respectively. In the two cases, we find that the new companion HIP39470 B is a stellar companion. The epoch astrometry of Gaia, expected with the fourth data release (end of 2025), or HCI observation, when possible, could constrain the exact orbital inclination of the companions and significantly improve the characterization of HIP39470 B.

The study of these stellar mass companions shows us the importance of properly determining the orbital inclination of a companion. Indeed, using only RV data, the minimum masses found for three of these companions were in the brown dwarfs' regime. Indeed, we find a minimum mass of about 37 and 52 \Mjup for the new companions HD56380 B and HD221638 B, and \cite{Moutou_2009} reported a minimum mass of about 18 \Mjupv for HD131664 B. However, coupling the RV data with absolute astrometry, we find an inclination of either $6.1 \pm 0.1$° or $173.6 \pm 0.1$°, either $31 \pm 1$° or $150 \pm 1$°, and either $9.3 \pm 0.3$° or $170.4 \pm 0.3$°, respectively, corresponding to a true mass of $0.39 \pm 0.02$, $0.132 \pm 0.006$, and $0.109 \pm 0.005$ \Msun ($410 \pm 19$, $138 \pm 6$, and $114 \pm 5$ \Mjup), respectively. These three companions are thus, in reality, low-mass stars. Note that \cite{2021MNRAS.507.2856F} had already identified HD131664 B as a stellar companion by coupling RV HARPS measurements with absolute astrometry.

Moreover, in most cases, when the period of the companion is much larger than the available RV baseline, the orbital parameters and the mass of the companion cannot be properly characterized. In these cases, the addition of relative astrometry significantly improves the constraint on the orbit of the companion and, consequently, its mass. Good examples are the cases of GJ680 B and HD111031 B. \cite{2022ApJS..262...21F} have combined RV data points, obtained with HARPS between 2004 and 2012, for GJ680, and with HIRES between 1997 and 2016, for HD111031, and absolute astrometry. They report a semi-major axis of $10.1_{-1.7}^{+1.8}$ and $13.1_{-1.1}^{+0.8}$ au and a true mass of $25.1_{-11.1}^{+6.2}$ and $54.2_{-6.1}^{+5.3}$ \Mjup respectively, corresponding to brown dwarfs. However, GJ680 B was detected in HCI in May 2005 and May 2007 with the NACO instrument \citep{2015MNRAS.449.2618W} and HD111031 B was detected in HCI in February 2018 with the NESSI instrument \citep{2021AJ....161..123D}. In this study, we consider the same RV dataset for GJ680 and we consider the RV HARPS measurements obtained between 2004 and 2013, the RV HIRES data obtained between 1997 and 2020, the RV APF measurements obtained between 2013 and 2020, and the RV LICK measurements obtained in 2001, for HD111031. Combining these RV data with absolute astrometry and relative astrometry data, we find significantly different results from those published by \cite{2022ApJS..262...21F} for the two companions. We find semi-major axes of $32_{-6}^{+9}$ and $21.1 \pm 0.6$ au and true masses of $0.178 \pm 0.004$ and $0.129 \pm 0.003$ \Msun ($186 \pm 4$ and $135 \pm 3$ \Mjup) respectively, corresponding to low mass stars instead of a brown dwarf.

As seen above, the algorithm used for this study, based on the MCMC sampling tool used by \cite{2023arXiv230500047L}, allows us to obtain a better exploration of the mass and orbital values compatible with the data. For example, in the case of GJ680 B, whose orbit is poorly constrained, we obtain, by considering only RV and absolute astrometry data, solutions corresponding to a $1\sigma$ (resp. $3\sigma$) confidence interval for the semi-major axis and the companion mass between 16 and 28 au (resp. 12 and 35 au) and between 70 and 160 \Mjup (resp. 50 and 190 \Mjupv), respectively. We notice that the solutions found by adding the HCI observations are well explored by the MCMC, but it is not possible without the imaging to conclude on the orbital parameters and the mass of the companion.

Finally, thanks to the HCI observations of HD93351 obtained with NACO, we find that the companion is actually a stellar binary with a combined mass of $0.70 \pm 0.05$ \Msun. However, the combination of RV, HCI, and absolute astrometry data does not allow us to properly determine the orbital parameters of the binary. We can only obtain an estimation for the semi-major axis with a solution at $83_{-19}^{+18}$ au.

\section{Conclusion}

Using the PMa of the stars as given by Gaia DR3 and Hipparcos observed with HARPS RV data points spread over at least 1000 days, we identified good candidates to host long-period sub-stellar companions, which allowed us to optimize our search for sub-stellar companions. We detected at least one companion, stellar or sub-stellar, for 86\% of the selected stars, including systems hosting multiple companions as well as those with companion's orbit poorly constrained. Moreover, 23\% of the systems are composed of at least one confirmed sub-stellar companion (41\% considering companions detected in multiple systems with a minimum mass smaller than 80 \Mjup). This is significantly higher than the detection rates obtained in the case of the large RV surveys (\cite{2011arXiv1109.2497M}, \cite{2020MNRAS.492..377W}, \cite{2021ApJS..255....8R}), which do not use selection criteria to optimize the detection of sub-stellar companions.

Among our sample, we have detected three new brown dwarfs (GJ660.1 C, HD73256 C, and HD165131 B) and six new planets (HD75302 b, HD108202 b, HD135625 b, HD185283 b, HIP10337 b, and HIP54597 b). As already demonstrated in the direct imaging study (\cite{2022MNRAS.513.5588B}, \cite{2022AJ....164..152S}, \cite{2022ApJ...934L..18K}), we confirm that the study of the PMa allows optimizing the selection of targets for the search of sub-stellar companions. However, this method remains sensitive only to massive objects (typically > 1 \Mjupv) and is not adapted for the detection of low-mass planets (super-Earths, mini-Neptune).

\begin{figure*}[h!]
 \centering
\includegraphics[width=0.9\textwidth]{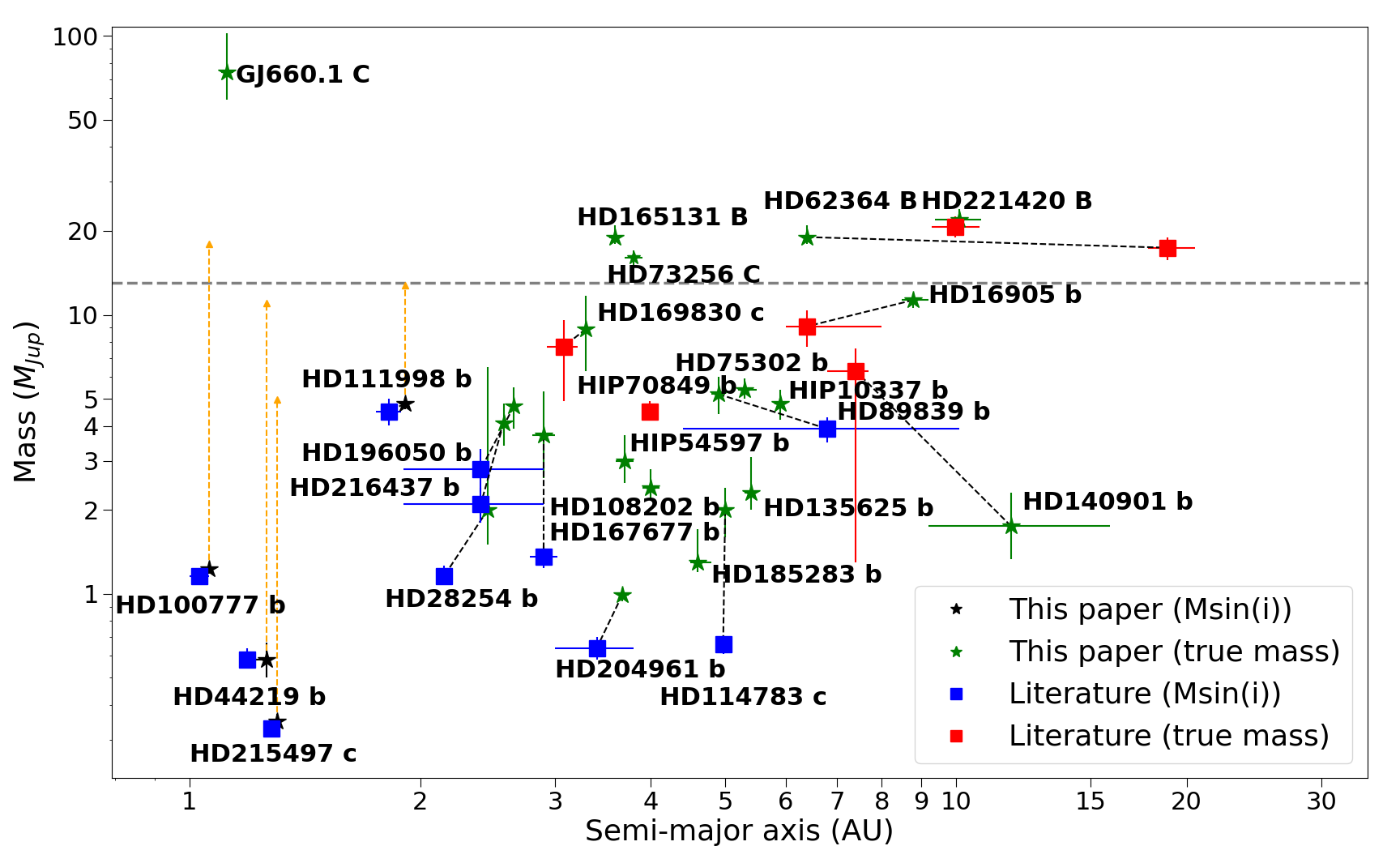}
\caption{Semi-major axis and masses of the different sub-stellar companions. For known companions, the black dotted lines allow comparing the solutions obtained in this study, represented by a star, with those obtained in previous studies, represented by a square. The orange arrow corresponds to the maximum mass of the planet at 3$\sigma$ found with Gaston. The gray dotted horizontal line corresponds to the deuterium burning limit ($\sim$13 \Mjupv).
\label{sma_mass}} 
\end{figure*}

The combination of RV data with absolute astrometry data allows us to properly characterize the orbital parameters of these ten new companions and in particular their orbital inclination. We thus determine their true mass and thus their exact nature. Moreover, by applying the same method, we are able to improve the characterization and determine precisely the true mass of 10 known sub-stellar companions and, using GASTON, we are able to confirm the planetary nature of three known candidate planets (Fig. \ref{sma_mass}). We determined the properties of 25 stellar companions, including 12 new detections. We show also that five companions previously reported as brown dwarfs are indeed low-mass stars: GJ680 B, HD56380 B, HD111031 B, HD131664 B, and HD221638 B.

The coupling of Hipparcos/Gaia measurements with RV data does not always allow for accurately characterizing the orbit of a companion, in particular when its period is not completely covered by the RV data. The addition of HCI data allows for better constraining the orbit of the companion and, consequently, measuring its mass. However, most of the companions revealed in this study have not been detected or observed in imaging. Thus, the orbital parameters nor the exact nature of 27 of the 80 single companions found in our sample cannot be determined.

In summary, we conclude that studying the PMa of stars allows optimizing the search for sub-stellar companions in RV that can then be properly characterized by combining RV, absolute astrometry, and, when available, relative astrometry data. However, this method is not sensitive to low-mass planets and therefore does not allow searching for Earth-like planets.

\begin{acknowledgements}
 This study was funded by a grant from PSL/OCAV. 
 This project has also received funding from the European Research Council (ERC) under the European Union's Horizon 2020 research and innovation programme (COBREX; grant agreement n° 885593).
 This work presents results from the European Space Agency (ESA) space mission Gaia. Gaia data are being processed by the Gaia Data Processing and Analysis Consortium (DPAC). Funding for the DPAC is provided by national institutions, in particular the institutions participating in the Gaia MultiLateral Agreement (MLA). The Gaia mission website is https://www.cosmos.esa.int/gaia. The Gaia archive website is https://archives.esac.esa.int/gaia.
 This publications makes use of the The Data \& Analysis Center for Exoplanets (DACE), which is a facility based at the University of Geneva (CH) dedicated to extrasolar planets data visualisation, exchange and analysis. DACE is a platform of the Swiss National Centre of Competence in Research (NCCR) PlanetS, federating the Swiss expertise in Exoplanet research. The DACE platform is available at https://dace.unige.ch.
 This work has made use of the SPHERE Data Center, jointly operated by OSUG/IPAG (Grenoble), PYTHEAS/LAM/CeSAM (Marseille), OCA/Lagrange (Nice), Observatoire de Paris/LESIA (Paris), and Observatoire de Lyon (OSUL/CRAL).
 This research has made use of the SIMBAD database and VizieR catalogue access tool, operated at CDS, Strasbourg, France.
 Based on data retrieved from the SOPHIE archive at Observatoire de Haute-Provence (OHP), available at atlas.obs-hp.fr/sophie.
 Based on spectral data retrieved from the ELODIE archive at Observatoire de Haute-Provence (OHP).
 Based on observations collected at the European Southern Observatory under ESO programmes.
 We finally thank our anonymous referee, whose feedback helped us to improve the quality of this study.
\end{acknowledgements}

\bibliographystyle{aa}
\bibliography{CB}

\begin{thebibliography}{129}
\expandafter\ifx\csname natexlab\endcsname\relax\def\natexlab#1{#1}\fi

\bibitem[{{Adibekyan} {et~al.}(2021){Adibekyan}, {Santos}, {Demangeon},
  {Faria}, {Barros}, {Oshagh}, {Figueira}, {Delgado Mena}, {Sousa},
  {Israelian}, {Campante}, \& {Hakobyan}}]{2021A&A...649A.111A}
{Adibekyan}, V., {Santos}, N.~C., {Demangeon}, O.~D.~S., {et~al.} 2021, \aap,
  649, A111

\bibitem[{{Aganze} {et~al.}(2016){Aganze}, {Burgasser}, {Faherty}, {Choban},
  {Escala}, {Lopez}, {Jin}, {Tamiya}, {Tallis}, \&
  {Rockward}}]{2016AJ....151...46A}
{Aganze}, C., {Burgasser}, A.~J., {Faherty}, J.~K., {et~al.} 2016, \aj, 151, 46

\bibitem[{{Ammler-von Eiff} \& {Reiners}(2012)}]{2012A&A...542A.116A}
{Ammler-von Eiff}, M. \& {Reiners}, A. 2012, \aap, 542, A116

\bibitem[{{Astudillo-Defru} {et~al.}(2017){Astudillo-Defru}, {Delfosse},
  {Bonfils}, {Forveille}, {Lovis}, \& {Rameau}}]{2017A&A...600A..13A}
{Astudillo-Defru}, N., {Delfosse}, X., {Bonfils}, X., {et~al.} 2017, \aap, 600,
  A13

\bibitem[{{Bailey} {et~al.}(2009){Bailey}, {Butler}, {Tinney}, {Jones},
  {O'Toole}, {Carter}, \& {Marcy}}]{2009ApJ...690..743B}
{Bailey}, J., {Butler}, R.~P., {Tinney}, C.~G., {et~al.} 2009, \apj, 690, 743

\bibitem[{{Baranne} {et~al.}(1996){Baranne}, {Queloz}, {Mayor}, {Adrianzyk},
  {Knispel}, {Kohler}, {Lacroix}, {Meunier}, {Rimbaud}, \&
  {Vin}}]{1996A&AS..119..373B}
{Baranne}, A., {Queloz}, D., {Mayor}, M., {et~al.} 1996, \aaps, 119, 373

\bibitem[{{Bernstein} {et~al.}(2003){Bernstein}, {Shectman}, {Gunnels},
  {Mochnacki}, \& {Athey}}]{2003SPIE.4841.1694B}
{Bernstein}, R., {Shectman}, S.~A., {Gunnels}, S.~M., {Mochnacki}, S., \&
  {Athey}, A.~E. 2003, in Society of Photo-Optical Instrumentation Engineers
  (SPIE) Conference Series, Vol. 4841, Instrument Design and Performance for
  Optical/Infrared Ground-based Telescopes, ed. M.~{Iye} \& A.~F.~M.
  {Moorwood}, 1694--1704

\bibitem[{{Beuzit} {et~al.}(2019){Beuzit}, {Vigan}, {Mouillet}, {Dohlen},
  {Gratton}, {Boccaletti}, {Sauvage}, {Schmid}, {Langlois}, {Petit},
  {Baruffolo}, {Feldt}, {Milli}, {Wahhaj}, {Abe}, {Anselmi}, {Antichi},
  {Barette}, {Baudrand}, {Baudoz}, {Bazzon}, {Bernardi}, {Blanchard}, {Brast},
  {Bruno}, {Buey}, {Carbillet}, {Carle}, {Cascone}, {Chapron}, {Charton},
  {Chauvin}, {Claudi}, {Costille}, {De Caprio}, {de Boer}, {Delboulb{\'e}},
  {Desidera}, {Dominik}, {Downing}, {Dupuis}, {Fabron}, {Fantinel}, {Farisato},
  {Feautrier}, {Fedrigo}, {Fusco}, {Gigan}, {Ginski}, {Girard}, {Giro},
  {Gisler}, {Gluck}, {Gry}, {Henning}, {Hubin}, {Hugot}, {Incorvaia}, {Jaquet},
  {Kasper}, {Lagadec}, {Lagrange}, {Le Coroller}, {Le Mignant}, {Le Ruyet},
  {Lessio}, {Lizon}, {Llored}, {Lundin}, {Madec}, {Magnard}, {Marteaud},
  {Martinez}, {Maurel}, {M{\'e}nard}, {Mesa}, {M{\"o}ller-Nilsson}, {Moulin},
  {Moutou}, {Orign{\'e}}, {Parisot}, {Pavlov}, {Perret}, {Pragt}, {Puget},
  {Rabou}, {Ramos}, {Reess}, {Rigal}, {Rochat}, {Roelfsema}, {Rousset}, {Roux},
  {Saisse}, {Salasnich}, {Santambrogio}, {Scuderi}, {Segransan}, {Sevin},
  {Siebenmorgen}, {Soenke}, {Stadler}, {Suarez}, {Tiph{\`e}ne}, {Turatto},
  {Udry}, {Vakili}, {Waters}, {Weber}, {Wildi}, {Zins}, \&
  {Zurlo}}]{2019A&A...631A.155B}
{Beuzit}, J.~L., {Vigan}, A., {Mouillet}, D., {et~al.} 2019, \aap, 631, A155

\bibitem[{{Biller} {et~al.}(2022){Biller}, {Grandjean}, {Messina}, {Desidera},
  {Delorme}, {Lagrange}, {Hambsch}, {Mesa}, {Janson}, {Gratton}, {D'Orazi},
  {Langlois}, {Maire}, {Schlieder}, {Henning}, {Zurlo}, {Hagelberg},
  {Brown-Sevilla}, {Romero}, {Bonnefoy}, {Chauvin}, {Feldt}, {Meyer}, {Vigan},
  {Pavlov}, {Soenke}, {LeMignant}, \& {Roux}}]{Biller_2022}
{Biller}, B.~A., {Grandjean}, A., {Messina}, S., {et~al.} 2022, \aap, 658, A145

\bibitem[{{Bonavita} {et~al.}(2022{\natexlab{a}}){Bonavita}, {Fontanive},
  {Gratton}, {Mu{\v{z}}i{\'c}}, {Desidera}, {Mesa}, {Biller}, {Scholz},
  {Sozzetti}, \& {Squicciarini}}]{2022MNRAS.513.5588B}
{Bonavita}, M., {Fontanive}, C., {Gratton}, R., {et~al.} 2022{\natexlab{a}},
  \mnras, 513, 5588

\bibitem[{{Bonavita} {et~al.}(2022{\natexlab{b}}){Bonavita}, {Gratton},
  {Desidera}, {Squicciarini}, {D'Orazi}, {Zurlo}, {Biller}, {Chauvin},
  {Fontanive}, {Janson}, {Messina}, {Menard}, {Meyer}, {Vigan}, {Avenhaus},
  {Asensio Torres}, {Beuzit}, {Boccaletti}, {Bonnefoy}, {Brandner},
  {Cantalloube}, {Cheetham}, {Cudel}, {Daemgen}, {Delorme}, {Desgrange},
  {Dominik}, {Engler}, {Feautrier}, {Feldt}, {Galicher}, {Garufi}, {Gasparri},
  {Ginski}, {Girard}, {Grandjean}, {Hagelberg}, {Henning}, {Hunziker},
  {Kasper}, {Keppler}, {Lagadec}, {Lagrange}, {Langlois}, {Lannier}, {Lazzoni},
  {Le Coroller}, {Ligi}, {Lombart}, {Maire}, {Mazevet}, {Mesa}, {Mouillet},
  {Moutou}, {M{\"u}ller}, {Peretti}, {Perrot}, {Petrus}, {Potier}, {Ramos},
  {Rickman}, {Rouan}, {Salter}, {Samland}, {Schmidt}, {Sissa}, {Stolker},
  {Szul{\'a}gyi}, {Turatto}, {Udry}, \& {Wildi}}]{Bonavita_2022}
{Bonavita}, M., {Gratton}, R., {Desidera}, S., {et~al.} 2022{\natexlab{b}},
  \aap, 663, A144

\bibitem[{{Borgniet} {et~al.}(2017){Borgniet}, {Lagrange}, {Meunier}, \&
  {Galland}}]{Borgniet_2017}
{Borgniet}, S., {Lagrange}, A.~M., {Meunier}, N., \& {Galland}, F. 2017, \aap,
  599, A57

\bibitem[{{Boro Saikia} {et~al.}(2018){Boro Saikia}, {Marvin}, {Jeffers},
  {Reiners}, {Cameron}, {Marsden}, {Petit}, {Warnecke}, \&
  {Yadav}}]{2018A&A...616A.108B}
{Boro Saikia}, S., {Marvin}, C.~J., {Jeffers}, S.~V., {et~al.} 2018, \aap, 616,
  A108

\bibitem[{{Bouchy} {et~al.}(2013){Bouchy}, {D{\'\i}az}, {H{\'e}brard},
  {Arnold}, {Boisse}, {Delfosse}, {Perruchot}, \&
  {Santerne}}]{2013A&A...549A..49B}
{Bouchy}, F., {D{\'\i}az}, R.~F., {H{\'e}brard}, G., {et~al.} 2013, \aap, 549,
  A49

\bibitem[{{Bowler} {et~al.}(2021){Bowler}, {Cochran}, {Endl}, {Franson},
  {Brandt}, {Dupuy}, {MacQueen}, {Kratter}, {Mawet}, \&
  {Ruane}}]{2021AJ....161..106B}
{Bowler}, B.~P., {Cochran}, W.~D., {Endl}, M., {et~al.} 2021, \aj, 161, 106

\bibitem[{{Brandt} {et~al.}(2021{\natexlab{a}}){Brandt}, {Brandt}, {Dupuy},
  {Li}, \& {Michalik}}]{2021AJ....161..179B}
{Brandt}, G.~M., {Brandt}, T.~D., {Dupuy}, T.~J., {Li}, Y., \& {Michalik}, D.
  2021{\natexlab{a}}, \aj, 161, 179

\bibitem[{{Brandt} {et~al.}(2021{\natexlab{b}}){Brandt}, {Dupuy}, {Li}, {Chen},
  {Brandt}, {Wong}, {Currie}, {Bowler}, {Liu}, {Best}, \&
  {Phillips}}]{2021AJ....162..301B}
{Brandt}, G.~M., {Dupuy}, T.~J., {Li}, Y., {et~al.} 2021{\natexlab{b}}, \aj,
  162, 301

\bibitem[{Brandt {et~al.}(2021)Brandt, Michalik, Brandt, Li, Dupuy, \&
  Zeng}]{Brandt_2021_htof}
Brandt, G.~M., Michalik, D., Brandt, T.~D., {et~al.} 2021, The Astronomical
  Journal, 162, 230

\bibitem[{{Brandt}(2021)}]{2021ApJS..254...42B}
{Brandt}, T.~D. 2021, \apjs, 254, 42

\bibitem[{{Brandt} {et~al.}(2019){Brandt}, {Dupuy}, \&
  {Bowler}}]{2019AJ....158..140B}
{Brandt}, T.~D., {Dupuy}, T.~J., \& {Bowler}, B.~P. 2019, \aj, 158, 140

\bibitem[{{Brandt} {et~al.}(2021){Brandt}, {Dupuy}, {Li}, {Brandt}, {Zeng},
  {Michalik}, {Bardalez Gagliuffi}, \& {Raposo-Pulido}}]{2021AJ....162..186B}
{Brandt}, T.~D., {Dupuy}, T.~J., {Li}, Y., {et~al.} 2021, \aj, 162, 186

\bibitem[{Brandt {et~al.}(2021)Brandt, Dupuy, Li, Brandt, Zeng, Michalik,
  Gagliuffi, \& Raposo-Pulido}]{Brandt_2021_orvara}
Brandt, T.~D., Dupuy, T.~J., Li, Y., {et~al.} 2021, The Astronomical Journal,
  162, 186

\bibitem[{{Bryan} {et~al.}(2016){Bryan}, {Knutson}, {Howard}, {Ngo}, {Batygin},
  {Crepp}, {Fulton}, {Hinkley}, {Isaacson}, {Johnson}, {Marcy}, \&
  {Wright}}]{2016ApJ...821...89B}
{Bryan}, M.~L., {Knutson}, H.~A., {Howard}, A.~W., {et~al.} 2016, \apj, 821, 89

\bibitem[{{Butler} {et~al.}(2017){Butler}, {Vogt}, {Laughlin}, {Burt},
  {Rivera}, {Tuomi}, {Teske}, {Arriagada}, {Diaz}, {Holden}, \&
  {Keiser}}]{2017AJ....153..208B}
{Butler}, R.~P., {Vogt}, S.~S., {Laughlin}, G., {et~al.} 2017, \aj, 153, 208

\bibitem[{{Butler} {et~al.}(2006){Butler}, {Wright}, {Marcy}, {Fischer},
  {Vogt}, {Tinney}, {Jones}, {Carter}, {Johnson}, {McCarthy}, \&
  {Penny}}]{2006ApJ...646..505B}
{Butler}, R.~P., {Wright}, J.~T., {Marcy}, G.~W., {et~al.} 2006, \apj, 646, 505

\bibitem[{{Costa Silva} {et~al.}(2020){Costa Silva}, {Delgado Mena}, \&
  {Tsantaki}}]{2020A&A...634A.136C}
{Costa Silva}, A.~R., {Delgado Mena}, E., \& {Tsantaki}, M. 2020, \aap, 634,
  A136

\bibitem[{{Crane} {et~al.}(2010){Crane}, {Shectman}, {Butler}, {Thompson},
  {Birk}, {Jones}, \& {Burley}}]{2010SPIE.7735E..53C}
{Crane}, J.~D., {Shectman}, S.~A., {Butler}, R.~P., {et~al.} 2010, in Society
  of Photo-Optical Instrumentation Engineers (SPIE) Conference Series, Vol.
  7735, Ground-based and Airborne Instrumentation for Astronomy III, ed. I.~S.
  {McLean}, S.~K. {Ramsay}, \& H.~{Takami}, 773553

\bibitem[{{Dalba} {et~al.}(2021){Dalba}, {Kane}, {Howell}, {Horch}, {Li},
  {Hirsch}, {Burt}, {Brandt}, {Mo{\v{c}}nik}, {Henry}, {Everett}, {Rosenthal},
  \& {Howard}}]{2021AJ....161..123D}
{Dalba}, P.~A., {Kane}, S.~R., {Howell}, S.~B., {et~al.} 2021, \aj, 161, 123

\bibitem[{{Damasso} {et~al.}(2020){Damasso}, {Sozzetti}, {Lovis}, {Barros},
  {Sousa}, {Demangeon}, {Faria}, {Lillo-Box}, {Cristiani}, {Pepe}, {Rebolo},
  {Santos}, {Zapatero Osorio}, {Gonz{\'a}lez Hern{\'a}ndez}, {Amate},
  {Pasquini}, {Zerbi}, {Adibekyan}, {Abreu}, {Affolter}, {Alibert}, {Aliverti},
  {Allart}, {Allende Prieto}, {{\'A}lvarez}, {Alves}, {Avila}, {Baldini},
  {Bandy}, {Benz}, {Bianco}, {Borsa}, {Bossini}, {Bourrier}, {Bouchy}, {Broeg},
  {Cabral}, {Calderone}, {Cirami}, {Coelho}, {Conconi}, {Coretti}, {Cumani},
  {Cupani}, {D'Odorico}, {Deiries}, {Dekker}, {Delabre}, {Di Marcantonio},
  {Dumusque}, {Ehrenreich}, {Figueira}, {Fragoso}, {Genolet}, {Genoni},
  {G{\'e}nova Santos}, {Hughes}, {Iwert}, {Kerber}, {Knudstrup}, {Landoni},
  {Lavie}, {Lizon}, {Lo Curto}, {Maire}, {Martins}, {M{\'e}gevand}, {Mehner},
  {Micela}, {Modigliani}, {Molaro}, {Monteiro}, {Monteiro}, {Moschetti},
  {Mueller}, {Murphy}, {Nunes}, {Oggioni}, {Oliveira}, {Oshagh}, {Pall{\'e}},
  {Pariani}, {Poretti}, {Rasilla}, {Rebord{\~a}o}, {Redaelli}, {Riva}, {Santana
  Tschudi}, {Santin}, {Santos}, {S{\'e}gransan}, {Schmidt}, {Segovia},
  {Sosnowska}, {Span{\`o}}, {Su{\'a}rez Mascare{\~n}o}, {Tabernero}, {Tenegi},
  {Udry}, \& {Zanutta}}]{2020A&A...642A..31D}
{Damasso}, M., {Sozzetti}, A., {Lovis}, C., {et~al.} 2020, \aap, 642, A31

\bibitem[{{Delgado Mena} {et~al.}(2019){Delgado Mena}, {Moya}, {Adibekyan},
  {Tsantaki}, {Gonz{\'a}lez Hern{\'a}ndez}, {Israelian}, {Davies}, {Chaplin},
  {Sousa}, {Ferreira}, \& {Santos}}]{2019A&A...624A..78D}
{Delgado Mena}, E., {Moya}, A., {Adibekyan}, V., {et~al.} 2019, \aap, 624, A78

\bibitem[{{Delorme} {et~al.}(2017){Delorme}, {Meunier}, {Albert}, {Lagadec},
  {Le Coroller}, {Galicher}, {Mouillet}, {Boccaletti}, {Mesa}, {Meunier},
  {Beuzit}, {Lagrange}, {Chauvin}, {Sapone}, {Langlois}, {Maire},
  {Montarg{\`e}s}, {Gratton}, {Vigan}, \& {Surace}}]{delorme17sphere}
{Delorme}, P., {Meunier}, N., {Albert}, D., {et~al.} 2017, in SF2A-2017:
  Proceedings of the Annual meeting of the French Society of Astronomy and
  Astrophysics, ed. C.~{Reyl{\'e}}, P.~{Di Matteo}, F.~{Herpin}, E.~{Lagadec},
  A.~{Lan{\c{c}}on}, Z.~{Meliani}, \& F.~{Royer}, Di

\bibitem[{{Diego} {et~al.}(1990){Diego}, {Charalambous}, {Fish}, \&
  {Walker}}]{1990SPIE.1235..562D}
{Diego}, F., {Charalambous}, A., {Fish}, A.~C., \& {Walker}, D.~D. 1990, in
  Society of Photo-Optical Instrumentation Engineers (SPIE) Conference Series,
  Vol. 1235, Instrumentation in Astronomy VII, ed. D.~L. {Crawford}, 562--576

\bibitem[{{dos Santos} {et~al.}(2017){dos Santos}, {Mel{\'e}ndez}, {Bedell},
  {Bean}, {Spina}, {Alves-Brito}, {Dreizler}, {Ram{\'\i}rez}, \&
  {Asplund}}]{2017MNRAS.472.3425D}
{dos Santos}, L.~A., {Mel{\'e}ndez}, J., {Bedell}, M., {et~al.} 2017, \mnras,
  472, 3425

\bibitem[{{Eggenberger} {et~al.}(2007){Eggenberger}, {Udry}, {Chauvin},
  {Beuzit}, {Lagrange}, {S{\'e}gransan}, \& {Mayor}}]{2007A&A...474..273E}
{Eggenberger}, A., {Udry}, S., {Chauvin}, G., {et~al.} 2007, \aap, 474, 273

\bibitem[{{Ehrenreich} {et~al.}(2010){Ehrenreich}, {Lagrange}, {Montagnier},
  {Chauvin}, {Galland}, {Beuzit}, \& {Rameau}}]{Ehrenreich_2010}
{Ehrenreich}, D., {Lagrange}, A.~M., {Montagnier}, G., {et~al.} 2010, \aap,
  523, A73

\bibitem[{{Emsenhuber} {et~al.}(2021){Emsenhuber}, {Mordasini}, {Burn},
  {Alibert}, {Benz}, \& {Asphaug}}]{2021A&A...656A..69E}
{Emsenhuber}, A., {Mordasini}, C., {Burn}, R., {et~al.} 2021, \aap, 656, A69

\bibitem[{{Enard}(1982)}]{1982SPIE..331..232E}
{Enard}, D. 1982, in Society of Photo-Optical Instrumentation Engineers (SPIE)
  Conference Series, Vol. 331, Instrumentation in Astronomy IV, 232--242

\bibitem[{{Feng} {et~al.}(2021){Feng}, {Butler}, {Jones}, {Phillips}, {Vogt},
  {Oppenheimer}, {Holden}, {Burt}, \& {Boss}}]{2021MNRAS.507.2856F}
{Feng}, F., {Butler}, R.~P., {Jones}, H. R.~A., {et~al.} 2021, \mnras, 507,
  2856

\bibitem[{{Feng} {et~al.}(2022){Feng}, {Butler}, {Vogt}, {Clement}, {Tinney},
  {Cui}, {Aizawa}, {Jones}, {Bailey}, {Burt}, {Carter}, {Crane}, {Dotti},
  {Holden}, {Ma}, {Ogihara}, {Oppenheimer}, {O'Toole}, {Shectman},
  {Wittenmyer}, {Wang}, {Wright}, \& {Xuan}}]{2022ApJS..262...21F}
{Feng}, F., {Butler}, R.~P., {Vogt}, S.~S., {et~al.} 2022, \apjs, 262, 21

\bibitem[{{Fischer} {et~al.}(2014){Fischer}, {Marcy}, \&
  {Spronck}}]{2014ApJS..210....5F}
{Fischer}, D.~A., {Marcy}, G.~W., \& {Spronck}, J. F.~P. 2014, \apjs, 210, 5

\bibitem[{Foreman-Mackey {et~al.}(2013)Foreman-Mackey, Hogg, Lang, \&
  Goodman}]{Foreman_Mackey_2013}
Foreman-Mackey, D., Hogg, D.~W., Lang, D., \& Goodman, J. 2013, Publications of
  the Astronomical Society of the Pacific, 125, 306

\bibitem[{{Franson} {et~al.}(2023){Franson}, {Bowler}, {Bonavita}, {Brandt},
  {Chen}, {Samland}, {Zhang}, {Lueber}, {Heng}, {Kitzmann}, {Wolf}, {Jones},
  {Tran}, {Bardalez Gagliuffi}, {Biller}, {Chilcote}, {Crepp}, {Dupuy},
  {Faherty}, {Fontanive}, {Groff}, {Gratton}, {Guyon}, {Jensen-Clem},
  {Jovanovic}, {Kasdin}, {Lozi}, {Magnier}, {Mu{\v{z}}i{\'c}}, {Sanghi}, \&
  {Theissen}}]{2023AJ....165...39F}
{Franson}, K., {Bowler}, B.~P., {Bonavita}, M., {et~al.} 2023, \aj, 165, 39

\bibitem[{{Gaia Collaboration}(2020)}]{2020yCat.1350....0G}
{Gaia Collaboration}. 2020, VizieR Online Data Catalog, I/350

\bibitem[{{Gaia Collaboration} {et~al.}(2021){Gaia Collaboration}, {Brown},
  {Vallenari}, {Prusti}, {de Bruijne}, {Babusiaux}, {Biermann}, {Creevey},
  {Evans}, {Eyer}, {Hutton}, {Jansen}, {Jordi}, {Klioner}, {Lammers},
  {Lindegren}, {Luri}, {Mignard}, {Panem}, {Pourbaix}, {Randich}, {Sartoretti},
  {Soubiran}, {Walton}, {Arenou}, {Bailer-Jones}, {Bastian}, {Cropper},
  {Drimmel}, {Katz}, {Lattanzi}, {van Leeuwen}, {Bakker}, {Cacciari},
  {Casta{\~n}eda}, {De Angeli}, {Ducourant}, {Fabricius}, {Fouesneau},
  {Fr{\'e}mat}, {Guerra}, {Guerrier}, {Guiraud}, {Jean-Antoine Piccolo},
  {Masana}, {Messineo}, {Mowlavi}, {Nicolas}, {Nienartowicz}, {Pailler},
  {Panuzzo}, {Riclet}, {Roux}, {Seabroke}, {Sordo}, {Tanga}, {Th{\'e}venin},
  {Gracia-Abril}, {Portell}, {Teyssier}, {Altmann}, {Andrae}, {Bellas-Velidis},
  {Benson}, {Berthier}, {Blomme}, {Brugaletta}, {Burgess}, {Busso}, {Carry},
  {Cellino}, {Cheek}, {Clementini}, {Damerdji}, {Davidson}, {Delchambre},
  {Dell'Oro}, {Fern{\'a}ndez-Hern{\'a}ndez}, {Galluccio}, {Garc{\'\i}a-Lario},
  {Garcia-Reinaldos}, {Gonz{\'a}lez-N{\'u}{\~n}ez}, {Gosset}, {Haigron},
  {Halbwachs}, {Hambly}, {Harrison}, {Hatzidimitriou}, {Heiter},
  {Hern{\'a}ndez}, {Hestroffer}, {Hodgkin}, {Holl}, {Jan{\ss}en}, {Jevardat de
  Fombelle}, {Jordan}, {Krone-Martins}, {Lanzafame}, {L{\"o}ffler}, {Lorca},
  {Manteiga}, {Marchal}, {Marrese}, {Moitinho}, {Mora}, {Muinonen}, {Osborne},
  {Pancino}, {Pauwels}, {Petit}, {Recio-Blanco}, {Richards}, {Riello},
  {Rimoldini}, {Robin}, {Roegiers}, {Rybizki}, {Sarro}, {Siopis}, {Smith},
  {Sozzetti}, {Ulla}, {Utrilla}, {van Leeuwen}, {van Reeven}, {Abbas}, {Abreu
  Aramburu}, {Accart}, {Aerts}, {Aguado}, {Ajaj}, {Altavilla}, {{\'A}lvarez},
  {{\'A}lvarez Cid-Fuentes}, {Alves}, {Anderson}, {Anglada Varela}, {Antoja},
  {Audard}, {Baines}, {Baker}, {Balaguer-N{\'u}{\~n}ez}, {Balbinot}, {Balog},
  {Barache}, {Barbato}, {Barros}, {Barstow}, {Bartolom{\'e}}, {Bassilana},
  {Bauchet}, {Baudesson-Stella}, {Becciani}, {Bellazzini}, {Bernet}, {Bertone},
  {Bianchi}, {Blanco-Cuaresma}, {Boch}, {Bombrun}, {Bossini}, {Bouquillon},
  {Bragaglia}, {Bramante}, {Breedt}, {Bressan}, {Brouillet}, {Bucciarelli},
  {Burlacu}, {Busonero}, {Butkevich}, {Buzzi}, {Caffau}, {Cancelliere},
  {C{\'a}novas}, {Cantat-Gaudin}, {Carballo}, {Carlucci}, {Carnerero},
  {Carrasco}, {Casamiquela}, {Castellani}, {Castro-Ginard}, {Castro Sampol},
  {Chaoul}, {Charlot}, {Chemin}, {Chiavassa}, {Cioni}, {Comoretto}, {Cooper},
  {Cornez}, {Cowell}, {Crifo}, {Crosta}, {Crowley}, {Dafonte}, {Dapergolas},
  {David}, {David}, {de Laverny}, {De Luise}, {De March}, {De Ridder}, {de
  Souza}, {de Teodoro}, {de Torres}, {del Peloso}, {del Pozo}, {Delbo},
  {Delgado}, {Delgado}, {Delisle}, {Di Matteo}, {Diakite}, {Diener},
  {Distefano}, {Dolding}, {Eappachen}, {Edvardsson}, {Enke}, {Esquej}, {Fabre},
  {Fabrizio}, {Faigler}, {Fedorets}, {Fernique}, {Fienga}, {Figueras},
  {Fouron}, {Fragkoudi}, {Fraile}, {Franke}, {Gai}, {Garabato},
  {Garcia-Gutierrez}, {Garc{\'\i}a-Torres}, {Garofalo}, {Gavras}, {Gerlach},
  {Geyer}, {Giacobbe}, {Gilmore}, {Girona}, {Giuffrida}, {Gomel}, {Gomez},
  {Gonzalez-Santamaria}, {Gonz{\'a}lez-Vidal}, {Granvik},
  {Guti{\'e}rrez-S{\'a}nchez}, {Guy}, {Hauser}, {Haywood}, {Helmi}, {Hidalgo},
  {Hilger}, {H{\l}adczuk}, {Hobbs}, {Holland}, {Huckle}, {Jasniewicz},
  {Jonker}, {Juaristi Campillo}, {Julbe}, {Karbevska}, {Kervella}, {Khanna},
  {Kochoska}, {Kontizas}, {Kordopatis}, {Korn}, {Kostrzewa-Rutkowska},
  {Kruszy{\'n}ska}, {Lambert}, {Lanza}, {Lasne}, {Le Campion}, {Le Fustec},
  {Lebreton}, {Lebzelter}, {Leccia}, {Leclerc}, {Lecoeur-Taibi}, {Liao},
  {Licata}, {Lindstr{\o}m}, {Lister}, {Livanou}, {Lobel}, {Madrero Pardo},
  {Managau}, {Mann}, {Marchant}, {Marconi}, {Marcos Santos}, {Marinoni},
  {Marocco}, {Marshall}, {Martin Polo}, {Mart{\'\i}n-Fleitas}, {Masip},
  {Massari}, {Mastrobuono-Battisti}, {Mazeh}, {McMillan}, {Messina},
  {Michalik}, {Millar}, {Mints}, {Molina}, {Molinaro}, {Moln{\'a}r},
  {Montegriffo}, {Mor}, {Morbidelli}, {Morel}, {Morris}, {Mulone}, {Munoz},
  {Muraveva}, {Murphy}, {Musella}, {Noval}, {Ord{\'e}novic}, {Orr{\`u}},
  {Osinde}, {Pagani}, {Pagano}, {Palaversa}, {Palicio}, {Panahi}, {Pawlak},
  {Pe{\~n}alosa Esteller}, {Penttil{\"a}}, {Piersimoni}, {Pineau}, {Plachy},
  {Plum}, {Poggio}, {Poretti}, {Poujoulet}, {Pr{\v{s}}a}, {Pulone}, {Racero},
  {Ragaini}, {Rainer}, {Raiteri}, {Rambaux}, {Ramos}, {Ramos-Lerate}, {Re
  Fiorentin}, {Regibo}, {Reyl{\'e}}, {Ripepi}, {Riva}, {Rixon}, {Robichon},
  {Robin}, {Roelens}, {Rohrbasser}, {Romero-G{\'o}mez}, {Rowell}, {Royer},
  {Rybicki}, {Sadowski}, {Sagrist{\`a} Sell{\'e}s}, {Sahlmann}, {Salgado},
  {Salguero}, {Samaras}, {Sanchez Gimenez}, {Sanna}, {Santove{\~n}a},
  {Sarasso}, {Schultheis}, {Sciacca}, {Segol}, {Segovia}, {S{\'e}gransan},
  {Semeux}, {Shahaf}, {Siddiqui}, {Siebert}, {Siltala}, {Slezak}, {Smart},
  {Solano}, {Solitro}, {Souami}, {Souchay}, {Spagna}, {Spoto}, {Steele},
  {Steidelm{\"u}ller}, {Stephenson}, {S{\"u}veges}, {Szabados}, {Szegedi-Elek},
  {Taris}, {Tauran}, {Taylor}, {Teixeira}, {Thuillot}, {Tonello}, {Torra},
  {Torra}, {Turon}, {Unger}, {Vaillant}, {van Dillen}, {Vanel}, {Vecchiato},
  {Viala}, {Vicente}, {Voutsinas}, {Weiler}, {Wevers}, {Wyrzykowski}, {Yoldas},
  {Yvard}, {Zhao}, {Zorec}, {Zucker}, {Zurbach}, \&
  {Zwitter}}]{2021A&A...649A...1G}
{Gaia Collaboration}, {Brown}, A.~G.~A., {Vallenari}, A., {et~al.} 2021, \aap,
  649, A1

\bibitem[{{Gaia Collaboration} {et~al.}(2016){Gaia Collaboration}, {Prusti},
  {de Bruijne}, {Brown}, {Vallenari}, {Babusiaux}, {Bailer-Jones}, {Bastian},
  {Biermann}, {Evans}, {Eyer}, {Jansen}, {Jordi}, {Klioner}, {Lammers},
  {Lindegren}, {Luri}, {Mignard}, {Milligan}, {Panem}, {Poinsignon},
  {Pourbaix}, {Randich}, {Sarri}, {Sartoretti}, {Siddiqui}, {Soubiran},
  {Valette}, {van Leeuwen}, {Walton}, {Aerts}, {Arenou}, {Cropper}, {Drimmel},
  {H{\o}g}, {Katz}, {Lattanzi}, {O'Mullane}, {Grebel}, {Holland}, {Huc},
  {Passot}, {Bramante}, {Cacciari}, {Casta{\~n}eda}, {Chaoul}, {Cheek}, {De
  Angeli}, {Fabricius}, {Guerra}, {Hern{\'a}ndez}, {Jean-Antoine-Piccolo},
  {Masana}, {Messineo}, {Mowlavi}, {Nienartowicz}, {Ord{\'o}{\~n}ez-Blanco},
  {Panuzzo}, {Portell}, {Richards}, {Riello}, {Seabroke}, {Tanga},
  {Th{\'e}venin}, {Torra}, {Els}, {Gracia-Abril}, {Comoretto},
  {Garcia-Reinaldos}, {Lock}, {Mercier}, {Altmann}, {Andrae}, {Astraatmadja},
  {Bellas-Velidis}, {Benson}, {Berthier}, {Blomme}, {Busso}, {Carry},
  {Cellino}, {Clementini}, {Cowell}, {Creevey}, {Cuypers}, {Davidson}, {De
  Ridder}, {de Torres}, {Delchambre}, {Dell'Oro}, {Ducourant}, {Fr{\'e}mat},
  {Garc{\'\i}a-Torres}, {Gosset}, {Halbwachs}, {Hambly}, {Harrison}, {Hauser},
  {Hestroffer}, {Hodgkin}, {Huckle}, {Hutton}, {Jasniewicz}, {Jordan},
  {Kontizas}, {Korn}, {Lanzafame}, {Manteiga}, {Moitinho}, {Muinonen},
  {Osinde}, {Pancino}, {Pauwels}, {Petit}, {Recio-Blanco}, {Robin}, {Sarro},
  {Siopis}, {Smith}, {Smith}, {Sozzetti}, {Thuillot}, {van Reeven}, {Viala},
  {Abbas}, {Abreu Aramburu}, {Accart}, {Aguado}, {Allan}, {Allasia},
  {Altavilla}, {{\'A}lvarez}, {Alves}, {Anderson}, {Andrei}, {Anglada Varela},
  {Antiche}, {Antoja}, {Ant{\'o}n}, {Arcay}, {Atzei}, {Ayache}, {Bach},
  {Baker}, {Balaguer-N{\'u}{\~n}ez}, {Barache}, {Barata}, {Barbier}, {Barblan},
  {Baroni}, {Barrado y Navascu{\'e}s}, {Barros}, {Barstow}, {Becciani},
  {Bellazzini}, {Bellei}, {Bello Garc{\'\i}a}, {Belokurov}, {Bendjoya},
  {Berihuete}, {Bianchi}, {Bienaym{\'e}}, {Billebaud}, {Blagorodnova},
  {Blanco-Cuaresma}, {Boch}, {Bombrun}, {Borrachero}, {Bouquillon}, {Bourda},
  {Bouy}, {Bragaglia}, {Breddels}, {Brouillet}, {Br{\"u}semeister},
  {Bucciarelli}, {Budnik}, {Burgess}, {Burgon}, {Burlacu}, {Busonero}, {Buzzi},
  {Caffau}, {Cambras}, {Campbell}, {Cancelliere}, {Cantat-Gaudin}, {Carlucci},
  {Carrasco}, {Castellani}, {Charlot}, {Charnas}, {Charvet}, {Chassat},
  {Chiavassa}, {Clotet}, {Cocozza}, {Collins}, {Collins}, {Costigan}, {Crifo},
  {Cross}, {Crosta}, {Crowley}, {Dafonte}, {Damerdji}, {Dapergolas}, {David},
  {David}, {De Cat}, {de Felice}, {de Laverny}, {De Luise}, {De March}, {de
  Martino}, {de Souza}, {Debosscher}, {del Pozo}, {Delbo}, {Delgado},
  {Delgado}, {di Marco}, {Di Matteo}, {Diakite}, {Distefano}, {Dolding}, {Dos
  Anjos}, {Drazinos}, {Dur{\'a}n}, {Dzigan}, {Ecale}, {Edvardsson}, {Enke},
  {Erdmann}, {Escolar}, {Espina}, {Evans}, {Eynard Bontemps}, {Fabre},
  {Fabrizio}, {Faigler}, {Falc{\~a}o}, {Farr{\`a}s Casas}, {Faye}, {Federici},
  {Fedorets}, {Fern{\'a}ndez-Hern{\'a}ndez}, {Fernique}, {Fienga}, {Figueras},
  {Filippi}, {Findeisen}, {Fonti}, {Fouesneau}, {Fraile}, {Fraser}, {Fuchs},
  {Furnell}, {Gai}, {Galleti}, {Galluccio}, {Garabato}, {Garc{\'\i}a-Sedano},
  {Gar{\'e}}, {Garofalo}, {Garralda}, {Gavras}, {Gerssen}, {Geyer}, {Gilmore},
  {Girona}, {Giuffrida}, {Gomes}, {Gonz{\'a}lez-Marcos},
  {Gonz{\'a}lez-N{\'u}{\~n}ez}, {Gonz{\'a}lez-Vidal}, {Granvik}, {Guerrier},
  {Guillout}, {Guiraud}, {G{\'u}rpide}, {Guti{\'e}rrez-S{\'a}nchez}, {Guy},
  {Haigron}, {Hatzidimitriou}, {Haywood}, {Heiter}, {Helmi}, {Hobbs},
  {Hofmann}, {Holl}, {Holland}, {Hunt}, {Hypki}, {Icardi}, {Irwin}, {Jevardat
  de Fombelle}, {Jofr{\'e}}, {Jonker}, {Jorissen}, {Julbe}, {Karampelas},
  {Kochoska}, {Kohley}, {Kolenberg}, {Kontizas}, {Koposov}, {Kordopatis},
  {Koubsky}, {Kowalczyk}, {Krone-Martins}, {Kudryashova}, {Kull}, {Bachchan},
  {Lacoste-Seris}, {Lanza}, {Lavigne}, {Le Poncin-Lafitte}, {Lebreton},
  {Lebzelter}, {Leccia}, {Leclerc}, {Lecoeur-Taibi}, {Lemaitre}, {Lenhardt},
  {Leroux}, {Liao}, {Licata}, {Lindstr{\o}m}, {Lister}, {Livanou}, {Lobel},
  {L{\"o}ffler}, {L{\'o}pez}, {Lopez-Lozano}, {Lorenz}, {Loureiro},
  {MacDonald}, {Magalh{\~a}es Fernandes}, {Managau}, {Mann}, {Mantelet},
  {Marchal}, {Marchant}, {Marconi}, {Marie}, {Marinoni}, {Marrese},
  {Marschalk{\'o}}, {Marshall}, {Mart{\'\i}n-Fleitas}, {Martino}, {Mary},
  {Matijevi{\v{c}}}, {Mazeh}, {McMillan}, {Messina}, {Mestre}, {Michalik},
  {Millar}, {Miranda}, {Molina}, {Molinaro}, {Molinaro}, {Moln{\'a}r},
  {Moniez}, {Montegriffo}, {Monteiro}, {Mor}, {Mora}, {Morbidelli}, {Morel},
  {Morgenthaler}, {Morley}, {Morris}, {Mulone}, {Muraveva}, {Musella},
  {Narbonne}, {Nelemans}, {Nicastro}, {Noval}, {Ord{\'e}novic},
  {Ordieres-Mer{\'e}}, {Osborne}, {Pagani}, {Pagano}, {Pailler}, {Palacin},
  {Palaversa}, {Parsons}, {Paulsen}, {Pecoraro}, {Pedrosa}, {Pentik{\"a}inen},
  {Pereira}, {Pichon}, {Piersimoni}, {Pineau}, {Plachy}, {Plum}, {Poujoulet},
  {Pr{\v{s}}a}, {Pulone}, {Ragaini}, {Rago}, {Rambaux}, {Ramos-Lerate},
  {Ranalli}, {Rauw}, {Read}, {Regibo}, {Renk}, {Reyl{\'e}}, {Ribeiro},
  {Rimoldini}, {Ripepi}, {Riva}, {Rixon}, {Roelens}, {Romero-G{\'o}mez},
  {Rowell}, {Royer}, {Rudolph}, {Ruiz-Dern}, {Sadowski}, {Sagrist{\`a}
  Sell{\'e}s}, {Sahlmann}, {Salgado}, {Salguero}, {Sarasso}, {Savietto},
  {Schnorhk}, {Schultheis}, {Sciacca}, {Segol}, {Segovia}, {Segransan},
  {Serpell}, {Shih}, {Smareglia}, {Smart}, {Smith}, {Solano}, {Solitro},
  {Sordo}, {Soria Nieto}, {Souchay}, {Spagna}, {Spoto}, {Stampa}, {Steele},
  {Steidelm{\"u}ller}, {Stephenson}, {Stoev}, {Suess}, {S{\"u}veges}, {Surdej},
  {Szabados}, {Szegedi-Elek}, {Tapiador}, {Taris}, {Tauran}, {Taylor},
  {Teixeira}, {Terrett}, {Tingley}, {Trager}, {Turon}, {Ulla}, {Utrilla},
  {Valentini}, {van Elteren}, {Van Hemelryck}, {van Leeuwen}, {Varadi},
  {Vecchiato}, {Veljanoski}, {Via}, {Vicente}, {Vogt}, {Voss}, {Votruba},
  {Voutsinas}, {Walmsley}, {Weiler}, {Weingrill}, {Werner}, {Wevers},
  {Whitehead}, {Wyrzykowski}, {Yoldas}, {{\v{Z}}erjal}, {Zucker}, {Zurbach},
  {Zwitter}, {Alecu}, {Allen}, {Allende Prieto}, {Amorim},
  {Anglada-Escud{\'e}}, {Arsenijevic}, {Azaz}, {Balm}, {Beck}, {Bernstein},
  {Bigot}, {Bijaoui}, {Blasco}, {Bonfigli}, {Bono}, {Boudreault}, {Bressan},
  {Brown}, {Brunet}, {Bunclark}, {Buonanno}, {Butkevich}, {Carret}, {Carrion},
  {Chemin}, {Ch{\'e}reau}, {Corcione}, {Darmigny}, {de Boer}, {de Teodoro}, {de
  Zeeuw}, {Delle Luche}, {Domingues}, {Dubath}, {Fodor}, {Fr{\'e}zouls},
  {Fries}, {Fustes}, {Fyfe}, {Gallardo}, {Gallegos}, {Gardiol}, {Gebran},
  {Gomboc}, {G{\'o}mez}, {Grux}, {Gueguen}, {Heyrovsky}, {Hoar}, {Iannicola},
  {Isasi Parache}, {Janotto}, {Joliet}, {Jonckheere}, {Keil}, {Kim},
  {Klagyivik}, {Klar}, {Knude}, {Kochukhov}, {Kolka}, {Kos}, {Kutka}, {Lainey},
  {LeBouquin}, {Liu}, {Loreggia}, {Makarov}, {Marseille}, {Martayan},
  {Martinez-Rubi}, {Massart}, {Meynadier}, {Mignot}, {Munari}, {Nguyen},
  {Nordlander}, {Ocvirk}, {O'Flaherty}, {Olias Sanz}, {Ortiz}, {Osorio},
  {Oszkiewicz}, {Ouzounis}, {Palmer}, {Park}, {Pasquato}, {Peltzer}, {Peralta},
  {P{\'e}turaud}, {Pieniluoma}, {Pigozzi}, {Poels}, {Prat}, {Prod'homme},
  {Raison}, {Rebordao}, {Risquez}, {Rocca-Volmerange}, {Rosen}, {Ruiz-Fuertes},
  {Russo}, {Sembay}, {Serraller Vizcaino}, {Short}, {Siebert}, {Silva},
  {Sinachopoulos}, {Slezak}, {Soffel}, {Sosnowska}, {Strai{\v{z}}ys}, {ter
  Linden}, {Terrell}, {Theil}, {Tiede}, {Troisi}, {Tsalmantza}, {Tur},
  {Vaccari}, {Vachier}, {Valles}, {Van Hamme}, {Veltz}, {Virtanen}, {Wallut},
  {Wichmann}, {Wilkinson}, {Ziaeepour}, \& {Zschocke}}]{2016A&A...595A...1G}
{Gaia Collaboration}, {Prusti}, T., {de Bruijne}, J.~H.~J., {et~al.} 2016,
  \aap, 595, A1

\bibitem[{Galicher {et~al.}(2018)Galicher, Boccaletti, Mesa, Delorme, Gratton,
  Langlois, Lagrange, Maire, Le~Coroller, Chauvin,
  {et~al.}}]{galicher2018specal}
Galicher, R., Boccaletti, A., Mesa, D., {et~al.} 2018, Astronomy \&
  Astrophysics, 615, A92

\bibitem[{{Galland} {et~al.}(2005){Galland}, {Lagrange}, {Udry}, {Chelli},
  {Pepe}, {Queloz}, {Beuzit}, \& {Mayor}}]{2005A&A...443..337G}
{Galland}, F., {Lagrange}, A.~M., {Udry}, S., {et~al.} 2005, \aap, 443, 337

\bibitem[{{Ghazaryan} {et~al.}(2019){Ghazaryan}, {Alecian}, \&
  {Hakobyan}}]{2019MNRAS.487.5922G}
{Ghazaryan}, S., {Alecian}, G., \& {Hakobyan}, A.~A. 2019, \mnras, 487, 5922

\bibitem[{{Gomes da Silva} {et~al.}(2021){Gomes da Silva}, {Santos},
  {Adibekyan}, {Sousa}, {Campante}, {Figueira}, {Bossini}, {Delgado-Mena},
  {Monteiro}, {de Laverny}, {Recio-Blanco}, \& {Lovis}}]{2021A&A...646A..77G}
{Gomes da Silva}, J., {Santos}, N.~C., {Adibekyan}, V., {et~al.} 2021, \aap,
  646, A77

\bibitem[{{Gondoin}(2020)}]{2020A&A...641A.110G}
{Gondoin}, P. 2020, \aap, 641, A110

\bibitem[{{Gorrini} {et~al.}(2022){Gorrini}, {Astudillo-Defru}, {Dreizler},
  {Damasso}, {D{\'\i}az}, {Bonfils}, {Jeffers}, {Barnes}, {Del Sordo},
  {Almenara}, {Artigau}, {Bouchy}, {Charbonneau}, {Delfosse}, {Doyon},
  {Figueira}, {Forveille}, {Haswell}, {L{\'o}pez-Gonz{\'a}lez}, {Melo},
  {Mennickent}, {Gaisn{\'e}}, {Morales Morales}, {Murgas}, {Pepe},
  {Rodr{\'\i}guez}, {Santos}, {Tal-Or}, {Tsapras}, \&
  {Udry}}]{2022A&A...664A..64G}
{Gorrini}, P., {Astudillo-Defru}, N., {Dreizler}, S., {et~al.} 2022, \aap, 664,
  A64

\bibitem[{{Grandjean} {et~al.}(2019){Grandjean}, {Lagrange}, {Beust}, {Rodet},
  {Milli}, {Rubini}, {Babusiaux}, {Meunier}, {Delorme}, {Aigrain}, {Zicher},
  {Bonnefoy}, {Biller}, {Baudino}, {Bonavita}, {Boccaletti}, {Cheetham},
  {Girard}, {Hagelberg}, {Janson}, {Lannier}, {Lazzoni}, {Ligi}, {Maire},
  {Mesa}, {Perrot}, {Rouan}, \& {Zurlo}}]{2019A&A...629C...1G}
{Grandjean}, A., {Lagrange}, A.~M., {Beust}, H., {et~al.} 2019, \aap, 629, C1

\bibitem[{{Grieves} {et~al.}(2018){Grieves}, {Ge}, {Thomas}, {Willis}, {Ma},
  {Lorenzo-Oliveira}, {Queiroz}, {Ghezzi}, {Chiappini}, {Anders},
  {Dutra-Ferreira}, {Porto de Mello}, {Santiago}, {da Costa}, {Ogando}, {del
  Peloso}, {Tan}, {Schneider}, {Pepper}, {Stassun}, {Zhao}, {Bizyaev}, \&
  {Pan}}]{2018MNRAS.481.3244G}
{Grieves}, N., {Ge}, J., {Thomas}, N., {et~al.} 2018, \mnras, 481, 3244

\bibitem[{{Hartmann} \& {Hatzes}(2015)}]{Hartmann_2015}
{Hartmann}, M. \& {Hatzes}, A.~P. 2015, \aap, 582, A84

\bibitem[{{H{\'e}brard} {et~al.}(2016){H{\'e}brard}, {Arnold}, {Forveille},
  {Correia}, {Laskar}, {Bonfils}, {Boisse}, {D{\'\i}az}, {Hagelberg},
  {Sahlmann}, {Santos}, {Astudillo-Defru}, {Borgniet}, {Bouchy}, {Bourrier},
  {Courcol}, {Delfosse}, {Deleuil}, {Demangeon}, {Ehrenreich}, {Gregorio},
  {Jovanovic}, {Labrevoir}, {Lagrange}, {Lovis}, {Lozi}, {Moutou},
  {Montagnier}, {Pepe}, {Rey}, {Santerne}, {S{\'e}gransan}, {Udry},
  {Vanhuysse}, {Vigan}, \& {Wilson}}]{2016A&A...588A.145H}
{H{\'e}brard}, G., {Arnold}, L., {Forveille}, T., {et~al.} 2016, \aap, 588,
  A145

\bibitem[{{Hojjatpanah} {et~al.}(2019){Hojjatpanah}, {Figueira}, {Santos},
  {Adibekyan}, {Sousa}, {Delgado-Mena}, {Alibert}, {Cristiani}, {Gonz{\'a}lez
  Hern{\'a}ndez}, {Lanza}, {Di Marcantonio}, {Martins}, {Micela}, {Molaro},
  {Neves}, {Oshagh}, {Pepe}, {Poretti}, {Rojas-Ayala}, {Rebolo}, {Su{\'a}rez
  Mascare{\~n}o}, \& {Zapatero Osorio}}]{2019A&A...629A..80H}
{Hojjatpanah}, S., {Figueira}, P., {Santos}, N.~C., {et~al.} 2019, \aap, 629,
  A80

\bibitem[{{Hojjatpanah} {et~al.}(2020){Hojjatpanah}, {Oshagh}, {Figueira},
  {Santos}, {Amazo-G{\'o}mez}, {Sousa}, {Adibekyan}, {Akinsanmi}, {Demangeon},
  {Faria}, {Gomes da Silva}, \& {Meunier}}]{2020A&A...639A..35H}
{Hojjatpanah}, S., {Oshagh}, M., {Figueira}, P., {et~al.} 2020, \aap, 639, A35

\bibitem[{{Jenkins} {et~al.}(2017){Jenkins}, {Jones}, {Tuomi}, {D{\'\i}az},
  {Cordero}, {Aguayo}, {Pantoja}, {Arriagada}, {Mahu}, {Brahm}, {Rojo}, {Soto},
  {Ivanyuk}, {Becerra Yoma}, {Day-Jones}, {Ruiz}, {Pavlenko}, {Barnes},
  {Murgas}, {Pinfield}, {Jones}, {L{\'o}pez-Morales}, {Shectman}, {Butler}, \&
  {Minniti}}]{2017MNRAS.466..443J}
{Jenkins}, J.~S., {Jones}, H.~R.~A., {Tuomi}, M., {et~al.} 2017, \mnras, 466,
  443

\bibitem[{{Jones} {et~al.}(2002){Jones}, {Paul Butler}, {Marcy}, {Tinney},
  {Penny}, {McCarthy}, \& {Carter}}]{2002MNRAS.337.1170J}
{Jones}, H. R.~A., {Paul Butler}, R., {Marcy}, G.~W., {et~al.} 2002, \mnras,
  337, 1170

\bibitem[{{Kane} {et~al.}(2019){Kane}, {Dalba}, {Li}, {Horch}, {Hirsch},
  {Horner}, {Wittenmyer}, {Howell}, {Everett}, {Butler}, {Tinney}, {Carter},
  {Wright}, {Jones}, {Bailey}, \& {O'Toole}}]{2019AJ....157..252K}
{Kane}, S.~R., {Dalba}, P.~A., {Li}, Z., {et~al.} 2019, \aj, 157, 252

\bibitem[{{Kervella} {et~al.}(2019){Kervella}, {Arenou}, {Mignard}, \&
  {Th{\'e}venin}}]{2019A&A...623A..72K}
{Kervella}, P., {Arenou}, F., {Mignard}, F., \& {Th{\'e}venin}, F. 2019, \aap,
  623, A72

\bibitem[{{Kervella} {et~al.}(2022){Kervella}, {Arenou}, \&
  {Th{\'e}venin}}]{2022A&A...657A...7K}
{Kervella}, P., {Arenou}, F., \& {Th{\'e}venin}, F. 2022, \aap, 657, A7

\bibitem[{{Kiefer} {et~al.}(2021){Kiefer}, {H{\'e}brard}, {Lecavelier des
  Etangs}, {Martioli}, {Dalal}, \& {Vidal-Madjar}}]{2021A&A...645A...7K}
{Kiefer}, F., {H{\'e}brard}, G., {Lecavelier des Etangs}, A., {et~al.} 2021,
  \aap, 645, A7

\bibitem[{{Kiefer} {et~al.}(2019){Kiefer}, {H{\'e}brard}, {Sahlmann}, {Sousa},
  {Forveille}, {Santos}, {Mayor}, {Deleuil}, {Wilson}, {Dalal}, {D{\'\i}az},
  {Henry}, {Hagelberg}, {Hobson}, {Demangeon}, {Bourrier}, {Delfosse},
  {Arnold}, {Astudillo-Defru}, {Beuzit}, {Boisse}, {Bonfils}, {Borgniet},
  {Bouchy}, {Courcol}, {Ehrenreich}, {Hara}, {Lagrange}, {Lovis}, {Montagnier},
  {Moutou}, {Pepe}, {Perrier}, {Rey}, {Santerne}, {S{\'e}gransan}, {Udry}, \&
  {Vidal-Madjar}}]{2019A&A...631A.125K}
{Kiefer}, F., {H{\'e}brard}, G., {Sahlmann}, J., {et~al.} 2019, \aap, 631, A125

\bibitem[{{Kuzuhara} {et~al.}(2022){Kuzuhara}, {Currie}, {Takarada}, {Brandt},
  {Sato}, {Uyama}, {Janson}, {Chilcote}, {Tobin}, {Lawson}, {Hori}, {Guyon},
  {Groff}, {Lozi}, {Vievard}, {Sahoo}, {Deo}, {Jovanovic}, {Ahn}, {Martinache},
  {Skaf}, {Akiyama}, {Norris}, {Bonnefoy}, {He{\l}miniak}, {Kudo}, {McElwain},
  {Samland}, {Wagner}, {Wisniewski}, {Knapp}, {Kwon}, {Nishikawa}, {Serabyn},
  {Hayashi}, \& {Tamura}}]{2022ApJ...934L..18K}
{Kuzuhara}, M., {Currie}, T., {Takarada}, T., {et~al.} 2022, \apjl, 934, L18

\bibitem[{{Lagrange} {et~al.}(2019){Lagrange}, {Meunier}, {Rubini}, {Keppler},
  {Galland}, {Chapellier}, {Michel}, {Balona}, {Beust}, {Guillot}, {Grandjean},
  {Borgniet}, {M{\'e}karnia}, {Wilson}, {Kiefer}, {Bonnefoy}, {Lillo-Box},
  {Pantoja}, {Jones}, {Iglesias}, {Rodet}, {Diaz}, {Zapata}, {Abe}, \&
  {Schmider}}]{2019NatAs...3.1135L}
{Lagrange}, A.~M., {Meunier}, N., {Rubini}, P., {et~al.} 2019, Nature
  Astronomy, 3, 1135

\bibitem[{{Lagrange} {et~al.}(2023){Lagrange}, {Philipot}, {Rubini}, {Meunier},
  {Kiefer}, {Kervella}, {Delorme}, \& {Beust}}]{2023arXiv230500047L}
{Lagrange}, A.~M., {Philipot}, F., {Rubini}, P., {et~al.} 2023, arXiv e-prints,
  arXiv:2305.00047

\bibitem[{{Lenzen} {et~al.}(2003){Lenzen}, {Hartung}, {Brandner}, {Finger},
  {Hubin}, {Lacombe}, {Lagrange}, {Lehnert}, {Moorwood}, \&
  {Mouillet}}]{2003SPIE.4841..944L}
{Lenzen}, R., {Hartung}, M., {Brandner}, W., {et~al.} 2003, in Society of
  Photo-Optical Instrumentation Engineers (SPIE) Conference Series, Vol. 4841,
  Instrument Design and Performance for Optical/Infrared Ground-based
  Telescopes, ed. M.~{Iye} \& A.~F.~M. {Moorwood}, 944--952

\bibitem[{{Li} {et~al.}(2021){Li}, {Brandt}, {Brandt}, {Dupuy}, {Michalik},
  {Jensen-Clem}, {Zeng}, {Faherty}, \& {Mitra}}]{2021AJ....162..266L}
{Li}, Y., {Brandt}, T.~D., {Brandt}, G.~M., {et~al.} 2021, \aj, 162, 266

\bibitem[{{Liebing} {et~al.}(2021){Liebing}, {Jeffers}, {Reiners}, \&
  {Zechmeister}}]{2021A&A...654A.168L}
{Liebing}, F., {Jeffers}, S.~V., {Reiners}, A., \& {Zechmeister}, M. 2021,
  \aap, 654, A168

\bibitem[{{Lindegren} {et~al.}(2021){Lindegren}, {Klioner}, {Hern{\'a}ndez},
  {Bombrun}, {Ramos-Lerate}, {Steidelm{\"u}ller}, {Bastian}, {Biermann}, {de
  Torres}, {Gerlach}, {Geyer}, {Hilger}, {Hobbs}, {Lammers}, {McMillan},
  {Stephenson}, {Casta{\~n}eda}, {Davidson}, {Fabricius}, {Gracia-Abril},
  {Portell}, {Rowell}, {Teyssier}, {Torra}, {Bartolom{\'e}}, {Clotet},
  {Garralda}, {Gonz{\'a}lez-Vidal}, {Torra}, {Abbas}, {Altmann}, {Anglada
  Varela}, {Balaguer-N{\'u}{\~n}ez}, {Balog}, {Barache}, {Becciani}, {Bernet},
  {Bertone}, {Bianchi}, {Bouquillon}, {Brown}, {Bucciarelli}, {Busonero},
  {Butkevich}, {Buzzi}, {Cancelliere}, {Carlucci}, {Charlot}, {Cioni},
  {Crosta}, {Crowley}, {del Peloso}, {del Pozo}, {Drimmel}, {Esquej}, {Fienga},
  {Fraile}, {Gai}, {Garcia-Reinaldos}, {Guerra}, {Hambly}, {Hauser},
  {Jan{\ss}en}, {Jordan}, {Kostrzewa-Rutkowska}, {Lattanzi}, {Liao}, {Licata},
  {Lister}, {L{\"o}ffler}, {Marchant}, {Masip}, {Mignard}, {Mints}, {Molina},
  {Mora}, {Morbidelli}, {Murphy}, {Pagani}, {Panuzzo}, {Pe{\~n}alosa Esteller},
  {Poggio}, {Re Fiorentin}, {Riva}, {Sagrist{\`a} Sell{\'e}s}, {Sanchez
  Gimenez}, {Sarasso}, {Sciacca}, {Siddiqui}, {Smart}, {Souami}, {Spagna},
  {Steele}, {Taris}, {Utrilla}, {van Reeven}, \& {Vecchiato}}]{Lindegren2021}
{Lindegren}, L., {Klioner}, S.~A., {Hern{\'a}ndez}, J., {et~al.} 2021, \aap,
  649, A2

\bibitem[{{Llorente de Andr{\'e}s} {et~al.}(2021){Llorente de Andr{\'e}s},
  {Chavero}, {de la Reza}, {Roca-F{\`a}brega}, \&
  {Cifuentes}}]{2021A&A...654A.137L}
{Llorente de Andr{\'e}s}, F., {Chavero}, C., {de la Reza}, R.,
  {Roca-F{\`a}brega}, S., \& {Cifuentes}, C. 2021, \aap, 654, A137

\bibitem[{{Lo Curto} {et~al.}(2010{\natexlab{a}}){Lo Curto}, {Mayor}, {Benz},
  {Bouchy}, {Lovis}, {Moutou}, {Naef}, {Pepe}, {Queloz}, {Santos}, {Segransan},
  \& {Udry}}]{2010A&A...512A..48L}
{Lo Curto}, G., {Mayor}, M., {Benz}, W., {et~al.} 2010{\natexlab{a}}, \aap,
  512, A48

\bibitem[{{Lo Curto} {et~al.}(2010{\natexlab{b}}){Lo Curto}, {Mayor}, {Benz},
  {Bouchy}, {Lovis}, {Moutou}, {Naef}, {Pepe}, {Queloz}, {Santos}, {Segransan},
  \& {Udry}}]{Locurto_2010}
{Lo Curto}, G., {Mayor}, M., {Benz}, W., {et~al.} 2010{\natexlab{b}}, \aap,
  512, A48

\bibitem[{{Lo Curto} {et~al.}(2015){Lo Curto}, {Pepe}, {Avila}, {Boffin},
  {Bovay}, {Chazelas}, {Coffinet}, {Fleury}, {Hughes}, {Lovis}, {Maire},
  {Manescau}, {Pasquini}, {Rihs}, {Sinclaire}, \& {Udry}}]{2015Msngr.162....9L}
{Lo Curto}, G., {Pepe}, F., {Avila}, G., {et~al.} 2015, The Messenger, 162, 9

\bibitem[{{Maldonado} {et~al.}(2020){Maldonado}, {Micela}, {Baratella},
  {D'Orazi}, {Affer}, {Biazzo}, {Lanza}, {Maggio}, {Gonz{\'a}lez
  Hern{\'a}ndez}, {Perger}, {Pinamonti}, {Scandariato}, {Sozzetti}, {Locci},
  {Di Maio}, {Bignamini}, {Claudi}, {Molinari}, {Rebolo}, {Ribas},
  {Toledo-Padr{\'o}n}, {Covino}, {Desidera}, {Herrero}, {Morales},
  {Su{\'a}rez-Mascare{\~n}o}, {Pagano}, {Petralia}, {Piotto}, \&
  {Poretti}}]{2020A&A...644A..68M}
{Maldonado}, J., {Micela}, G., {Baratella}, M., {et~al.} 2020, \aap, 644, A68

\bibitem[{{Mayor} {et~al.}(2011){Mayor}, {Marmier}, {Lovis}, {Udry},
  {S{\'e}gransan}, {Pepe}, {Benz}, {Bertaux}, {Bouchy}, {Dumusque}, {Lo Curto},
  {Mordasini}, {Queloz}, \& {Santos}}]{2011arXiv1109.2497M}
{Mayor}, M., {Marmier}, M., {Lovis}, C., {et~al.} 2011, arXiv e-prints,
  arXiv:1109.2497

\bibitem[{{Mayor} {et~al.}(2003){Mayor}, {Pepe}, {Queloz}, {Bouchy},
  {Rupprecht}, {Lo Curto}, {Avila}, {Benz}, {Bertaux}, {Bonfils}, {Dall},
  {Dekker}, {Delabre}, {Eckert}, {Fleury}, {Gilliotte}, {Gojak}, {Guzman},
  {Kohler}, {Lizon}, {Longinotti}, {Lovis}, {Megevand}, {Pasquini}, {Reyes},
  {Sivan}, {Sosnowska}, {Soto}, {Udry}, {van Kesteren}, {Weber}, \&
  {Weilenmann}}]{2003Msngr.114...20M}
{Mayor}, M., {Pepe}, F., {Queloz}, D., {et~al.} 2003, The Messenger, 114, 20

\bibitem[{{Mayor} \& {Queloz}(1995)}]{1995Natur.378..355M}
{Mayor}, M. \& {Queloz}, D. 1995, \nat, 378, 355

\bibitem[{{Mayor} {et~al.}(2004{\natexlab{a}}){Mayor}, {Udry}, {Naef}, {Pepe},
  {Queloz}, {Santos}, \& {Burnet}}]{Mayor_2004}
{Mayor}, M., {Udry}, S., {Naef}, D., {et~al.} 2004{\natexlab{a}}, \aap, 415,
  391

\bibitem[{{Mayor} {et~al.}(2004{\natexlab{b}}){Mayor}, {Udry}, {Naef}, {Pepe},
  {Queloz}, {Santos}, \& {Burnet}}]{2004A&A...415..391M}
{Mayor}, M., {Udry}, S., {Naef}, D., {et~al.} 2004{\natexlab{b}}, \aap, 415,
  391

\bibitem[{{Melo} {et~al.}(2007){Melo}, {Santos}, {Gieren}, {Pietrzynski},
  {Ruiz}, {Sousa}, {Bouchy}, {Lovis}, {Mayor}, {Pepe}, {Queloz}, {da Silva}, \&
  {Udry}}]{2007A&A...467..721M}
{Melo}, C., {Santos}, N.~C., {Gieren}, W., {et~al.} 2007, \aap, 467, 721

\bibitem[{{Meunier} \& {Lagrange}(2019)}]{2019A&A...628A.125M}
{Meunier}, N. \& {Lagrange}, A.~M. 2019, \aap, 628, A125

\bibitem[{{Meunier} {et~al.}(2019){Meunier}, {Lagrange}, {Boulet}, \&
  {Borgniet}}]{2019A&A...627A..56M}
{Meunier}, N., {Lagrange}, A.~M., {Boulet}, T., \& {Borgniet}, S. 2019, \aap,
  627, A56

\bibitem[{{Mordasini}(2018)}]{2018haex.bookE.143M}
{Mordasini}, C. 2018, in Handbook of Exoplanets, ed. H.~J. {Deeg} \& J.~A.
  {Belmonte}, 143

\bibitem[{{Moutou} {et~al.}(2011){Moutou}, {Mayor}, {Lo Curto},
  {S{\'e}gransan}, {Udry}, {Bouchy}, {Benz}, {Lovis}, {Naef}, {Pepe}, {Queloz},
  {Santos}, \& {Sousa}}]{2011A&A...527A..63M}
{Moutou}, C., {Mayor}, M., {Lo Curto}, G., {et~al.} 2011, \aap, 527, A63

\bibitem[{{Moutou} {et~al.}(2009){Moutou}, {Mayor}, {Lo Curto}, {Udry},
  {Bouchy}, {Benz}, {Lovis}, {Naef}, {Pepe}, {Queloz}, \&
  {Santos}}]{Moutou_2009}
{Moutou}, C., {Mayor}, M., {Lo Curto}, G., {et~al.} 2009, \aap, 496, 513

\bibitem[{{Naef} {et~al.}(2007){Naef}, {Mayor}, {Benz}, {Bouchy}, {Lo Curto},
  {Lovis}, {Moutou}, {Pepe}, {Queloz}, {Santos}, \&
  {Udry}}]{2007A&A...470..721N}
{Naef}, D., {Mayor}, M., {Benz}, W., {et~al.} 2007, \aap, 470, 721

\bibitem[{{Naef} {et~al.}(2010){Naef}, {Mayor}, {Lo Curto}, {Bouchy}, {Lovis},
  {Moutou}, {Benz}, {Pepe}, {Queloz}, {Santos}, {S{\'e}gransan}, {Udry},
  {Bonfils}, {Delfosse}, {Forveille}, {H{\'e}brard}, {Mordasini}, {Perrier},
  {Boisse}, \& {Sosnowska}}]{2010A&A...523A..15N}
{Naef}, D., {Mayor}, M., {Lo Curto}, G., {et~al.} 2010, \aap, 523, A15

\bibitem[{{Naef} {et~al.}(2001){Naef}, {Mayor}, {Pepe}, {Queloz}, {Santos},
  {Udry}, \& {Burnet}}]{2001A&A...375..205N}
{Naef}, D., {Mayor}, M., {Pepe}, F., {et~al.} 2001, \aap, 375, 205

\bibitem[{{Pavlov} {et~al.}(2008){Pavlov}, {M{\"o}ller-Nilsson}, {Feldt},
  {Henning}, {Beuzit}, \& {Mouillet}}]{Pavlolv_drh}
{Pavlov}, A., {M{\"o}ller-Nilsson}, O., {Feldt}, M., {et~al.} 2008, in Society
  of Photo-Optical Instrumentation Engineers (SPIE) Conference Series, Vol.
  7019, Advanced Software and Control for Astronomy II, ed. A.~{Bridger} \&
  N.~M. {Radziwill}, 701939

\bibitem[{{Perruchot} {et~al.}(2008){Perruchot}, {Kohler}, {Bouchy}, {Richaud},
  {Richaud}, {Moreaux}, {Merzougui}, {Sottile}, {Hill}, {Knispel}, {Regal},
  {Meunier}, {Ilovaisky}, {Le Coroller}, {Gillet}, {Schmitt}, {Pepe}, {Fleury},
  {Sosnowska}, {Vors}, {M{\'e}gevand}, {Blanc}, {Carol}, {Point}, {Laloge}, \&
  {Brunel}}]{2008SPIE.7014E..0JP}
{Perruchot}, S., {Kohler}, D., {Bouchy}, F., {et~al.} 2008, in Society of
  Photo-Optical Instrumentation Engineers (SPIE) Conference Series, Vol. 7014,
  Ground-based and Airborne Instrumentation for Astronomy II, ed. I.~S.
  {McLean} \& M.~M. {Casali}, 70140J

\bibitem[{{Perryman} {et~al.}(1997){Perryman}, {Lindegren}, {Kovalevsky},
  {Hoeg}, {Bastian}, {Bernacca}, {Cr{\'e}z{\'e}}, {Donati}, {Grenon},
  {Grewing}, {van Leeuwen}, {van der Marel}, {Mignard}, {Murray}, {Le Poole},
  {Schrijver}, {Turon}, {Arenou}, {Froeschl{\'e}}, \&
  {Petersen}}]{1997A&A...323L..49P}
{Perryman}, M.~A.~C., {Lindegren}, L., {Kovalevsky}, J., {et~al.} 1997, \aap,
  323, L49

\bibitem[{{Philipot} {et~al.}(2023){Philipot}, {Lagrange}, {Rubini}, {Kiefer},
  \& {Chomez}}]{2023A&A...670A..65P}
{Philipot}, F., {Lagrange}, A.~M., {Rubini}, P., {Kiefer}, F., \& {Chomez}, A.
  2023, \aap, 670, A65

\bibitem[{{Queloz} {et~al.}(1999){Queloz}, {Mayor}, {Weber}, {Blecha},
  {Burnet}, {Confino}, {Naef}, {Pepe}, {Santos}, \&
  {Udry}}]{1999astro.ph.10223Q}
{Queloz}, D., {Mayor}, M., {Weber}, L., {et~al.} 1999, arXiv e-prints, astro

\bibitem[{{Reiners} \& {Zechmeister}(2020)}]{2020ApJS..247...11R}
{Reiners}, A. \& {Zechmeister}, M. 2020, \apjs, 247, 11

\bibitem[{{Rice} \& {Brewer}(2020)}]{2020ApJ...898..119R}
{Rice}, M. \& {Brewer}, J.~M. 2020, \apj, 898, 119

\bibitem[{{Rigaut} {et~al.}(1998){Rigaut}, {Salmon}, {Arsenault}, {Thomas},
  {Lai}, {Rouan}, {V{\'e}ran}, {Gigan}, {Crampton}, {Fletcher}, {Stilburn},
  {Boyer}, \& {Jagourel}}]{1998PASP..110..152R}
{Rigaut}, F., {Salmon}, D., {Arsenault}, R., {et~al.} 1998, \pasp, 110, 152

\bibitem[{{Rosenthal} {et~al.}(2021){Rosenthal}, {Fulton}, {Hirsch},
  {Isaacson}, {Howard}, {Dedrick}, {Sherstyuk}, {Blunt}, {Petigura}, {Knutson},
  {Behmard}, {Chontos}, {Crepp}, {Crossfield}, {Dalba}, {Fischer}, {Henry},
  {Kane}, {Kosiarek}, {Marcy}, {Rubenzahl}, {Weiss}, \&
  {Wright}}]{2021ApJS..255....8R}
{Rosenthal}, L.~J., {Fulton}, B.~J., {Hirsch}, L.~A., {et~al.} 2021, \apjs,
  255, 8

\bibitem[{{Rousset} {et~al.}(2003){Rousset}, {Lacombe}, {Puget}, {Hubin},
  {Gendron}, {Fusco}, {Arsenault}, {Charton}, {Feautrier}, {Gigan}, {Kern},
  {Lagrange}, {Madec}, {Mouillet}, {Rabaud}, {Rabou}, {Stadler}, \&
  {Zins}}]{2003SPIE.4839..140R}
{Rousset}, G., {Lacombe}, F., {Puget}, P., {et~al.} 2003, in Society of
  Photo-Optical Instrumentation Engineers (SPIE) Conference Series, Vol. 4839,
  Adaptive Optical System Technologies II, ed. P.~L. {Wizinowich} \&
  D.~{Bonaccini}, 140--149

\bibitem[{{Sahlmann} {et~al.}(2011){Sahlmann}, {S{\'e}gransan}, {Queloz},
  {Udry}, {Santos}, {Marmier}, {Mayor}, {Naef}, {Pepe}, \&
  {Zucker}}]{2011A&A...525A..95S}
{Sahlmann}, J., {S{\'e}gransan}, D., {Queloz}, D., {et~al.} 2011, \aap, 525,
  A95

\bibitem[{{Santos} {et~al.}(2011){Santos}, {Mayor}, {Bonfils}, {Dumusque},
  {Bouchy}, {Figueira}, {Lovis}, {Melo}, {Pepe}, {Queloz}, {S{\'e}gransan},
  {Sousa}, \& {Udry}}]{Santos_2011}
{Santos}, N.~C., {Mayor}, M., {Bonfils}, X., {et~al.} 2011, \aap, 526, A112

\bibitem[{{Schneider} {et~al.}(2011){Schneider}, {Melis}, {Song}, \&
  {Zuckerman}}]{2011ApJ...743..109S}
{Schneider}, A., {Melis}, C., {Song}, I., \& {Zuckerman}, B. 2011, \apj, 743,
  109

\bibitem[{{Scott} {et~al.}(2018){Scott}, {Howell}, {Horch}, \&
  {Everett}}]{2018PASP..130e4502S}
{Scott}, N.~J., {Howell}, S.~B., {Horch}, E.~P., \& {Everett}, M.~E. 2018,
  \pasp, 130, 054502

\bibitem[{{S{\'e}gransan} {et~al.}(2011){S{\'e}gransan}, {Mayor}, {Udry},
  {Lovis}, {Benz}, {Bouchy}, {Lo Curto}, {Mordasini}, {Moutou}, {Naef}, {Pepe},
  {Queloz}, \& {Santos}}]{2011A&A...535A..54S}
{S{\'e}gransan}, D., {Mayor}, M., {Udry}, S., {et~al.} 2011, \aap, 535, A54

\bibitem[{{S{\'e}gransan} {et~al.}(2010){S{\'e}gransan}, {Udry}, {Mayor},
  {Naef}, {Pepe}, {Queloz}, {Santos}, {Demory}, {Figueira}, {Gillon},
  {Marmier}, {M{\'e}gevand}, {Sosnowska}, {Tamuz}, \&
  {Triaud}}]{2010A&A...511A..45S}
{S{\'e}gransan}, D., {Udry}, S., {Mayor}, M., {et~al.} 2010, \aap, 511, A45

\bibitem[{{Service} {et~al.}(2016){Service}, {Lu}, {Campbell}, {Sitarski},
  {Ghez}, \& {Anderson}}]{2016PASP..128i5004S}
{Service}, M., {Lu}, J.~R., {Campbell}, R., {et~al.} 2016, \pasp, 128, 095004

\bibitem[{{Sousa} {et~al.}(2018){Sousa}, {Adibekyan}, {Delgado-Mena}, {Santos},
  {Andreasen}, {Ferreira}, {Tsantaki}, {Barros}, {Demangeon}, {Israelian},
  {Faria}, {Figueira}, {Mortier}, {Brand{\~a}o}, {Montalto}, {Rojas-Ayala}, \&
  {Santerne}}]{2018A&A...620A..58S}
{Sousa}, S.~G., {Adibekyan}, V., {Delgado-Mena}, E., {et~al.} 2018, \aap, 620,
  A58

\bibitem[{{Stanford-Moore} {et~al.}(2020){Stanford-Moore}, {Nielsen}, {De
  Rosa}, {Macintosh}, \& {Czekala}}]{2020ApJ...898...27S}
{Stanford-Moore}, S.~A., {Nielsen}, E.~L., {De Rosa}, R.~J., {Macintosh}, B.,
  \& {Czekala}, I. 2020, \apj, 898, 27

\bibitem[{{Stelzer} {et~al.}(2013){Stelzer}, {Marino}, {Micela},
  {L{\'o}pez-Santiago}, \& {Liefke}}]{2013MNRAS.431.2063S}
{Stelzer}, B., {Marino}, A., {Micela}, G., {L{\'o}pez-Santiago}, J., \&
  {Liefke}, C. 2013, \mnras, 431, 2063

\bibitem[{{Swimmer} {et~al.}(2022){Swimmer}, {Currie}, {Steiger}, {Brandt},
  {Brandt}, {Guyon}, {Kuzuhara}, {Chilcote}, {Tobin}, {Groff}, {Lozi},
  {Bailey}, {Walter}, {Fruitwala}, {Zobrist}, {Smith}, {Coiffard}, {Dodkins},
  {Davis}, {Daal}, {Bumble}, {Vievard}, {Skaf}, {Deo}, {Jovanovic},
  {Martinache}, {Tamura}, {Kasdin}, \& {Mazin}}]{2022AJ....164..152S}
{Swimmer}, N., {Currie}, T., {Steiger}, S., {et~al.} 2022, \aj, 164, 152

\bibitem[{{Tal-Or} {et~al.}(2019){Tal-Or}, {Trifonov}, {Zucker}, {Mazeh}, \&
  {Zechmeister}}]{2019MNRAS.484L...8T}
{Tal-Or}, L., {Trifonov}, T., {Zucker}, S., {Mazeh}, T., \& {Zechmeister}, M.
  2019, \mnras, 484, L8

\bibitem[{{Tran} {et~al.}(2021){Tran}, {Bowler}, {Cochran}, {Endl},
  {Stef{\'a}nsson}, {Mahadevan}, {Ninan}, {Bender}, {Halverson}, {Roy}, \&
  {Terrien}}]{2021AJ....161..173T}
{Tran}, Q.~H., {Bowler}, B.~P., {Cochran}, W.~D., {et~al.} 2021, \aj, 161, 173

\bibitem[{{Trifonov} {et~al.}(2020){Trifonov}, {Tal-Or}, {Zechmeister},
  {Kaminski}, {Zucker}, \& {Mazeh}}]{2020A&A...636A..74T}
{Trifonov}, T., {Tal-Or}, L., {Zechmeister}, M., {et~al.} 2020, \aap, 636, A74

\bibitem[{{Tull}(1998)}]{1998SPIE.3355..387T}
{Tull}, R.~G. 1998, in Society of Photo-Optical Instrumentation Engineers
  (SPIE) Conference Series, Vol. 3355, Optical Astronomical Instrumentation,
  ed. S.~{D'Odorico}, 387--398

\bibitem[{{Udry} {et~al.}(2003{\natexlab{a}}){Udry}, {Mayor}, {Clausen},
  {Freyhammer}, {Helt}, {Lovis}, {Naef}, {Olsen}, {Pepe}, {Queloz}, \&
  {Santos}}]{2003A&A...407..679U}
{Udry}, S., {Mayor}, M., {Clausen}, J.~V., {et~al.} 2003{\natexlab{a}}, \aap,
  407, 679

\bibitem[{{Udry} {et~al.}(2003{\natexlab{b}}){Udry}, {Mayor}, {Clausen},
  {Freyhammer}, {Helt}, {Lovis}, {Naef}, {Olsen}, {Pepe}, {Queloz}, \&
  {Santos}}]{Udry_2003}
{Udry}, S., {Mayor}, M., {Clausen}, J.~V., {et~al.} 2003{\natexlab{b}}, \aap,
  407, 679

\bibitem[{{van Leeuwen}(2007)}]{2007A&A...474..653V}
{van Leeuwen}, F. 2007, \aap, 474, 653

\bibitem[{{Venner} {et~al.}(2021){Venner}, {Vanderburg}, \&
  {Pearce}}]{2021AJ....162...12V}
{Venner}, A., {Vanderburg}, A., \& {Pearce}, L.~A. 2021, \aj, 162, 12

\bibitem[{{Vogt}(1987)}]{1987PASP...99.1214V}
{Vogt}, S.~S. 1987, \pasp, 99, 1214

\bibitem[{{Vogt} {et~al.}(1994){Vogt}, {Allen}, {Bigelow}, {Bresee}, {Brown},
  {Cantrall}, {Conrad}, {Couture}, {Delaney}, {Epps}, {Hilyard}, {Hilyard},
  {Horn}, {Jern}, {Kanto}, {Keane}, {Kibrick}, {Lewis}, {Osborne},
  {Pardeilhan}, {Pfister}, {Ricketts}, {Robinson}, {Stover}, {Tucker}, {Ward},
  \& {Wei}}]{1994SPIE.2198..362V}
{Vogt}, S.~S., {Allen}, S.~L., {Bigelow}, B.~C., {et~al.} 1994, in Society of
  Photo-Optical Instrumentation Engineers (SPIE) Conference Series, Vol. 2198,
  Instrumentation in Astronomy VIII, ed. D.~L. {Crawford} \& E.~R. {Craine},
  362

\bibitem[{{Vogt} {et~al.}(2002){Vogt}, {Butler}, {Marcy}, {Fischer},
  {Pourbaix}, {Apps}, \& {Laughlin}}]{2002ApJ...568..352V}
{Vogt}, S.~S., {Butler}, R.~P., {Marcy}, G.~W., {et~al.} 2002, \apj, 568, 352

\bibitem[{{Vogt} {et~al.}(2014){Vogt}, {Radovan}, {Kibrick}, {Butler},
  {Alcott}, {Allen}, {Arriagada}, {Bolte}, {Burt}, {Cabak}, {Chloros},
  {Cowley}, {Deich}, {Dupraw}, {Earthman}, {Epps}, {Faber}, {Fischer}, {Gates},
  {Hilyard}, {Holden}, {Johnston}, {Keiser}, {Kanto}, {Katsuki}, {Laiterman},
  {Lanclos}, {Laughlin}, {Lewis}, {Lockwood}, {Lynam}, {Marcy}, {McLean},
  {Miller}, {Misch}, {Peck}, {Pfister}, {Phillips}, {Rivera}, {Sandford},
  {Saylor}, {Stover}, {Thompson}, {Walp}, {Ward}, {Wareham}, {Wei}, \&
  {Wright}}]{2014PASP..126..359V}
{Vogt}, S.~S., {Radovan}, M., {Kibrick}, R., {et~al.} 2014, \pasp, 126, 359

\bibitem[{{Ward-Duong} {et~al.}(2015){Ward-Duong}, {Patience}, {De Rosa},
  {Bulger}, {Rajan}, {Goodwin}, {Parker}, {McCarthy}, \&
  {Kulesa}}]{2015MNRAS.449.2618W}
{Ward-Duong}, K., {Patience}, J., {De Rosa}, R.~J., {et~al.} 2015, \mnras, 449,
  2618

\bibitem[{{Wittenmyer} {et~al.}(2009){Wittenmyer}, {Endl}, {Cochran},
  {Levison}, \& {Henry}}]{2009ApJS..182...97W}
{Wittenmyer}, R.~A., {Endl}, M., {Cochran}, W.~D., {Levison}, H.~F., \&
  {Henry}, G.~W. 2009, \apjs, 182, 97

\bibitem[{{Wittenmyer} {et~al.}(2014){Wittenmyer}, {Tuomi}, {Butler}, {Jones},
  {Anglada-Escud{\'e}}, {Horner}, {Tinney}, {Marshall}, {Carter}, {Bailey},
  {Salter}, {O'Toole}, {Wright}, {Crane}, {Schectman}, {Arriagada}, {Thompson},
  {Minniti}, {Jenkins}, \& {Diaz}}]{2014ApJ...791..114W}
{Wittenmyer}, R.~A., {Tuomi}, M., {Butler}, R.~P., {et~al.} 2014, \apj, 791,
  114

\bibitem[{{Wittenmyer} {et~al.}(2020){Wittenmyer}, {Wang}, {Horner}, {Butler},
  {Tinney}, {Carter}, {Wright}, {Jones}, {Bailey}, {O'Toole}, \&
  {Johns}}]{2020MNRAS.492..377W}
{Wittenmyer}, R.~A., {Wang}, S., {Horner}, J., {et~al.} 2020, \mnras, 492, 377

\bibitem[{{Yee} {et~al.}(2017){Yee}, {Petigura}, \& {von
  Braun}}]{2017ApJ...836...77Y}
{Yee}, S.~W., {Petigura}, E.~A., \& {von Braun}, K. 2017, \apj, 836, 77

\bibitem[{{Zechmeister} {et~al.}(2013){Zechmeister}, {K{\"u}rster}, {Endl}, {Lo
  Curto}, {Hartman}, {Nilsson}, {Henning}, {Hatzes}, \&
  {Cochran}}]{Zechmeister_2013}
{Zechmeister}, M., {K{\"u}rster}, M., {Endl}, M., {et~al.} 2013, \aap, 552, A78

\end{thebibliography}

\onecolumn

\appendix

\section{Target list}

\begin{table*}[h!]
\centering
\caption{Proper motion values from HGCA of the \textbf{116} selected stars.}
\begin{adjustbox}{width=\textwidth}
\begin{tabular}[h!]{cccccccccccc}
\hline
 Star & $\sigma$ PMa$^a$ & $M_{b}$ min at 10au & Parallax$^b$ & $\mu_{Hip}^{\alpha}$ & $\mu_{Hip}^{\delta}$ & $\mu_{EDR3}^{\alpha}$ & $\mu_{EDR3}^{\delta}$ & $\mu_{Hip-EDR3}^{\alpha}$ & $\mu_{Hip-EDR3}^{\delta}$ & No. of \\ 
 & & ($M_{Jup}$) & (pc) & (mas/yr) & (mas/yr) & (mas/yr) & (mas/yr) & (mas/yr) & (mas/yr) & companion(s) \\ 
\hline
 GJ660.1 & > 4 & 11.8 & $43.297 \pm 0.397$ & $192.695 \pm 0.525$ & $-704.419 \pm 0.354$ & $184.702 \pm 4.626$ & $-693.716 \pm 3.139$ & $190.922 \pm 0.129$ & $-702.635 \pm 0.080$ & 1 \\ [0.1cm]
 GJ676 & > 4 & 3.4 & $62.609 \pm 0.034$ & $-258.759 \pm 0.046$ & $-185.119 \pm 0.034$ & $-260.087 \pm 1.371$ & $-185.157 \pm 0.831$ & $-258.838 \pm 0.048$ & $-185.880 \pm 0.033$ & 4 \\ [0.1cm]
 GJ680 & > 4 & 16.4 & $103.345 \pm 0.024$ & $74.069 \pm 0.033$ & $470.166 \pm 0.020$ & $83.508 \pm 3.523$ & $454.994 \pm 2.580$ & $77.747 \pm 0.082$ & $464.975 \pm 0.057$ & 1 \\ [0.1cm]
 HD142 & > 4 & $11.7$ & $38.213 \pm 0.041$ & $575.099 \pm 0.024$ & $-40.874 \pm 0.033$ & $575.297 \pm 0.447$ & $-39.223 \pm 0.469$ & $575.093 \pm 0.012$ & $-39.845 \pm 0.013$ & 2 \\ [0.1cm]
 HD457 & > 4 & 6.1 & $18.102 \pm 0.025$ & $113.112 \pm 0.025$ & $-18.873 \pm 0.028$ & $112.462 \pm 0.592$ & $-19.548 \pm 0.678$ & $112.791 \pm 0.018$ & $-18.826 \pm 0.018$ & $1^{*}$ \\ [0.1cm]
 HD1388 & > 4 & $28.5$ & $37.176 \pm 0.029$ & $401.642 \pm 0.043$ & $-0.238 \pm 0.024$ & $396.154 \pm 0.824$ & $-0.481 \pm 0.457$ & $398.610 \pm 0.021$ & $-0.437 \pm 0.016$ & 1 \\ [0.1cm]
 HD7449 & > 4 & 37.4 & $25.9132 \pm 0.0287$ & $-164.544 \pm 0.053$ & $-134.382 \pm 0.039$ & $-161.045 \pm 0.621$ & $-138.692 \pm 0.464$ & $-163.019 \pm 0.022$ & $-136.855 \pm 0.017$ & 2 \\ [0.1cm]
 HD11195 & $3.4$ & $4.2$ & $17.4998 \pm 0.0184$ & $-6.488 \pm 0.028$ & $14.994 \pm 0.025$ & $-6.833 \pm 0.508$ & $15.066 \pm 0.502$ & $-6.512 \pm 0.018$ & $14.763 \pm 0.018$ & 0 \\ [0.1cm]
 HD11397 & > 4 & $119$ & $18.4075 \pm 0.0196$ & $132.287 \pm 0.027$ & $-352.861 \pm 0.020$ & $126.848 \pm 1.338$ & $-363.072 \pm 1.007$ & $131.014 \pm 0.045$ & $-358.017 \pm 0.033$ & $1^{*}$ \\ [0.1cm]
 HD13350 & 2.0 & $0.1$ & $54.6825 \pm 0.0354$ & $310.909 \pm 0.064$ & $1049.061 \pm 0.082$ & $311.667 \pm 0.397$ & $1050.236 \pm 0.444$ & $310.720 \pm 0.013$ & $1049.658 \pm 0.012$ & 0 \\ [0.1cm]
 HD13357 & $2.0$ & $0.1$ & $12.1502 \pm 0.0157$ & $-30.155 \pm 0.022$ & $-195.200 \pm 0.020$ & $-30.826 \pm 1.055$ & $-195.625 \pm 0.919$ & $-30.082 \pm 0.029$ & $-195.058 \pm 0.031$ & 0 \\ [0.1cm]
 HD13612 & > 4 & $17.3$ & $26.1678 \pm 0.0274$ & $366.258 \pm 0.043$ & $-67.275 \pm 0.030$ & $376.253 \pm 9.264$ & $-72.567 \pm 6.630$ & $367.067 \pm 0.684$ & $-66.783 \pm 0.411$ & $1^{*}$ \\ [0.1cm]
 HD15337 & > 4 & 13.2 & $22.2922 \pm 0.0160$ & $-73.581 \pm 0.021$ & $-211.935 \pm 0.020$ & $-72.014 \pm 1.023$ & $-210.732 \pm 0.920$ & $-72.829 \pm 0.034$ & $-211.474 \pm 0.031$ & $1^{*}$ \\ [0.1cm]
 HD16160 & $> 4$ & 75.3 & $35.8970 \pm 0.1225$ & $114.381 \pm 0.209$ & $-199.348 \pm 0.131$ & $99.098 \pm 2.307$ & $-187.790 \pm 1.360$ & $105.221 \pm 0.079$ & $-190.355 \pm 0.046$ & 2 \\ [0.1cm]
 HD16548 & > 4 & $108.2$ & $18.9061 \pm 0.0226$ & $261.625 \pm 0.042$ & $-62.568 \pm 0.028$ & $273.099 \pm 0.730$ & $-51.351 \pm 0.795$ & $266.536 \pm 0.025$ & $-61.001 \pm 0.023$ & 1 \\ [0.1cm]
 HD16905 & > 4 & $5.1$ & $25.1176 \pm 0.0123$ & $36.720 \pm 0.024$ & $-184.705 \pm 0.018$ & $37.464 \pm 0.966$ & $-183.811 \pm 0.898$ & $37.153 \pm 0.029$ & $-184.468 \pm 0.029$ & 1 \\ [0.1cm]
 HD19467 & $3.0$ & 2.3 & $22.0344 \pm 0.0177$ & $36.061 \pm 0.026$ & $-80.180 \pm 0.020$ & $36.537 \pm 1.713$ & $-78.766 \pm 1.067$ & $36.267 \pm 0.045$ & $-80.069 \pm 0.028$ & $1^{*}$ \\ [0.1cm]
 HD19641 & > 4 & 20 & $19.0259 \pm 0.0216$ & $-25.024 \pm 0.024$ & $114.795 \pm 0.030$ & $-25.973 \pm 0.673$ & $117.150 \pm 0.749$ & $-25.519 \pm 0.020$ & $115.886 \pm 0.021$ & $1^{*}$ \\ [0.1cm]
 HD20010 & > 4 & 43 & $71.4337 \pm 0.1320$ & $359.565 \pm 0.112$ & $619.186 \pm 0.161$ & $371.142 \pm 0.427$ & $611.897 \pm 0.550$ & $364.729 \pm 0.015$ & $616.214 \pm 0.017$ & $1^{*}$ \\ [0.1cm]
 HD21019 & 2.8 & 0.1 & $26.7913 \pm 0.0912$ & $1.295 \pm 0.246$ & $-219.292 \pm 0.170$ & $2.035 \pm 0.751$ & $-219.220 \pm 0.511$ & $1.799 \pm 0.022$ & $-219.483 \pm 0.014$ & $1^{*}$ \\ [0.1cm]
 HD21175 & > 4 & 59 & $56.5570 \pm 0.0270$ & $63.985 \pm 0.033$ & $35.175 \pm 0.049$ & $40.646 \pm 0.780$ & $42.888 \pm 0.778$ & $53.036 \pm 0.026$ & $39.153 \pm 0.025$ & 1 \\ [0.1cm]
 HD24633 & > 4 & 32.4 & $18.7655 \pm 0.0884$ & $4.290 \pm 0.147$ & $64.156 \pm 0.111$ & $4.088 \pm 1.494$ & $64.266 \pm 1.419$ & $5.886 \pm 0.042$ & $65.529 \pm 0.035$ & $1^{*}$ \\ [0.1cm]
 HD27894 & > 4 & 7 & $22.8888 \pm 0.0121$ & $182.473 \pm 0.017$ & $270.012 \pm 0.023$ & $182.047 \pm 0.915$ & $272.210 \pm 1.075$ & $182.032 \pm 0.030$ & $270.376 \pm 0.033$ & 0 \\ [0.1cm]
 HD28254 & 2.2 & 0.1 & $15.2528 \pm 0.0183$ & $181.376 \pm 0.027$ & $-164.130 \pm 0.026$ & $182.195 \pm 1.058$ & $-163.440 \pm 0.488$ & $181.367 \pm 0.032$ & $-164.082 \pm 0.019$ & 1 \\ [0.1cm]
 HD38459 & > 4 & 2.9 & $26.5541 \pm 0.0114$ & $-19.556 \pm 0.021$ & $-39.926 \pm 0.021$ & $-20.643 \pm 0.894$ & $-40.680 \pm 0.962$ & $-19.706 \pm 0.030$ & $-40.138 \pm 0.030$ & 0 \\ [0.1cm]
 HD39091 & 2.7 & 2.3 & $18.2833 \pm 0.0216$ & $74.236 \pm 0.026$ & $-214.258 \pm 0.023$ & $75.863 \pm 1.357$ & $-213.220 \pm 0.847$ & $74.058 \pm 0.042$ & $-214.357 \pm 0.024$ & 0 \\ [0.1cm]
 HD40397 & > 4 & 20 & $41.5366 \pm 0.0210$ & $74.849 \pm 0.029$ & $-200.354 \pm 0.027$ & $70.996 \pm 0.789$ & $-203.208 \pm 0.541$ & $73.753 \pm 0.024$ & $-202.096 \pm 0.017$ & 1 \\ [0.1cm]
 HD42581 & > 4 & 7.7 & $138.3400 \pm 0.3177$ & $1778.585 \pm 0.589$ & $1477.306 \pm 0.340$ & $1807.221 \pm 1.454$ & $1443.037 \pm 0.713$ & $1792.759 \pm 0.033$ & $1450.768 \pm 0.022$ & 0 \\ [0.1cm]
 HD42659 & > 4 & 42 & $7.7597 \pm 0.0308$ & $-0.118 \pm 0.033$ & $-18.771 \pm 0.036$ & $0.465 \pm 0.617$ & $-17.998 \pm 0.513$ & $0.284 \pm 0.017$ & $-18.558 \pm 0.016$ & 1 \\ [0.1cm]
 HD42936 & > 4 & 0.9 & $17.8589 \pm 0.0131$ & $323.791 \pm 0.013$ & $-281.239 \pm 0.017$ & $324.151 \pm 0.515$ & $-281.307 \pm 0.860$ & $323.830 \pm 0.017$ & $-280.993 \pm 0.026$ & 2 \\ [0.1cm]
 HD43197 & 3.7 & 5.7 & $16.0171 \pm 0.0106$ & $147.648 \pm 0.013$ & $16.600 \pm 0.015$ & $148.028 \pm 1.432$ & $13.923 \pm 1.615$ & $147.878 \pm 0.044$ & $16.381 \pm 0.052$ & 2 \\ [0.1cm]
 HD44219 & 2.2 & 0.1 & $31.2191 \pm 0.0240$ & $-8.694 \pm 0.036$ & $-260.642 \pm 0.035$ & $-7.903 \pm 0.734$ & $-260.163 \pm 0.791$ & $-8.445 \pm 0.022$ & $-260.600 \pm 0.017$ & 1 \\ [0.1cm]
 HD46569 & > 4 & 87 & $27.1217 \pm 0.0476$ & $102.014 \pm 0.074$ & $104.577 \pm 0.074$ & $102.415 \pm 0.418$ & $104.925 \pm 0.423$ & $107.769 \pm 0.014$ & $103.583 \pm 0.014$ & 1 \\ [0.1cm]
 HD49095 & > 4 & 74 & $41.7838 \pm 0.0270$ & $-217.138 \pm 0.038$ & $-317.283 \pm 0.038$ & $-204.905 \pm 0.450$ & $-304.276 \pm 0.461$ & $-212.357 \pm 0.013$ & $-310.692 \pm 0.015$ & 0 \\ [0.1cm]
 HD56380 & > 4 & 162 & $18.7261 \pm 0.2264$ & $48.032 \pm 0.304$ & $113.332 \pm 0.388$ & $66.521 \pm 2.491$ & $105.182 \pm 2.474$ & $57.329 \pm 0.080$ & $104.799 \pm 0.083$ & 1 \\ [0.1cm]
 HD61383 & > 4 & 136 & $18.8219 \pm 0.0241$ & $226.754 \pm 0.034$ & $-126.880 \pm 0.026$ & $228.806 \pm 0.875$ & $-122.982 \pm 0.509$ & $230.681 \pm 0.025$ & $-122.299 \pm 0.015$ & 1 \\ [0.1cm]
 HD61986 & > 4 & 5.7 & $32.9777 \pm 0.0239$ & $-148.238 \pm 0.034$ & $60.209 \pm 0.028$ & $-147.092 \pm 0.945$ & $59.709 \pm 0.581$ & $-148.613 \pm 0.028$ & $59.841 \pm 0.014$ & 0 \\ [0.1cm]
 HD62364 & > 4 & 16 & $18.8811 \pm 0.0229$ & $-123.632 \pm 0.045$ & $-8.469 \pm 0.039$ & $-124.392 \pm 0.536$ & $-9.721 \pm 0.541$ & $-123.989 \pm 0.018$ & $-9.190 \pm 0.017$ & 1 \\ [0.1cm]
 HD63454 & > 4 & 2.0 & $23.3044 \pm 0.0283$ & $114.008 \pm 0.043$ & $-72.063 \pm 0.037$ & $112.003 \pm 1.660$ & $-73.320 \pm 1.260$ & $114.225 \pm 0.059$ & $-72.085 \pm 0.044$ & 2 \\ [0.1cm]
 HD65277 & > 4 & 5.0 & $56.6412 \pm 0.0254$ & $-157.142 \pm 0.032$ & $7.580 \pm 0.022$ & $-155.772 \pm 1.537$ & $3.088 \pm 1.089$ & $-158.025 \pm 0.043$ & $6.923 \pm 0.032$ & $1^{*}$ \\ [0.1cm]
 HD67200 & 2.2 & 0.1 & $17.6470 \pm 0.0197$ & $-88.625 \pm 0.028$ & $-39.919 \pm 0.024$ & $-89.442 \pm 0.818$ & $-39.482 \pm 0.872$ & $-88.698 \pm 0.028$ & $-39.832 \pm 0.025$ & $1^{*}$ \\ [0.1cm]
 HD68161 & > 4 & 152 & $3.7118 \pm 0.0994$ & $-9.131 \pm 0.143$ & $-1.733 \pm 0.154$ & $-9.992 \pm 0.443$ & $-1.443 \pm 0.379$ & $-9.854 \pm 0.014$ & $-1.243 \pm 0.013$ & 0 \\ [0.1cm]
 HD72659 & > 4 & 11 & $19.2582 \pm 0.0271$ & $-112.313 \pm 0.036$ & $-96.386 \pm 0.024$ & $-112.566 \pm 0.689$ & $-98.187 \pm 0.601$ & $-112.625 \pm 0.024$ & $-97.048 \pm 0.016$ & 2 \\ [0.1cm]
 HD73256 & > 4 & 14 & $27.2441 \pm 0.0217$ & $-182.193 \pm 0.024$ & $67.373 \pm 0.029$ & $-180.349 \pm 0.430$ & $66.238 \pm 0.542$ & $-181.090 \pm 0.013$ & $67.182 \pm 0.019$ & 2 \\ [0.1cm]
 HD74957 & > 4 & 27 & $20.5730 \pm 0.0153$ & $64.486 \pm 0.025$ & $78.146 \pm 0.022$ & $59.966 \pm 0.706$ & $76.040 \pm 0.669$ & $62.862 \pm 0.023$ & $77.328 \pm 0.022$ & $1^{*}$ \\ [0.1cm]
 HD75302 & > 4 & 5.7 & $34.6094 \pm 0.0239$ & $27.283 \pm 0.029$ & $112.918 \pm 0.034$ & $27.126 \pm 0.665$ & $112.245 \pm 0.590$ & $27.068 \pm 0.021$ & $112.434 \pm 0.020$ & 1 \\ [0.1cm]
 HD78612 & > 4 & 101 & $24.4502 \pm 0.0210$ & $-368.061 \pm 0.028$ & $-108.564 \pm 0.021$ & $-355.978 \pm 0.782$ & $-108.083 \pm 0.621$ & $-362.803 \pm 0.024$ & $-109.072 \pm 0.020$ & 1 \\ [0.1cm]
 HD85390 & 2.4 & 0.1 & $65.5889 \pm 0.0342$ & $-415.507 \pm 0.053$ & $-213.991 \pm 0.032$ & $-415.899 \pm 0.630$ & $-214.685 \pm 0.548$ & $-415.721 \pm 0.019$ & $-213.898 \pm 0.014$ & 2 \\ [0.1cm]
 HD88072 & 2.4 & 2.2 & $29.9081 \pm 0.0377$ & $-262.971 \pm 0.080$ & $-3.649 \pm 0.093$ & $-263.271 \pm 0.511$ & $-3.995 \pm 0.392$ & $-263.333 \pm 0.018$ & $-3.496 \pm 0.014$ & $1^{*}$ \\ [0.1cm]
 HD88218 & > 4 & 30 & $31.5137 \pm 0.0307$ & $-445.632 \pm 0.036$ & $18.774 \pm 0.040$ & $-443.157 \pm 0.447$ & $16.019 \pm 0.521$ & $-444.675 \pm 0.014$ & $16.690 \pm 0.016$ & $1^{*}$ \\ [0.1cm]
 HD89839 & 3.3 & 2.4 & $21.9392 \pm 0.0275$ & $-39.173 \pm 0.037$ & $-120.068 \pm 0.031$ & $-38.414 \pm 1.088$ & $-119.640 \pm 0.458$ & $-39.236 \pm 0.033$ & $-120.267 \pm 0.018$ & 1 \\ [0.1cm]
 HD93351 & > 4 & 10 & $17.9129 \pm 0.0168$ & $88.176 \pm 0.021$ & $-142.442 \pm 0.021$ & $87.116 \pm 1.324$ & $-143.170 \pm 0.942$ & $87.655 \pm 0.045$ & $-142.596 \pm 0.029$ & 2 \\ [0.1cm]
 HD96116 & > 4 & 128 & $17.5165 \pm 0.0164$ & $18.249 \pm 0.022$ & $-39.264 \pm 0.019$ & $30.843 \pm 0.823$ & $-35.226 \pm 0.686$ & $23.825 \pm 0.026$ & $-38.921 \pm 0.025$ & 1 \\ [0.1cm]
 HD97037 & > 4 & 15 & $30.8043 \pm 0.0212$ & $-238.340 \pm 0.039$ & $-173.313 \pm 0.025$ & $-240.783 \pm 0.558$ & $-171.223 \pm 0.559$ & $-239.437 \pm 0.020$ & $-172.792 \pm 0.017$ & $1^{*}$ \\ [0.1cm]
 HD100777 & > 4 & 7.8 & $20.1590 \pm 0.0285$ & $-11.535 \pm 0.038$ & $35.586 \pm 0.031$ & $-11.575 \pm 0.990$ & $34.643 \pm 0.852$ & $-11.268 \pm 0.033$ & $35.034 \pm 0.030$ & 1 \\ [0.1cm]
 HD101198 & > 4 & 42 & $37.2450 \pm 0.0730$ & $91.849 \pm 0.104$ & $125.335 \pm 0.069$ & $99.625 \pm 0.453$ & $125.765 \pm 0.337$ & $95.408 \pm 0.018$ & $126.203 \pm 0.012$ & 1 \\ [0.1cm]
 HD103743 & 2.2 & 0.1 & $28.2336 \pm 0.0231$ & $-171.869 \pm 0.038$ & $-8.144 \pm 0.026$ & $-171.494 \pm 7.626$ & $-11.607 \pm 4.054$ & $-171.894 \pm 0.382$ & $-8.165 \pm 0.241$ & 0 \\ [0.1cm]
 HD104731 & > 4 & 33 & $40.4345 \pm 0.0884$ & $318.554 \pm 0.079$ & $-111.120 \pm 0.076$ & $323.541 \pm 0.384$ & $-111.825 \pm 0.371$ & $321.144 \pm 0.012$ & $-111.265 \pm 0.011$ & 0 \\ [0.1cm]
 HD106906 & 3.5 & 9.2 & $9.7673 \pm 0.0185$ & $-39.066 \pm 0.020$ & $-12.698 \pm 0.022$ & $-38.695 \pm 0.617$ & $-12.303 \pm 0.651$ & $-39.222 \pm 0.019$ & $-12.684 \pm 0.021$ & 2 \\ [0.1cm]
 HD107094 & > 4 & 19 & $18.6573 \pm 0.0234$ & $-3.315 \pm 0.025$ & $-45.172 \pm 0.022$ & $-8.639 \pm 0.741$ & $-44.534 \pm 0.643$ & $-2.702 \pm 0.027$ & $-46.030 \pm 0.022$ & 1 \\ [0.1cm]
 HD108063 & 2.0 & 0.1 & $19.0734 \pm 0.0231$ & $-135.319 \pm 0.027$ & $-22.538 \pm 0.018$ & $-135.004 \pm 0.451$ & $-22.583 \pm 0.432$ & $-135.170 \pm 0.014$ & $-22.539 \pm 0.014$ & $1^{*}$ \\ [0.1cm]
 HD108202 & 2.3 & 0.4 & $25.8379 \pm 0.0177$ & $-309.967 \pm 0.032$ & $-77.177 \pm 0.022$ & $-311.265 \pm 1.621$ & $-77.031 \pm 1.142$ & $-309.702 \pm 0.061$ & $-77.415 \pm 0.045$ & 1 \\ [0.1cm]
 HD110537 & > 4 & 4.3 & $21.2496 \pm 0.1143$ & $-0.716 \pm 0.182$ & $-382.755 \pm 0.176$ & $0.561 \pm 0.896$ & $-383.939 \pm 0.940$ & $-1.046 \pm 0.028$ & $-382.604 \pm 0.028$ & $1^{*}$ \\ [0.1cm]
 HD111031 & > 4 & 39 & $32.0310 \pm 0.0219$ & $-279.627 \pm 0.034$ & $46.791 \pm 0.026$ & $-282.073 \pm 0.633$ & $54.801 \pm 0.454$ & $-281.564 \pm 0.022$ & $49.571 \pm 0.016$ & 1 \\ [0.1cm]
 HD111232 & > 4 & 5.5 & $24.4620 \pm 0.0455$ & $108.284 \pm 0.077$ & $-50.040 \pm 0.081$ & $108.246 \pm 0.657$ & $-49.514 \pm 0.569$ & $108.551 \pm 0.025$ & $-49.692 \pm 0.021$ & 2 \\ [0.1cm]
 HD111998 & 2.4 & 1.8 & $19.7872 \pm 0.0205$ & $-191.118 \pm 0.019$ & $-65.020 \pm 0.021$ & $-190.718 \pm 0.513$ & $-65.308 \pm 0.649$ & $-191.179 \pm 0.017$ & $-65.216 \pm 0.021$ & 1 \\ [0.1cm]
 HD114174 & > 4 & 37 & $37.8677 \pm 0.0243$ & $85.533 \pm 0.037$ & $-680.324 \pm 0.037$ & $84.368 \pm 0.759$ & $-670.201 \pm 0.603$ & $84.993 \pm 0.024$ & $-676.775 \pm 0.018$ & $1^{*}$ \\ [0.1cm]
 HD114330 & 2.0 & 0.1 & $11.1834 \pm 0.4110$ & $-35.252 \pm 0.752$ & $-33.302 \pm 0.552$ & $-35.625 \pm 0.990$ & $-32.322 \pm 0.829$ & $-36.673 \pm 0.049$ & $-32.368 \pm 0.038$ & 0 \\ [0.1cm]
 HD114783 & 2.9 & 1.1 & $47.5529 \pm 0.0291$ & $-138.362 \pm 0.047$ & $10.284 \pm 0.030$ & $-138.541 \pm 0.817$ & $9.834 \pm 0.851$ & $-138.488 \pm 0.028$ & $10.508 \pm 0.025$ & 2 \\ [0.1cm]
 HD125276 & > 4 & 11 & $55.5900 \pm 0.0531$ & $-359.037 \pm 0.062$ & $364.426 \pm 0.056$ & $-356.395 \pm 0.887$ & $366.796 \pm 0.737$ & $-357.805 \pm 0.026$ & $365.366 \pm 0.019$ & 0 \\ [0.1cm]
 HD128571 & > 4 & 6.1 & $17.0083 \pm 0.0184$ & $-71.750 \pm 0.017$ & $19.160 \pm 0.025$ & $-71.718 \pm 0.503$ & $19.211 \pm 0.687$ & $-71.476 \pm 0.016$ & $18.924 \pm 0.021$ & $1^{*}$ \\ [0.1cm]
\hline
\end{tabular}
\end{adjustbox}
\textbf{Notes:} $^a$: Significance of the PMa. $^b$: Gaia DR3. $\mu_{Hip}$ corresponds to the proper motion obtained by Hipparcos. $^{*}$: Orbital parameters and mass of the companion poorly constrained. $\mu_{EDR3}$ correspond to the proper motion obtained by Gaia EDR3. $\mu_{Hip-EDR3}$ correspond to the proper motion obtained by the Hipparcos-Gaia EDR3 positional difference.
\label{PMvalues}
\end{table*}

\begin{table*}[h!]
\centering
\caption{Table A.1 continued.}
\begin{adjustbox}{width=\textwidth}
\begin{tabular}[h!]{cccccccccccc}
\hline
 Star & $\sigma$ PMa$^a$ & $M_{b}$ at 10au & Parallax$^b$ & $\mu_{Hip}^{\alpha}$ & $\mu_{Hip}^{\delta}$ & $\mu_{EDR3}^{\alpha}$ & $\mu_{EDR3}^{\delta}$ & $\mu_{Hip-EDR3}^{\alpha}$ & $\mu_{Hip-EDR3}^{\delta}$ & No. of \\ 
 & & ($M_{Jup}$) & (mas) & (mas/yr) & (mas/yr) & (mas/yr) & (mas/yr) & (mas/yr) & (mas/yr) & companion(s) \\ 
\hline
 HD129191 & > 4 & 11 & $18.4622 \pm 0.0242$ & $26.520 \pm 0.042$ & $-152.832 \pm 0.035$ & $26.278 \pm 1.394$ & $-152.293 \pm 1.540$ & $26.058 \pm 0.039$ & $-152.433 \pm 0.038$ & $1^{*}$ \\ [0.1cm]
 HD131664 & > 4 & 99 & $19.1360 \pm 0.0829$ & $8.008 \pm 0.105$ & $22.284 \pm 0.126$ & $14.979 \pm 0.710$ & $28.685 \pm 0.740$ & $11.370 \pm 0.022$ & $25.899 \pm 0.023$ & 1 \\ [0.1cm]
 HD131977 & > 4 & 3.7 & $169.8843 \pm 0.0653$ & $1031.472 \pm 0.093$ & $-1723.619 \pm 0.075$ & $1035.702 \pm 1.343$ & $-1725.446 \pm 1.007$ & $1032.968 \pm 0.051$ & $-1724.781 \pm 0.033$ & 0 \\ [0.1cm]
 HD135625 & 2.0 & 0.1 & $18.9018 \pm 0.0191$ & $72.385 \pm 0.026$ & $-16.574 \pm 0.027$ & $72.305 \pm 0.974$ & $-17.487 \pm 0.728$ & $72.283 \pm 0.023$ & $-16.433 \pm 0.016$ & 1 \\ [0.1cm]
 HD140901 & 2.0 & 0.1 & $26.2048 \pm 0.0179$ & $-4.040 \pm 0.019$ & $-176.046 \pm 0.025$ & $-3.664 \pm 0.508$ & $-176.463 \pm 0.808$ & $-4.053 \pm 0.015$ & $-176.132 \pm 0.022$ & 1 \\ [0.1cm]
 HD142070 & > 4 & 81 & $5.5771 \pm 0.0307$ & $-14.315 \pm 0.043$ & $-24.244 \pm 0.033$ & $-15.266 \pm 1.073$ & $-25.030 \pm 1.016$ & $-13.895 \pm 0.031$ & $-23.580 \pm 0.023$ & 0 \\ [0.1cm]
 HD143361 & 3.0 & 3.0 & $14.5456 \pm 0.0202$ & $-156.561 \pm 0.026$ & $-120.231 \pm 0.019$ & $-157.333 \pm 1.362$ & $-119.387 \pm 1.039$ & $-156.630 \pm 0.040$ & $-120.369 \pm 0.027$ & 2 \\ [0.1cm]
 HD150139 & 2.3 & 0.1 & $20.5407 \pm 0.0251$ & $139.201 \pm 0.044$ & $-143.615 \pm 0.037$ & $138.903 \pm 0.807$ & $-142.494 \pm 0.594$ & $139.101 \pm 0.031$ & $-143.648 \pm 0.023$ & $1^{*}$ \\ [0.1cm]
 HD156079 & > 4 & 4.4 & $17.5648 \pm 0.0239$ & $-69.179 \pm 0.031$ & $-165.076 \pm 0.023$ & $-69.736 \pm 0.909$ & $-164.554 \pm 0.561$ & $-68.923 \pm 0.028$ & $-165.238 \pm 0.015$ & 0 \\ [0.1cm]
 HD156274 & > 4 & 9.9 & $113.7513 \pm 0.0725$ & $1029.609 \pm 0.117$ & $106.935 \pm 0.101$ & $1036.484 \pm 1.039$ & $109.036 \pm 0.520$ & $1033.069 \pm 0.039$ & $107.932 \pm 0.021$ & $1^{*}$ \\ [0.1cm]
 HD165131 & > 4 & 15 & $17.2691 \pm 0.0317$ & $-61.918 \pm 0.043$ & $-124.587 \pm 0.030$ & $-59.570 \pm 1.126$ & $-125.994 \pm 0.717$ & $-62.476 \pm 0.036$ & $-124.074 \pm 0.023$ & 1 \\ [0.1cm]
 HD166724 & 3.2 & 1.6 & $37.4641 \pm 0.0253$ & $-43.140 \pm 0.040$ & $72.728 \pm 0.037$ & $-43.593 \pm 0.393$ & $73.244 \pm 0.361$ & $-43.414 \pm 0.013$ & $72.873 \pm 0.012$ & $1^{*}$ \\ [0.1cm]
 HD167677 & 3.2 & 3.0 & $16.7894 \pm 0.0166$ & $33.127 \pm 0.020$ & $12.219 \pm 0.017$ & $33.530 \pm 1.050$ & $12.868 \pm 0.828$ & $33.235 \pm 0.035$ & $12.392 \pm 0.027$ & 1 \\ [0.1cm]
 HD169830 & 2.0 & 0.1 & $27.1461 \pm 0.1469$ & $-0.341 \pm 0.221$ & $16.103 \pm 0.158$ & $-0.721 \pm 0.881$ & $15.434 \pm 0.537$ & $0.064 \pm 0.023$ & $16.374 \pm 0.014$ & 2 \\ [0.1cm]
 HD171587 & 2.0 & 0.1 & $28.5668 \pm 0.0217$ & $126.024 \pm 0.029$ & $-193.681 \pm 0.025$ & $126.315 \pm 0.938$ & $-194.794 \pm 0.733$ & $125.979 \pm 0.034$ & $-193.715 \pm 0.023$ & 0 \\ [0.1cm]
 HD174429 & 3.1 & 2.1 & $21.1621 \pm 0.0223$ & $16.272 \pm 0.025$ & $-85.519 \pm 0.024$ & $17.353 \pm 1.217$ & $-83.766 \pm 0.830$ & $16.081 \pm 0.035$ & $-85.590 \pm 0.025$ & 0 \\ [0.1cm]
 HD179346 & > 4 & 53 & $18.0318 \pm 0.0942$ & $-7.800 \pm 0.126$ & $-53.982 \pm 0.116$ & $-9.922 \pm 0.877$ & $-52.875 \pm 0.720$ & $-4.943 \pm 0.030$ & $-53.287 \pm 0.021$ & 0 \\ [0.1cm]
 HD181433 & > 4 & 2.1 & $37.0511 \pm 0.0211$ & $-230.723 \pm 0.018$ & $235.806 \pm 0.025$ & $-230.886 \pm 0.841$ & $235.382 \pm 0.996$ & $-230.958 \pm 0.030$ & $235.640 \pm 0.029$ & 3 \\ [0.1cm]
 HD183414 & 2.2 & 0.2 & $36.1158 \pm 0.0264$ & $1.440 \pm 0.037$ & $-174.330 \pm 0.042$ & $2.701 \pm 0.701$ & $-174.656 \pm 0.464$ & $1.268 \pm 0.023$ & $-174.389 \pm 0.015$ & 0 \\ [0.1cm]
 HD185283 & 2.3 & 0.8 & $31.9046 \pm 0.0217$ & $-163.887 \pm 0.022$ & $-53.650 \pm 0.017$ & $-162.481 \pm 1.218$ & $-53.718 \pm 1.109$ & $-163.982 \pm 0.033$ & $-53.707 \pm 0.025$ & 1 \\ [0.1cm]
 HD191797 & > 4 & 58 & $22.7336 \pm 0.0272$ & $-8.443 \pm 0.038$ & $51.173 \pm 0.033$ & $-12.621 \pm 2.324$ & $54.340 \pm 1.456$ & $-5.479 \pm 0.051$ & $48.047 \pm 0.033$ & 1 \\ [0.1cm]
 HD195145 & > 4 & 3.1 & $16.6547 \pm 0.0305$ & $4.132 \pm 0.052$ & $-4.200 \pm 0.050$ & $1.989 \pm 0.769$ & $-4.698 \pm 0.481$ & $4.050 \pm 0.021$ & $-4.430 \pm 0.012$ & 0 \\ [0.1cm]
 HD195564 & > 4 & 21 & $40.4241 \pm 0.0497$ & $309.736 \pm 0.068$ & $109.910 \pm 0.051$ & $307.694 \pm 0.651$ & $107.216 \pm 0.518$ & $308.366 \pm 0.021$ & $108.386 \pm 0.012$ & 1 \\ [0.1cm]
 HD196050 & > 4 & 6.1 & $22.2309 \pm 0.0234$ & $-219.316 \pm 0.041$ & $-179.612 \pm 0.039$ & $-219.148 \pm 0.675$ & $-180.370 \pm 0.547$ & $-219.413 \pm 0.026$ & $-179.984 \pm 0.023$ & 1 \\ [0.1cm]
 HD196385 & > 4 & 24 & $20.9103 \pm 0.0239$ & $73.036 \pm 0.031$ & $-16.412 \pm 0.024$ & $73.293 \pm 0.772$ & $-15.056 \pm 0.445$ & $73.633 \pm 0.022$ & $-15.383 \pm 0.012$ & $1^{*}$ \\ [0.1cm]
 HD202206 & 3.2 & 2.2 & $22.2420 \pm 0.0740$ & $352.541 \pm 0.107$ & $-31.590 \pm 0.064$ & $350.543 \pm 0.977$ & $-32.314 \pm 0.509$ & $352.754 \pm 0.031$ & $-31.388 \pm 0.022$ & 2 \\ [0.1cm]
 HD204313 & > 4 & 11 & $20.7705 \pm 0.0343$ & $42.751 \pm 0.054$ & $-270.443 \pm 0.034$ & $42.551 \pm 0.747$ & $-271.125 \pm 0.483$ & $42.234 \pm 0.027$ & $-270.727 \pm 0.018$ & 3 \\ [0.1cm]
 HD204961 & > 4 & 0.7 & $201.3252 \pm 0.0237$ & $-45.917 \pm 0.032$ & $-816.875 \pm 0.025$ & $-45.677 \pm 1.028$ & $-817.318 \pm 0.656$ & $-46.049 \pm 0.030$ & $-816.302 \pm 0.025$ & 1 \\ [0.1cm]
 HD207700 & > 4 & 11 & $26.0873 \pm 0.0172$ & $-196.563 \pm 0.025$ & $-437.609 \pm 0.030$ & $-195.605 \pm 0.674$ & $-438.053 \pm 0.573$ & $-195.753 \pm 0.019$ & $-437.915 \pm 0.018$ & 1 \\ [0.1cm]
 HD207869 & 3.1 & 1.6 & $19.5789 \pm 0.0118$ & $-156.358 \pm 0.014$ & $-248.956 \pm 0.017$ & $-157.311 \pm 0.951$ & $-249.431 \pm 0.807$ & $-156.228 \pm 0.032$ & $-248.970 \pm 0.030$ & 0 \\ [0.1cm]
 HD210797 & > 4 & 13 & $15.4128 \pm 0.0210$ & $258.946 \pm 0.027$ & $-122.008 \pm 0.024$ & $258.354 \pm 1.303$ & $-118.366 \pm 1.076$ & $258.487 \pm 0.037$ & $-121.500 \pm 0.036$ & $1^{*}$ \\ [0.1cm]
 HD212036 & > 4 & 69 & $19.1167 \pm 0.0803$ & $21.555 \pm 0.105$ & $-234.978 \pm 0.116$ & $29.075 \pm 1.182$ & $-224.715 \pm 1.505$ & $22.904 \pm 0.036$ & $-231.443 \pm 0.033$ & 1 \\ [0.1cm]
 HD215257 & > 4 & 6.0 & $23.6251 \pm 0.0251$ & $150.456 \pm 0.035$ & $333.617 \pm 0.033$ & $150.934 \pm 0.889$ & $331.705 \pm 0.653$ & $150.716 \pm 0.026$ & $333.292 \pm 0.020$ & $1^{*}$ \\ [0.1cm]
 HD215497 & 2.0 & 0.1 & $24.6569 \pm 0.0152$ & $-54.666 \pm 0.016$ & $-61.147 \pm 0.019$ & $-53.881 \pm 0.939$ & $-60.575 \pm 0.933$ & $-54.665 \pm 0.031$ & $-61.222 \pm 0.033$ & 2 \\ [0.1cm]
 HD216437 & 2.8 & 1.7 & $26.9269 \pm 0.0154$ & $52.792 \pm 0.025$ & $215.141 \pm 0.027$ & $51.858 \pm 0.767$ & $214.317 \pm 0.865$ & $52.863 \pm 0.026$ & $214.857 \pm 0.028$ & 1 \\ [0.1cm]
 HD218396 & 2.1 & 3.5 & $28.2075 \pm 0.0230$ & $-179.085 \pm 0.037$ & $-6.549 \pm 0.023$ & $-196.276 \pm 10.121$ & $-9.146 \pm 5.480$ & $-179.244 \pm 0.344$ & $-6.934 \pm 0.226$ & 0 \\ [0.1cm]
 HD219834 & > 4 & 0.1 & $173.5740 \pm 0.0170$ & $-135.692 \pm 0.015$ & $-719.178 \pm 0.023$ & $-137.136 \pm 0.575$ & $-713.779 \pm 0.834$ & $-136.426 \pm 0.018$ & $-715.027 \pm 0.028$ & 0 \\ [0.1cm]
 HD221146 & > 4 & 29 & $27.5348 \pm 0.0247$ & $-10.513 \pm 0.033$ & $-24.287 \pm 0.024$ & $-11.763 \pm 1.093$ & $-28.467 \pm 0.843$ & $-11.268 \pm 0.030$ & $-25.794 \pm 0.022$ & 1 \\ [0.1cm]
 HD221420 & > 4 & 16.3 & $32.1023 \pm 0.0325$ & $16.306 \pm 0.056$ & $0.736 \pm 0.058$ & $15.885 \pm 0.437$ & $1.647 \pm 0.385$ & $14.973 \pm 0.014$ & $0.455 \pm 0.012$ & 1 \\ [0.1cm]
 HD221638 & > 4 & 40 & $19.1599 \pm 0.0269$ & $56.624 \pm 0.041$ & $72.839 \pm 0.042$ & $52.210 \pm 0.623$ & $77.925 \pm 0.609$ & $55.789 \pm 0.019$ & $74.527 \pm 0.019$ & 1 \\ [0.1cm]
 HD222237 & > 4 & 3.7 & $87.3724 \pm 0.0187$ & $143.736 \pm 0.030$ & $-736.907 \pm 0.032$ & $141.756 \pm 0.524$ & $-736.300 \pm 0.507$ & $142.849 \pm 0.018$ & $-736.651 \pm 0.017$ & $1^{*}$ \\ [0.1cm]
 HD223238 & 2.5 & 0.1 & $17.8497 \pm 0.0171$ & $-1.916 \pm 0.016$ & $-202.994 \pm 0.018$ & $-2.808 \pm 1.100$ & $-202.424 \pm 1.016$ & $-1.814 \pm 0.037$ & $-203.140 \pm 0.035$ & 0 \\ [0.1cm]
 HIP9603 & > 4 & 109 & $18.8225 \pm 0.0866$ & $-111.175 \pm 0.143$ & $257.497 \pm 0.147$ & $-110.184 \pm 0.933$ & $245.727 \pm 0.990$ & $-119.933 \pm 0.033$ & $255.468 \pm 0.033$ & 1 \\ [0.1cm]
 HIP10337 & > 4 & 1.9 & $42.1225 \pm 0.0167$ & $375.454 \pm 0.019$ & $55.718 \pm 0.020$ & $377.499 \pm 1.622$ & $55.290 \pm 1.164$ & $375.597 \pm 0.051$ & $55.414 \pm 0.040$ & 1 \\ [0.1cm]
 HIP39470 & > 4 & 91 & $22.6284 \pm 0.0325$ & $16.231 \pm 0.044$ & $-82.449 \pm 0.047$ & $9.228 \pm 1.155$ & $-82.499 \pm 1.131$ & $18.669 \pm 0.039$ & $-76.357 \pm 0.041$ & 1 \\ [0.1cm]
 HIP54597 & 2.3 & 0.1 & $19.3611 \pm 0.0183$ & $40.676 \pm 0.019$ & $53.190 \pm 0.025$ & $41.256 \pm 0.462$ & $54.595 \pm 0.690$ & $40.793 \pm 0.016$ & $53.317 \pm 0.022$ & 1 \\ [0.1cm]
 HIP70849 & > 4 & 4.6 & $41.4618 \pm 0.0175$ & $-44.051 \pm 0.023$ & $-201.577 \pm 0.028$ & $-47.003 \pm 2.099$ & $-203.284 \pm 1.890$ & $-44.428 \pm 0.060$ & $-202.047 \pm 0.044$ & 1 \\ [0.1cm]
 HIP113201 & > 4 & 132 & $68.7356 \pm 0.0272$ & $341.147 \pm 0.037$ & $-219.108 \pm 0.034$ & $343.238 \pm 2.100$ & $-218.067 \pm 3.265$ & $341.100 \pm 0.055$ & $-219.083 \pm 0.063$ & 1 \\ [0.1cm]
\hline
\end{tabular}
\end{adjustbox}
\label{PMvalues_suite}
\end{table*}

\begin{table*}[h!]
\centering
\caption{Characteristics of the 53 stars with characterized companions.}
\begin{adjustbox}{width=1.0\textwidth}
\begin{tabular}[h!]{ccccccccccccc}
\hline
Name & HIP & Mass$^{1}$ & Distance$^{1}$ & Spectral$^{2}$ & V$^{2}$ & \rhk & Vsin(\textit{I}) & Age & $\chi^{2}$ $^{22}$ \\ 
 & & (M$_{\sun}$) & (pc) & type & (mag) &  & (km/s) & (Gyr) & \\ 
\hline
GJ660.1 & 84212 & $0.43 \pm 0.01$ & $23.10 \pm 0.21$ & M1V & 11.62 & -5.30$^{7}$ & - & > 2$^{23}$ & 69 \\ [0.1cm]
GJ680 & 86057 & $0.47 \pm 0.01$ & $9.676 \pm 0.003$ & M3V & 10.13 & -5.16$^{7}$ & 3.0$^{12}$ & $8.1 \pm 4.2^{24}$ & 8601 \\ [0.1cm]
HD1388 & 1444 & $1.18 \pm 0.06$ & $26.90 \pm 0.02$ & G0V & 6.50 & -4.98$^{6}$ & 3.1$^{13}$ & $4.6 \pm 1.7^{25}$ & 4141 \\ [0.1cm]
HD16160 & 12114 & $0.80 \pm 0.04$ & $7.23 \pm 0.02$ & K3V & 5.60$^{3}$ & -4.87$^{3}$ & 4.7$^{13}$ & $4.7 \pm 3.9^{13}$ & 6078 \\ [0.1cm]
HD16548 & 12350 & $1.31 \pm 0.07$ & $52.81 \pm 0.07$ & G5V & 6.98 & -5.13$^{3}$ & 5.2$^{13}$ & $4.9 \pm 0.3^{25}$ & 11460 \\ [0.1cm]
HD16905 & 12436 & $0.85 \pm	0.04$ & $39.80 \pm 0.02$ & K3V & 9.46 & -4.72$^{3}$ & 1.5$^{14}$ & $6.1 \pm 4.7^{25}$ & 184 \\ [0.1cm]
HD21175 & 15799 & $0.86 \pm 0.04$ & $17.68 \pm 0.01$ & K0V & 6.90 & -4.74$^{6}$ & 1.8$^{14}$ & $3.7_{-1.7}^{+3.3}$ $^{6}$ & 69960 \\ [0.1cm]
HD28254 & 20606 & $1.10 \pm 0.05$ & $55.16 \pm 0.08$ & G5V & 7.68 & -5.17$^{3}$ & 2.3$^{5}$ & $6.0 \pm 1.3^{25}$ & 10.3 \\ [0.1cm]
HD42659 & 29365 & $2.20 \pm 0.11$ & $128.5 \pm 0.6$ & A3 & 6.75 & - & 19.0$^{15}$ & - & 15 \\ [0.1cm]
HD44219 & 30114 & $1.20 \pm 0.06$ & $52.84 \pm 0.06$ & G3V & 7.70 & -5.04$^{3}$ & 2.2$^{13}$ & $8.4 \pm 0.9^{25}$ & 23 \\ [0.1cm]
HD46569 & 31079 & $1.50	\pm 0.07$ & $36.87 \pm 0.07$ & F8V & 5.57 & -4.93$^{6}$ & 5.9$^{16}$ & $7.6^{+3.2}_{-3.0}$$^{6}$ & 6232 \\ [0.1cm]
HD56380 & 34961 & $0.98 \pm 0.05$ & $53.1 \pm 0.3$ & G8V & 9.20 & -5.00$^{3}$ & 0.5$^{14}$ & $12.2 \pm 1.1^{25}$ & 3697 \\ [0.1cm]
HD61383 & 37233 & $1.17 \pm 0.06$ & $53.05 \pm 0.08$ & G3V & 7.57 & -5.02$^{3}$ & 2.7$^{13}$ & $11.4 \pm 0.2$$^{13}$ & 36110 \\ [0.1cm]
HD62364 & 36941 & $1.30 \pm	0.06$ & $52.95 \pm 0.08$ & F7V & 7.31 & -4.95$^{3}$ & 3.6$^{13}$ & $3.9 \pm 0.7^{25}$ & 364 \\ [0.1cm]
HD73256 & 42214 & $1.00 \pm	0.05$ & $36.68 \pm 0.04$ & G8IV & 8.06 & -4.46$^{5}$ & 3.50$^{13}$ & $2.2 \pm 2.1^{25}$ & 2040 \\ [0.1cm]
HD75302 & 43297 & $1.06 \pm 0.05$ & $30.29 \pm 0.03$ & G5V & 7.46$^{4}$ & -4.62$^{6}$ & 0.46$^{17}$ & $2.1_{-1.1}^{+2.6}$ $^{6}$ & 299 \\ [0.1cm]
HD78612 & 44896 & $1.20 \pm	0.06$ & $40.85 \pm 0.04$ & G3V & 7.15 & -4.98$^{3}$ & 1.6$^{13}$ & $10.2 \pm 0.8^{25}$ & 20060 \\ [0.1cm]
HD89839 & 50653 & $1.30 \pm 0.06$ & $50.10 \pm 0.07$ & F7V & 7.64 & -4.94$^{3}$ & 4.2$^{14}$ & $1.6 \pm 0.9^{25}$ & 56 \\ [0.1cm]
HD93351 & 52720 & $0.98 \pm 0.05$ & $55.72 \pm 0.05$ & G6V & 9.14 & -4.98$^{3}$ & - & $4.3 \pm 3.9$$^{13}$ & 143 \\ [0.1cm]
HD96116 & 54102 & $1.08 \pm 0.05$ & $57.05 \pm 0.06$ & G3V & 8.69 & -4.69$^{3}$ & 1.7$^{13}$ & $2.1 \pm 1.8^{25}$ & 28020 \\ [0.1cm]
HD100777 & 56572 & $1.08 \pm 0.05$ & $49.52 \pm 0.08$ & G8V & 8.42 & -5.09$^{3}$ & 1.8$^{13}$ & $5.7 \pm 3.3^{25}$ & 173 \\ [0.1cm]
HD101198 & 56802 & $1.33 \pm 0.07$ & $26.82 \pm 0.03$ & F6.5V & 5.48 & -4.87$^{6}$ & 5$^{18}$ & $6.2^{+3.2}_{-2.6}$$^{6}$ & 1529 \\ [0.1cm]
HD107094 & 60051 & $0.98 \pm 0.05$ & $53.54 \pm 0.07$ & G5V & 9.13 & -4.83$^{3}$ & 1.8$^{13}$ & $8.6 \pm 4.0^{25}$ & 1225 \\ [0.1cm]
HD108202 & 60648 & $0.74 \pm 0.04$ & $38.65 \pm 0.03$ & K4/5V & 10.22 & -4.50$^{3}$ & - & $6.5 \pm 4.0^{28}$ & 28 \\ [0.1cm]
HD111031 & 62345 & $1.17 \pm 0.06$ & $31.19 \pm 0.02$ & G5V & 6.87 & -5.09$^{3}$ & 1.9$^{13}$ & $4.1 \pm 1.6^{25}$ & 8191 \\ [0.1cm]
HD111998 & 62875 & $1.40 \pm 0.07$ & $33.39 \pm 0.05$ & F6V & 6.11 & -4.58$^{6}$ & 20.0$^{18}$ & $1.7_{-0.9}^{+3.4}$ $^{6}$ & 27 \\ [0.1cm]
HD114783 & 64457 & $0.90 \pm 0.04$ & $21.01 \pm 0.02$ & K1V & 7.55 & -4.97$^{3}$ & 2.0$^{13}$ & $6.0 \pm 4.9^{25}$ & 33 \\ [0.1cm]
HD131664 & 73408 & $1.16 \pm 0.06$ & $52.2 \pm 0.3$ & G3V & 8.13 & -4.85$^{5}$ & 3.0$^{13}$ & $1.2 \pm 1.0^{25}$ & 1823 \\ [0.1cm]
HD135625 & 74856 & $1.28 \pm 0.06$ & $56.58 \pm 0.07$ & G3IV/V & 7.71 & -4.99$^{6}$ & 4.5$^{13}$ & $4.4 \pm 0.9^{25}$ & 7 \\ [0.1cm]
HD140901 & 77358 & $1.06 \pm 0.05$ & $15.24 \pm 0.01$ & G7IV & 6.01 & -4.79$^{6}$ & 2.0$^{13}$ & $3.2 \pm 2.8^{25}$ & 18 \\ [0.1cm]
HD143361 & 78521 & $1.06 \pm 0.05$ & $68.62 \pm 0.10$ & G6V & 9.16 & -5.01$^{5}$ & 0.68$^{5}$ & $6.3 \pm 3.7^{29}$ & 21 \\ [0.1cm]
HD165131 & 88595 & $1.10 \pm 0.06$ & $57.8 \pm 0.1$ & G3/5V & 8.41 & -4.98$^{3}$ & 2.3$^{19}$ & $2.7 \pm 2.0^{25}$ & 264 \\ [0.1cm]
HD167677 & 89583 & $1.00 \pm 0.05$ & $54.59 \pm 0.07$ & G5V & 8.92 & -4.98$^{3}$ & 0.7$^{19}$ & $11.5 \pm 1.7^{25}$ & 22 \\ [0.1cm]
HD169830 & 90485 & $1.40 \pm 0.07$ & $36.8 \pm 0.2$ & F7V & 5.90 & -4.95$^{5}$ & 3.4$^{13}$ & $2.2 \pm 0.2^{25}$ & 12 \\ [0.1cm]
HD179346 & 94370 & $1.16 \pm 0.06$ & $55.4 \pm 0.3$ & G0V & 8.05 & -4.84$^{3}$ & 3.8$^{13}$ & $0.79 \pm 0.68^{25}$ & 496 \\ [0.1cm]
HD185283 & 96861 & $0.80 \pm 0.04$ & $31.32 \pm 0.02$ & K3V & 8.95 & -4.94$^{3}$ & 0.9$^{14}$ & $6.0 \pm 4.7^{25}$ & 9 \\ [0.1cm]
\hline
\end{tabular}
\end{adjustbox}
\textbf{Notes:} $^{1}$\cite{2022A&A...657A...7K}. $^{2}$\cite{2020A&A...636A..74T}. $^{3}$\cite{2021A&A...646A..77G}. $^{4}$\cite{2020ApJS..247...11R}. $^{5}$\cite{2021A&A...654A.168L}. $^{6}$\cite{2020ApJ...898...27S}. $^{7}$\cite{2017A&A...600A..13A}. $^{8}$\cite{2017ApJ...836...77Y}. $^{9}$\cite{2020A&A...641A.110G}. $^{10}$\cite{2018A&A...616A.108B}. $^{11}$\cite{2020A&A...639A..35H}. $^{12}$\cite{2013MNRAS.431.2063S}. $^{13}$\cite{2021A&A...654A.137L}. $^{14}$\cite{2020A&A...639A..35H}. $^{15}$\cite{2019MNRAS.487.5922G}. $^{16}$\cite{2012A&A...542A.116A}. $^{17}$\cite{2020ApJ...898..119R}. $^{18}$\cite{Borgniet_2017}. $^{19}$\cite{2020A&A...634A.136C}. $^{20}$\cite{2019A&A...629A..80H}. $^{21}$\cite{2018A&A...620A..58S}. $^{22}$\cite{2021ApJS..254...42B}. $^{23}$\cite{2011ApJ...743..109S}. $^{24}$\cite{2020A&A...644A..68M}. $^{25}$\cite{2019A&A...624A..78D}. $^{26}$\cite{2011A&A...535A..54S}. $^{27}$\cite{Biller_2022}. $^{28}$\cite{2018MNRAS.481.3244G}. $^{29}$\cite{2021A&A...649A.111A}. 
\label{table_carac_star_2}
\end{table*}

\begin{table*}[h!]
\centering
\caption{Table A.3 continued.}
\begin{adjustbox}{width=1.0\textwidth}
\begin{tabular}[h!]{ccccccccccccc}
\hline
Name & HIP & Mass$^{1}$ & Distance$^{1}$ & Spectral$^{2}$ & V$^{2}$ & \rhk & Vsin(\textit{I}) & Age & $\chi^{2}$ $^{22}$ \\ 
 & & (M$_{\sun}$) & (pc) & type & (mag) &  & (km/s) & (Gyr) & \\ 
\hline
HD191797 & 99695 & $0.85 \pm 0.04$ & $43.94 \pm 0.06$ & K0V & 9.50 & -4.37$^{3}$ & 3.6$^{13}$ & $4.6 \pm 4.3^{25}$ & 6461 \\ [0.1cm]
HD195564 & 101345 & $1.24 \pm 0.06$ & $24.71 \pm 0.04$ & G2.5IV & 5.65 & -5.18$^{3}$ & 1.9$^{13}$ & $7.4 \pm 0.3^{25}$ & 1054 \\ [0.1cm]
HD196050 & 101806 & $1.27 \pm 0.06$ & $50.49 \pm 0.05$ & G3V & 7.49 & -4.98$^{5}$ & 3.3$^{13}$ & $4.3 \pm 0.9^{25}$ & 655 \\ [0.1cm]
HD204961 & 106440 & $0.45 \pm 0.01$ & $4.966 \pm 0.001$ & M2/3V & 8.67 & -5.18$^{7}$ & 2.1$^{20}$ & $8.1 \pm 4.2^{24}$ & 278 \\ [0.1cm]
HD207700 & 108158 & $1.16 \pm 0.06$ & $38.31 \pm 0.03$ & G4V & 7.43 & -5.10$^{5}$ & 1.2$^{13}$ & $9.9 \pm 1.6^{25}$ & 655 \\ [0.1cm]
HD212036 & 110440 & $1.08 \pm 0.05$ & $52.2 \pm 0.3$ & G5V & 8.47 & -4.86$^{6}$ & 1.3$^{14}$ & $5.7 \pm 3.2^{25}$ & 894 \\ [0.1cm]
HD215497 & 112441 & $0.92 \pm 0.05$ & $40.53 \pm 0.02$ & K3V & 8.95 & -5.07$^{3}$ & - & $5.8 \pm 4.9^{29}$ & 4 \\ [0.1cm]
HD216437 & 113137 & $1.22 \pm 0.06$ & $26.68 \pm 0.02$ & G1V & 6.05 & -5.04$^{9}$ & 3.7$^{13}$ & $4.9 \pm 1.0$ $^{13}$ & 46 \\ [0.1cm]
HD221146 & 115951 & $1.24 \pm 0.06$ & $36.27 \pm 0.04$ & G1V & 6.89 & -4.98$^{5}$ & 2.9$^{13}$ & $6.4 \pm 1.1^{25}$ & 2220 \\ [0.1cm]
HD221420 & 116250 & $1.34 \pm 0.07$ & $31.14 \pm 0.04$ & G1V & 6.89 & -5.28$^{3}$ & 0.7$^{13}$ & $3.7 \pm 0.5^{25}$ & 541 \\ [0.1cm]
HD221638 & 116350 & $1.20 \pm 0.06$ & $52.15 \pm 0.09$ & F6V & 7.55 & -4.86$^{3}$ & 7.2$^{13}$ & $2.2 \pm 1.4^{25}$ & 1586 \\ [0.1cm]
HIP9603 & 9603 & $0.60 \pm 0.01$ & $27.8 \pm 0.1$ & K7V & 11.10 & - & - & - & 6339 \\ [0.1cm]
HIP10337 & 10337 & $0.69 \pm 0.01$ & $23.725 \pm 0.010$ & K7V & 9.83 & -4.43$^{10}$ & 3.2$^{5}$ & $4.0_{-2.8}^{+9.5}$ $^{8}$ & 62 \\ [0.1cm]
HIP39470 & 39470 & $0.69 \pm 0.01$ & $44.16 \pm 0.07$ & M0 & 10.83 & - & - & $5.1 \pm 4.6^{25}$ & 10660 \\ [0.1cm]
HIP54597 & 54597 & $0.80 \pm 0.04$ & $39.51 \pm 0.02$ & K5V & 9.82 & -5.04$^{11}$ & 0.5$^{14}$ & $4.3 \pm 4.0$ $^{13}$ & 9 \\ [0.1cm]
HIP70849 & 70849 & $0.65 \pm 0.01$ & $24.10 \pm 0.01$ & K7V & 10.38 & -4.48$^{10}$ & 0.3$^{21}$ & $1-5^{26}$ & 119 \\ [0.1cm]
HIP113201 & 113201 & $0.53 \pm 0.01$ & $23.5 \pm 0.2$ & M4V & 11.51 & -4.42$^{7}$ & - & $1.2 \pm 0.1^{27}$ & 4869 \\ [0.1cm]
\hline
\end{tabular}
\end{adjustbox}
\label{table_carac_star_2_suite}
\end{table*}

\newpage

\section{Data considered for orbital fits}

\begin{table*}[h!]
\centering
\caption{RV data considered for orbital fits.}
\begin{adjustbox}{width=1.0\textwidth}
\begin{tabular}[h!]{ccccccccccccc}
\hline
Star & Instrument & $\Delta$T (days) & N$_{obs}$ & Reference & Star & Instrument & $\Delta$T (days) & N$_obs$ & Reference \\ 
\hline
GJ660.1 & HARPS & 3649 & 59 & ESO archive & HD140901 & HARPS & 1293 & 52 & ESO archive \\ [0.1cm]
GJ680 & HARPS & 3008 & 39 & ESO archive & HD140901 & HARPS & 1293 & 52 & ESO archive \\ [0.1cm]
HD1388 & HARPS & 5082 & 212 & ESO archive & HD140901 & AAT & 5878 & 321 & \cite{2022ApJS..262...21F} \\ [0.1cm]
HD1388 & HIRES & 8357 & 53 & \cite{2021ApJS..255....8R} & HD140901 & PFS & 3994 & 70 & \cite{2022ApJS..262...21F} \\ [0.1cm]
HD16160 & HARPS & 1776 & 229 & ESO archive & HD143361 & HARPS & 4667 & 59 & ESO archive \\ [0.1cm]
HD16160 & HIRES & 4858 & 146 & \cite{2021ApJS..255....8R} & HD143361 & CORALIE & 2012 & 45 & \cite{2017MNRAS.466..443J} \\ [0.1cm]
HD16160 & APF & 2216 & 252 & \cite{2021ApJS..255....8R} & HD143361 & MIKE & 2155 & 17 & \cite{2017MNRAS.466..443J} \\ [0.1cm]
HD16160 & LICK & 8153 & 81 & \cite{2021ApJS..255....8R} & HD165131 & HARPS & 5596 & 69 & ESO archive \\ [0.1cm]
HD16548 & HARPS & 6333 & 52 & ESO archive & HD167677 & HARPS & 5205 & 44 & ESO archive \\ [0.1cm]
HD16548 & LICK & 1349 & 9 & \cite{2014ApJS..210....5F} & HD169830 & HARPS & 1920 & 86 & ESO archive \\ [0.1cm]
HD16905 & HARPS & 6263 & 52 & ESO archive & HD169830 & CORALIE & 1506 & 106 & \cite{Mayor_2004}$^{*}$ \\ [0.1cm]
HD21175 & HARPS & 1147 & 61 & ESO archive & HD169830 & HIRES & 6959 & 85 & \cite{2021ApJS..255....8R} \\ [0.1cm]
HD28254 & HARPS & 6338 & 72 & ESO archive & HD179346 & HARPS & 5861 & 44 & ESO archive \\ [0.1cm]
HD42659 & HARPS & 1535 & 49 & \cite{Hartmann_2015} & HD185283 & HARPS & 4827 & 78 & ESO archive \\ [0.1cm]
HD44219 & HARPS & 6353 & 65 & ESO archive & HD191797 & HARPS & 5523 & 42 & ESO archive \\ [0.1cm]
HD46569 & HARPS & 1297 & 63 & ESO archive & HD195564 & HARPS & 4363 & 95 & ESO archive \\ [0.1cm]
HD46569 & LC & 1925 & 53 & \cite{Zechmeister_2013} & HD195564 & HIRES & 7936 & 56 & \cite{2021ApJS..255....8R} \\ [0.1cm]
HD46569 & VLC & 2040 & 54 & \cite{Zechmeister_2013} & HD195564 & APF & 1856 & 299 & \cite{2021ApJS..255....8R} \\ [0.1cm]
HD56380 & HARPS & 6287 & 96 & ESO archive & HD196050 & HARPS & 6243 & 56 & ESO archive \\ [0.1cm]
HD61383 & HARPS & 6293 & 63 & ESO archive & HD196050 & CORALIE & 1256 & 31 & \cite{Mayor_2004}$^{*}$ \\ [0.1cm]
HD62364 & HARPS & 6233 & 87 & ESO archive & HD196050 & UCLES & 2513 & 44 & \cite{2006ApJ...646..505B}$^{*}$ \\ [0.1cm]
HD73256 & HARPS & 2986 & 26 & ESO archive & HD204961 & HARPS & 5864 & 183 & ESO archive \\ [0.1cm]
HD73256 & CORALIE & 755 & 38 & \cite{Udry_2003}$^{*}$ & HD204961 & UCLES & 3519 & 32 & \cite{2009ApJ...690..743B}$^{*}$ \\ [0.1cm]
HD73256 & HIRES & 4700 & 18 & \cite{2017AJ....153..208B}$^{*}$ & HD207700 & HARPS & 3260 & 36 & ESO archive \\ [0.1cm]
HD75302 & HARPS & 2160 & 35 & ESO archive & HD212036 & HARPS & 5152 & 51 & ESO archive \\ [0.1cm]
HD75302 & ELODIE & 3218 & 26 & OHP archive & HD215497 & HARPS & 1842 & 99 & \cite{Locurto_2010} \\ [0.1cm]
HD75302 & SOPHIE & 3227 & 20 & OHP archive & HD216437 & HARPS & 5031 & 100 & ESO archive \\ [0.1cm]
HD78612 & HARPS & 3073 & 31 & ESO archive & HD216437 & CORALIE & 1405 & 21 & \cite{Mayor_2004}$^{*}$ \\ [0.1cm]
HD89839 & HARPS & 6191 & 98 & ESO archive & HD216437 & UCLES & 2801 & 39 & \cite{2006ApJ...646..505B}$^{*}$ \\ [0.1cm]
HD93351 & HARPS & 6537 & 40 & ESO archive & HD221146 & HARPS & 3612 & 69 & ESO archive \\ [0.1cm]
HD96116 & HARPS & 6296 & 40 & ESO archive & HD221146 & HRS & 6957 & 50 & \cite{2021AJ....161..106B} \\ [0.1cm]
HD100777 & HARPS & 1231 & 37 & ESO archive & HD221420 & HARPS & 5493 & 77 & ESO archive \\ [0.1cm]
HD101198 & HARPS & 1988 & 91 & \cite{Borgniet_2017} & HD221420 & AAT & 6515 & 88 & \cite{2022ApJS..262...21F} \\ [0.1cm]
HD107094 & HARPS & 6566 & 98 & ESO archive & HD221420 & PFS & 3323 & 38 & \cite{2022ApJS..262...21F} \\ [0.1cm]
HD108202 & HARPS & 6088 & 42 & ESO archive & HD221638 & HARPS & 5670 & 57 & ESO archive \\ [0.1cm]
HD111031 & HARPS & 3337 & 40 & ESO archive & HIP9603 & HARPS & 5790 & 39 & ESO archive \\ [0.1cm]
HD111998 & HARPS & 2992 & 150 & \cite{Borgniet_2017} & HIP10337 & HARPS & 6435 & 90 & ESO archive \\ [0.1cm]
HD114783 & HARPS & 4415 & 91 & ESO archive & HIP10337 & HIRES & 2161 & 18 & \cite{2017AJ....153..208B}$^{*}$ \\ [0.1cm]
HD114783 & HIRES & 7849 & 157 & \cite{2021ApJS..255....8R} & HIP39470 & HARPS & 5554 & 32 & ESO archive \\ [0.1cm]
HD114783 & HRS & 826 & 34 & \cite{2009ApJS..182...97W} & HIP54597 & HARPS & 5836 & 69 & ESO archive \\ [0.1cm]
HD131664 & HARPS & 6100 & 65 & ESO archive & HIP70849 & HARPS & 5521 & 59 & ESO archive \\ [0.1cm]
HD135625 & HARPS & 6292 & 99 & ESO archive & HIP113201 & HARPS & 3259 & 66 & ESO archive \\ [0.1cm]
\hline
\end{tabular}
\end{adjustbox}
\textbf{Notes: $^{*}$Data retrieved from the DACE archives.} 
\label{table_RV_full}
\end{table*}

\begin{table*}[h!]
\centering
\caption{Relative astrometry considered for orbital fits.}
\begin{adjustbox}{width=1.0\textwidth}
\begin{tabular}[h!]{cccccc}
\hline
Companion & JD - 2400000 & SEP (mas) & PA ($\deg$) & Instrument & Reference \\ 
\hline
GJ680 B & 53491.03 & $3760 \pm 50$ & $325.4 \pm 0.3$ & NACO & \cite{2015MNRAS.449.2618W} \\ [0.1cm]
GJ680 B & 54235.64 & $3940 \pm 50$ & $323.7 \pm 0.3$ & NACO & \cite{2015MNRAS.449.2618W} \\ [0.1cm]
HD1388 B & 57999.50 & $1845.9 \pm 1.3$ & $85.8 \pm 0.6$ & NESSI$^{a}$ & \cite{2021AJ....161..123D} \\ [0.1cm]
HD1388 B & 53940.39 & $1490 \pm 10$ & $85.6 \pm 1$ & NACO & This paper \\ [0.1cm]
HD21175 B & 57665.35 & $2628.0 \pm 2.5$ & $122.75 \pm 0.08$ & SPHERE & This paper \\ [0.1cm]
HD78612 B & 53739.29 & $613 \pm 5$ & $272.3 \pm 1$ & NACO & This paper \\ [0.1cm]
HD93351 B/C & 56093.02 & $1684 \pm 30$ & $76.3 \pm 1$ & NACO & This paper \\ [0.1cm]
HD101198 B & 54421.70 & $891.2 \pm 13.6$ & $250.36 \pm 1.16$ & PUEO$^{b}$ & \cite{Ehrenreich_2010} \\ [0.1cm]
HD101198 B & 54948.30 & $926.4 \pm 6.3$ & $251.23 \pm 0.55$ & NACO & \cite{Ehrenreich_2010} \\ [0.1cm]
HD101198 B & 54948.30 & $927.0 \pm 10.6$ & $250.93 \pm 0.86$ & NACO & \cite{Ehrenreich_2010} \\ [0.1cm]
HD111031 B & 58151.50 & $1055.8 \pm 1.3$ & $121.3 \pm 0.6$ & NESSI & \cite{2021AJ....161..123D} \\ [0.1cm]
HD195564 B & 58006.70 & $1113.0 \pm 1.3$ & $49.4 \pm 0.6$ & NESSI & \cite{2021AJ....161..123D} \\ [0.1cm]
HD195564 B & 58006.70 & $1104.9 \pm 1.3$ & $49.8 \pm 0.6$ & NESSI & \cite{2021AJ....161..123D} \\ [0.1cm]
HD207700 B & 56092.36 & $1090 \pm 20$ & $284.2 \pm 1$ & NACO & This paper \\ [0.1cm]
HD221146 B & 53262.30 & $1723.0 \pm 5.0$ & $28.6 \pm 0.1$ & NIRC2$^{c}$ & \cite{2021AJ....161..106B} \\ [0.1cm]
HD221146 B & 53571.14 & $1720.0 \pm 5.0$ & $28.5 \pm 0.2$ & NIRC2 & \cite{2021AJ....161..106B} \\ [0.1cm]
HD221146 B & 58041.14 & $1623.0 \pm 5.0$ & $25.10 \pm 0.14$ & NIRC2 & \cite{2021AJ....161..106B} \\ [0.1cm]
HD221146 B & 58480.47 & $1600.0 \pm 5.0$ & $24.33 \pm 0.12$ & NIRC2 & \cite{2021AJ....161..106B} \\ [0.1cm]
HD221146 B & 58480.47 & $1604.0 \pm 5.0$ & $24.26 \pm 0.14$ & NIRC2 & \cite{2021AJ....161..106B} \\ [0.1cm]
HD221146 B & 58676.11 & $1592.0 \pm 5.0$ & $24.4 \pm 0.2$ & NIRC2 & \cite{2021AJ....161..106B} \\ [0.1cm]
HIP113201 B & 57649.13 & $161.21 \pm 0.38$ & $182.36 \pm 0.13$ & SPHERE & \cite{Bonavita_2022} \\ [0.1cm]
HIP113201 B & 57676.03 & $167.80 \pm 0.53$ & $180.90 \pm 0.53$ & SPHERE & \cite{Bonavita_2022} \\ [0.1cm]
HIP113201 B & 57709.01 & $172.61 \pm 1.36$ & $177.72 \pm 0.13$ & SPHERE & \cite{Bonavita_2022} \\ [0.1cm]
HIP113201 B & 57906.44 & $198.44 \pm 1.76$ & $162.72 \pm 0.51$ & SPHERE & \cite{Bonavita_2022} \\ [0.1cm]
HIP113201 B & 58621.39 & $285.49 \pm 1.29$ & $123.59 \pm 0.26$ & SPHERE & \cite{Bonavita_2022} \\ [0.1cm]
\hline
\end{tabular}
\end{adjustbox}
\textbf{Notes:} $^{a}$ \cite{2018PASP..130e4502S}. $^{b}$ \cite{1998PASP..110..152R}. $^{c}$ \cite{2016PASP..128i5004S}.
\label{table_relative_astrometry_full}
\end{table*}

\section{Additional results}

\begin{landscape}
\begin{table*}[h!]
\caption{Summary of posteriors obtained with our MCMC algorithm for known planets.}
\resizebox{1.3\textwidth}{!}{
\begin{tabular}[h!]{ccccccccccccccccccccc}
\hline
 Parameter & HD16905 b & HD28254 b & HD44219 b$^{*}$ & HD62364 B & HD89839 b & HD100777 b$^{*}$ & HD111998 b$^{*}$ & HD114783 c & HD140901 b \\ 
\hline
 \emph{a} (au) & $8.8_{-0.3}^{+0.4}$ & $2.45_{-0.04}^{+0.03}$ & $1.26_{-0.07}^{+0.02}$ & $6.4 \pm 0.1$ & $4.9 \pm 0.1$ & $1.06_{-0.01}^{+0.02}$ & $1.91_{-0.03}^{+0.02}$ & $5.0 \pm 0.1$ & $11.8_{-2.5}^{+4.1}$ \\ [0.1cm]
 P (days) & $10256_{-522}^{+618}$ & $1333 \pm 4$ & $472_{-4}^{+3}$ & $5172 \pm 30$ & $3448 \pm 26$ & $383 \pm 1$ & $810_{-5}^{+6}$ & $4352_{-76}^{+88}$ & $14386_{-4415}^{+8099}$ \\ [0.1cm]
 Ecc & $0.68_{-0.01}^{+0.02}$ & $0.95_{-0.04}^{+0.03}$ & $0.41_{-0.14}^{+0.12}$ & $0.610_{-0.005}^{+0.004}$ & $0.20 \pm 0.02$ & $0.38 \pm 0.02$ & $0.13_{-0.04}^{+0.05}$ & $0.05_{-0.03}^{+0.04}$ & $0.77_{-0.07}^{+0.06}$ \\ [0.1cm]
 \textit{I} (°)& $43 \pm 3$ or $136 \pm 3$ & $21_{-11}^{+38}$ or $162_{-27}^{+7}$ & fix to 90 & $42 \pm 2$ or $131 \pm 2$ & $53_{-9}^{+16}$ or $129_{-15}^{+9}$ & fix to 90 & fix to 90 & $21_{-4}^{+7}$ or $159_{-6}^{+4}$ & $40_{-11}^{+18}$ or $138_{-21}^{+12}$ \\ [0.1cm]
 Mass (\Mjupv) & $11.3_{-0.7}^{+0.6}$ & $3.8_{-2.2}^{+3.0}$ & $0.58_{-0.08}^{+0.09}$ (< 11.0) & $19_{-1}^{+2}$ & $5.2 \pm 0.8$ & $1.23 \pm 0.05$ (< 18.0) & $4.8 \pm 0.2$ (< 12.8) & $1.9_{-0.4}^{+0.5}$ & $1.8 \pm 0.5$ \\ [0.1cm]
 $\Omega$ (°) & $353_{-8}^{+4}$ or $125 \pm 5$ & $248_{-33}^{+53}$ or $163_{-36}^{+38}$ & fix to 0 & $136 \pm 3$ or $273 \pm 5$ & $195_{-10}^{+9}$ or $175 \pm 9$ & fix to 0 & fix to 0 & $307_{-11}^{+14}$ or $355_{-9}^{+11}$ & $153_{-16}^{+14}$ or $81_{-17}^{+22}$ \\ [0.1cm]
 $\omega$ (°) & $248 \pm 1$ & $289_{-18}^{+20}$ & $145_{-34}^{+23}$ & $0 \pm 1$ & $171_{-5}^{+4}$ & $204_{-3}^{+4}$ & $190_{-17}^{+22}$ & $153_{-48}^{+46}$ & $299_{-9}^{+8}$ \\ [0.1cm]
 $P_{time}$ & $1856 \pm 8$ & $246_{-10}^{+9}$ & $147_{-27}^{+42}$ & $3570 \pm 5$ & $1788_{-47}^{+45}$ & $224 \pm 3$ & $521_{-36}^{+45}$ & $1618_{-555}^{+583}$ & $1182_{-46}^{+47}$ \\ [0.1cm]
 $\sqrt{ecc}cos(\omega)$ & $-0.31 \pm 0.02$ & $0.32_{-0.29}^{+0.30}$ & $-0.33_{-0.17}^{+0.24}$ & $0.778 \pm 0.003$ & $-0.44_{-0.3}^{+0.2}$ & $-0.56 \pm 0.02$ & $-0.34_{-0.07}^{+0.09}$ & $-0.17_{-0.10}^{+0.17}$ & $0.42_{-0.13}^{+0.12}$ \\ [0.1cm]
 $\sqrt{ecc}sin(\omega)$ & $-0.766 \pm 0.007$ & $-0.90_{-0.06}^{+0.14}$ & $0.38_{-0.26}^{+0.16}$ & $0.01^{+0.01}_{-0.02}$ & $0.07_{-0.03}^{+0.04}$ & $-0.25_{-0.03}^{+0.04}$ & $-0.06_{-0.11}^{+0.09}$ & $0.08_{-0.13}^{+0.12}$ & $-0.76_{-0.06}^{+0.07}$ \\ [0.1cm]
 Jitter (m/s) & $2.9_{-0.3}^{+0.5}$ & $2.9_{-0.3}^{+0.4}$ & $9.7_{-0.9}^{+1.0}$ & $4.0_{-0.4}^{+0.5}$ & $4.3 \pm 0.4$ & $2.4 \pm 0.4$ & $18.7 \pm 1.2$ & $3.2_{-0.2}^{+0.1}$ & $9.2 \pm 0.3$ \\ [0.1cm]
 rms (m/s) & $3.6_{-0.2}^{+0.8}$ & $5.8_{-0.1}^{+0.3}$ & $10.8 \pm 0.3$ & $4.9_{-0.2}^{+0.5}$ & $5.7_{-0.1}^{+0.2}$ & $2.6 \pm 0.02$ & $19.4_{-0.3}^{+0.6}$ & $3.9 \pm 0.1$ & $9.4_{-0.1}^{+0.2}$ \\ [0.1cm]
\hline
 & H03 = $64804 \pm 2$ & H03 = $-9324 \pm 1$ & H03 = $-24155_{-2}^{+1}$ & H03 = $35510 \pm 1$ & H03 = $31758 \pm 1$ & H03 = $1250 \pm 1$ & H03 = $-3 \pm 3$ & H03 = $0 \pm 1$ & H03 = $-6701 \pm 2$ \\ [0.1cm]
 Instrumental & H15 = $64817 \pm 1$ & H15 = $-9308_{-1}^{+2}$ & H15 = $-24211_{-5}^{+6}$ & H15 = $35525 \pm 2$ & H15 = $31770_{-1}^{+2}$ & & & H15 = $0_{-1}^{+2}$ & AAT = $-3 \pm 1$ \\ [0.1cm]
 offset (m/s) & & & & & & & & HRS = $0 \pm 1$ & PFS = $-6 \pm 2$ \\ [0.1cm]
 & & & & & & & & Hir94 = $0 \pm 1$ & \\ [0.1cm]
 & & & & & & & & Hir04 = $0 \pm 1$ & \\ [0.1cm]
\hline
\end{tabular}}
\textbf{Notes:} The median and $1 \sigma$ confidence interval are given for each parameter. $^{*}$ corresponds to the solutions found considering only the RV data points. The reported masses are therefore the minimum masses and the maximum masses estimated with Gaston, with a confidence index of $3 \sigma$, are noted between parenthesis for these companions.
\label{table_summary_connu}
\end{table*}

\begin{table*}[h!]
\caption{Table C.1 continued.}
\resizebox{1.3\textwidth}{!}{
\begin{tabular}[h!]{ccccccccccccccccccccc}
\hline
 Parameter & HD143361 b & HD167677 b & HD169830 c & HD196050 b & HD204961 b & HD215497 c$^{*}$ & HD216437 b & HD221420 B & HIP70849 b \\ 
\hline
 \emph{a} (au) & $2.05 \pm 0.03$ & $2.9 \pm 0.1$ & $3.29_{-0.05}^{+0.04}$ & $2.65 \pm 0.03$ & $3.7 \pm 0.1$ & $1.31 \pm 0.02$ & $2.57 \pm 0.04$ & $10.1 \pm 0.7$ & $3.99_{-0.07}^{+0.06}$ \\ [0.1cm]
 P (days) & $1040 \pm 1$ & $1804 \pm 73$ & $1832 \pm 7$ & $1392 \pm 4$ & $3853^{+51}_{-47}$ & $566 \pm 4$ & $1359 \pm 1$ & $10066_{-1125}^{+929}$ & $3649 \pm 18$ \\ [0.1cm]
 Ecc & $0.198_{-0.006}^{+0.007}$ & $0.35_{-0.15}^{+0.25}$ & $0.23_{-0.02}^{+0.03}$ & $0.20 \pm 0.01$ & $0.05 \pm 0.03$ & $0.47_{-0.04}^{+0.03}$ & $0.31 \pm 0.01$ & $0.12_{-0.03}^{+0.04}$ & $0.65_{-0.01}^{+0.02}$ \\ [0.1cm]
 \textit{I} (°) & $50_{-12}^{+19}$ or $128_{-27}^{+18}$ & $20_{-6}^{+14}$ or $155_{-12}^{+6}$ & $23_{-6}^{+12}$ or $158_{-9}^{+5}$ & $42_{-7}^{+11}$ or $141_{-12}^{+7}$ & $51 \pm 3$ or $134 \pm 3$ & fix to 90 & $36_{-6}^{+8}$ or $148_{-7}^{+5}$ & $14 \pm 1$ or $162 \pm 2$ & $96 \pm 16$ \\ [0.1cm]
 Mass (\Mjupv) & $4.9_{-0.9}^{+1.4}$ & $3.7_{-1.1}^{+1.6}$ & $8.9_{-2.6}^{+2.8}$ & $4.7 \pm 0.8$ & $0.99_{-0.08}^{+0.09}$ & $0.35 \pm 0.02$ (< 5.0) & $4.2 \pm 0.7$ & $22 \pm 2$ & $4.5_{-0.3}^{+0.4}$ \\ [0.1cm]
 $\Omega$ (°) & $352_{-18}^{+30}$ or $3.91_{-16}^{+23}$ & $236 \pm 24$ & $158_{-29}^{+24}$ or $291_{-28}^{+25}$ & $5_{-8}^{+13}$ or $18_{-7}^{+8}$ & $61_{-13}^{+17}$ or $265_{-15}^{+12}$ & fix to 0 & $129 \pm 8$ or $101_{-6}^{+7}$ & $272 \pm 3$ or $241 \pm 3$ & $35 \pm 6$ \\ [0.1cm]
 $\omega$ (°) & $242 \pm 2$ & $283 \pm 25$ & $305 \pm 10$ & $178 \pm 6$ & $207_{-31}^{+22}$ & $42 \pm 4$ & $65 \pm 2$ & $217_{-10}^{+24}$ & $182 \pm 1$ \\ [0.1cm]
 $P_{time}$ (days) & $725_{-6}^{+7}$ & $488_{-150}^{+198}$ & $614_{-44}^{+40}$ & $962_{-26}^{+23}$ & $2696_{-338}^{+454}$ & $431_{-7}^{+6}$ & $667 \pm 8$ & $2624_{-220}^{+513}$ & $2719_{-13}^{+14}$ \\ [0.1cm]
 $\sqrt{ecc}cos(\omega)$ & $-0.21_{-0.01}^{+0.02}$ & $0.09_{-0.23}^{+0.30}$ & $0.36_{-0.07}^{+0.06}$ & $-0.44_{-0.02}^{+0.01}$ & $-0.16_{-0.08}^{+0.14}$ & $0.51 \pm 0.04$ & $0.23 \pm 0.02$ & $-0.28_{-0.07}^{+0.15}$ & $-0.81 \pm 0.01$ \\ [0.1cm]
 $\sqrt{ecc}sin(\omega)$ & $-0.39 \pm 0.01$ & $-0.56_{-0.12}^{+0.15}$ & $-0.35 \pm 0.05$ & $0.02_{-0.05}^{+0.04}$ & $-0.13_{-0.07}^{+0.11}$ & $0.46_{-0.04}^{+0.03}$ & $0.50_{-0.01}^{+0.02}$ & $-0.21_{-0.05}^{+0.03}$ & $-0.03_{-0.02}^{+0.01}$ \\ [0.1cm]
 Jitter (m/s) & $2.1_{-0.2}^{+0.3}$ & $9 \pm 1$ & $10.4_{-0.5}^{+0.6}$ & $3.9_{-0.4}^{+0.5}$ & $2.3 \pm 0.1$ & $1.5 \pm 0.2$ & $2.4 \pm 0.2$ & $3.4 \pm 0.2$ & $6.7_{-0.7}^{+0.8}$ \\ [0.1cm]
 rms (m/s) & $8.0 \pm 0.1$ & $9.3_{-0.7}^{+2.2}$ & $10.7_{-0.1}^{+0.2}$ & $6.7 \pm 0.1$ & $3.7_{-0.3}^{+0.8}$ & $2.1 \pm 0.1$ & $4.3 \pm 0.1$ & $4.5_{-0.5}^{+2.7}$ & $6.9_{-0.2}^{+0.3}$ \\ [0.1cm]
\hline
 & H03 = $444.4_{-0.5}^{+0.4}$ & H03 = $-57241_{-3}^{+5}$ & H03 = $-6 \pm 2$ & H03 = $61417 \pm 1$ & H03 = $13352 \pm 1$ & H03 = $49307 \pm 1$ & H03 = $-2227 \pm 1$ & H03 = $26582_{-5}^{+4}$ & H03 = $53 \pm 1$\\ [0.1cm]
 Instrumental & H15 = $434_{-2}^{+1}$ & H15 = $-57232 \pm 4$ & C98 = $-1_{-1}^{+2}$ & H15 = $61427_{-1}^{+2}$ & H15 = $13350 \pm 1$ & & H15 = $-2209 \pm 1$ & H15 = $26591_{-5}^{+4}$ & H15 = ${65}_{-1}^{+2}$ \\ [0.1cm]
 offset (m/s) & C07 = $3 \pm 2$ & & Hir94 = $-2 \pm 3$ & C98 = $61348_{-1}^{+2}$ & AAT = $0 \pm 1$ & & UCLES = $0 \pm 1$ & AAT = $43_{-5}^{+4}$ & \\ [0.1cm]
 & & $-35 \pm 1$ & & Hir04 = $-2_{-1}^{+2}$ & UCLES = $0 \pm 1$ & PFS = $-12 \pm 1$ & & C98 = $-2276 \pm 1$ & PFS = $25_{-6}^{+4}$ & \\ [0.1cm]
\hline
\end{tabular}}
\label{table_summary_connu2}
\end{table*}
\end{landscape}

\begin{landscape}
\begin{table*}[h!]
\centering
\caption{Orbital parameters and mass of single stellar companions.}
\resizebox{1.35\textwidth}{!}{
\begin{tabular}[h!]{cccccccccc}
\hline
 Star & \emph{a} (au) & Mass (\Msun) & Ecc & \textit{I} (°) & $\Omega$ (°) & $\omega$ (°) & $P_{time}$ (years) & Methods & Discovery reference \\ 
\hline
 GJ680 & $32_{-6}^{+9}$ & $0.178 \pm 0.004$ & $0.75_{-0.13}^{+0.11}$ & $167_{-10}^{+4}$ & $171_{-14}^{+18}$ & $60_{-20}^{+22}$ & $188_{-68}^{+102}$ & RV+HG+HCI & \cite{2015MNRAS.449.2618W} \\ [0.1cm]
 HD1388 & $77_{-10}^{+13}$ & $0.474_{-0.005}^{+0.006}$ & $0.43_{-0.05}^{+0.07}$ & $91 \pm 2$ & $86 \pm 1$ & $283_{-13}^{+10}$ & $501_{-103}^{+149}$ & RV+HG+HCI & \cite{2021AJ....161..123D} \\ [0.1cm]
 HD16160 & $17.1 \pm 0.3$ & $0.099 \pm 0.004$ & $0.635 \pm 0.003$ & $47 \pm 1$ & $307 \pm 1$ & $133.5 \pm 0.3$ & $11.54 \pm 0.01$ & RV+HG & \cite{2021MNRAS.507.2856F} \\ [0.1cm]
 HD16548 & $30_{-8}^{+11}$ & $0.51_{-0.06}^{+0.07}$ & $0.66 \pm 0.06$ & $24_{-4}^{+5}$ & $125_{-10}^{+13}$ & $4_{-10}^{+9}$ & $105_{-43}^{+73}$ & RV+HG & This paper \\ [0.1cm]
 HD21175 & $43_{-4}^{+6}$ & $1.05 \pm 0.05$ & $0.23 \pm 0.13$ & $27_{-6}^{+9}$ & $93_{-61}^{+248}$ & $284_{-247}^{+48}$ & $134_{-24}^{+60}$ & RV+HG+HCI & This paper \\ [0.1cm]
 HD42659 & $0.561_{-0.008}^{+0.007}$ & $0.50 \pm 0.02$ & $0.15_{-0.01}^{+0.02}$ & fix to 90 & fix to 0 & $227_{-8}^{+9}$ & $0.05 \pm 0.01$ & RV & \cite{Hartmann_2015} \\ [0.1cm]
 HD46569 & $10.4 \pm 0.2$ & $0.29 \pm 0.01$ & $0.580 \pm 0.004$ & $45 \pm 1$ or $138 \pm 1$ & $252 \pm 3$ or $313 \pm 3$ & $278.7 \pm 0.3$ & $19.4_{-0.6}^{+0.5}$ & RV+HG & \cite{Zechmeister_2013} \\ [0.1cm]
 HD56380 & $4.79 \pm 0.07$ & $0.40 \pm 0.01$ & $0.650 \pm 0.001$ & $6.1 \pm 0.1$ & $314 \pm 1$ & $24.0 \pm 0.1$ & $4.758_{-0.001}^{+0.002}$ & RV+HG & This paper \\ [0.1cm]
 HD61383 & $12.7 \pm 0.2$ & $0.227_{-0.006}^{+0.007}$ & $0.33 \pm 0.02$ & $87 \pm 4$ & $222 \pm 2$ & $169_{-1}^{+2}$ & $28.4 \pm 0.2$ & RV+HG & This paper \\ [0.1cm]
 HD78612 & $18.2_{-1.6}^{+3.5}$ & $0.29 \pm 0.01$ & $0.42_{-0.17}^{+0.13}$ & $48_{-9}^{+11}$ & $95.9 \pm 0.8$ & $11_{-9}^{+22}$ & $37_{-8}^{+17}$ & RV+HG+HCI & This paper \\ [0.1cm]
 HD96116 & $23_{-3}^{+5}$ & $0.41_{-0.05}^{+0.08}$ & $0.40_{-0.08}^{+0.07}$ & $55 \pm 10$ & $86_{-3}^{+2}$ & $67_{-16}^{+22}$ & $73_{-16}^{+26}$ & RV+HG & \cite{2017MNRAS.472.3425D} \\ [0.1cm]
 HD101198 & $26_{-1}^{+2}$ & $0.50 \pm 0.02$ & $0.64_{-0.12}^{+0.09}$ & $108 \pm 3$ & $227 \pm 5$ & $147 \pm 6$ & $70_{-6}^{+8}$ & RV+HG+HCI & \cite{Borgniet_2017} \\ [0.1cm]
 HD107094 & $54_{-8}^{+9}$ & $0.28 \pm 0.1$ & $0.81 \pm 0.03$ & $95.4 \pm 0.4$ & $278_{-6}^{+5}$ & $257 \pm 1$ & $357_{-72}^{+93}$ & RV+HG & \cite{Santos_2011} \\ [0.1cm]
 HD111031 & $21.1 \pm 0.6$ & $0.129 \pm 0.003$ & $0.71 \pm 0.01$ & $137_{-3}^{+4}$ & $325_{-1}^{+2}$ & $42_{-2}^{+3}$ & $53_{-3}^{+4}$ & RV+HG+HCI & \cite{2022ApJS..262...21F} \\ [0.1cm]
 HD131664 & $3.4 \pm 0.1$ & $0.132 \pm 0.006$ & $0.69 \pm 0.01$ & $9.3 \pm 0.3$ or $170.4 \pm 0.3$ & $268 \pm 1$ or $182 \pm 2$ & $151.7_{-0.4}^{+0.5}$ & $0.01 \pm 0.01$ & RV+HG & \cite{Moutou_2009} \\ [0.1cm]
 HD179346 & $9.4 \pm 0.2$ & $0.192 \pm 0.007$ & $0.624 \pm 0.002$ & $67 \pm 1$ & $150_{-1}^{+2}$ & $126.3_{-0.1}^{+0.2}$ & $7.413 \pm 0.003$ & RV+HG & This paper \\ [0.1cm]
 HD191797 & $5.4 \pm 0.1$ & $0.139 \pm 0.005$ & $0.78 \pm 0.01$ & $91 \pm 6$ & $315 \pm 1$ & $212 \pm 1$ & $0.46 \pm 0.01$ & RV+HG & This paper \\ [0.1cm]
 HD195564 & $73_{-12}^{+17}$ & $0.42_{-0.01}^{+0.02}$ & $0.94_{-0.01}^{+0.02}$ & $61_{-9}^{+7}$ & $263_{-11}^{+24}$ & $270 \pm 10$ & $29.0_{-0.7}^{+1.0}$ & RV+HG+HCI & \cite{2021AJ....161..123D} \\ [0.1cm]
 HD207700 & $40_{-9}^{+7}$ & $0.13 \pm 0.01$ & $0.28_{-0.19}^{+0.29}$ & $154_{-16}^{+10}$ & $79_{-30}^{24}$ & $318_{-66}^{+75}$ & $77_{-32}^{+90}$ & RV+HG+HCI & This paper \\ [0.1cm]
 HD212036 & $4.1 \pm 0.1$ & $0.164_{-0.005}^{+0.006}$ & $0.51 \pm 0.01$ & $104_{-7}^{+4}$ & $35_{-7}^{+4}$ & $181.11_{-0.05}^{+0.06}$ & $6.886 \pm 0.001$ & RV+HG & This paper \\ [0.1cm]
 HD221146 & $38.0_{-1.3}^{+3.3}$ & $0.633_{-0.008}^{+0.009}$ & $0.89_{-0.07}^{+0.02}$ & $142_{-15}^{+12}$ & $350_{-69}^{+18}$ & $126_{-66}^{+10}$ & $172_{-10}^{+23}$ & RV+HG+HCI & \cite{2021AJ....161..106B} \\ [0.1cm]
 HD221638 & $7.2 \pm 0.2$ & $0.108 \pm 0.005$ & $0.68 \pm 0.02$ & $149 \pm 1$ & $124_{-1}^{+2}$ & $196.7_{-0.8}^{+0.9}$ & $16.71 \pm 0.06$ & RV+HG & This paper \\ [0.1cm]
 HIP9603 & $15 \pm 3$ & $0.21 \pm 0.02$ & $0.63 \pm 0.05$ & $45 \pm 1$ & $71 \pm 5$ & $188 \pm 3$ & $12.80 \pm 0.04$ & RV+HG & This paper \\ [0.1cm]
 HIP39470 & $8.8 \pm 0.2$ or $7.6 \pm 0.1$ & $0.138_{-0.008}^{+0.007}$ or $0.147 \pm 0.005$ & $0.33_{-0.06}^{+0.05}$ or $0.26 \pm 0.04$ & $22_{-3}^{+2}$ or $170 \pm 1$ & $260 \pm 2$ or $142 \pm 2$ & $151 \pm 3$ or $150 \pm 4$ & $18.2 \pm 0.5$ or $16.4 \pm 0.4$ & RV+HG & This paper \\ [0.1cm]
 HIP113201 & $9.7 \pm 0.2$ & $0.118 \pm 0.002$ & $0.69 \pm 0.01$ & $144.8 \pm 0.4$ & $201 \pm 2$ & $317 \pm 1$ & $8.96 \pm 0.02$ & RV+HG+HCI & \cite{Biller_2022} \\ [0.1cm]
\hline
\end{tabular}}
\textbf{Notes: }HG corresponds to (\textit{Hipparcos/Gaia}) absolute astrometry data. For HD42659 B, the reported mass is the minimum mass.
\label{table_summary_stellar}
\end{table*}
\end{landscape}

\begin{figure}[h!]
 \centering
\includegraphics[width=0.95\textwidth]{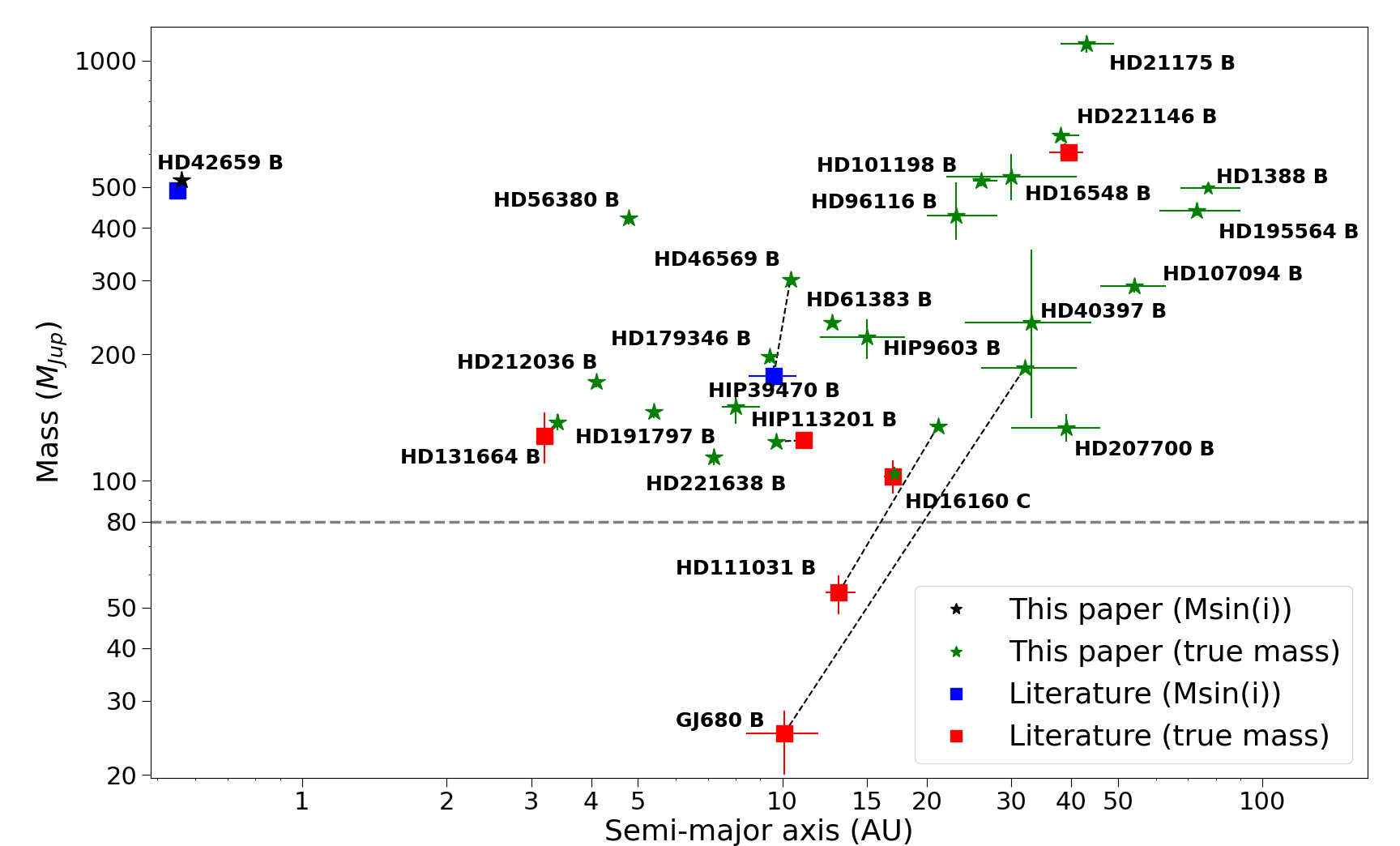}
\caption{Semi-major axis and masses of the different stellar companions. For known companions, when previous characterization is available, the dotted lines allow comparing the solutions obtained in this study with those obtained in previous studies. The grey dotted line corresponds to the hydrogen burning limit ($\sim$80 \Mjupv).
\label{sma_mass_stellar}} 
\end{figure}

\newpage

\section{additional figures}

\subsection{new companions}

\subsubsection{DPASS results}

\begin{figure}[h!]
 \centering
 \includegraphics[width=0.4\textwidth]{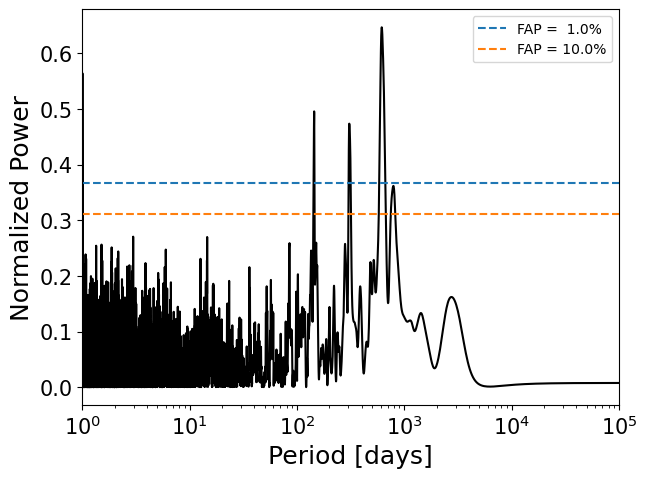}
 \includegraphics[width=0.4\textwidth]{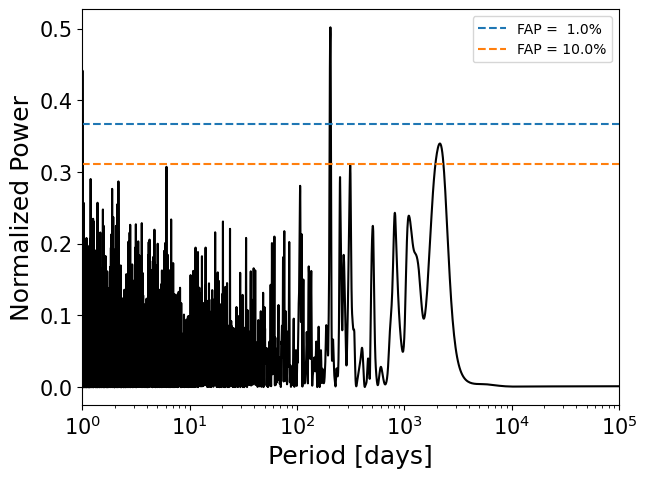}
 \includegraphics[width=0.4\textwidth]{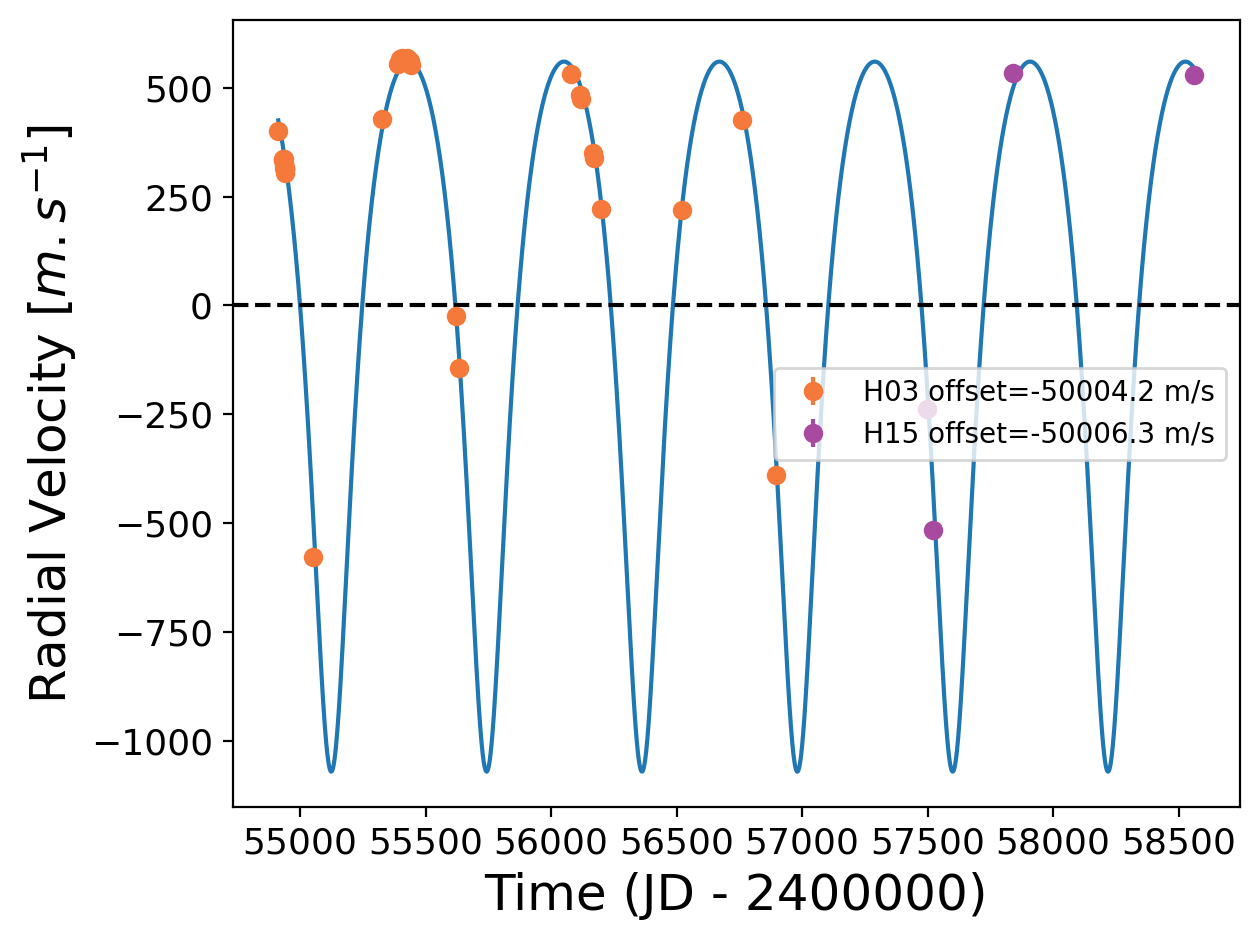}
 \includegraphics[width=0.4\textwidth]{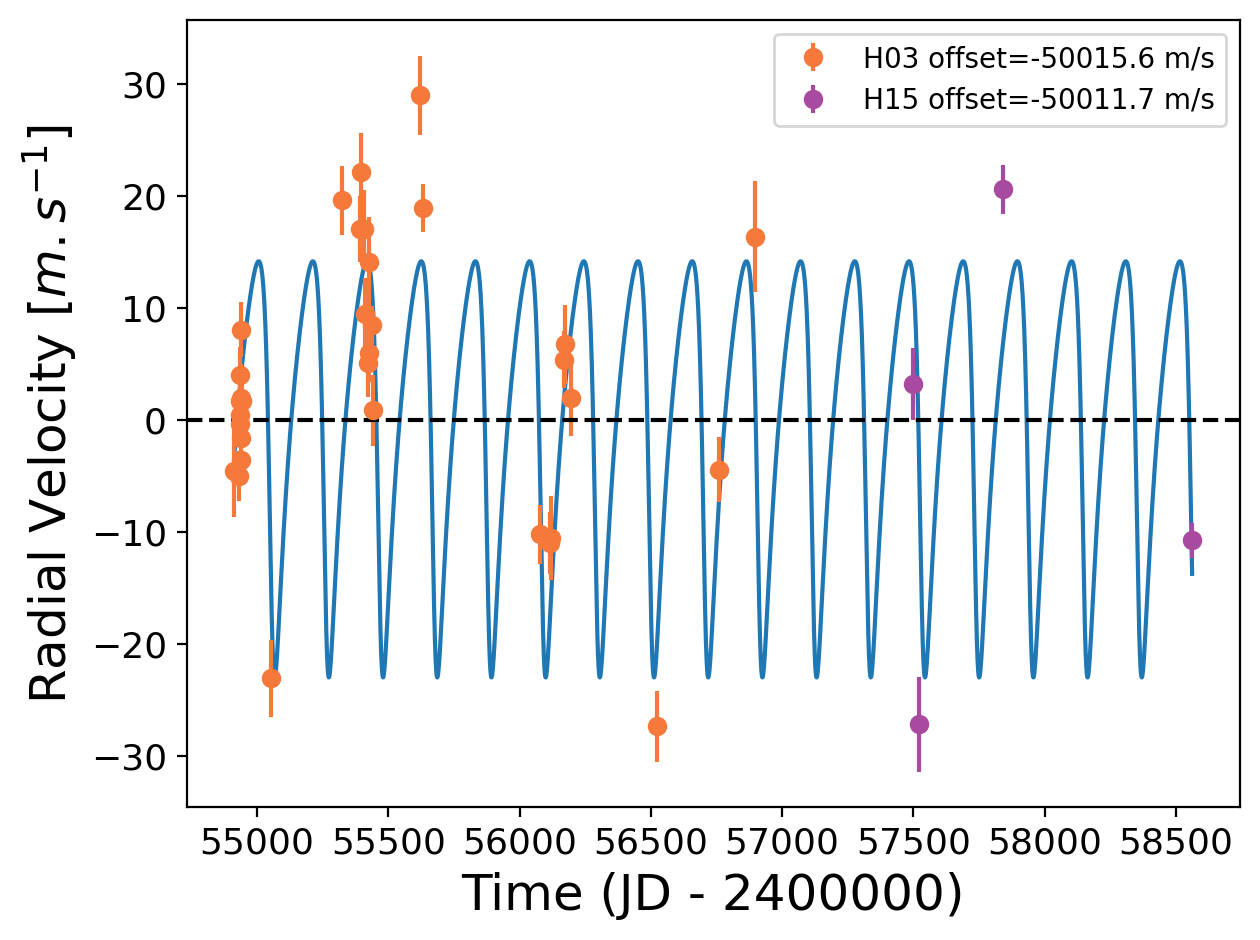}
 \caption{Periodograms of GJ660.1. \textit{Top left}: Periodogram of the GJ660.1 A RV measurements. \textit{Top right}: Periodogram of the RV residuals. \textit{Bottom left}: Best fit of the GJ660.1 A RV measurements obtained with DPASS. \textit{Bottom left}: Best fit of the GJ660.1 A RV measurements, corrected for the signal of GJ660.1 C, obtained with DPASS.
 \label{RV_DPASS_GJ660_1}} 
 \end{figure}

\begin{figure}[h!]
 \centering
\includegraphics[width=0.4\textwidth]{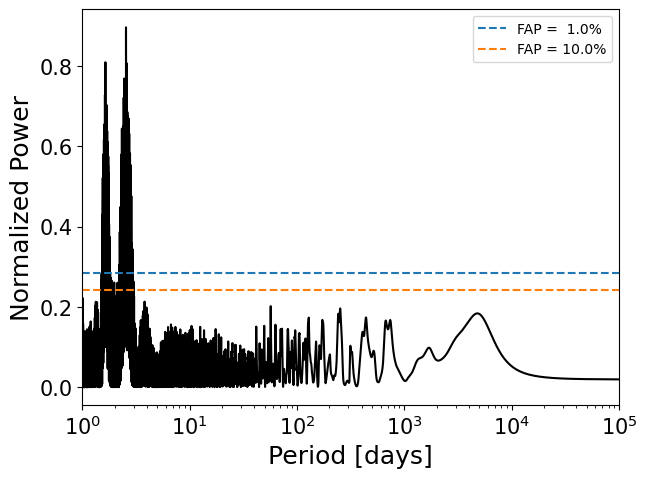}
\includegraphics[width=0.4\textwidth]{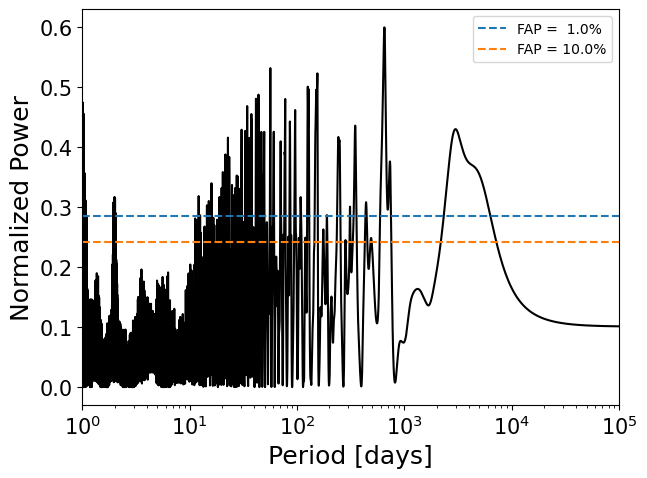}
\includegraphics[width=0.4\textwidth]{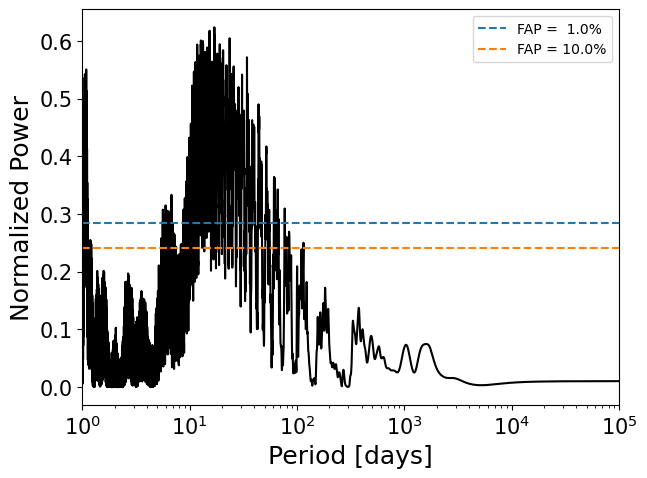}
\includegraphics[width=0.4\textwidth]{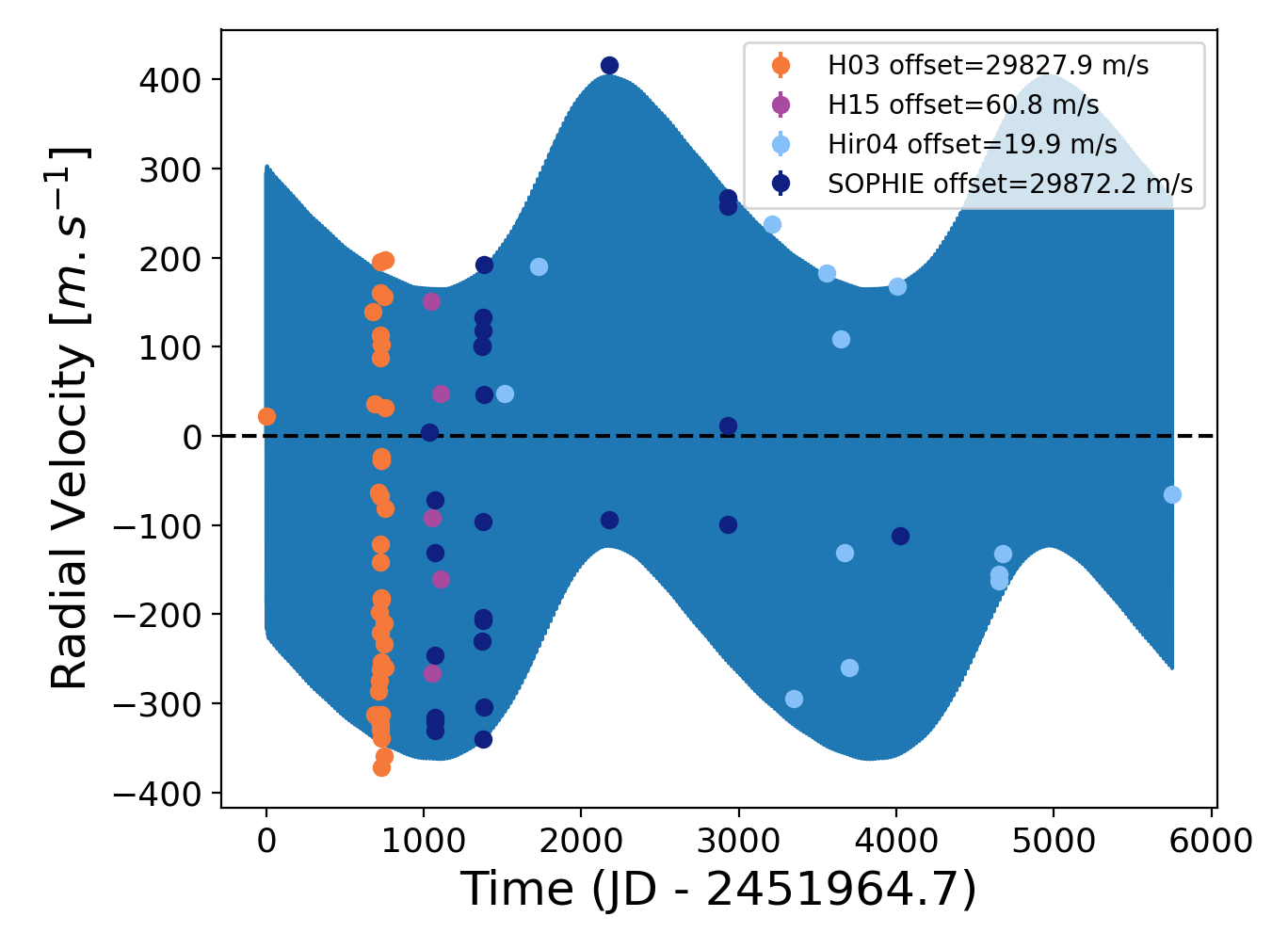}
\caption{Periodograms of HD73256. \textit{Top left}: Periodogram of the HD73256 RV measurements. \textit{Top right}: Periodogram of the HD73256 RV measurements, corrected for the signal of HD73256 b. \textit{Bottom left}: Periodogram of the RV residuals. \textit{Bottom right}: Fit of the HD73256 RV measurements obtained with DPASS, considering a two companions orbital fit with a period of 2.5 d and 2785 d, respectively.
\label{RV_DPASS_HD73256}} 
\end{figure}

\begin{figure}[h!]
 \centering
\includegraphics[width=0.32\textwidth]{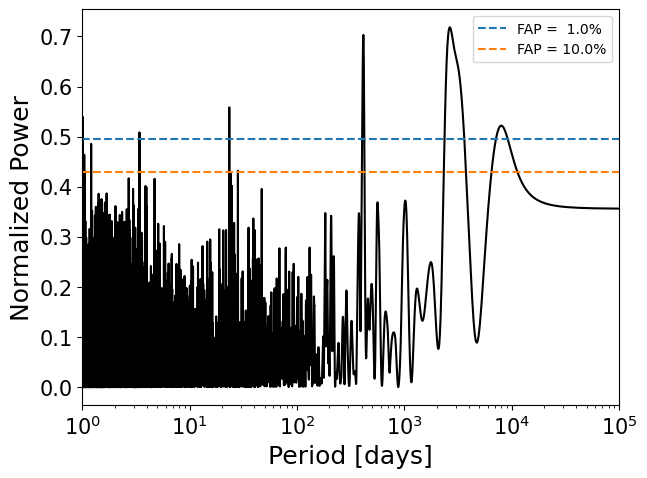}
\includegraphics[width=0.32\textwidth]{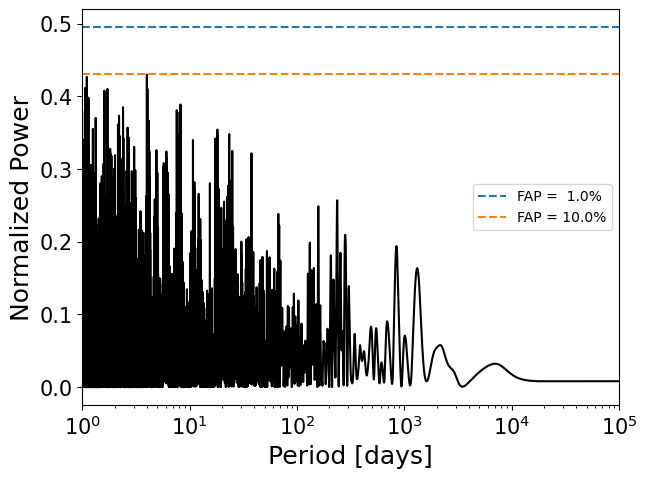}
\includegraphics[width=0.32\textwidth]{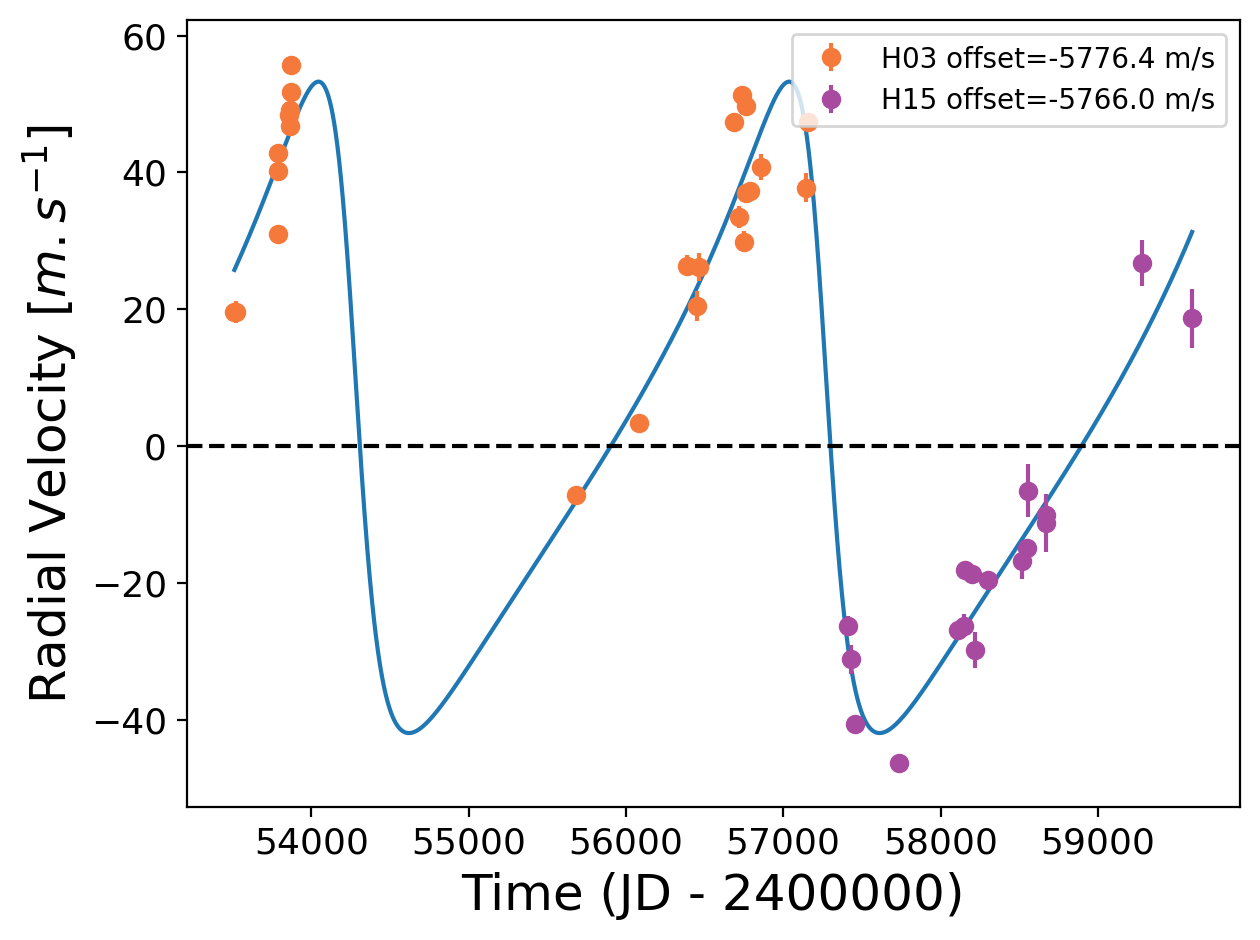}
\caption{Periodograms of HD108202. \textit{Left}: Periodogram of the HD108202 RV measurements. \textit{Middle}: Periodogram of the RV residuals. \textit{Right}: Best fit of the HD108202 RV measurements obtained with DPASS.
\label{RV_DPASS_HD108202}} 
\end{figure}

\begin{figure}[h!]
 \centering
\includegraphics[width=0.32\textwidth]{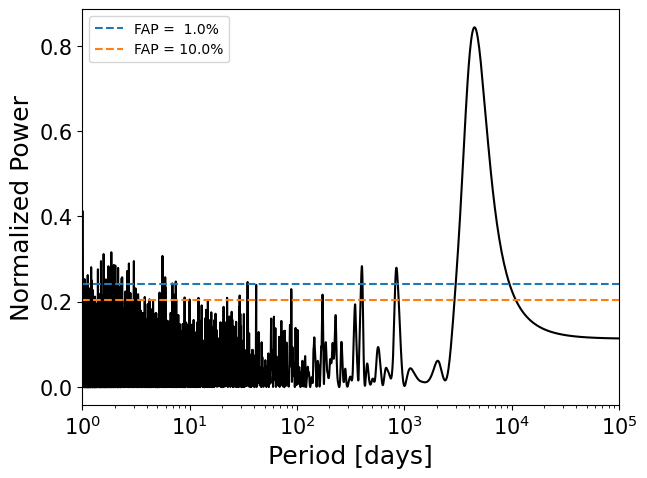}
\includegraphics[width=0.32\textwidth]{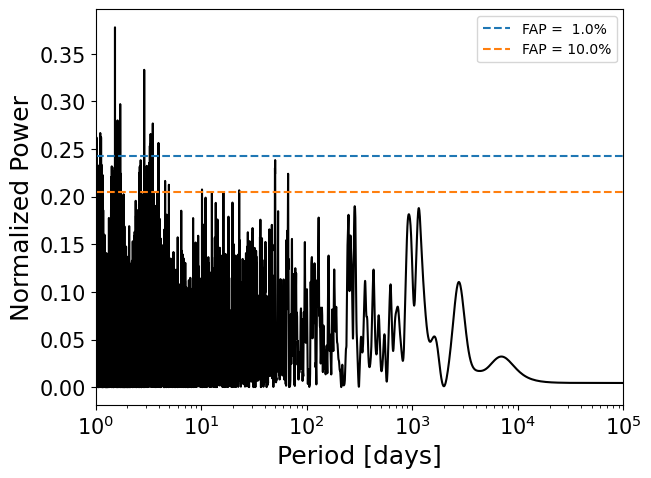}
\includegraphics[width=0.32\textwidth]{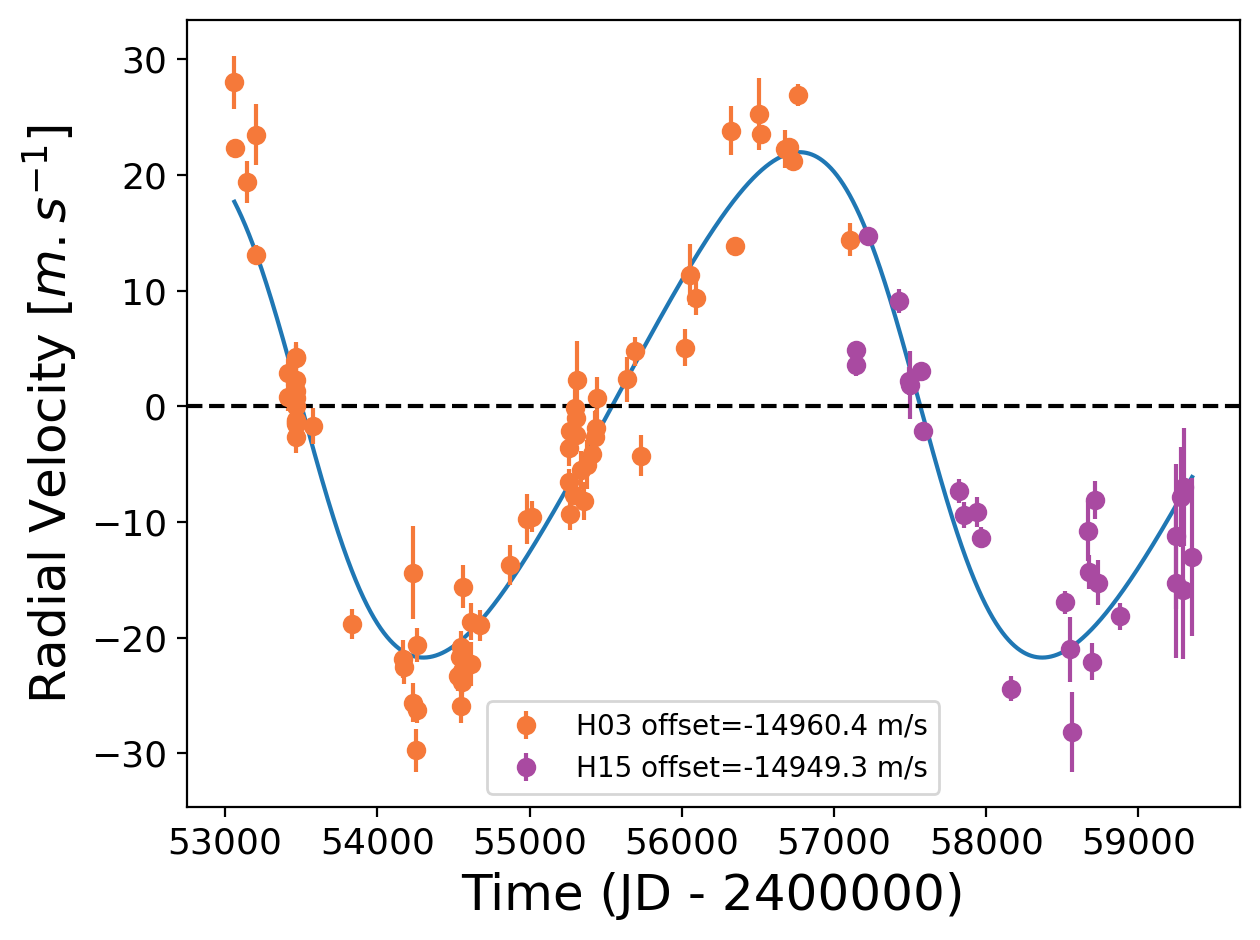}
\caption{Periodograms of HD135625. \textit{Left}: Periodogram of the HD135625 RV measurements. \textit{Middle}: Periodogram of the RV residuals. \textit{Right}: Best fit of the HD135625 RV measurements obtained with DPASS.
\label{RV_DPASS_HD135625}} 
\end{figure}

\begin{figure}[h!]
 \centering
\includegraphics[width=0.32\textwidth]{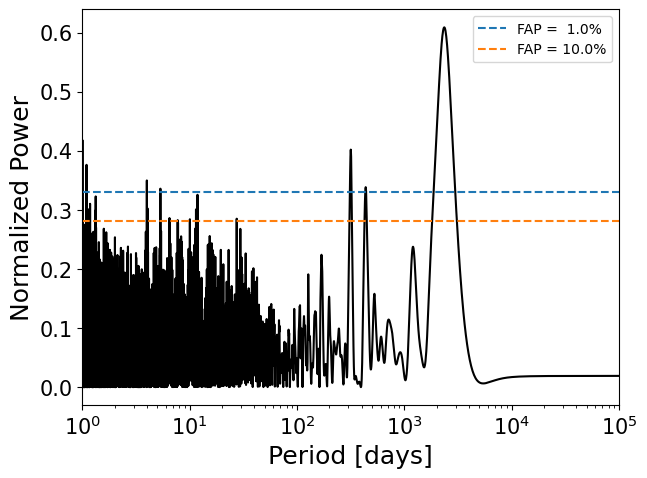}
\includegraphics[width=0.32\textwidth]{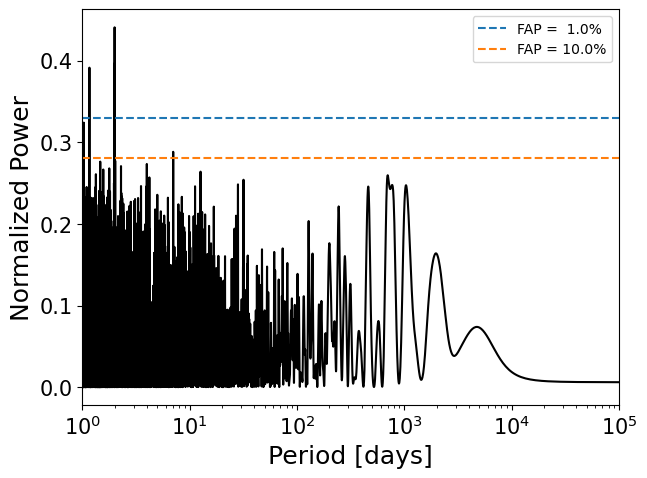}
\includegraphics[width=0.32\textwidth]{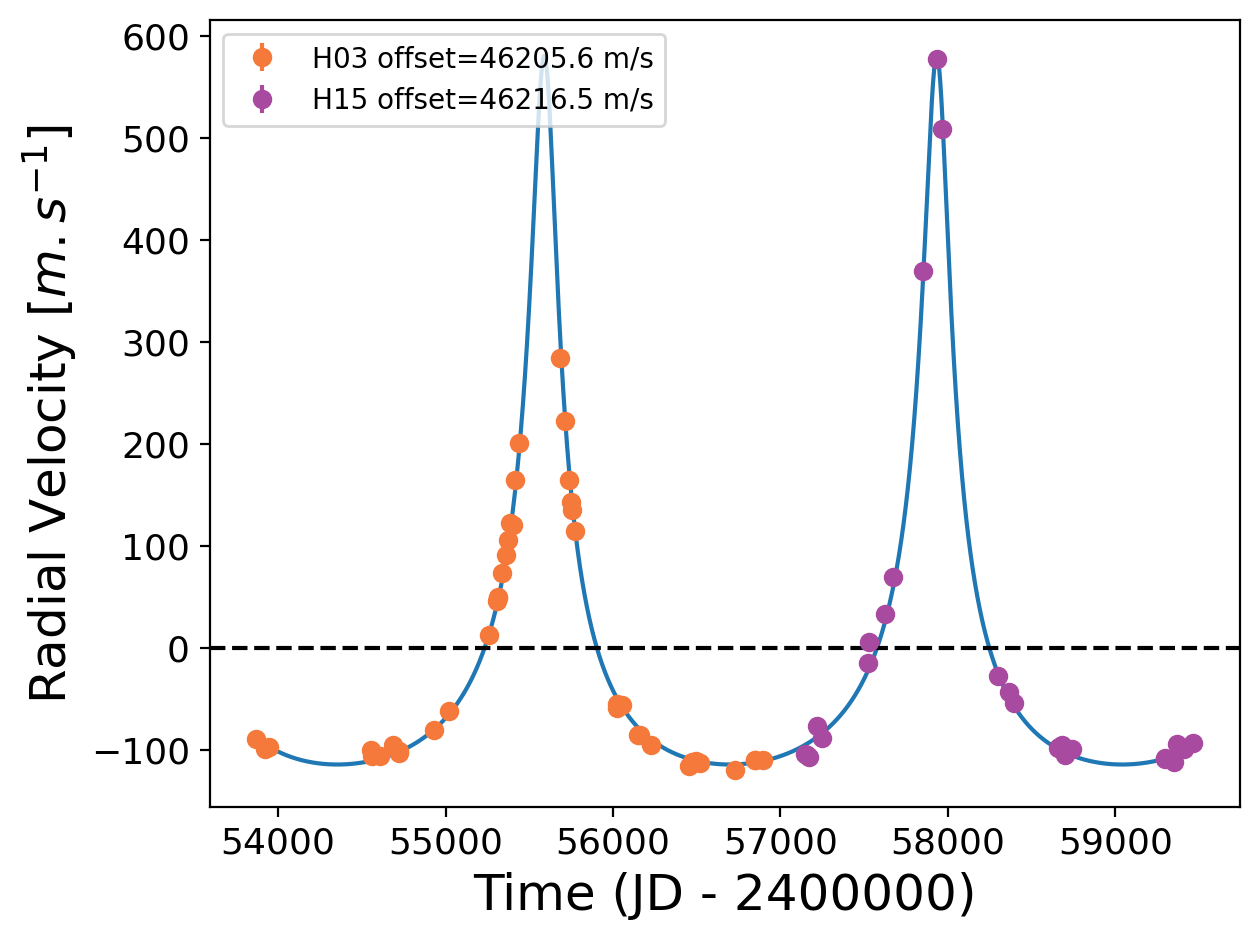}
\caption{Periodograms of HD165131. \textit{Left}: Periodogram of the HD165131 RV measurements. \textit{Middle}: Periodogram of the RV residuals. \textit{Right}: Best fit of the HD165131 RV measurements obtained with DPASS.
\label{RV_DPASS_HD165131}} 
\end{figure}

\begin{figure}[h!]
 \centering
\includegraphics[width=0.32\textwidth]{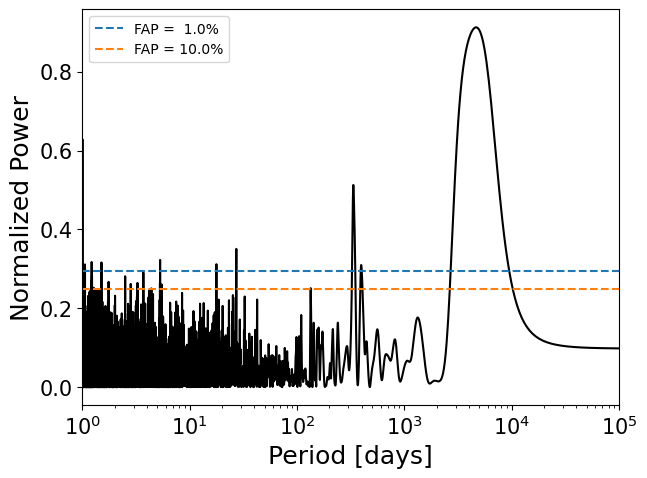}
\includegraphics[width=0.32\textwidth]{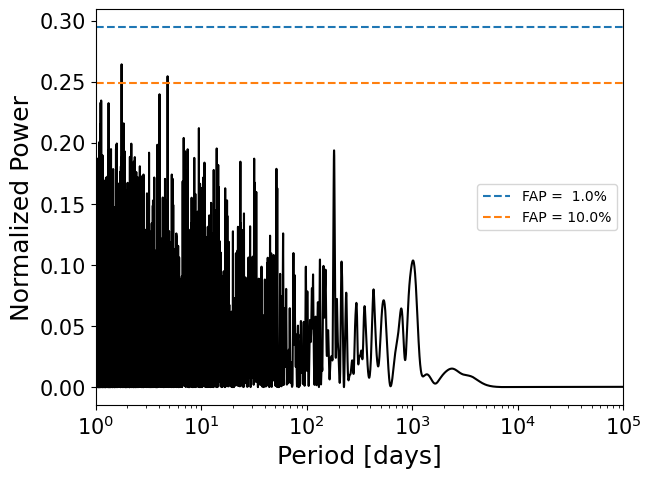}
\includegraphics[width=0.32\textwidth]{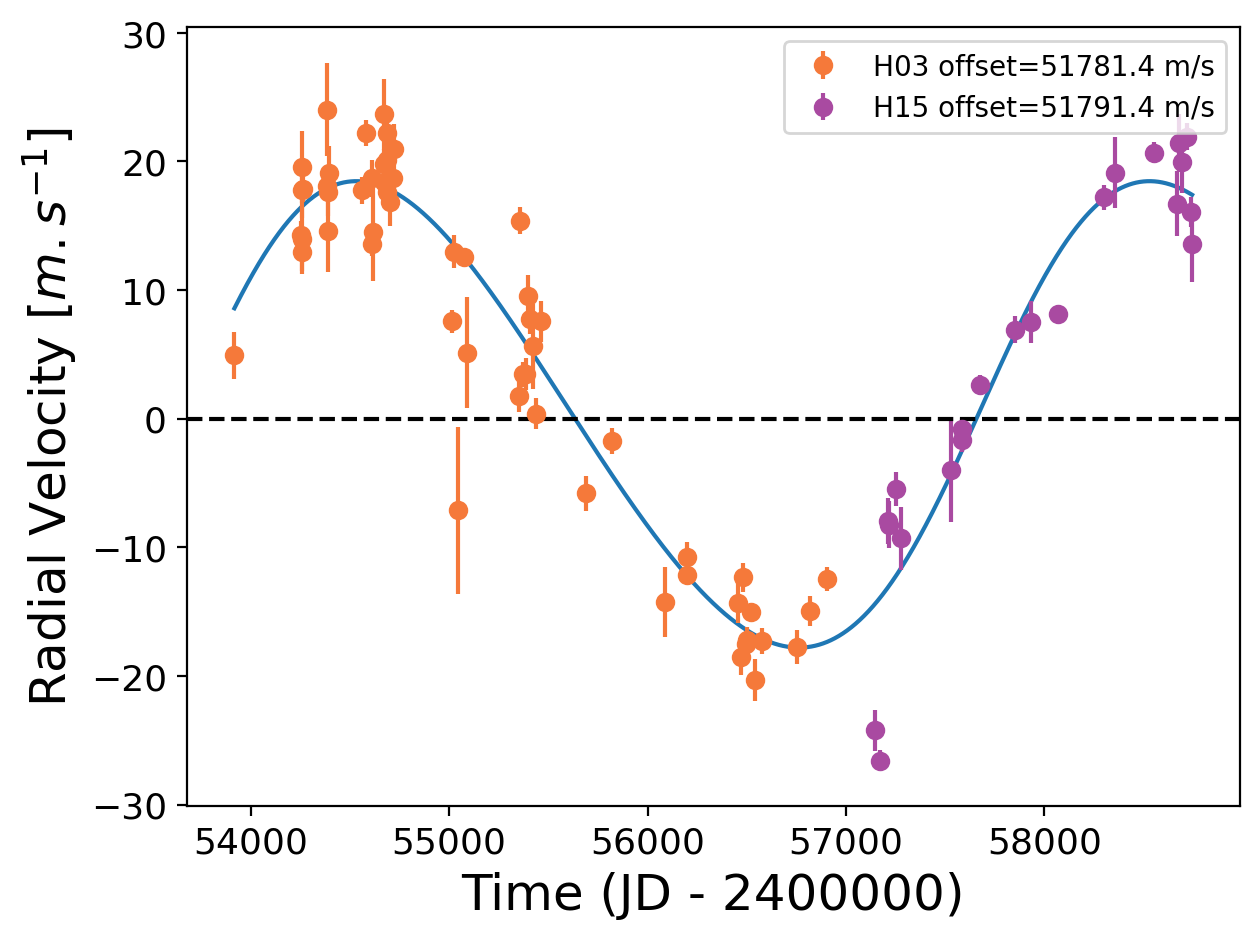}
\caption{Periodograms of HD185283. \textit{Left}: Periodogram of the HD185283 RV measurements. \textit{Middle}: Periodogram of the RV residuals. \textit{Right}: Best fit of the HD185283 RV measurements obtained with DPASS. 
\label{RV_DPASS_HD185283}} 
\end{figure}

\begin{figure}[h!]
 \centering
\includegraphics[width=0.32\textwidth]{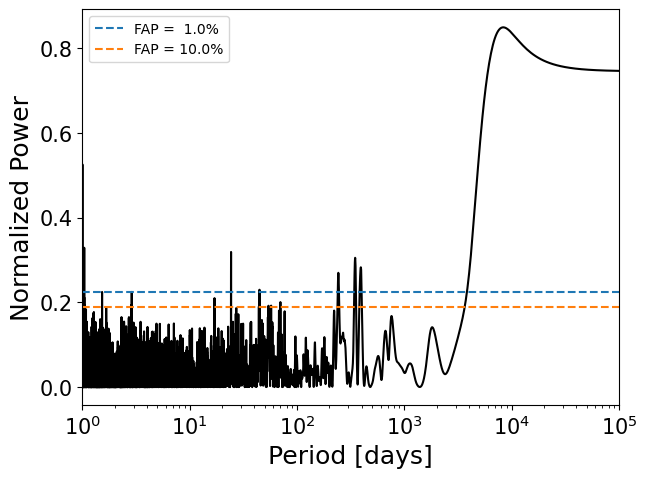}
\includegraphics[width=0.32\textwidth]{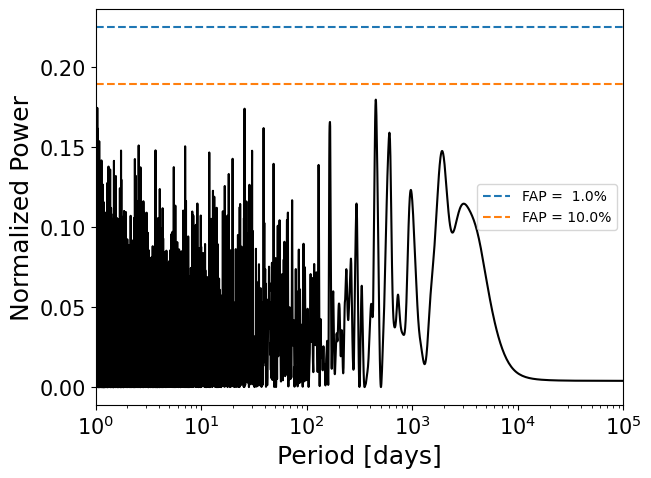}
\includegraphics[width=0.32\textwidth]{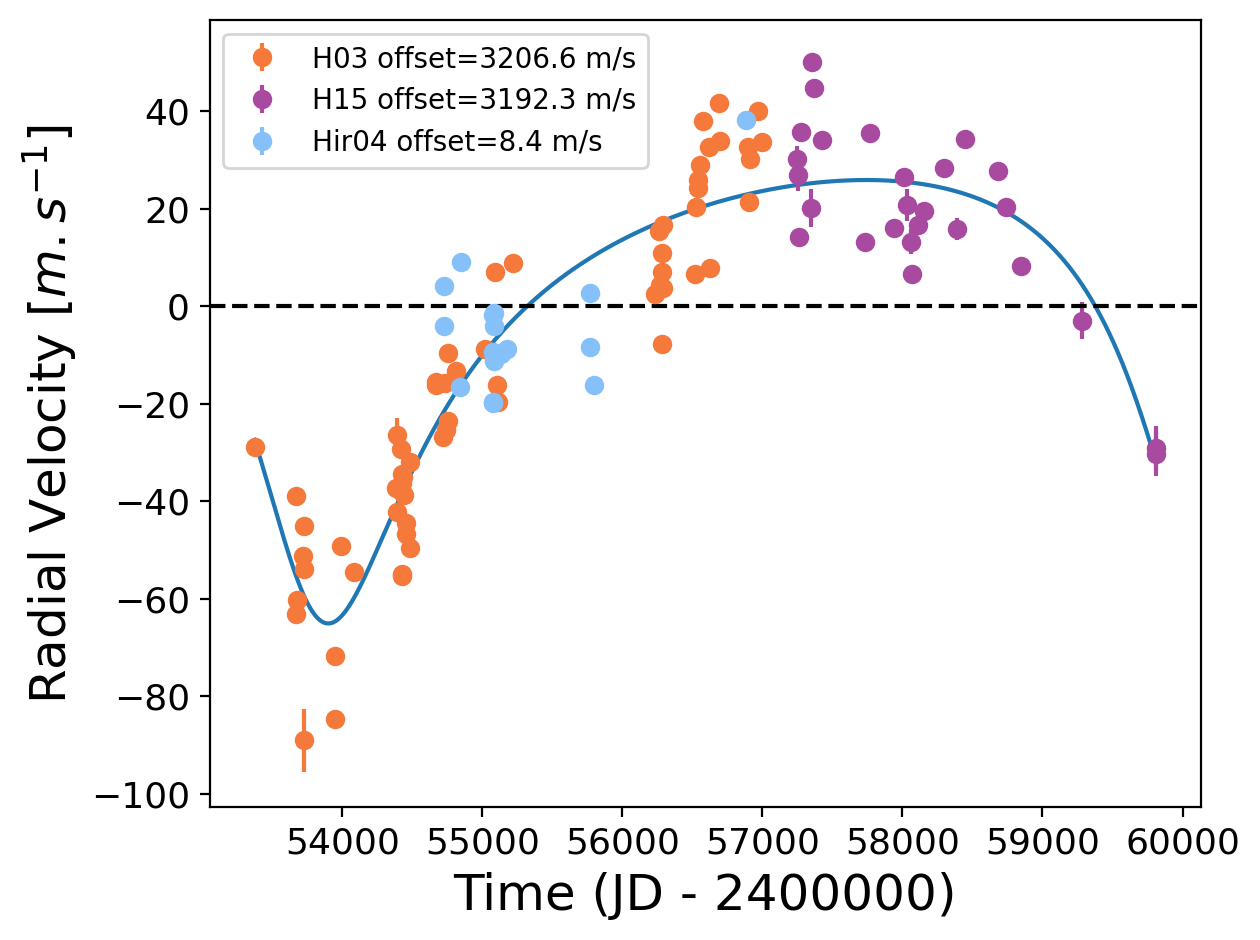}
\caption{Periodograms of HIP10337. \textit{Left}: Periodogram of the HIP10337 RV measurements. \textit{Middle}: Periodogram of the RV residuals. \textit{Right}: Best fit of the HIP10337 RV measurements obtained with DPASS. 
\label{RV_DPASS_HIP10337}} 
\end{figure}

\begin{figure}[h!]
 \centering
\includegraphics[width=0.32\textwidth]{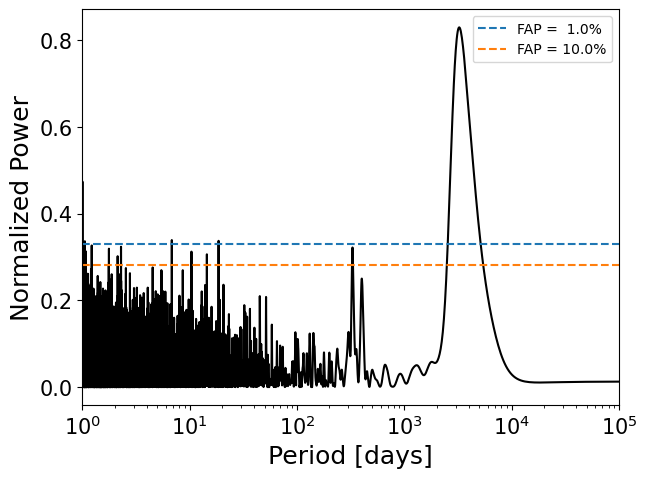}
\includegraphics[width=0.32\textwidth]{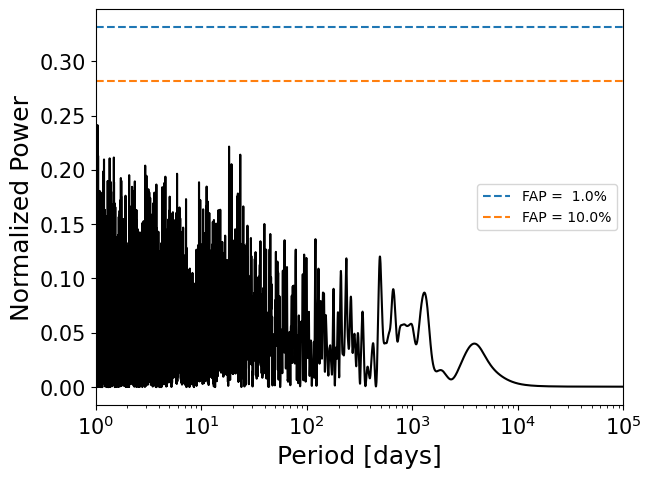}
\includegraphics[width=0.32\textwidth]{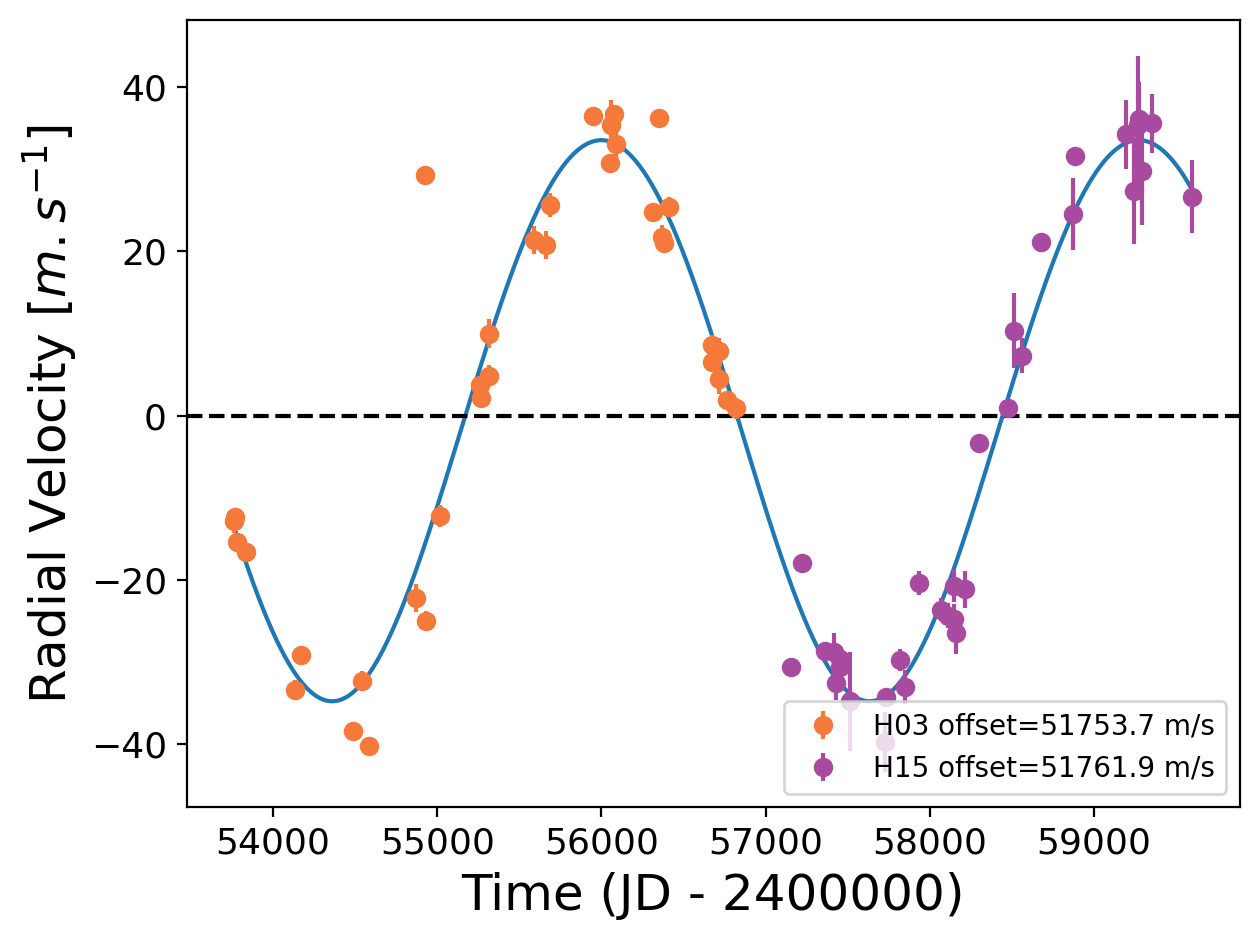}
\caption{Periodograms of HIP54597. \textit{Left}: Periodogram of the HIP54597 RV measurements. \textit{Middle}: Periodogram of the RV residuals. \textit{Right}: Best fit of the HIP54597 RV measurements obtained with DPASS. 
\label{RV_DPASS_HIP54597}} 
\end{figure}

\subsubsection{HD75302}

\begin{figure}[h!]
 \centering
\includegraphics[width=0.35\textwidth]{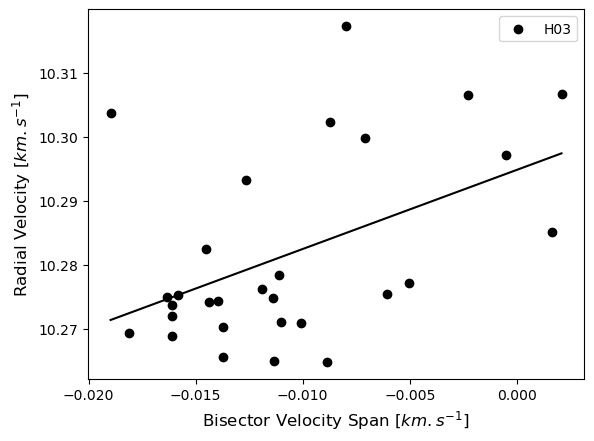}
\includegraphics[width=0.35\textwidth]{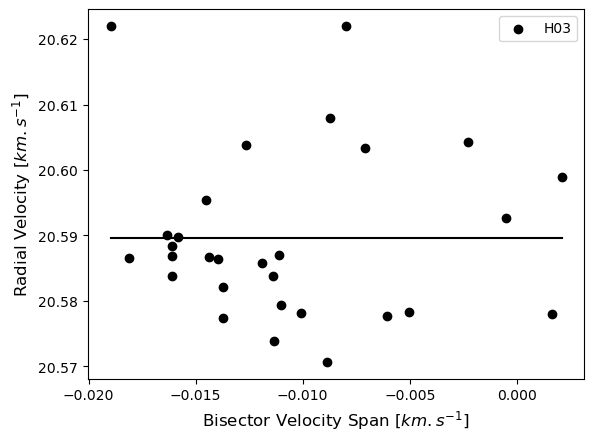}
\includegraphics[width=0.35\textwidth]{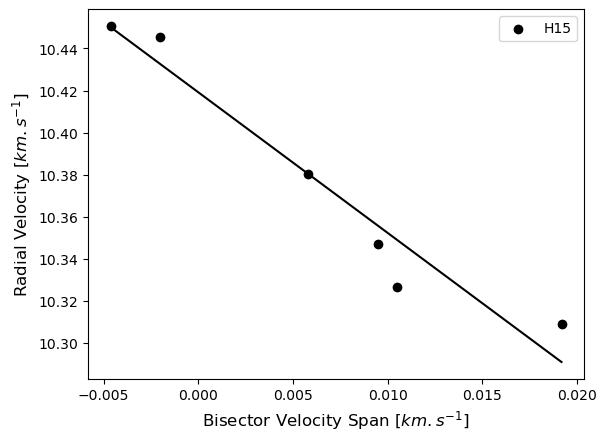}
\includegraphics[width=0.35\textwidth]{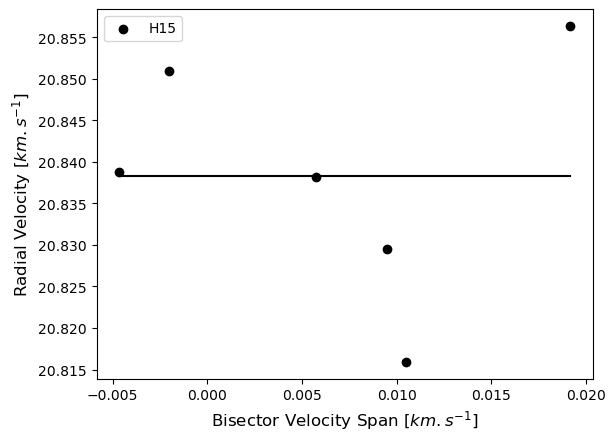}
\includegraphics[width=0.35\textwidth]{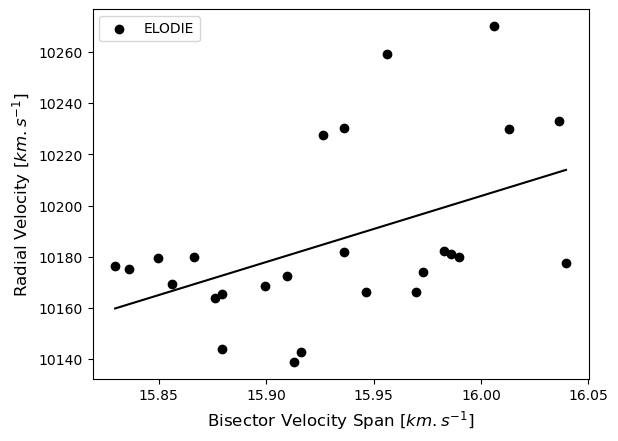}
\includegraphics[width=0.35\textwidth]{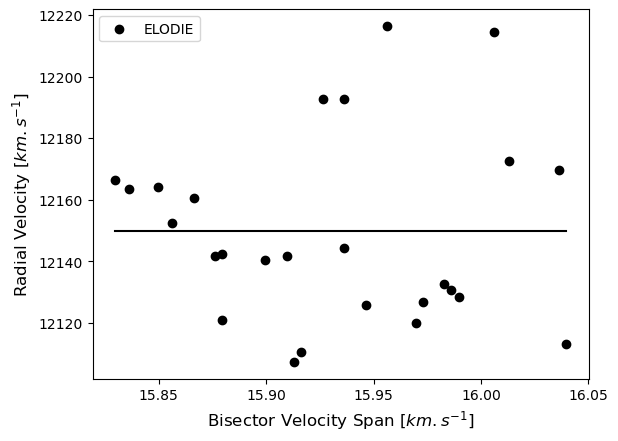}
\includegraphics[width=0.35\textwidth]{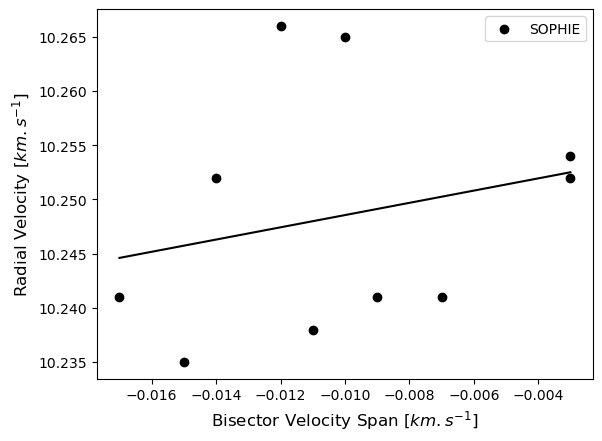}
\includegraphics[width=0.35\textwidth]{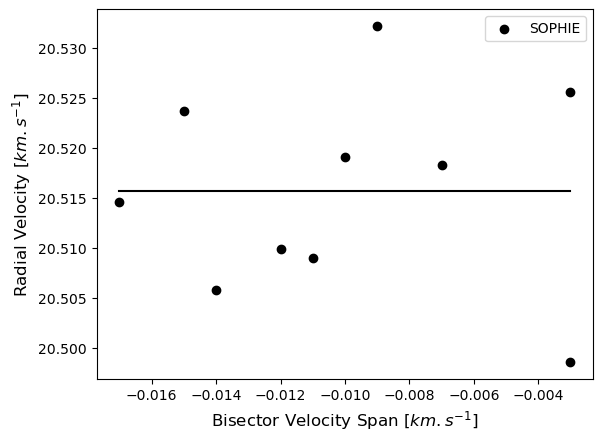}
\includegraphics[width=0.35\textwidth]{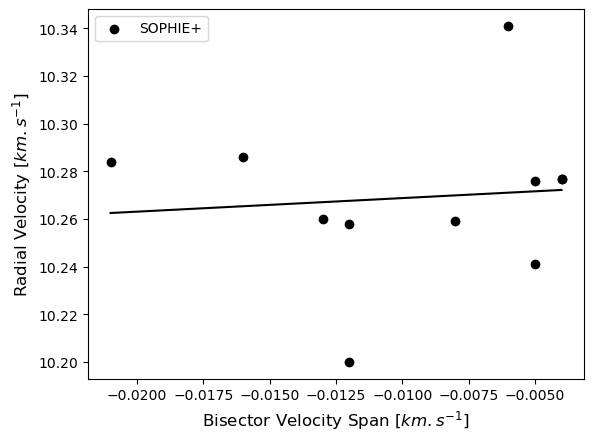}
\includegraphics[width=0.35\textwidth]{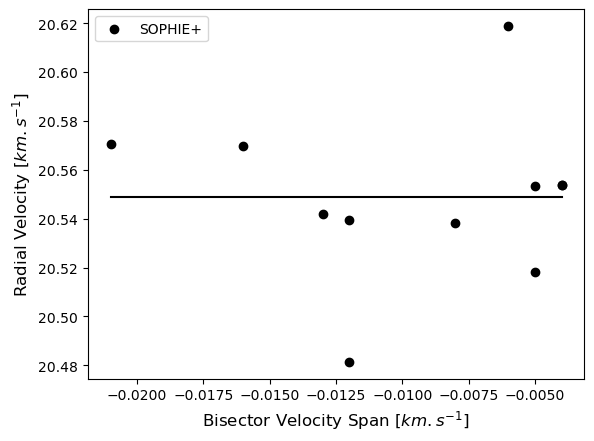}
\caption{Radial velocity vs. bisector velocity span of HD75302 before (\textit{left}) and after (\textit{right}) RV correction of the star's short-term activity. The black line corresponds to the best linear fit. The RV and BVS data, from \textit{top} to \textit{bottom}, correspond to the data obtained with the instruments H03, H15, ELODIE, SOPHIE, and SOPHIE+.
\label{RV_BVS_HD75302}} 
\end{figure}

\begin{figure}[h!]
 \centering
\includegraphics[width=0.45\textwidth]{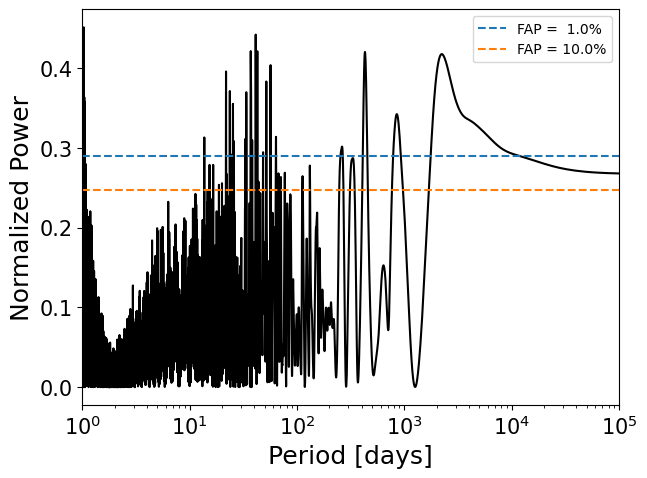}
\includegraphics[width=0.45\textwidth]{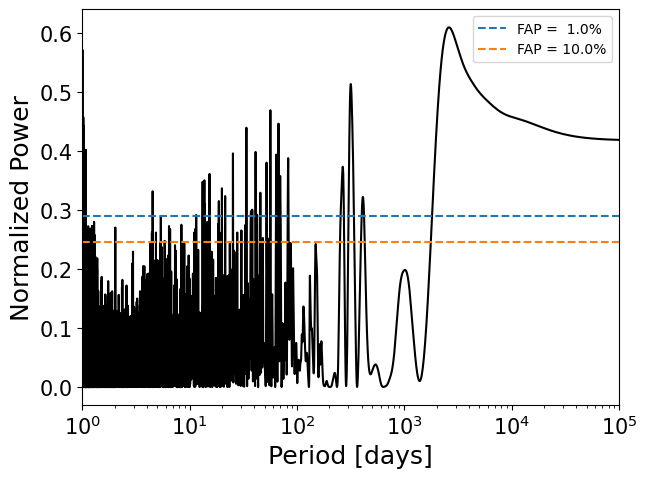}
\includegraphics[width=0.45\textwidth]{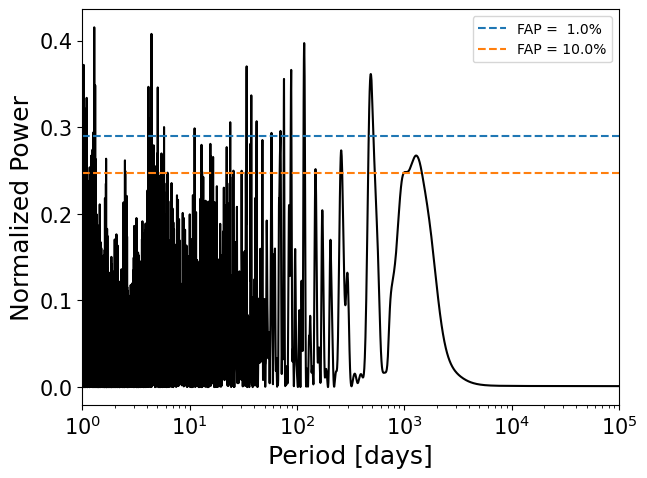}
\includegraphics[width=0.45\textwidth]{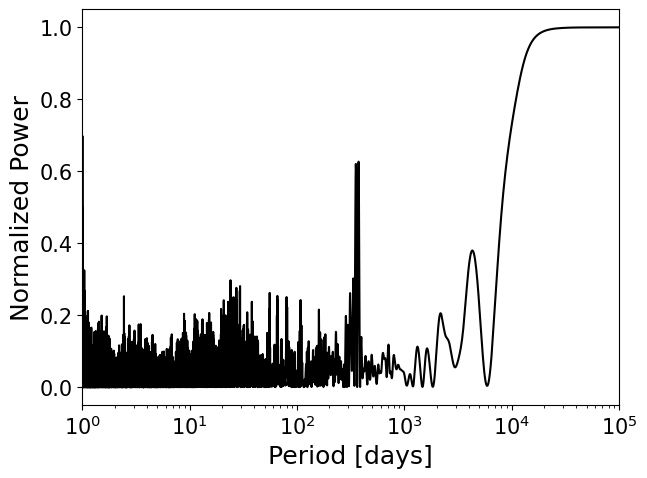}
\caption{Periodograms of HD75302. \textit{Top left}: Periodogram of the HD75302 RV measurements. \textit{Top right}: Periodogram of the HD75302 RV measurements after correction of the star's short-term activity.\textit{Bottom left}: Periodogram of the RV corrected residuals. \textit{Bottom right}: Sampling window periodogram of the RV measurements. 
\label{Perio_DPASS_HD75302}} 
\end{figure}

\begin{figure}[h!]
 \centering
\includegraphics[width=0.45\textwidth]{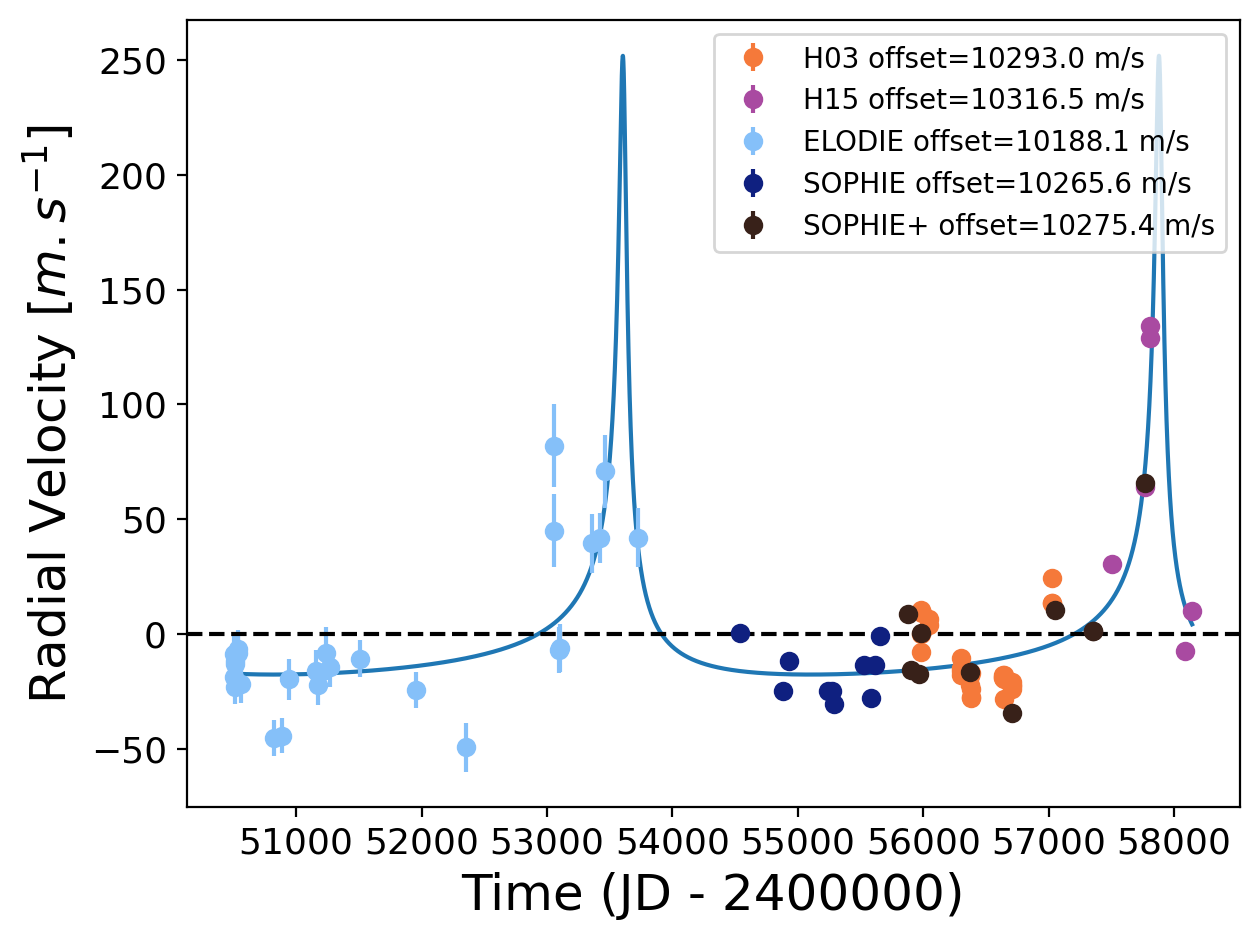}
\includegraphics[width=0.45\textwidth]{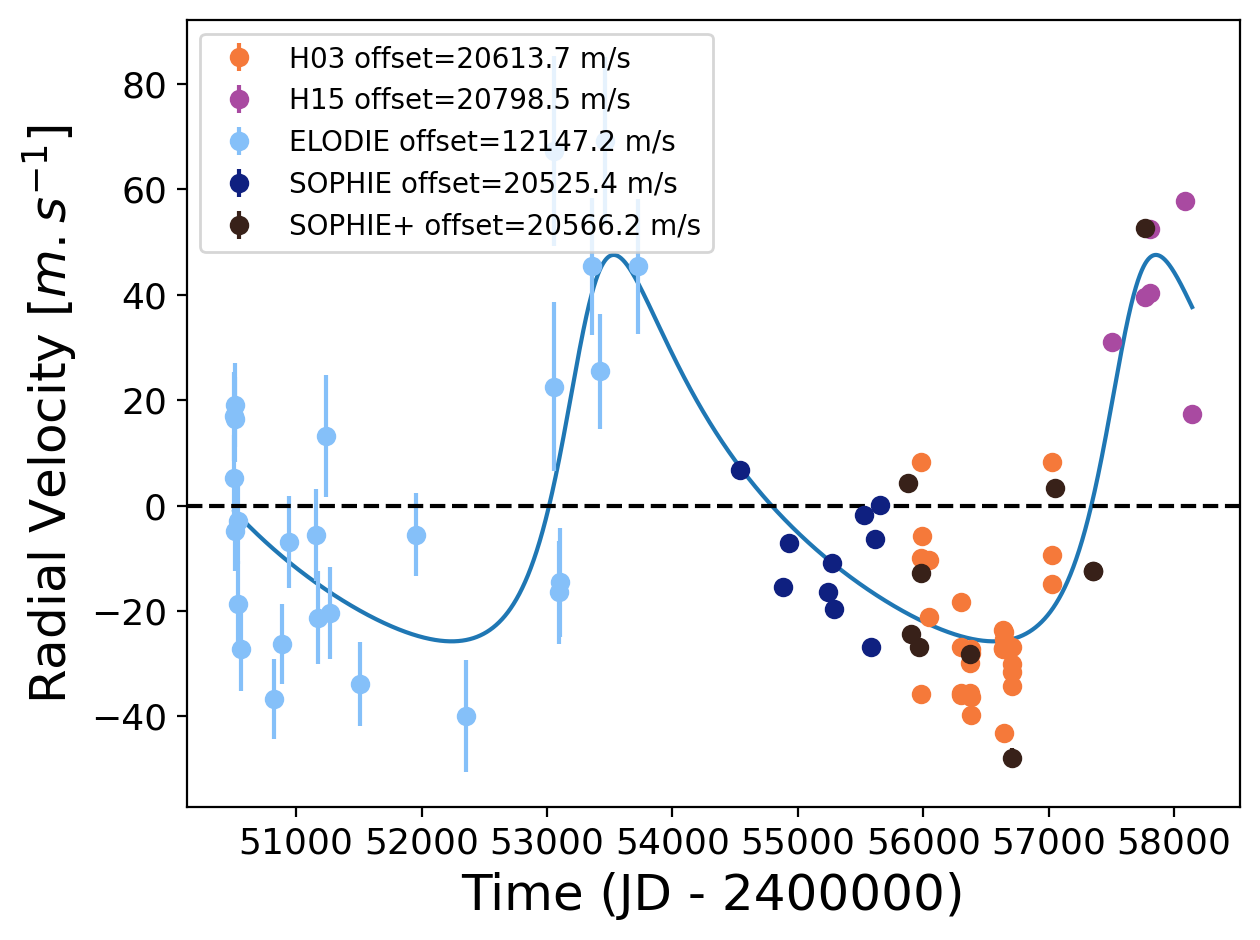}
\caption{Fit of the HD75302 RV measurements before (\textit{left}) and after (\textit{right}) correction of the star's short-term activity.
\label{RV_DPASS_HD75302}} 
\end{figure}

\newpage

\subsubsection{Corner plots}

\begin{figure}[h!]
 \centering
\includegraphics[width=1\textwidth]{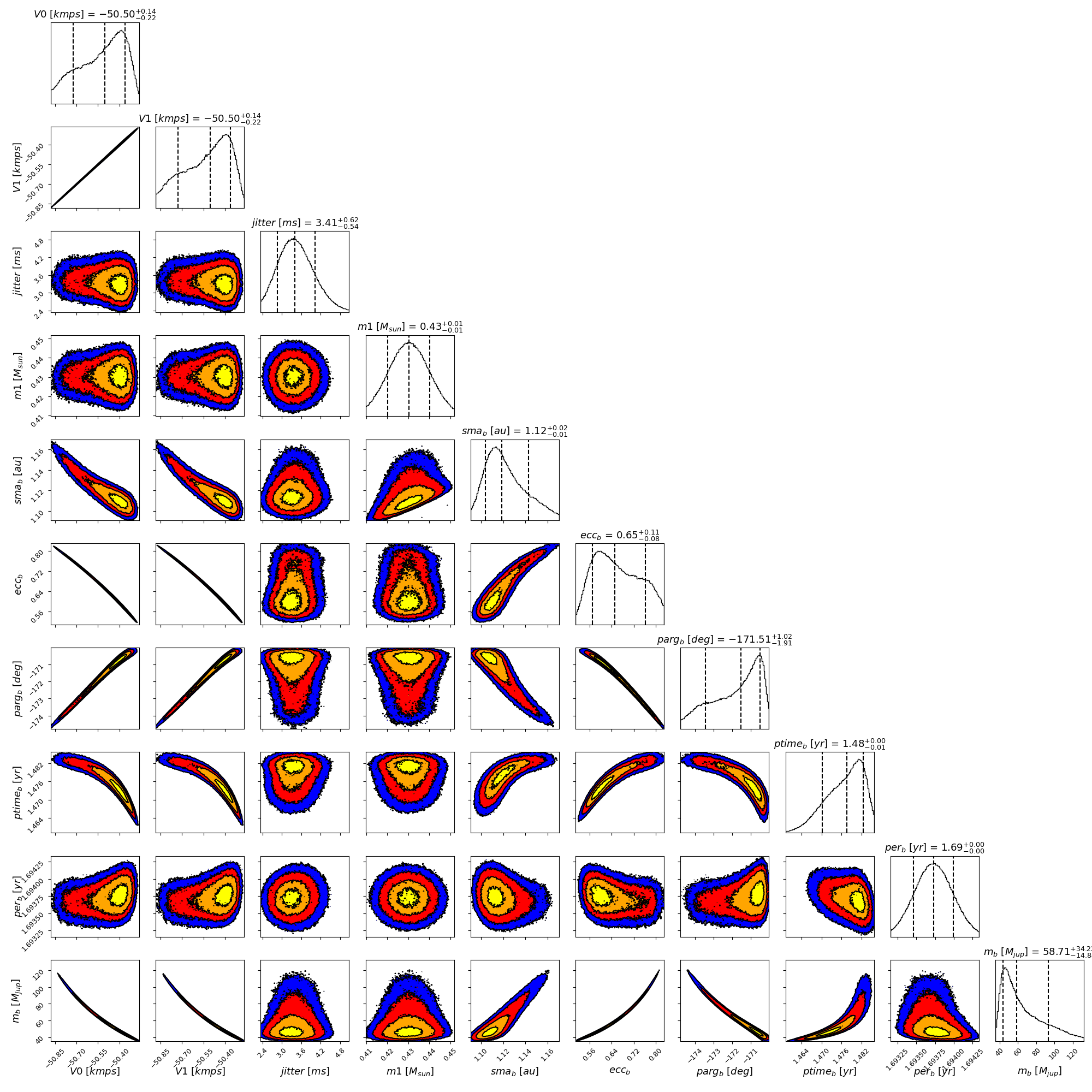}
\caption{Corner plot of the fitted parameters of GJ660.1 C. V0 and V1 correspond to H03 and H15 dataset.
\label{Corner_GJ660_1}} 
\end{figure}

\begin{figure}[h!]
 \centering
\includegraphics[width=1\textwidth]{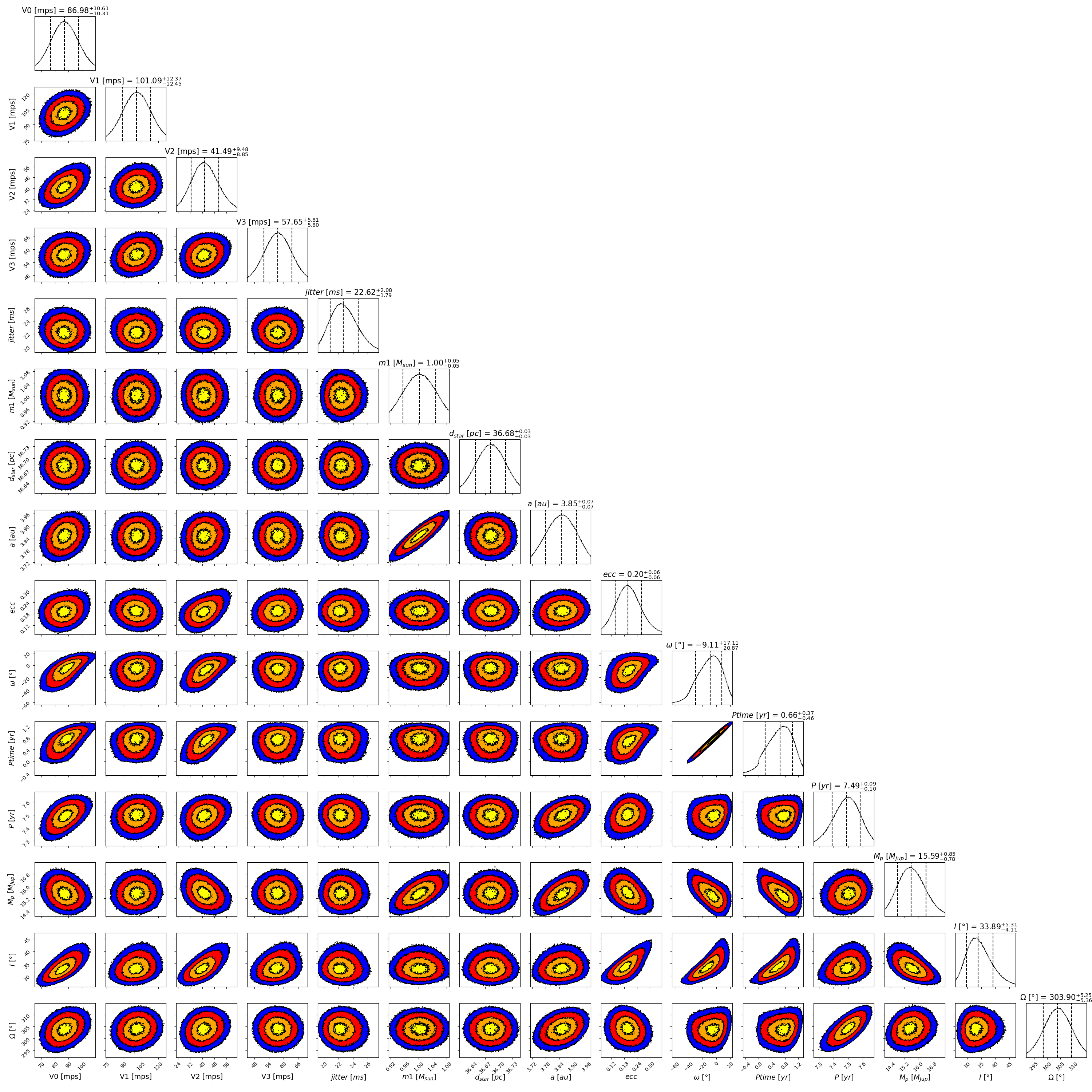}
\caption{Corner plot of the fitted parameters of HD73256 C, considering a prior on \textit{I} between 0 and 90°. V0, V1, V2, and V3 correspond to C98, Hir94, Hir04, and H03 dataset, respectively.
\label{Corner_HD73256}} 
\end{figure}

\begin{figure}[h!]
 \centering
\includegraphics[width=1\textwidth]{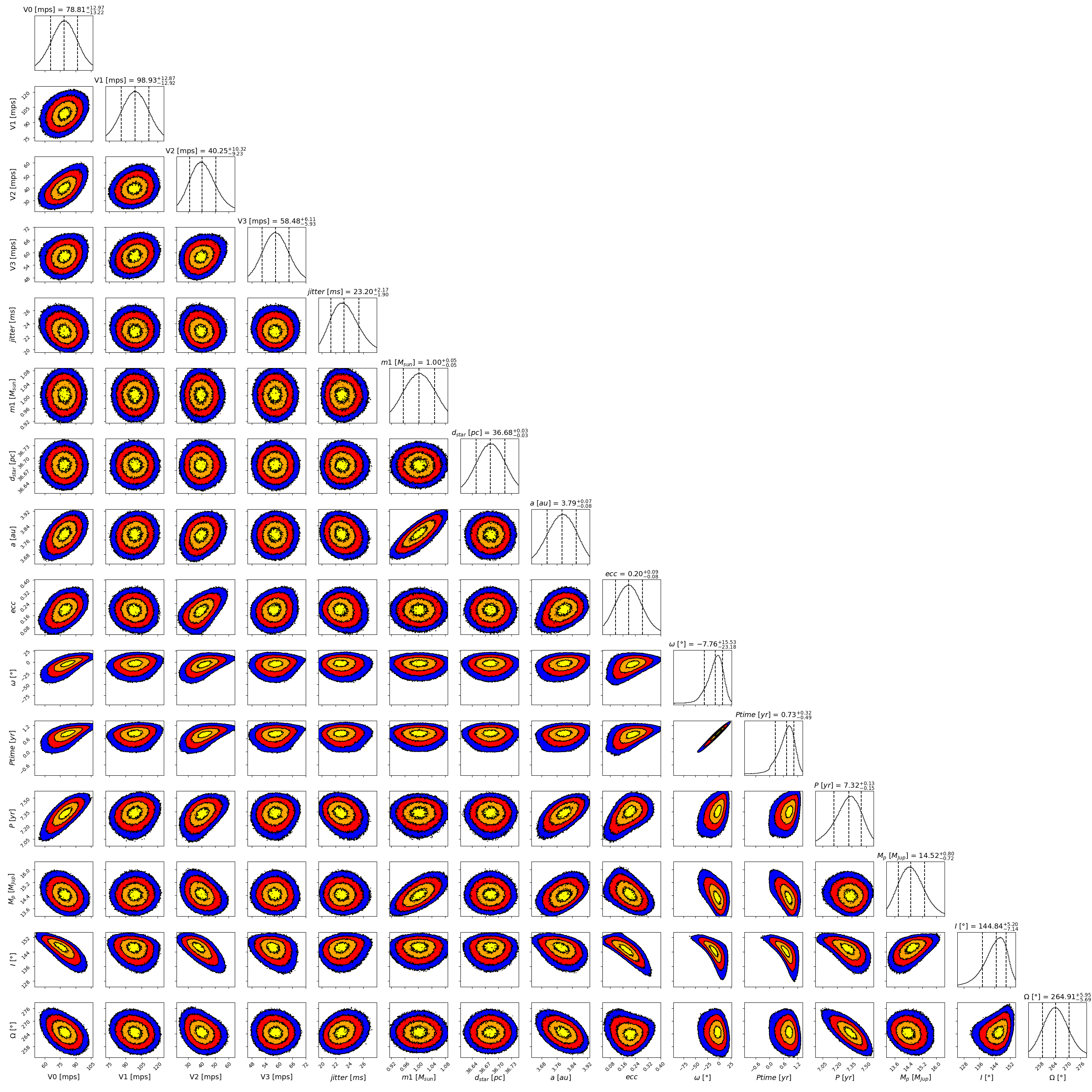}
\caption{Corner plot of the fitted parameters of HD73256 C, considering a prior on \textit{I} between 90 and 180°. V0, V1, V2, and V3 correspond to C98, Hir94, Hir04, and H03 dataset, respectively.
\label{Corner_HD73256_sup}} 
\end{figure}

\begin{figure}[h!]
 \centering
\includegraphics[width=1\textwidth]{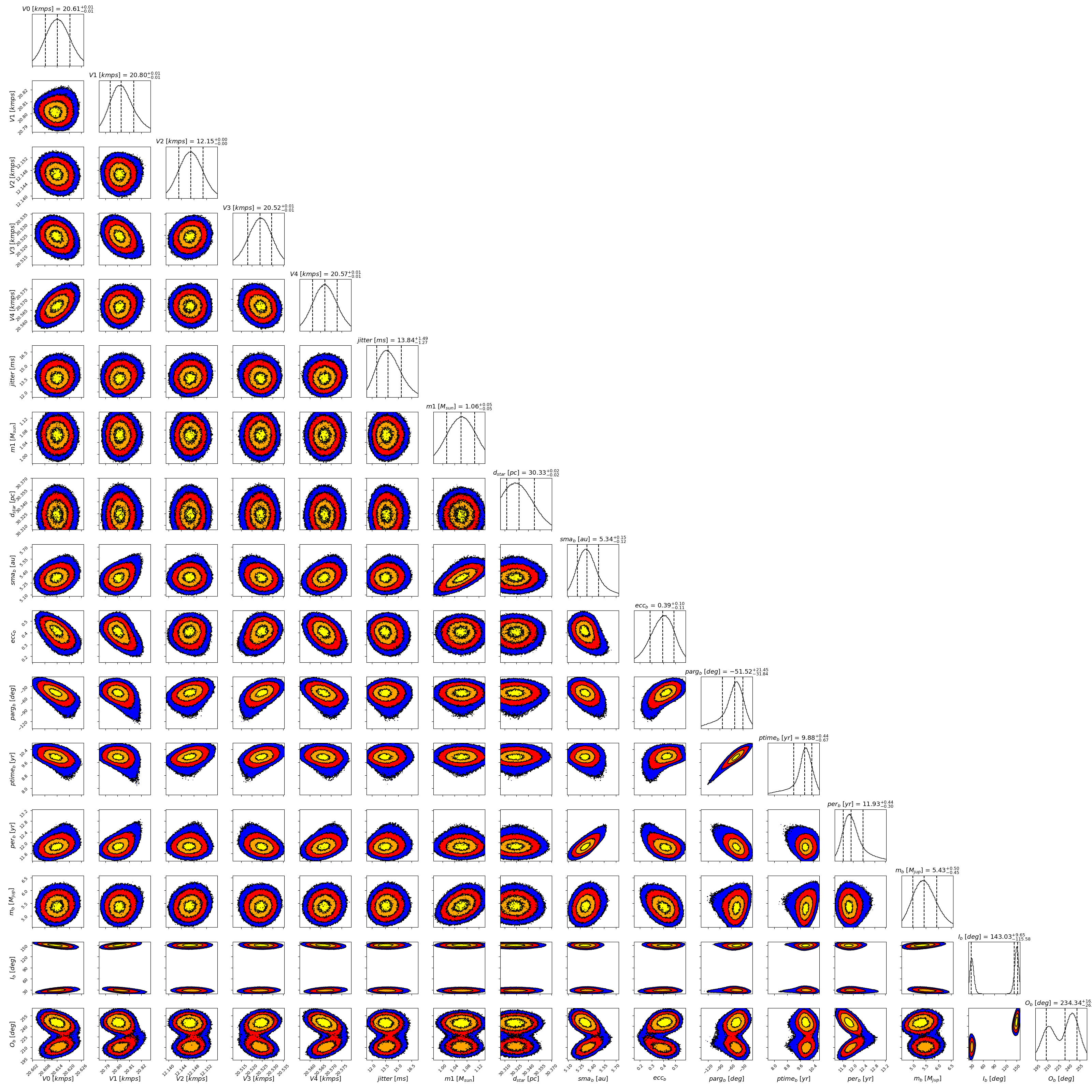}
\caption{Corner plot of the fitted parameters of HD75302 b. V0, V1, V2, V3 and V4 correspond to H03, H15, ELODIE, SOPHIE, and SOPHIE+ dataset, respectively.
\label{Corner_HD75302}} 
\end{figure}

\begin{figure}[h!]
 \centering
\includegraphics[width=1\textwidth]{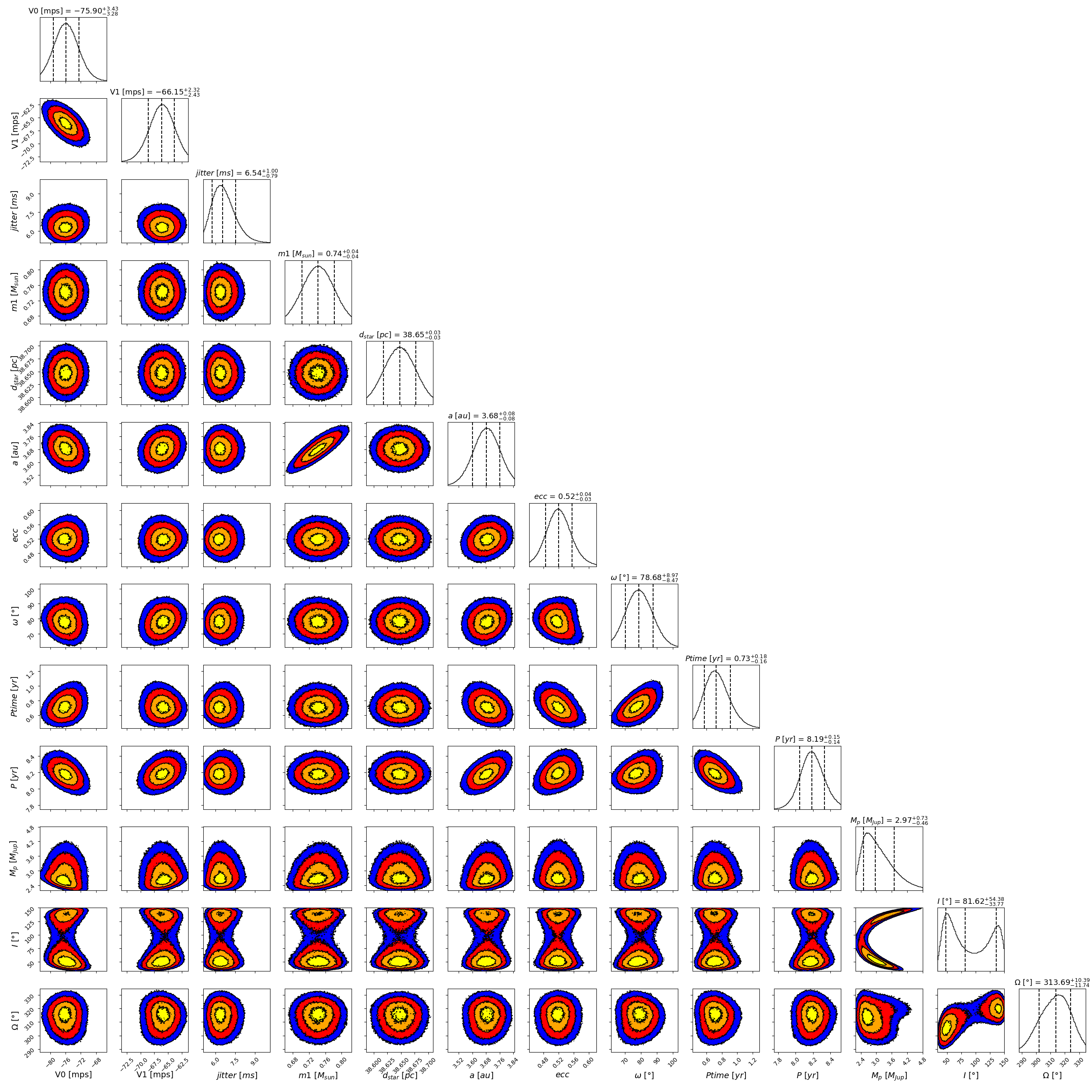}
\caption{Corner plot of the fitted parameters of HD108202 b. An offset of 5.7 km/s is added to V0 (H03) and V1 (H15) to improve readability.
\label{Corner_HD108202}} 
\end{figure}

\begin{figure}[h!]
 \centering
\includegraphics[width=1\textwidth]{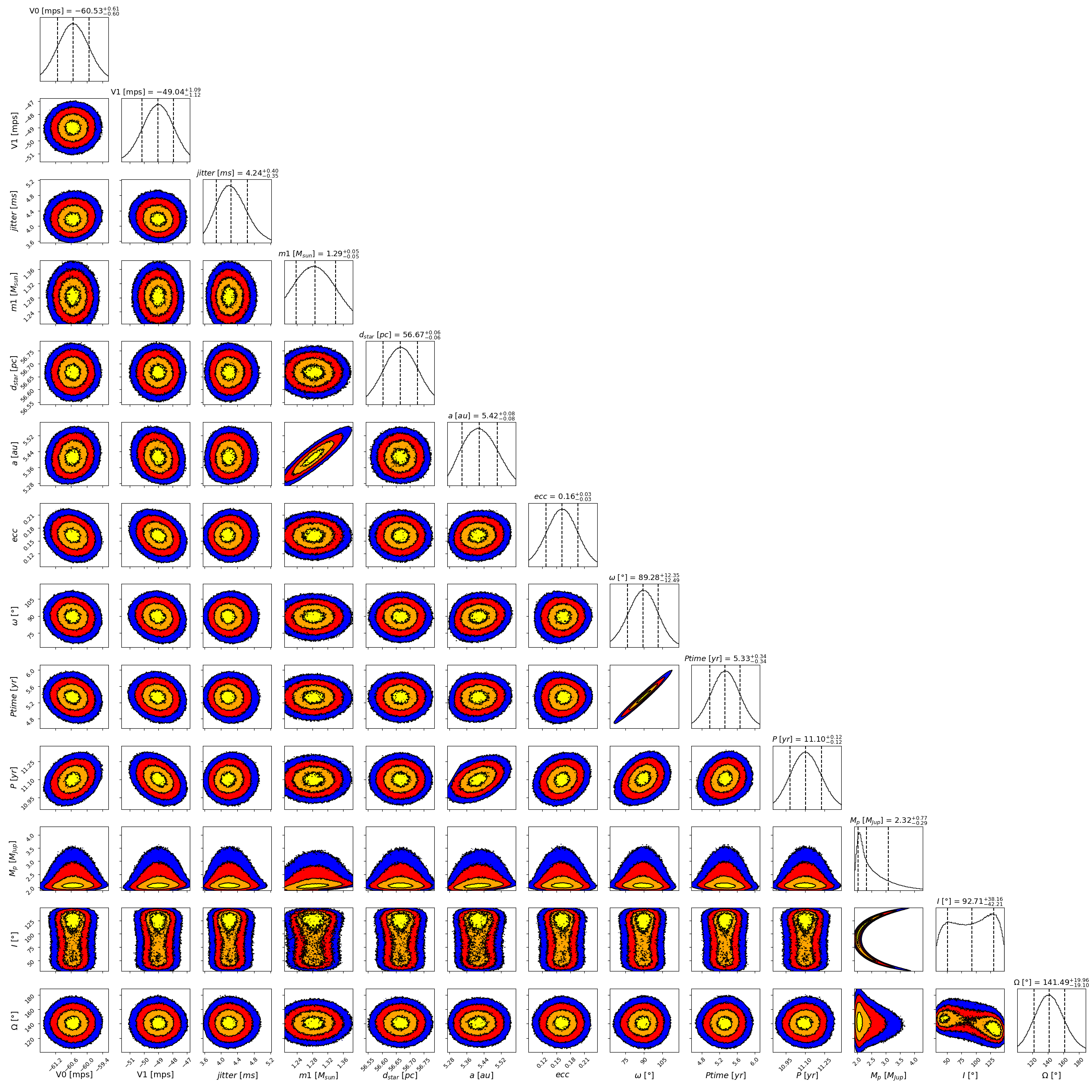}
\caption{Corner plot of the fitted parameters of HD135625 b. An offset of 14.9 km/s is added to V0 (H03) and V1 (H15) to improve readability.
\label{Corner_HD135625}} 
\end{figure}

\begin{figure}[h!]
 \centering
\includegraphics[width=1\textwidth]{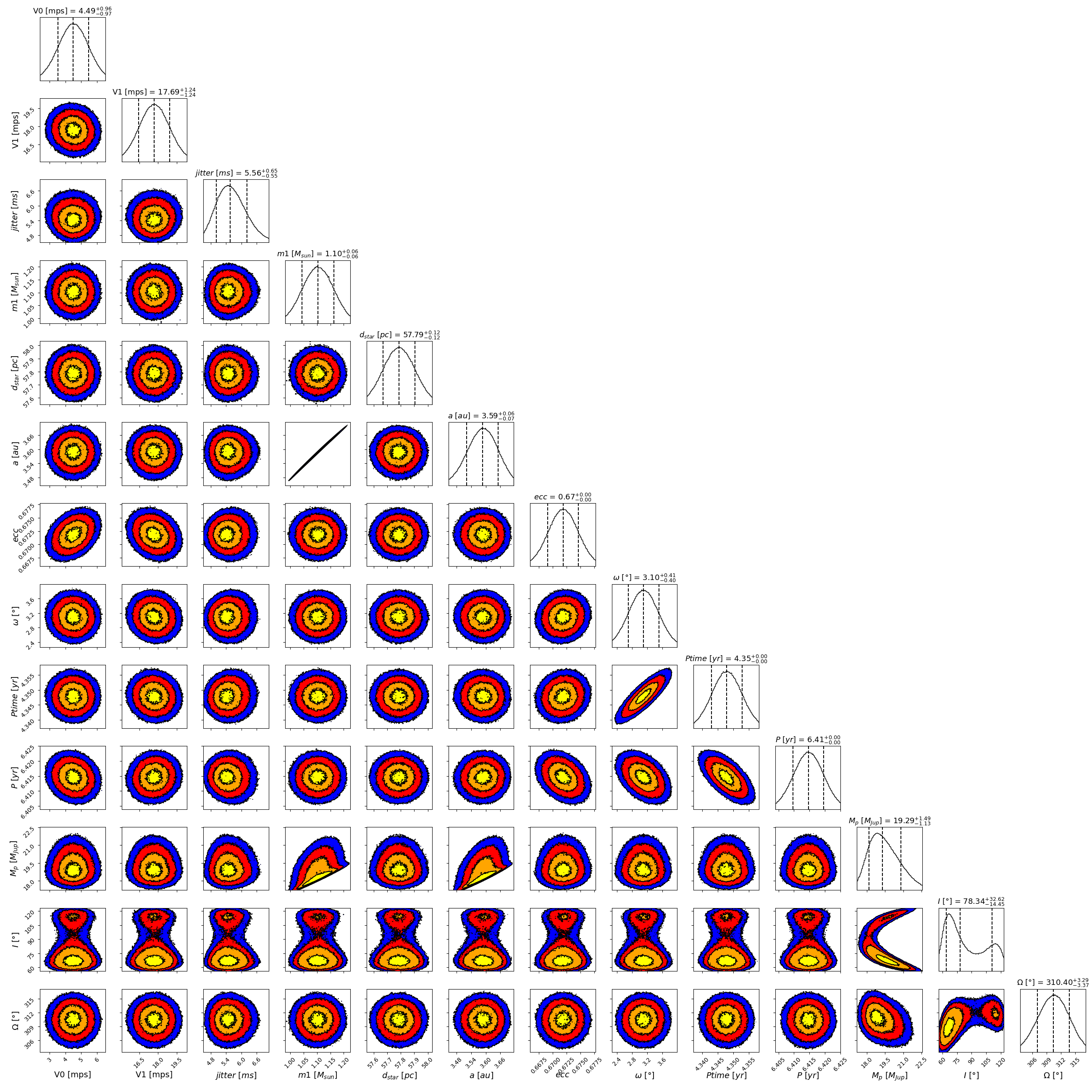}
\caption{Corner plot of the fitted parameters of HD165131 B. An offset of 46.2 km/s is subtracted to V0 (H03) and V1 (H15) to improve readability.
\label{Corner_HD165131}} 
\end{figure}

\begin{figure}[h!]
 \centering
\includegraphics[width=1\textwidth]{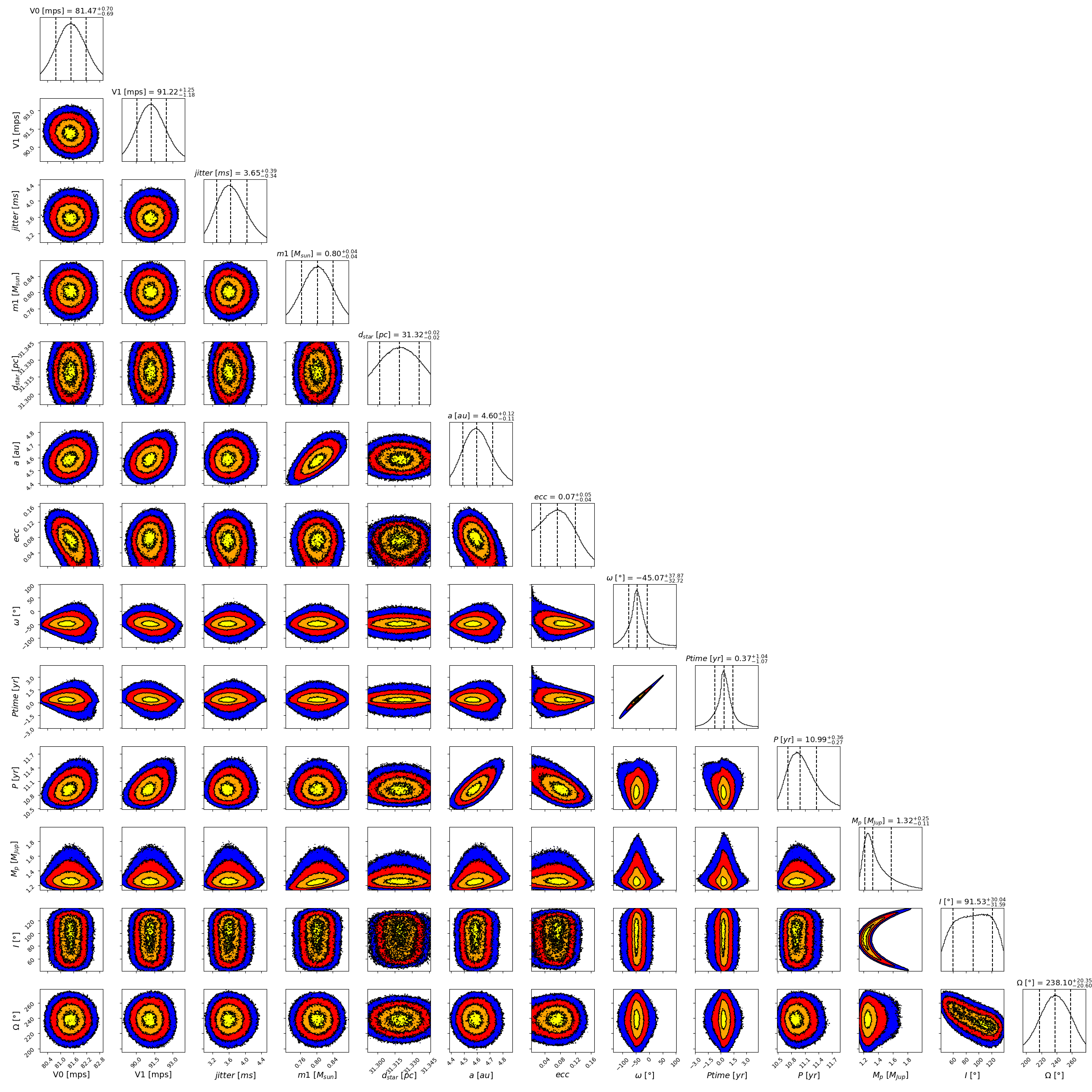}
\caption{Corner plot of the fitted parameters of HD185283 b. An offset of 51.7 km/s is subtracted to V0 (H03) and V1 (H15) to improve readability.
\label{Corner_HD185283}} 
\end{figure}

\begin{figure}[h!]
 \centering
\includegraphics[width=1\textwidth]{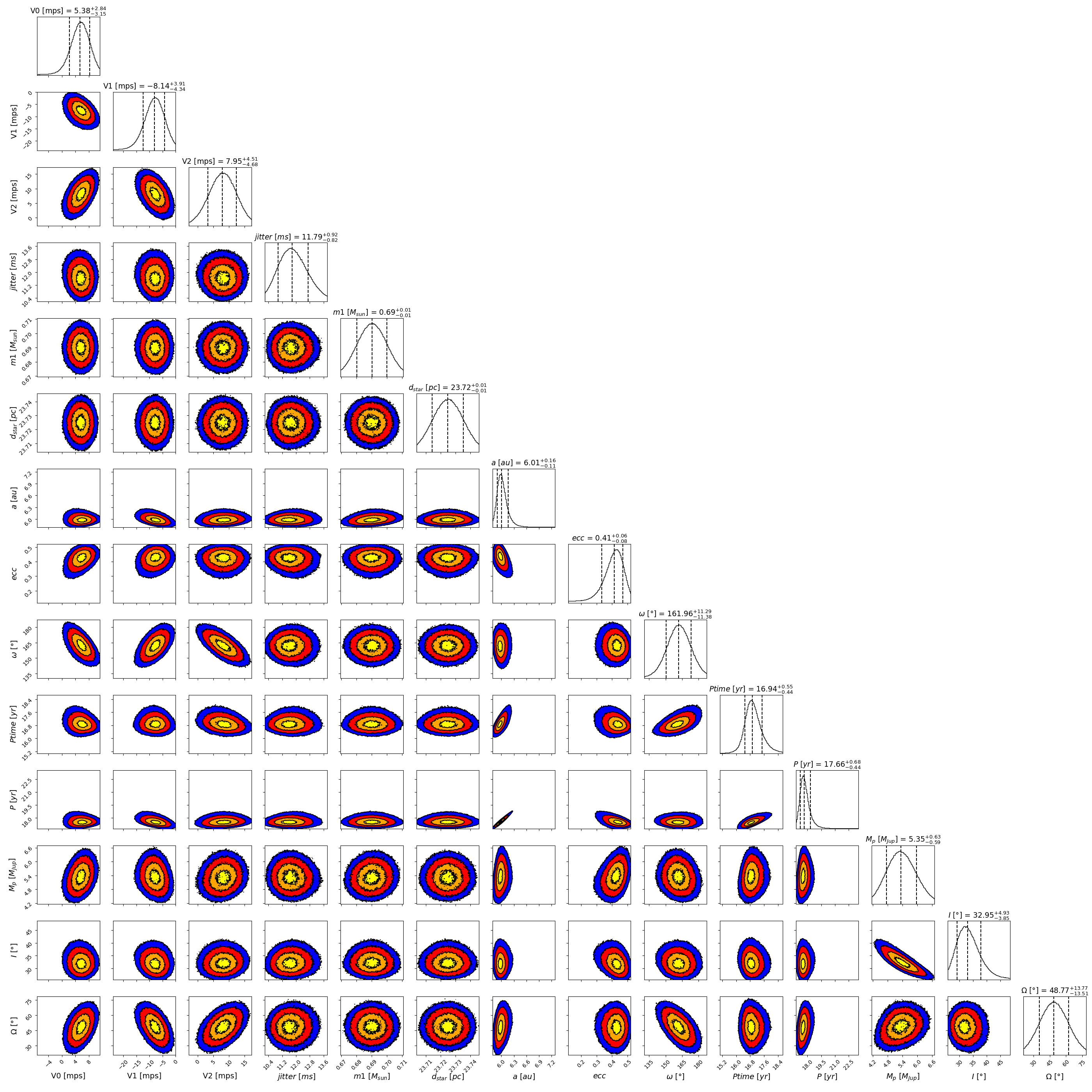}
\caption{Corner plot of the fitted parameters of HIP10337 b, considering a prior on \textit{I} between 0 and 90°. An offset of 3.2 km/s is subtracted to V0 (H03) and V1 (H15) to improve readability. V2 corresponds to Hir04 dataset.
\label{Corner_HIP10337}} 
\end{figure}

\begin{figure}[h!]
 \centering
\includegraphics[width=1\textwidth]{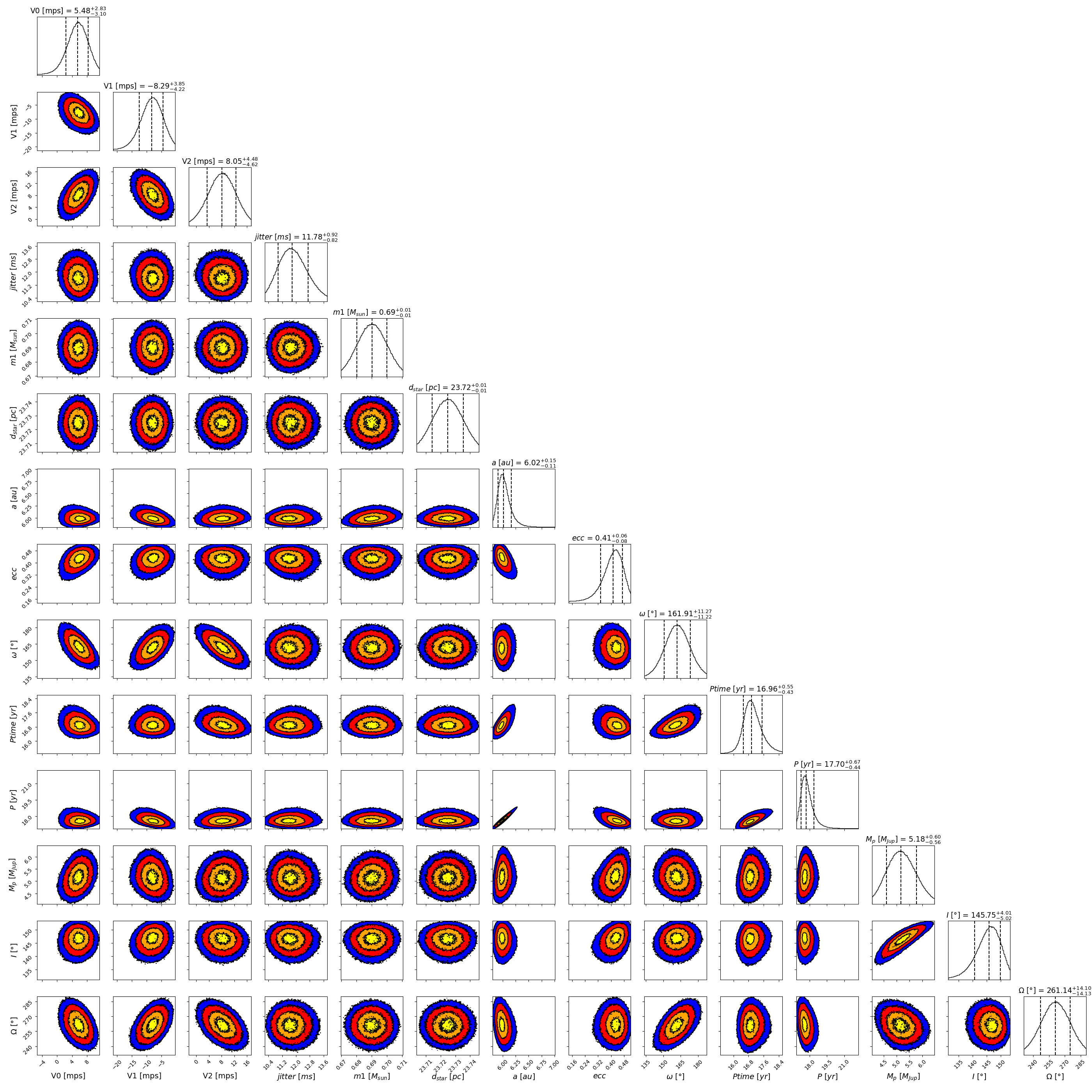}
\caption{Corner plot of the fitted parameters of HIP10337 b, considering a prior on \textit{I} between 90 and 180°. An offset of 3.2 km/s is subtracted to V0 (H03) and V1 (H15) to improve readability. V2 corresponds to Hir04 dataset.
\label{Corner_HIP10337_sup}} 
\end{figure}

\begin{figure}[h!]
 \centering
\includegraphics[width=1\textwidth]{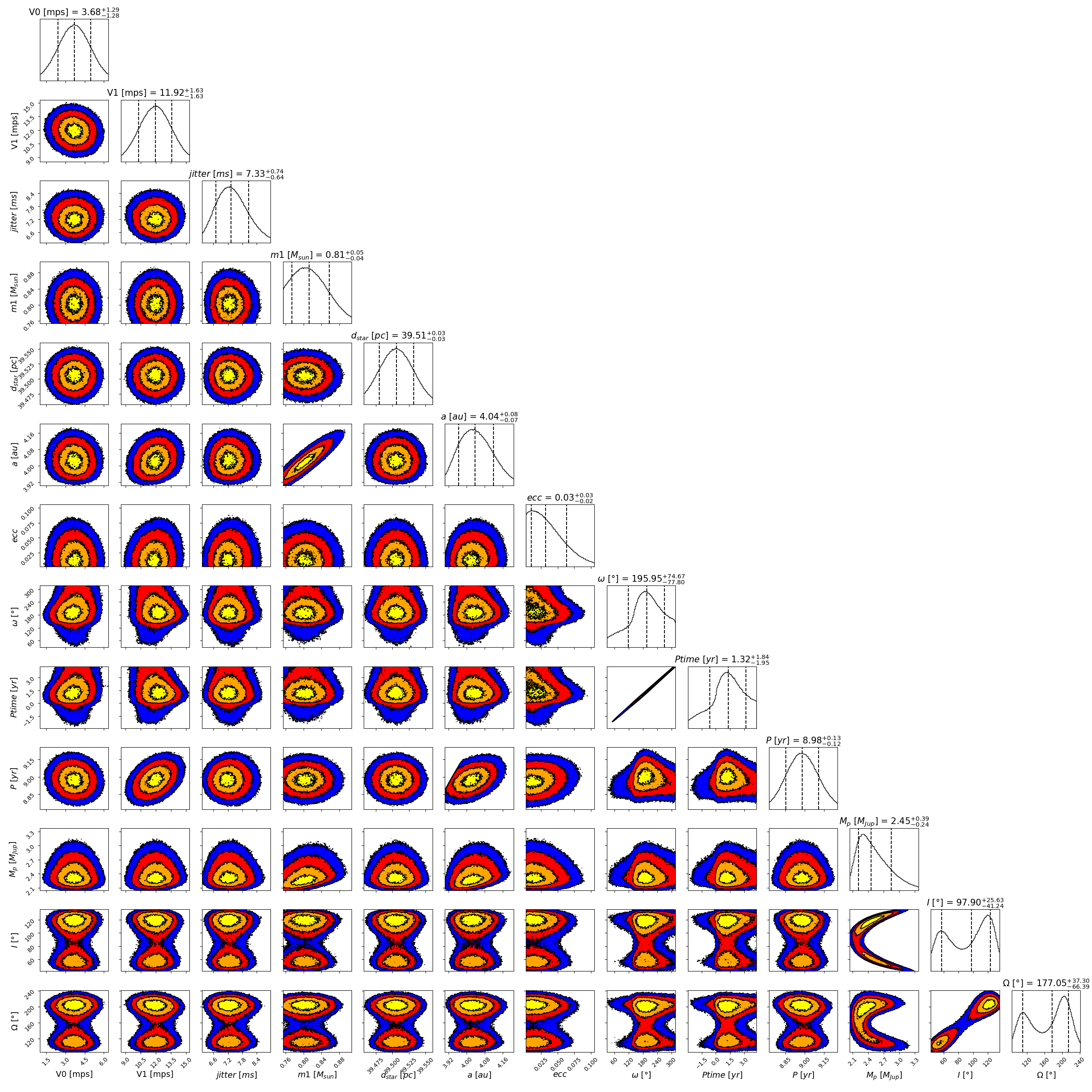}
\caption{Corner plot of the fitted parameters of HIP54597 b. An offset of 51.75 km/s is subtracted to V0 (H03) and V1 (H15) to improve readability.
\label{Corner_HIP54597}} 
\end{figure}

\newpage

\subsection{Plots for known sub-stellar companions}

\subsubsection{Long-period companions}

 \begin{figure}[h!]
 \centering
 \includegraphics[width=0.33\textwidth]{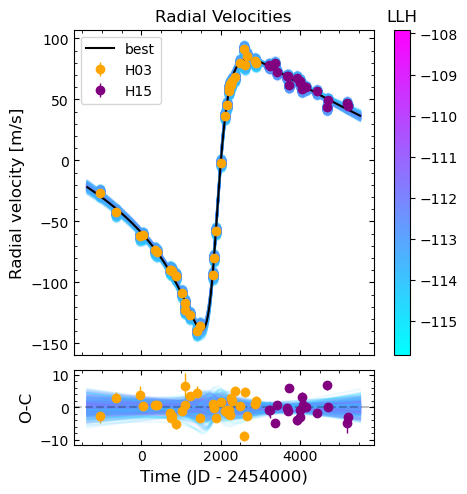}
 \includegraphics[width=0.6\textwidth]{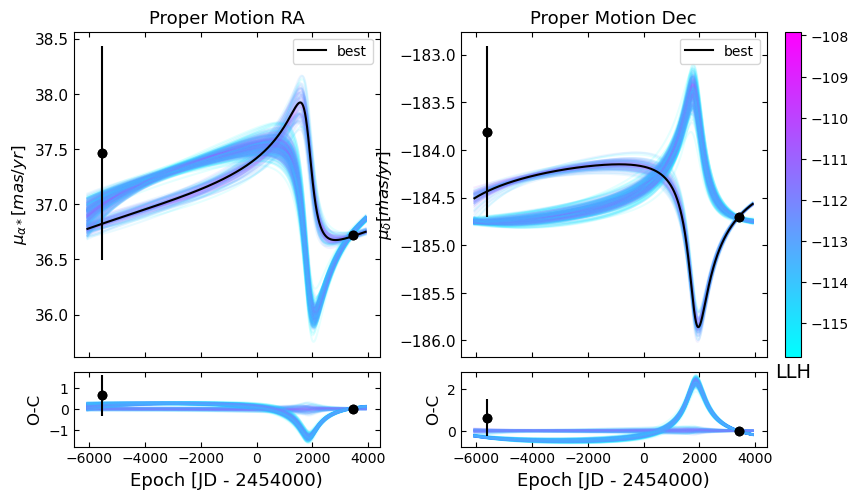}
 \includegraphics[width=0.88\textwidth]{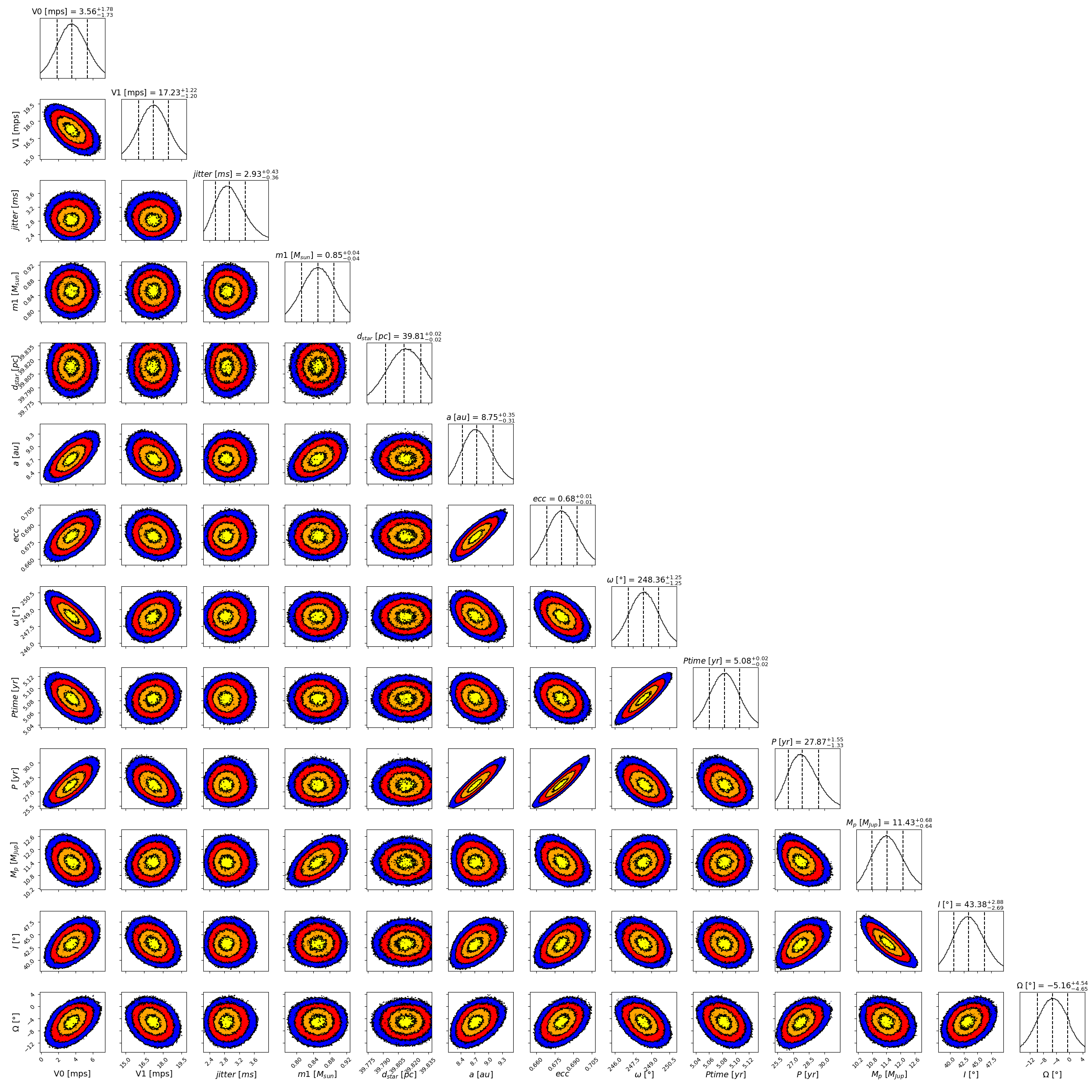}
 \caption{Fit of the orbit of HD16905. \textit{Top left}: Fit of the HD16905 RV measurements. \textit{Top right}: Fit of the HD16905 astrometric acceleration in right ascension (left) and declination (right). The black points correspond to the Hipparcos and Gaia EDR3 proper motion measurements. In each plot, the black curve corresponds to the best fit. The color bar indicates the log-likelihood corresponding to the different fits plotted. \textit{Bottom}: Corner plot of the fitted parameters of HD16905 b, considering a prior on \textit{I} between 0 and 90°. An offset of 64.8 km/s is subtracted to V0 (H03) and V1 (H15) to improve readability.
 \label{RV_AA_HD16905}}
 \end{figure}

 \begin{figure}[h!]
 \centering
 \includegraphics[width=1.0\textwidth]{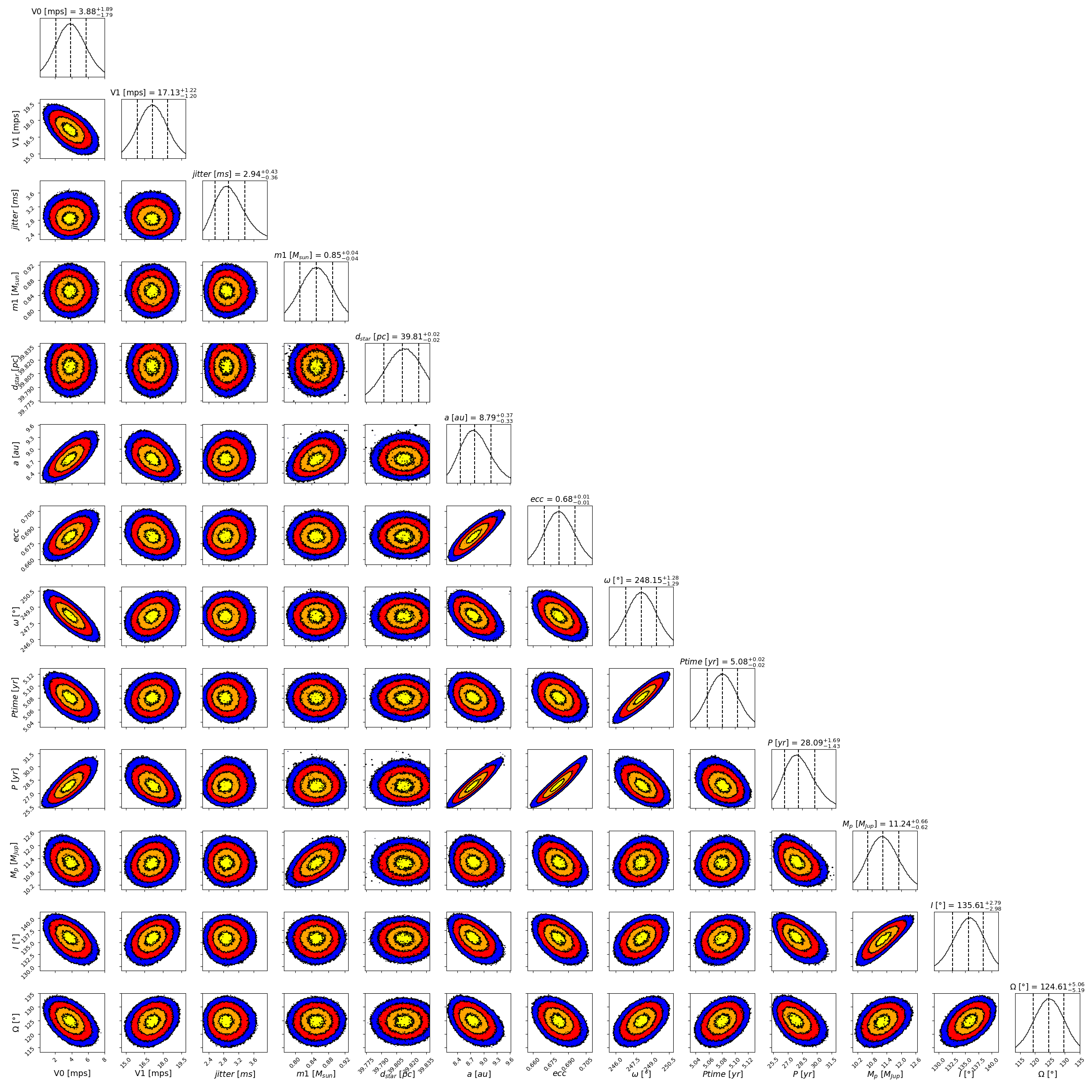}
 \caption{Corner plot of the fitted parameters of HD16905 b, considering a prior on \textit{I} between 90 and 180°. An offset of 64.8 km/s is subtracted to V0 (H03) and V1 (H15) to improve readability.
 \label{RV_AA_HD16905_sup}}
 \end{figure}

\begin{figure}[h!]
\centering
\includegraphics[width=0.33\textwidth]{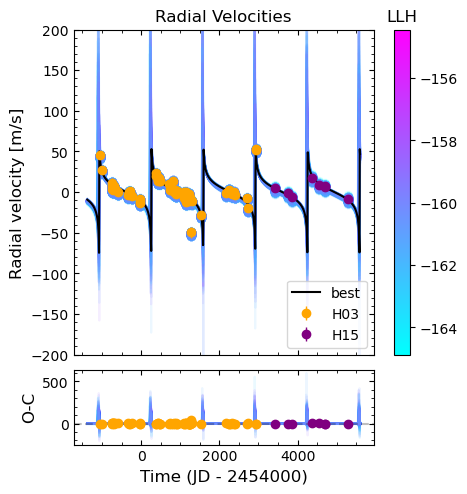}
\includegraphics[width=0.6\textwidth]{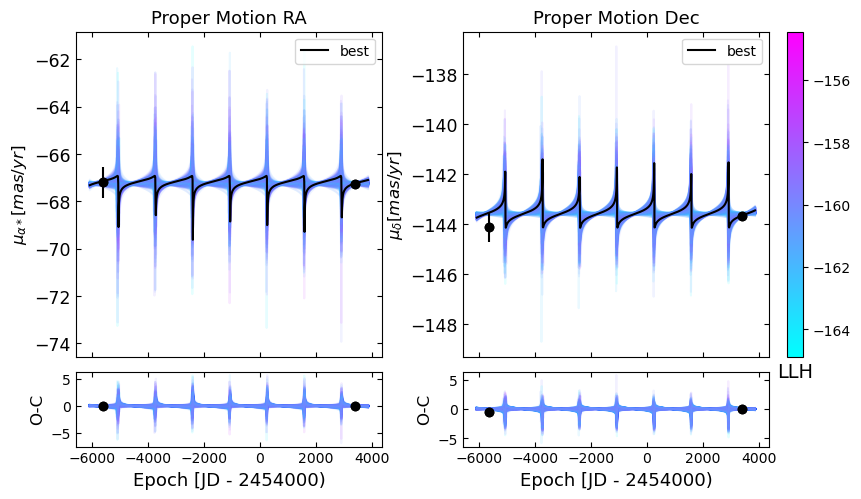}
\includegraphics[width=0.88\textwidth]{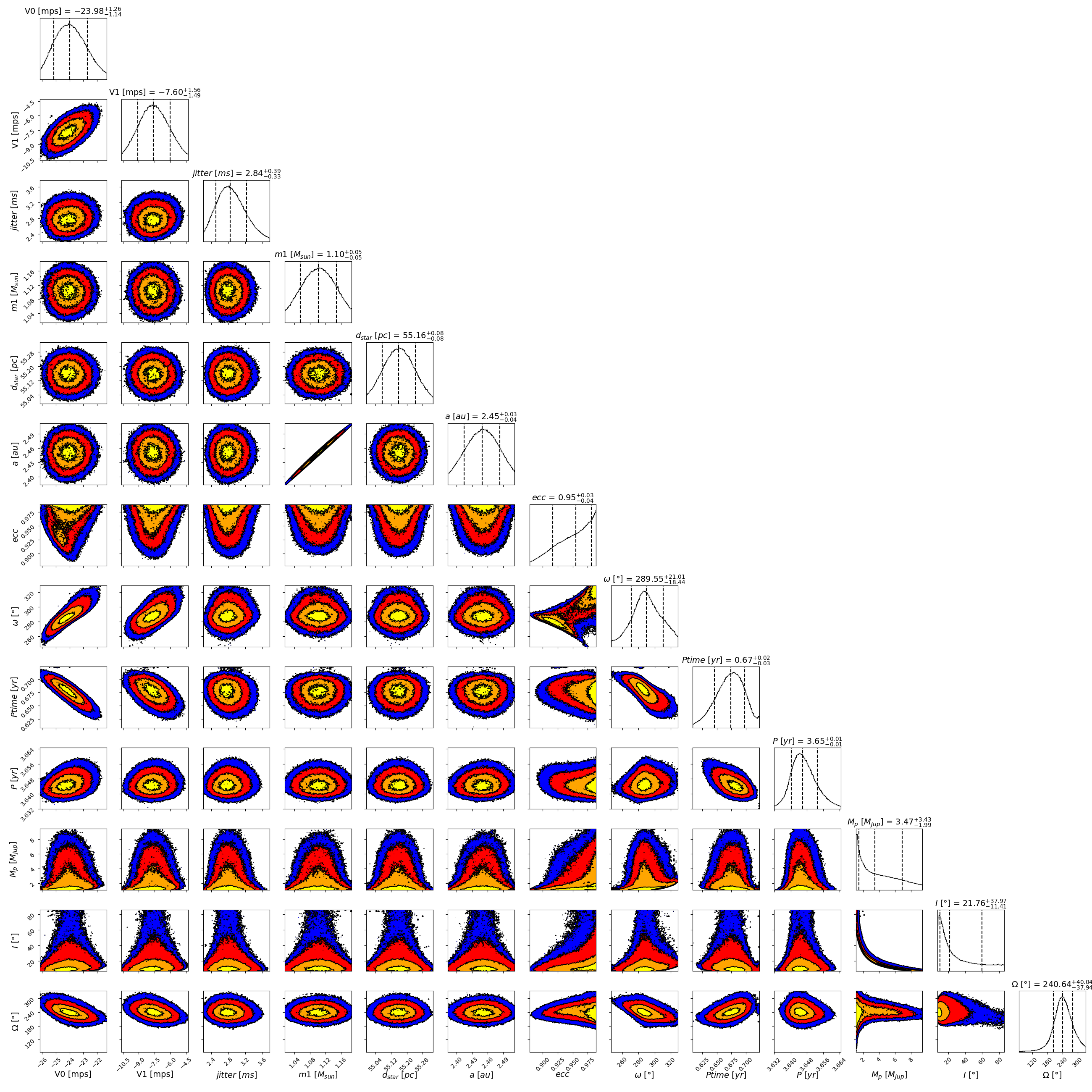}
\caption{Fit of the orbit of HD28254. \textit{Top left}: Fit of the HD28254 RV measurements. \textit{Top right}: Fit of the HD28254 astrometric acceleration in right ascension (left) and declination (right). The black points correspond to the Hipparcos and Gaia EDR3 proper motion measurements. In each plot, the black curve corresponds to the best fit. The color bar indicates the log-likelihood corresponding to the different fits plotted. \textit{Bottom}: Corner plot of the fitted parameters of HD28254 b, considering a prior on \textit{I} between 0 and 90°. An offset of 9.3 km/s is added to V0 (H03) and V1 (H15) to improve readability.
\label{RV_AA_HD28254}}
\end{figure}

\begin{figure}[h!]
\centering
\includegraphics[width=1.0\textwidth]{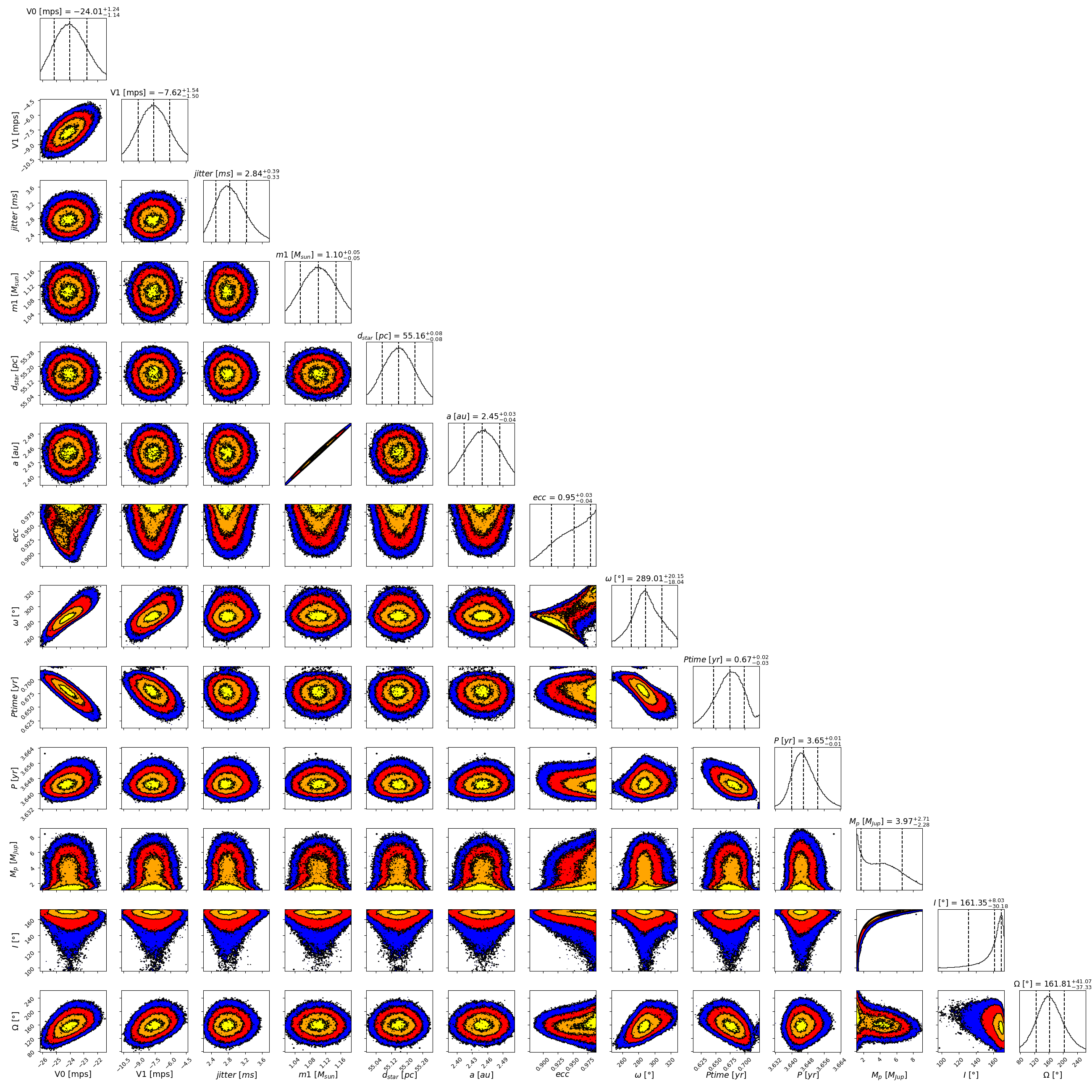}
\caption{Corner plot of the fitted parameters of HD28254 b, considering a prior on \textit{I} between 90 and 180°. An offset of 9.3 km/s is added to V0 (H03) and V1 (H15) to improve readability.
\label{RV_AA_HD28254_sup}}
\end{figure}

 \begin{figure}[h!]
 \centering
 \includegraphics[width=0.33\textwidth]{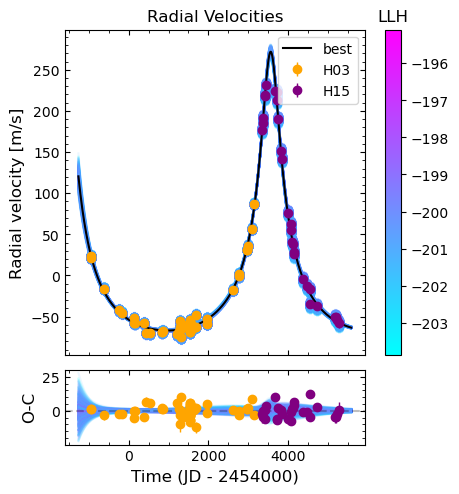}
 \includegraphics[width=0.6\textwidth]{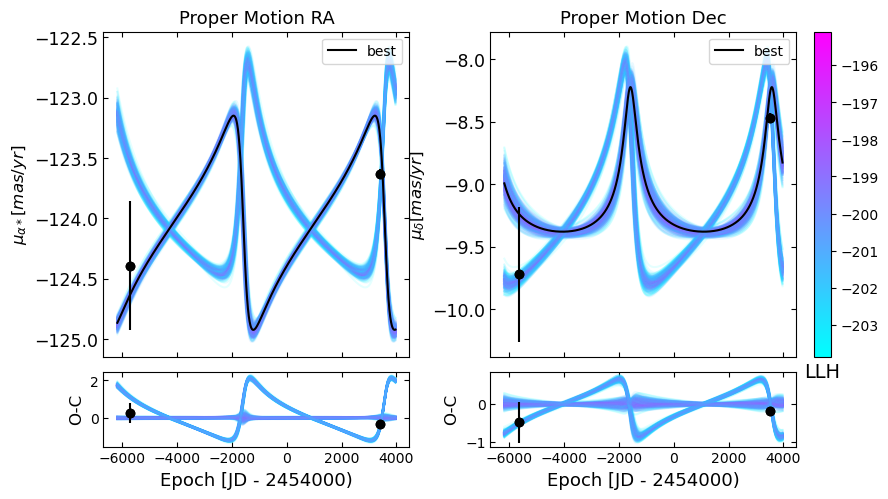}
 \includegraphics[width=0.88\textwidth]{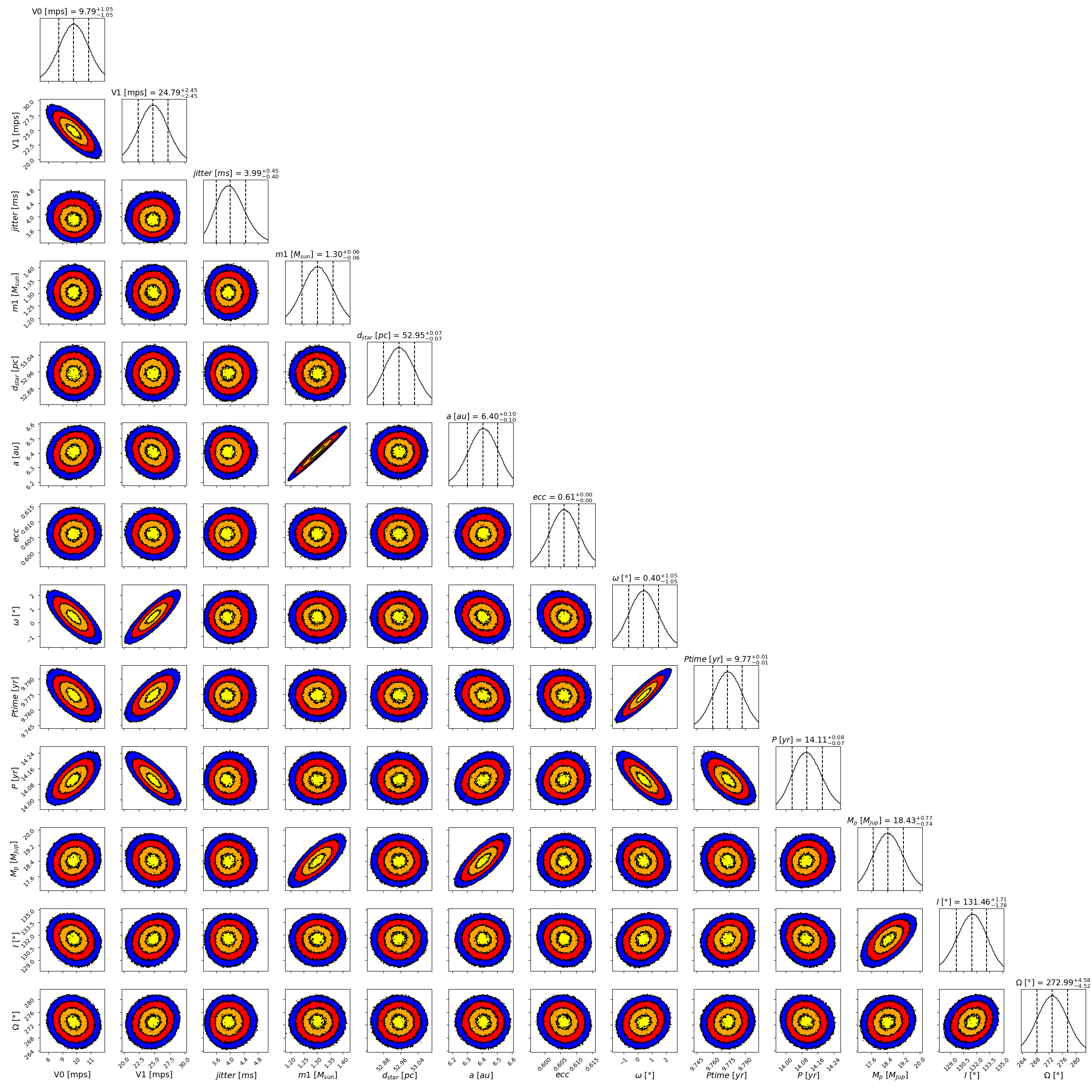}
 \caption{Fit of the orbit of HD62364. \textit{Top left}: Fit of the HD62364 RV measurements. \textit{Top right}: Fit of the HD62364 astrometric acceleration in right ascension (left) and declination (right). The black points correspond to the Hipparcos and Gaia EDR3 proper motion measurements. In each plot, the black curve corresponds to the best fit. The color bar indicates the log-likelihood corresponding to the different fits plotted. \textit{Bottom}: Corner plot of the fitted parameters of HD62364 B, considering a prior on \textit{I} between 0 and 90°. An offset of 35.5 km/s is subtracted to V0 (H03) and V1 (H15) to improve readability.
 \label{RV_AA_HD62364}} 
 \end{figure}

 \begin{figure}[h!]
 \centering
 \includegraphics[width=1.0\textwidth]{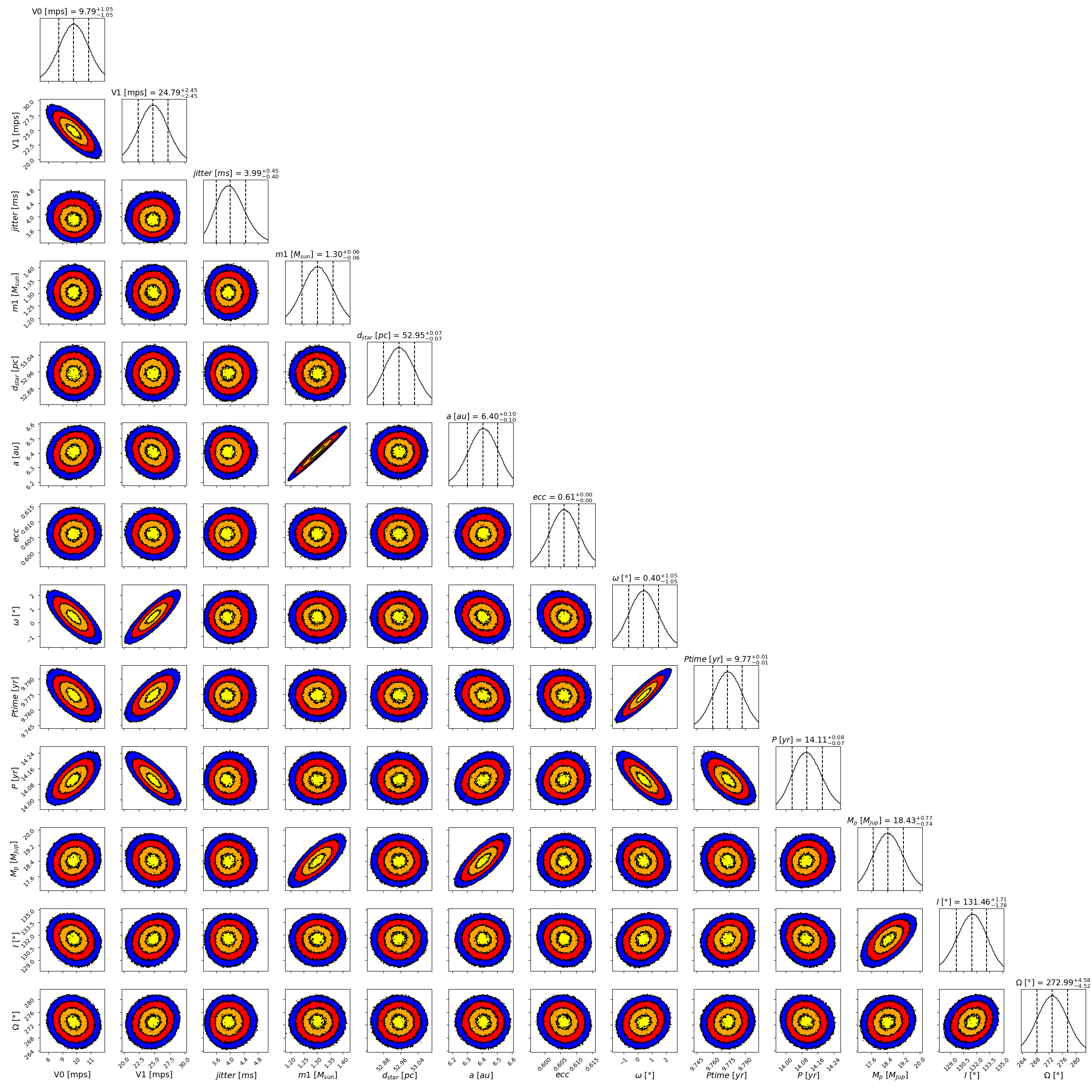}
 \caption{Corner plot of the fitted parameters of HD62364 B, considering a prior on \textit{I} between 90 and 180°. An offset of 35.5 km/s is subtracted to V0 (H03) and V1 (H15) to improve readability.
 \label{RV_AA_HD62364_sup}} 
 \end{figure}

 \begin{figure}[h!]
 \centering
 \includegraphics[width=0.33\textwidth]{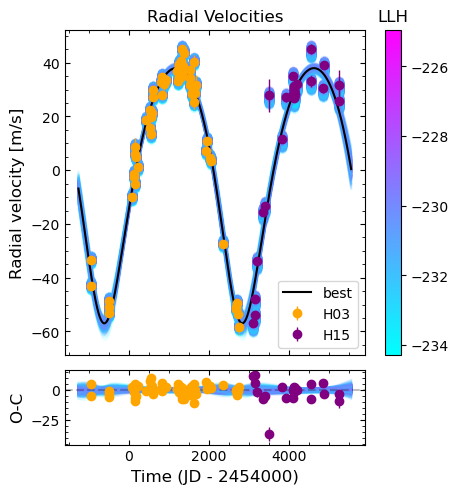}
 \includegraphics[width=0.6\textwidth]{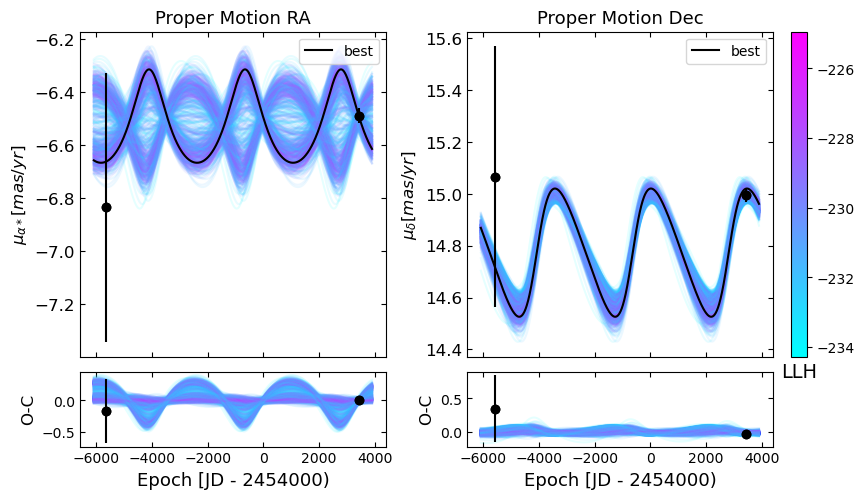}
 \includegraphics[width=0.88\textwidth]{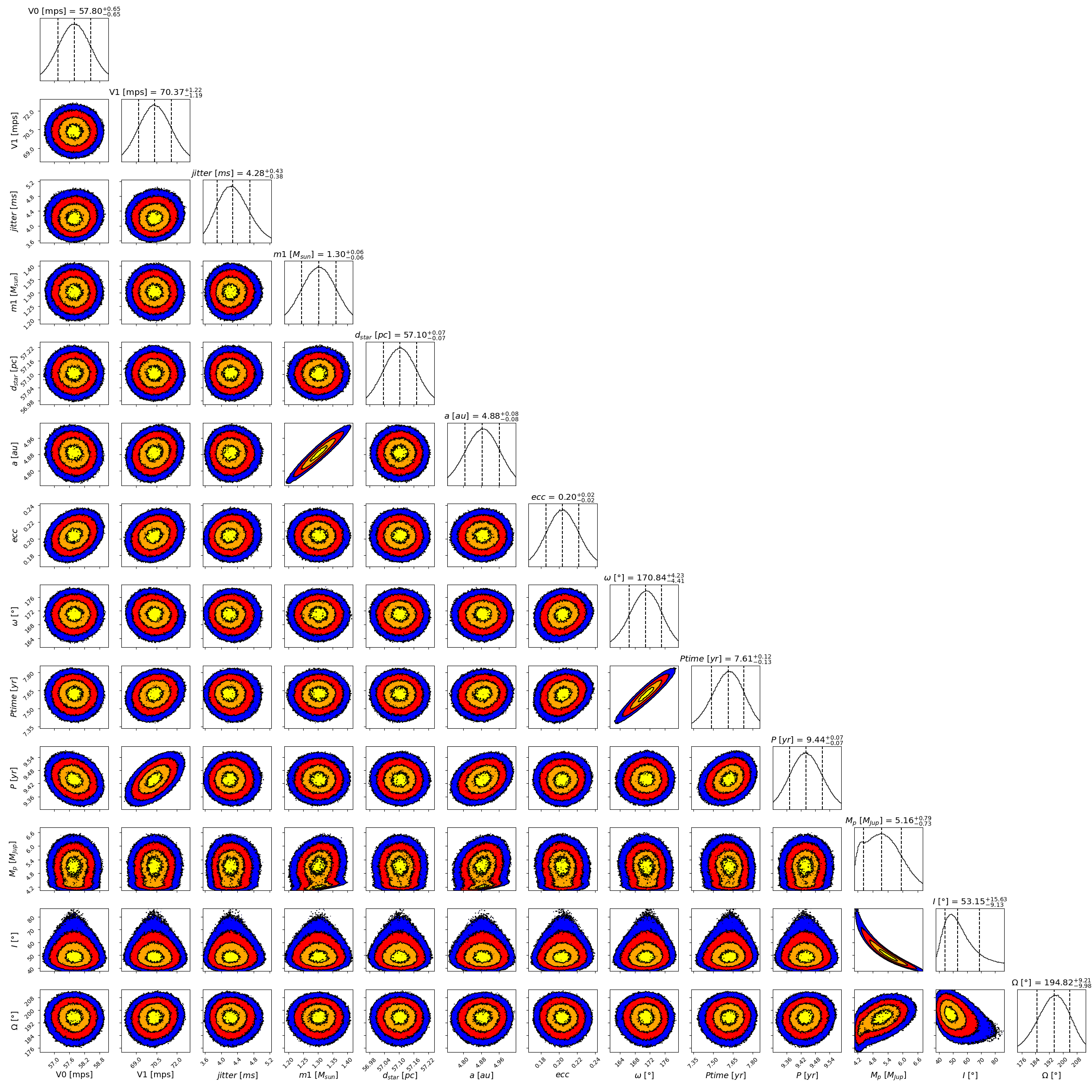}
 \caption{Fit of the orbit of HD89839. \textit{Top left}: Fit of the HD89839 RV measurements. \textit{Top right}: Fit of the HD89839 astrometric acceleration in right ascension (left) and declination (right). The black points correspond to the Hipparcos and Gaia EDR3 proper motion measurements. In each plot, the black curve corresponds to the best fit. The color bar indicates the log-likelihood corresponding to the different fits plotted. \textit{Bottom}: Corner plot of the fitted parameters of HD89839 b, considering a prior on \textit{I} between 0 and 90°. An offset of 31.7 km/s is subtracted to V0 (H03) and V1 (H15) to improve readability.
 \label{RV_AA_HD89839}} 
 \end{figure}

 \begin{figure}[h!]
 \centering
 \includegraphics[width=1.0\textwidth]{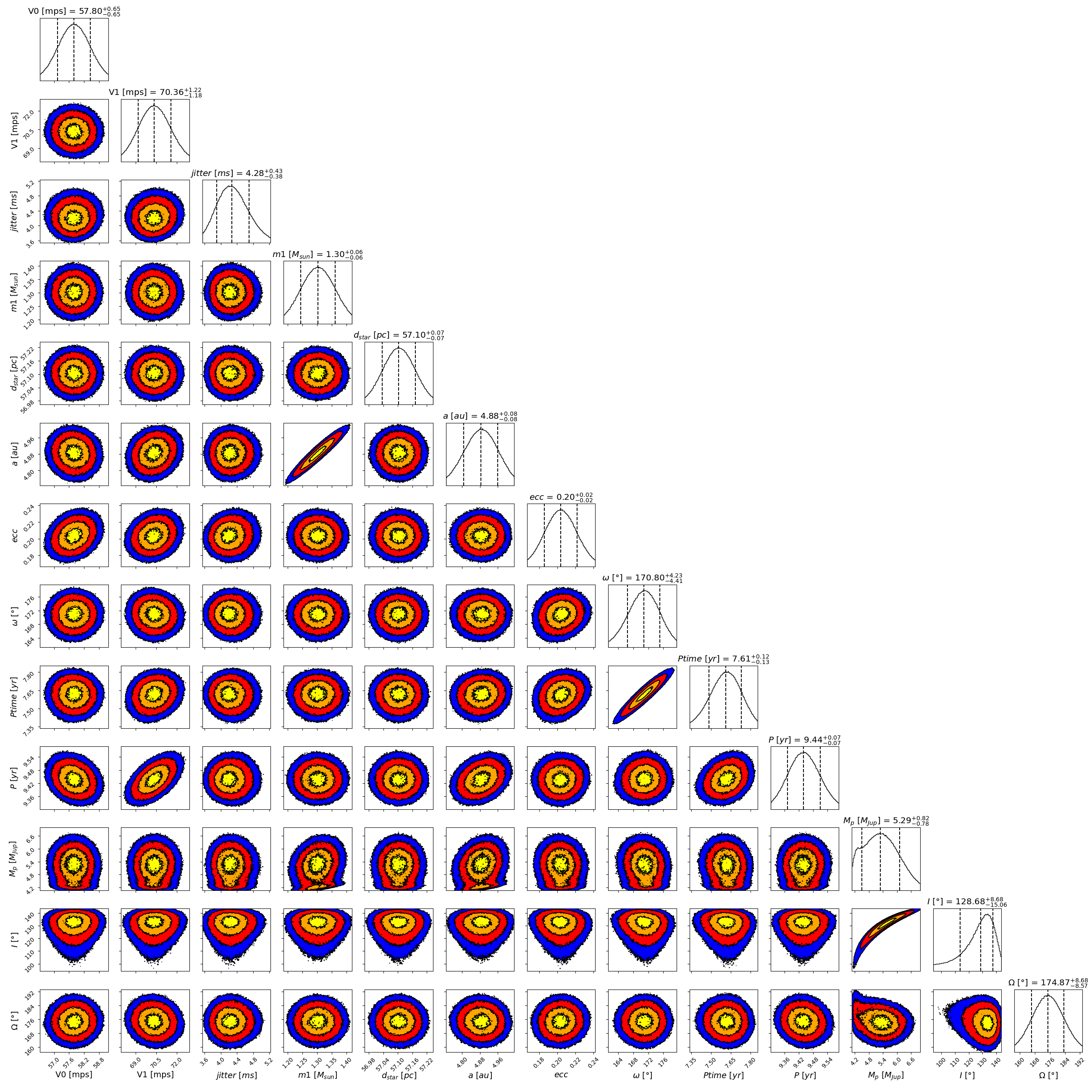}
 \caption{Corner plot of the fitted parameters of HD89839 b, considering a prior on \textit{I} between 90 and 180°. An offset of 31.7 km/s is subtracted to V0 (H03) and V1 (H15) to improve readability.
 \label{RV_AA_HD89839_sup}} 
 \end{figure}

 \begin{figure}[h!]
 \centering
 \includegraphics[width=0.33\textwidth]{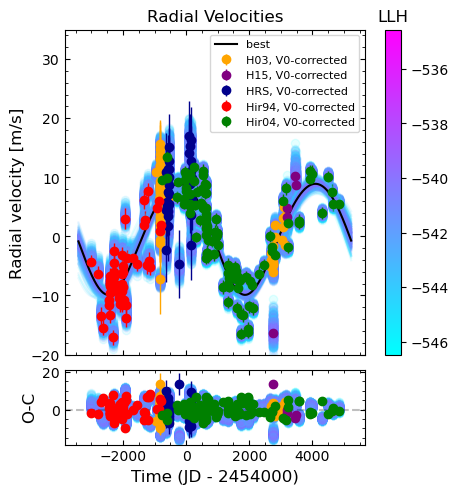}
 \includegraphics[width=0.6\textwidth]{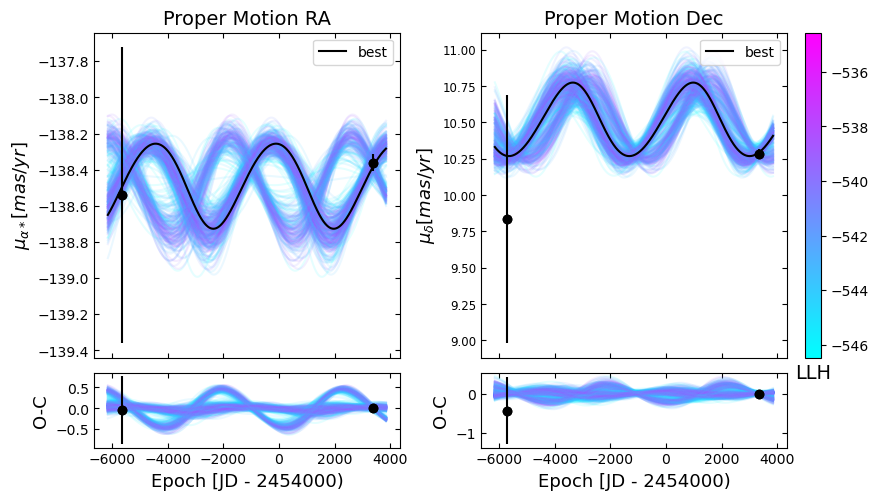}
 \includegraphics[width=0.88\textwidth]{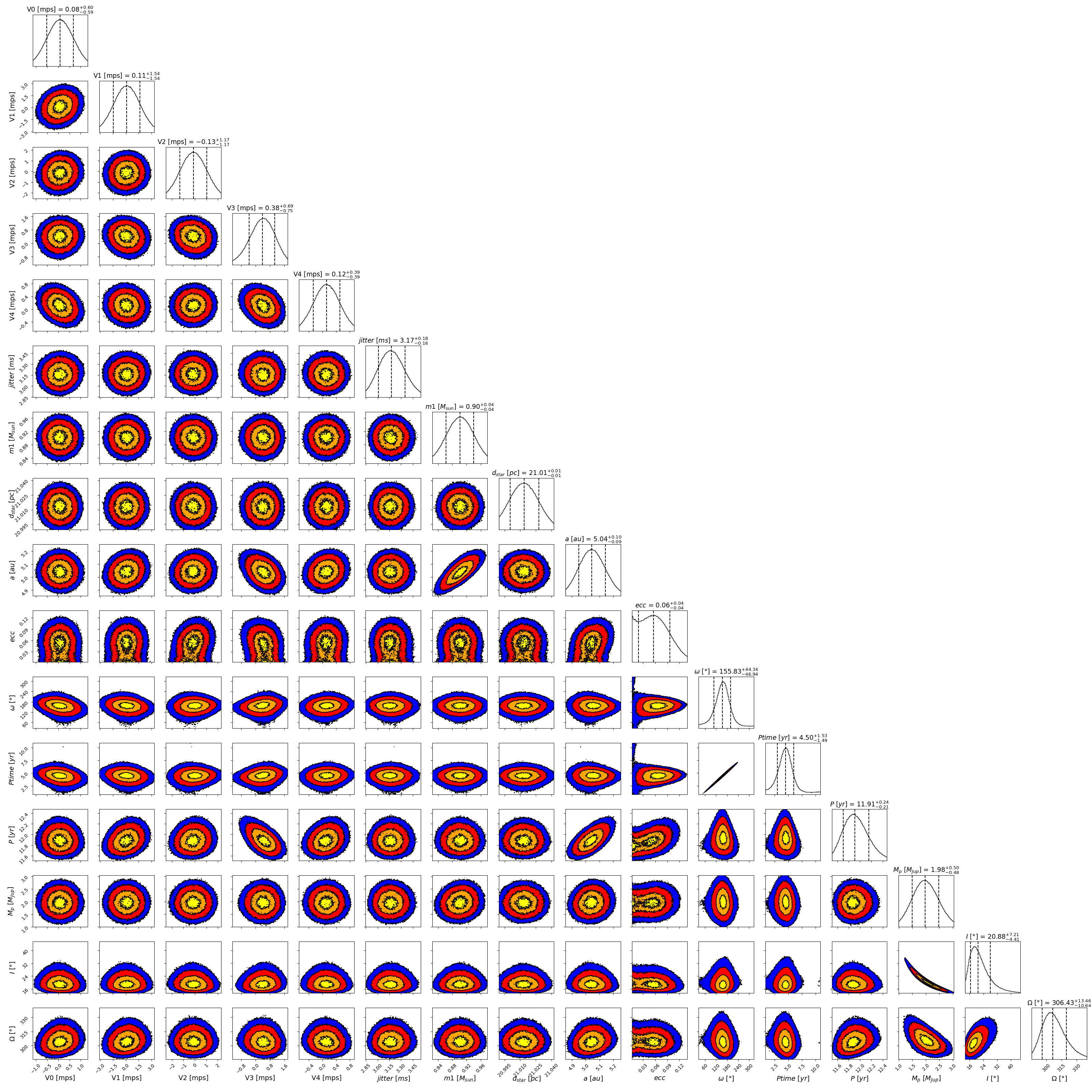}
 \caption{Fit of the orbit of HD114783. \textit{Top left}: Fit of the HD114783 RV measurements after subtracted the signal of the inner planet. \textit{Top right}: Fit of the HD114783 astrometric acceleration in right ascension (left) and declination (right). The black points correspond to the Hipparcos and Gaia EDR3 proper motion measurements. In each plot, the black curve corresponds to the best fit. The color bar indicates the log-likelihood corresponding to the different fits plotted. \textit{Bottom}: Corner plot of the fitted parameters of HD114783 c, considering a prior on \textit{I} between 0 and 90°. V0, V1, V2, V3, and V4 correspond to H03, H15, HRS, Hir94, and Hir04, respectively.
 \label{RV_AA_HD114783}} 
 \end{figure}

 \begin{figure}[h!]
 \centering
 \includegraphics[width=1.0\textwidth]{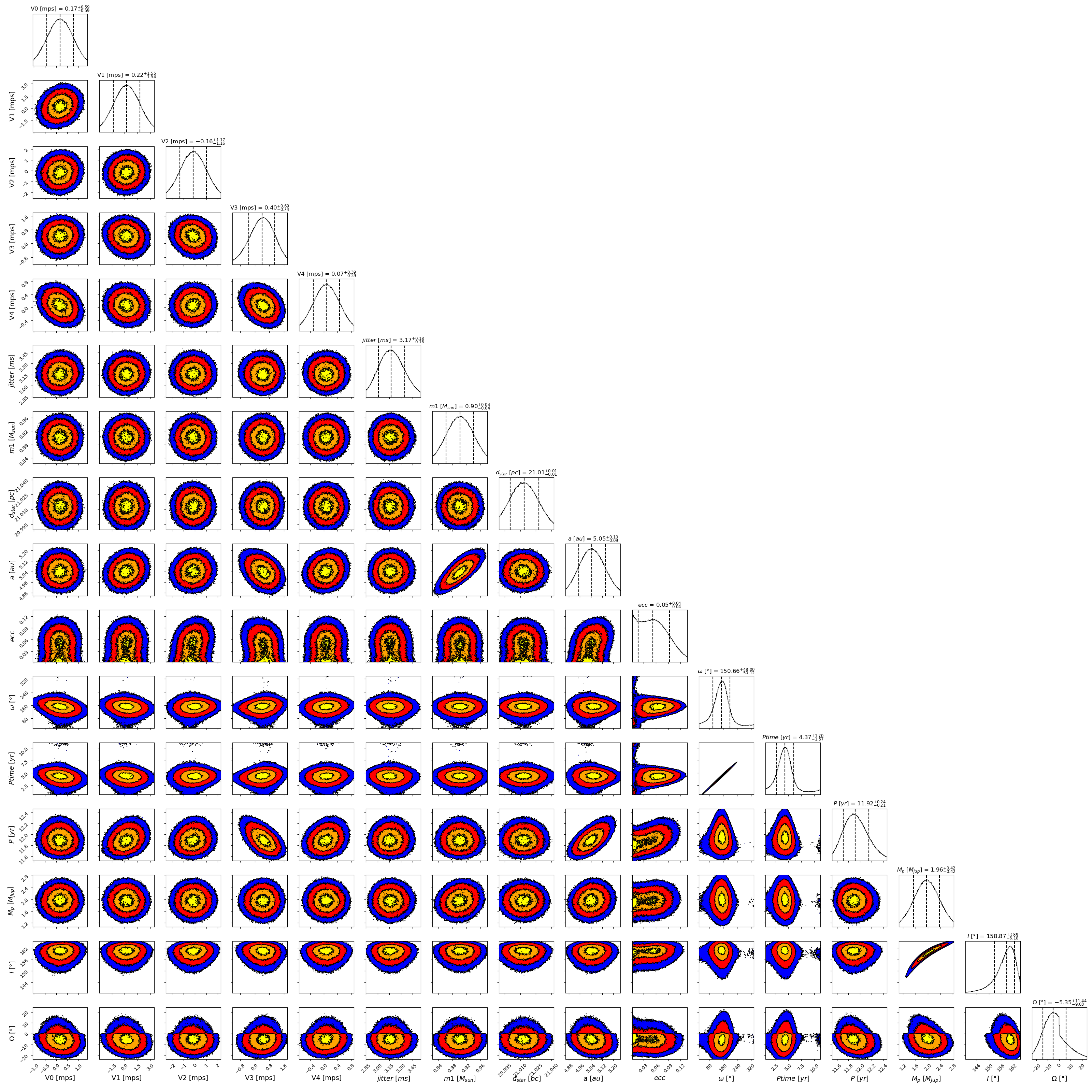}
 \caption{Corner plot of the fitted parameters of HD114783 c, considering a prior on \textit{I} between 90 and 180°. V0, V1, V2, V3, and V4 correspond to H03, H15, HRS, Hir94, and Hir04, respectively.
 \label{RV_AA_HD114783_sup}} 
 \end{figure}

 \begin{figure}[h!]
 \centering
 \includegraphics[width=0.33\textwidth]{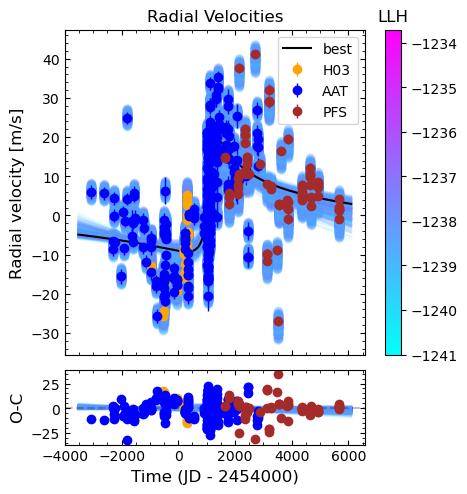}
 \includegraphics[width=0.6\textwidth]{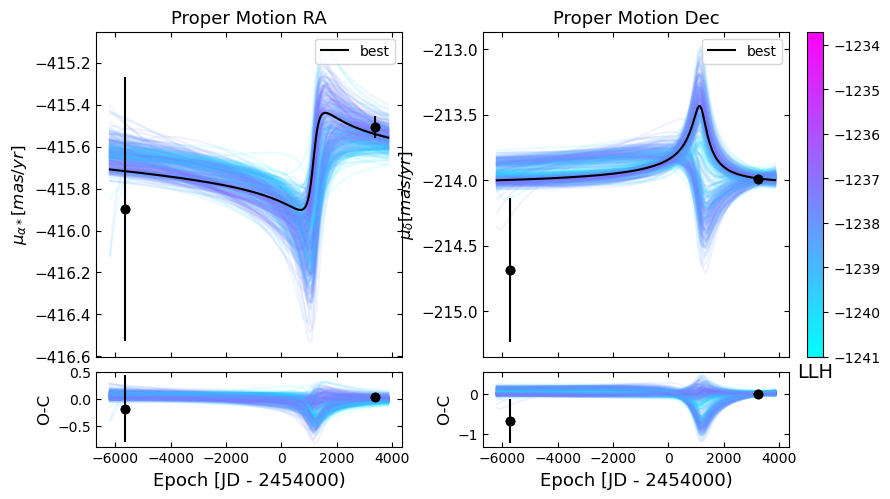}
 \includegraphics[width=0.88\textwidth]{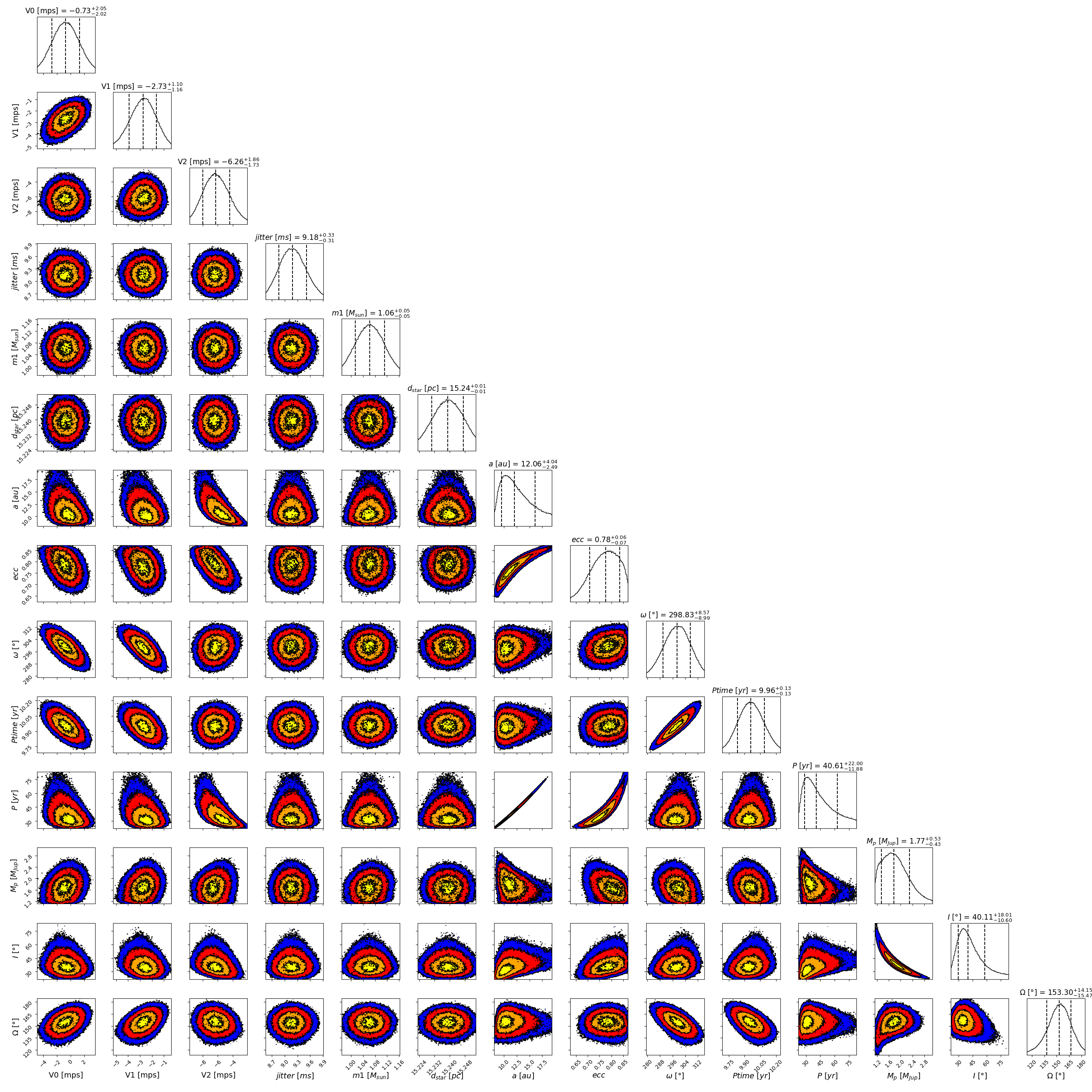}
 \caption{Fit of the orbit of HD140901. \textit{Top left}: Fit of the HD140901 RV measurements. \textit{Top right}: Fit of the HD140901 astrometric acceleration in right ascension (left) and declination (right). The black points correspond to the Hipparcos and Gaia EDR3 proper motion measurements. In each plot, the black curve corresponds to the best fit. The color bar indicates the log-likelihood corresponding to the different fits plotted. \textit{Bottom}: Corner plot of the fitted parameters of HD140901 b, considering a prior on \textit{I} between 0 and 90°. V0, V1, V2, and V3 correspond to H03, H15, C07, and MIKE, respectively.
 \label{RV_AA_HD140901}} 
 \end{figure}

 \begin{figure}[h!]
 \centering
 \includegraphics[width=1.0\textwidth]{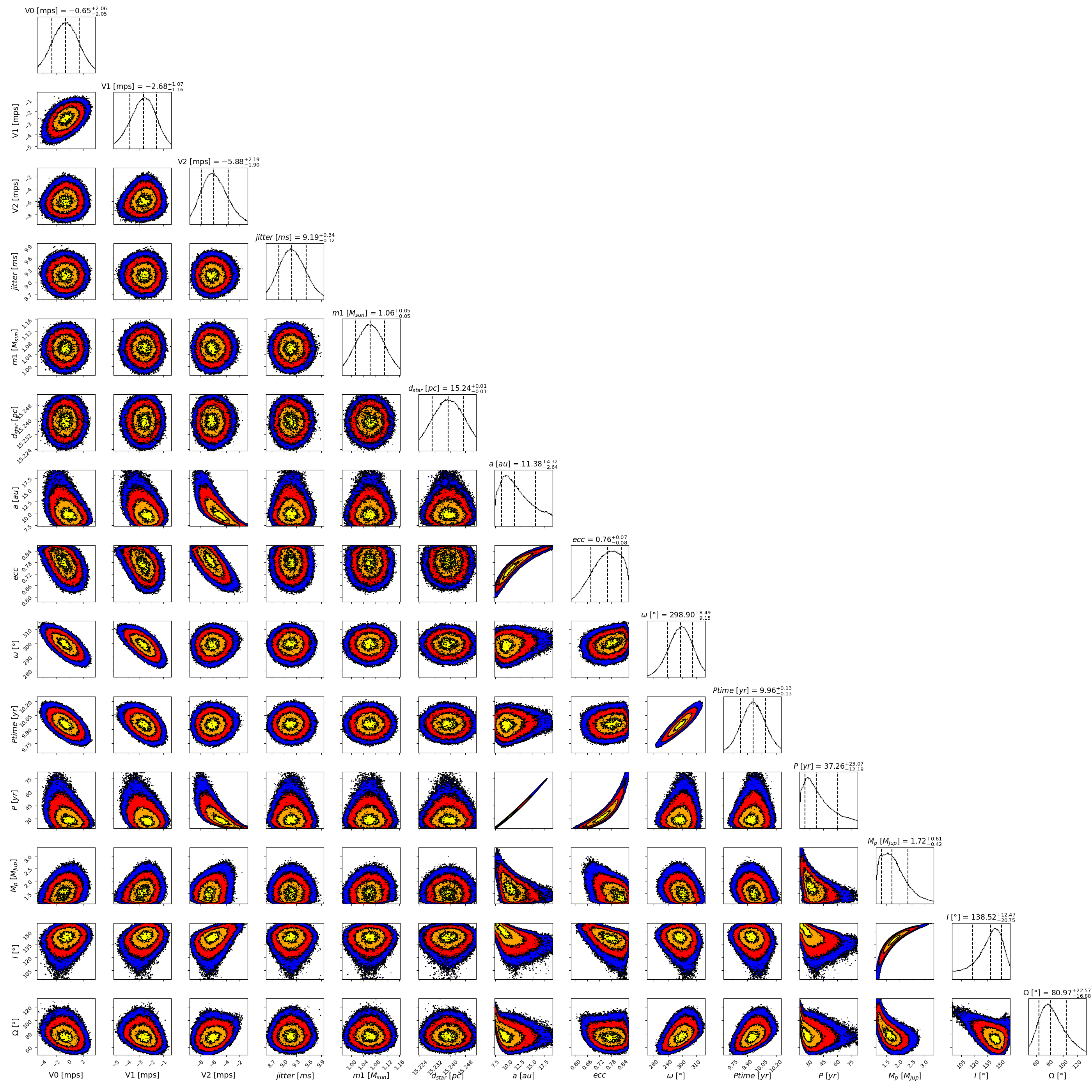}
 \caption{Corner plot of the fitted parameters of HD140901 b, considering a prior on \textit{I} between 90 and 180°. V0, V1, V2, and V3 correspond to H03, H15, C07, and MIKE, respectively.
 \label{RV_AA_HD140901_sup}} 
 \end{figure}

 \begin{figure}[h!]
 \centering
 \includegraphics[width=0.33\textwidth]{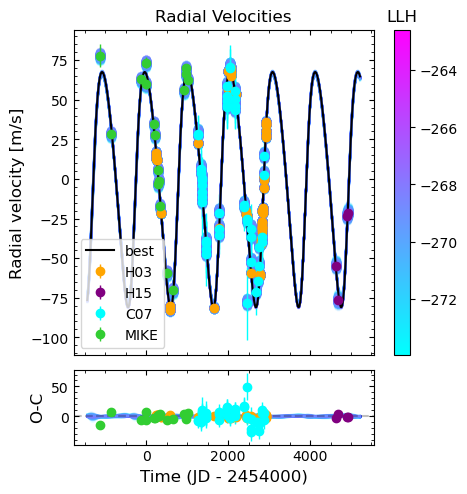}
 \includegraphics[width=0.6\textwidth]{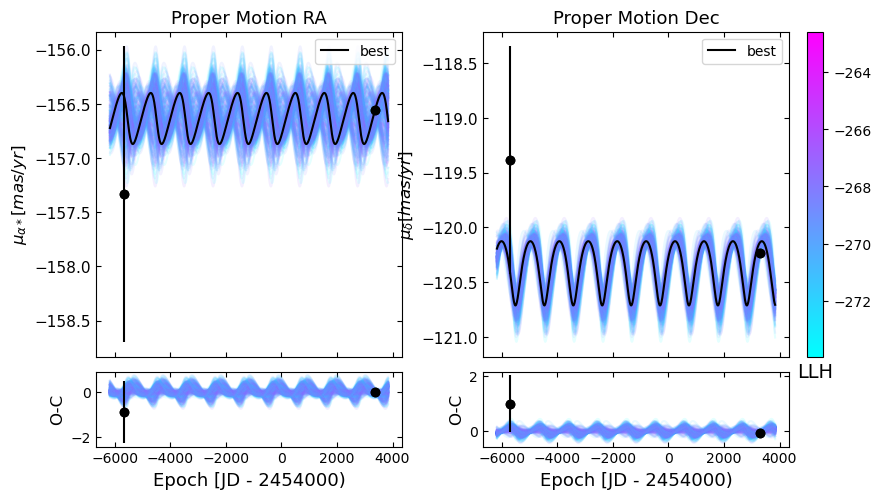}
 \includegraphics[width=0.88\textwidth]{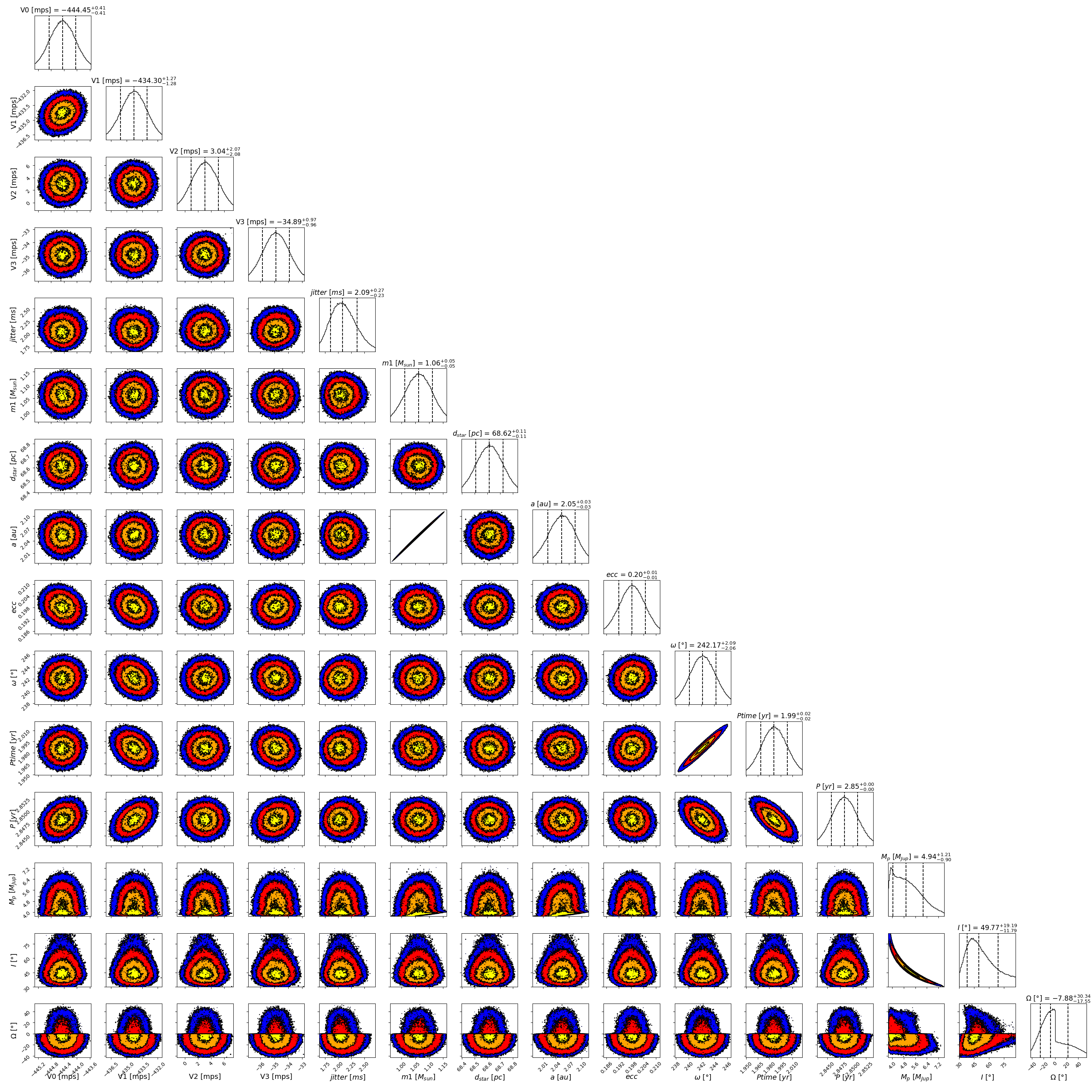}
 \caption{Fit of the orbit of HD143361. \textit{Top left}: Fit of the HD143361 RV measurements. \textit{Top right}: Fit of the HD143361 astrometric acceleration in right ascension (left) and declination (right). The black points correspond to the Hipparcos and Gaia EDR3 proper motion measurements. In each plot, the black curve corresponds to the best fit. The color bar indicates the log-likelihood corresponding to the different fits plotted. \textit{Bottom}: Corner plot of the fitted parameters of HD143361 b, considering a prior on \textit{I} between 0 and 90°. V0, V1, V2, and V3 correspond to H03, H15, C07, and MIKE, respectively.
 \label{RV_AA_HD143361}} 
 \end{figure}

 \begin{figure}[h!]
 \centering
 \includegraphics[width=1.0\textwidth]{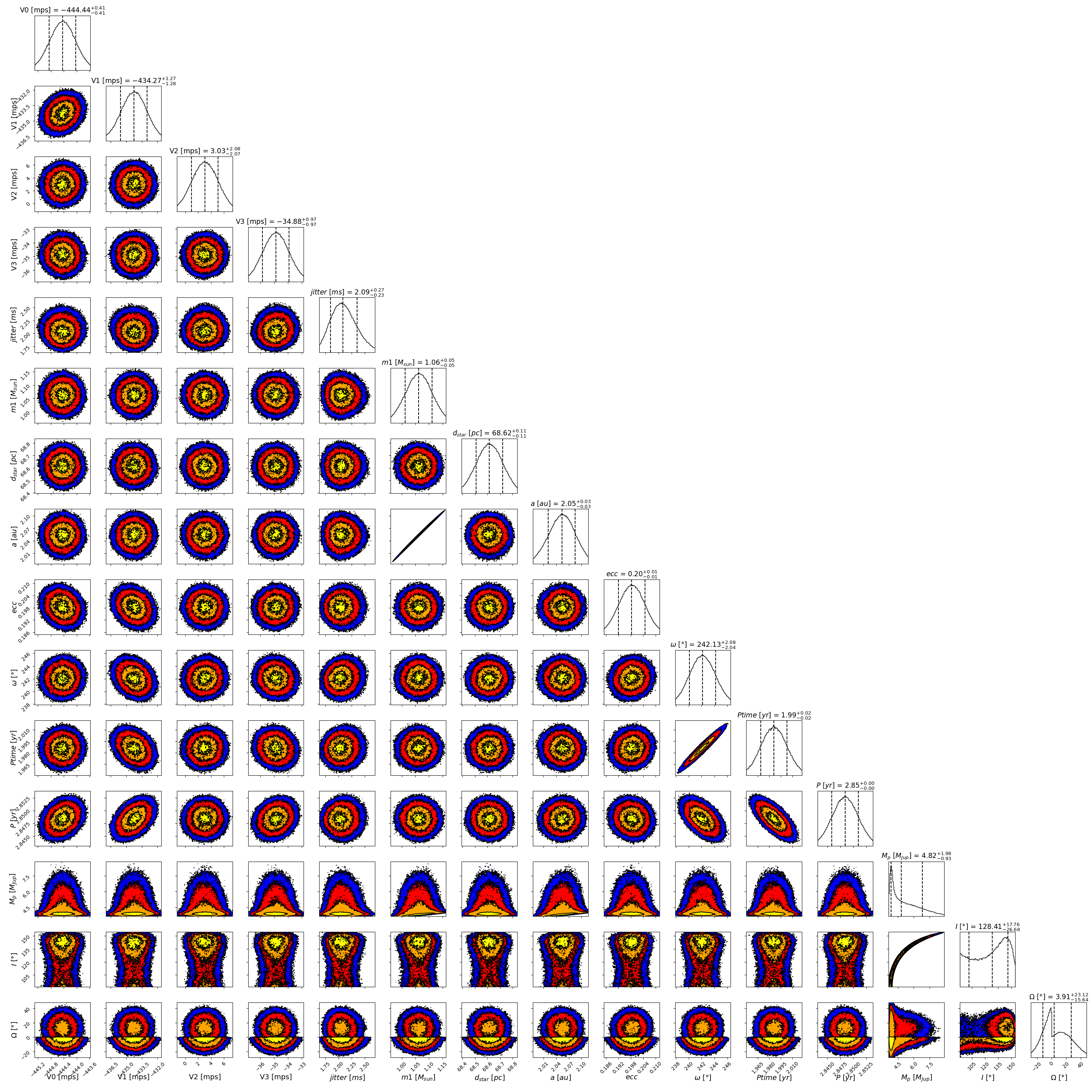}
 \caption{Corner plot of the fitted parameters of HD143361 b, considering a prior on \textit{I} between 90 and 180°. V0, V1, V2, and V3 correspond to H03, H15, C07, and MIKE, respectively.
 \label{RV_AA_HD143361_sup}} 
 \end{figure}

 \begin{figure}[h!]
 \centering
 \includegraphics[width=0.33\textwidth]{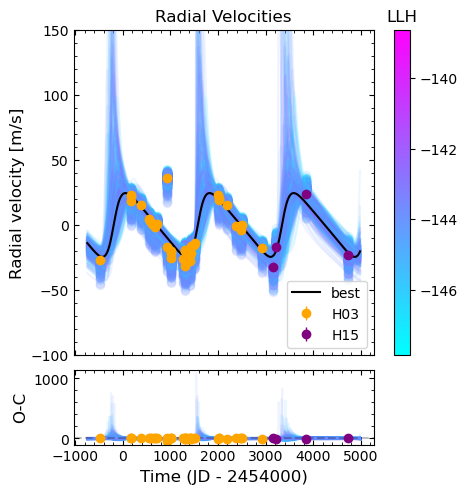}
 \includegraphics[width=0.6\textwidth]{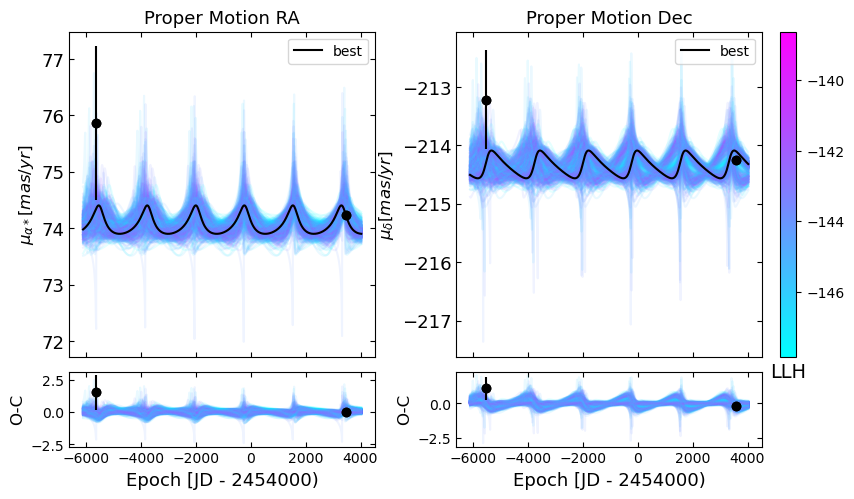}
 \includegraphics[width=0.88\textwidth]{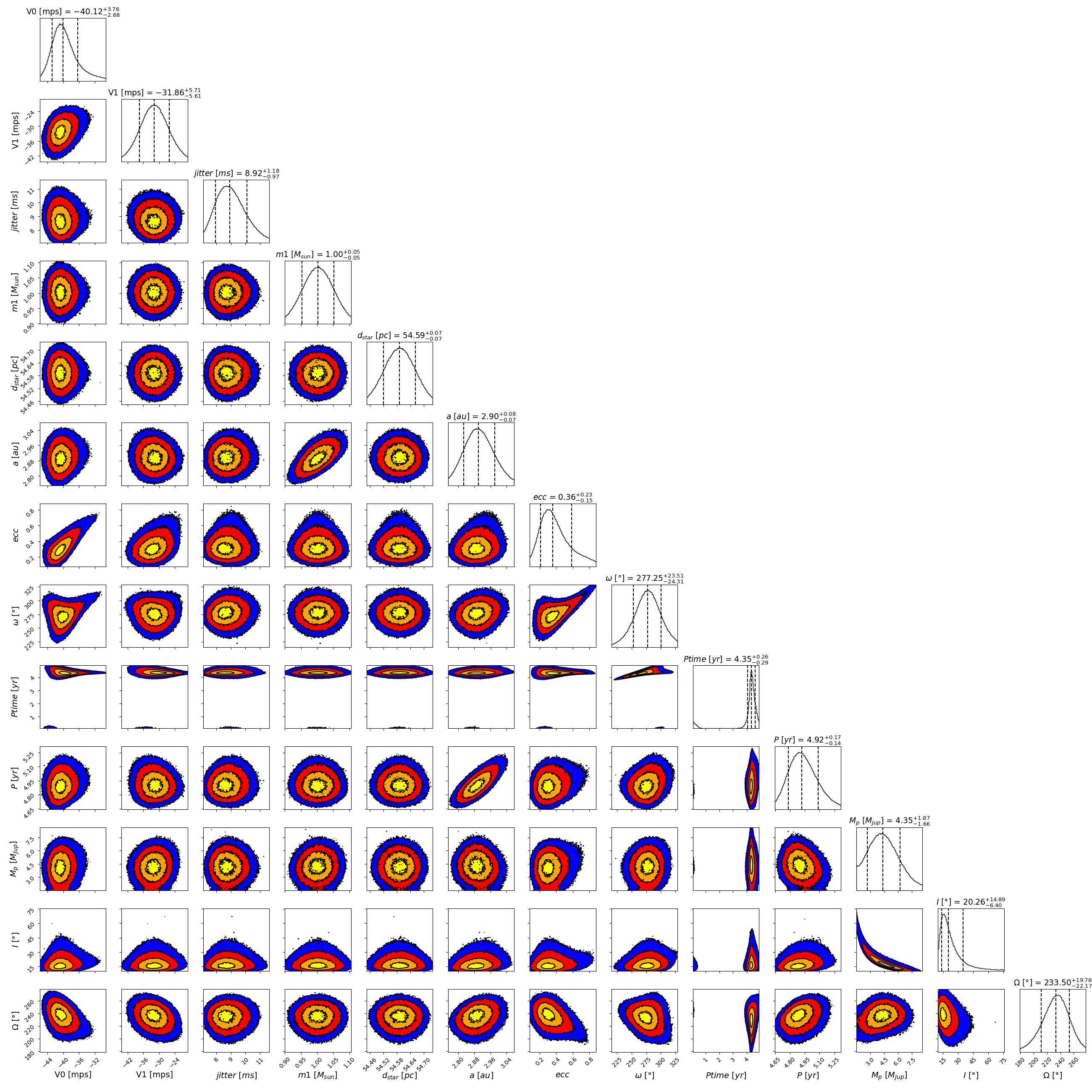}
 \caption{Fit of the orbit of HD167677. \textit{Top left}: Fit of the HD167677 RV measurements. \textit{Top right}: Fit of the HD167677 astrometric acceleration in right ascension (left) and declination (right). The black points correspond to the Hipparcos and Gaia EDR3 proper motion measurements. In each plot, the black curve corresponds to the best fit. The color bar indicates the log-likelihood corresponding to the different fits plotted. \textit{Bottom}: Corner plot of the fitted parameters of HD167677 b, considering a prior on \textit{I} between 0 and 90°. An offset of 57.2 km/s is added to V0 (H03) and V1 (H15) to improve readability.
 \label{RV_AA_HD167677}} 
 \end{figure}

 \begin{figure}[h!]
 \centering
 \includegraphics[width=1.0\textwidth]{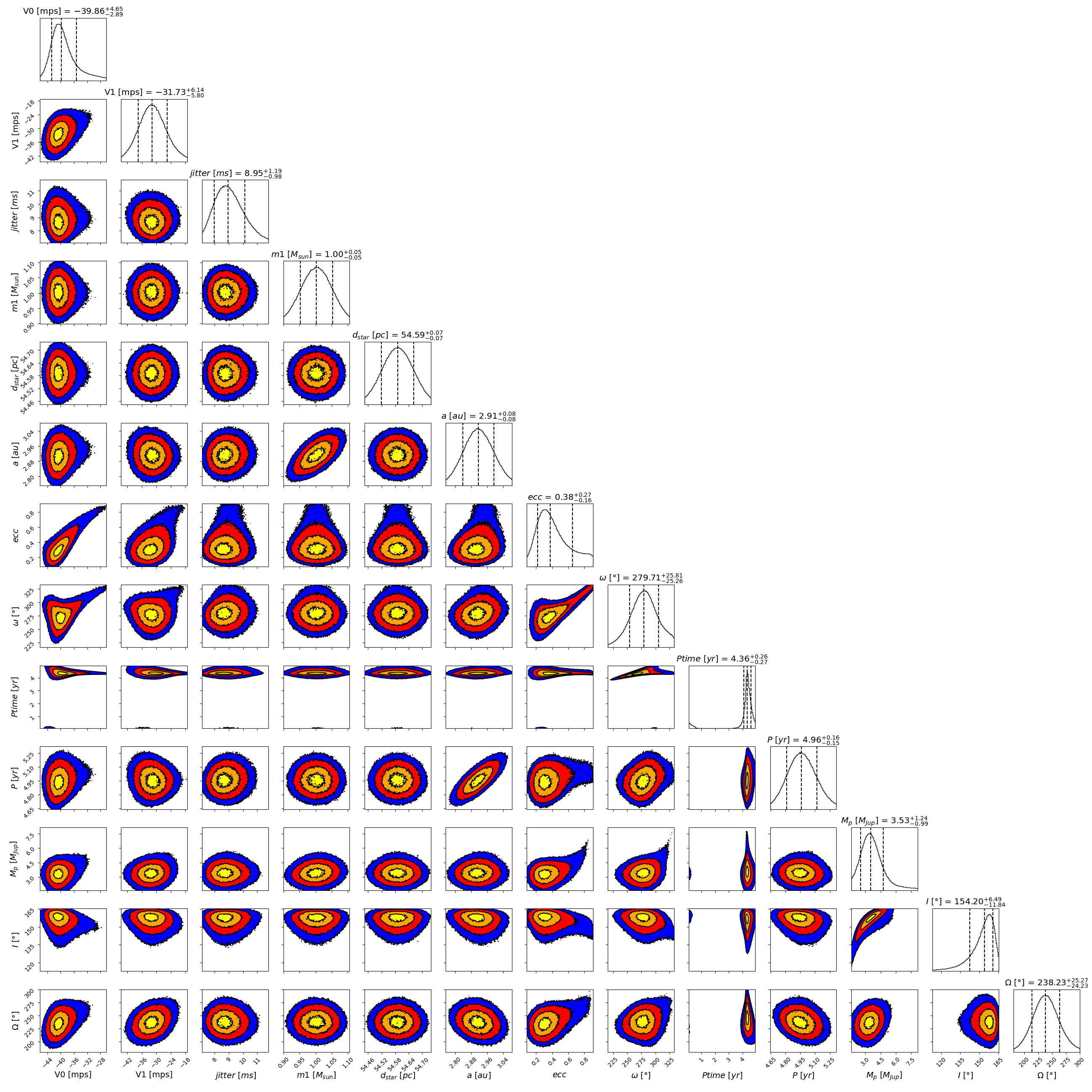}
 \caption{Corner plot of the fitted parameters of HD167677 b, considering a prior on \textit{I} between 90 and 180°. An offset of 57.2 km/s is added to V0 (H03) and V1 (H15) to improve readability.
 \label{RV_AA_HD167677_sup}} 
 \end{figure}

\begin{figure}[h!]
\centering
\includegraphics[width=0.33\textwidth]{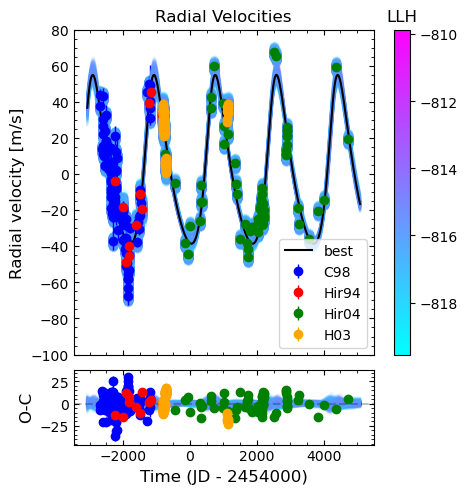}
\includegraphics[width=0.6\textwidth]{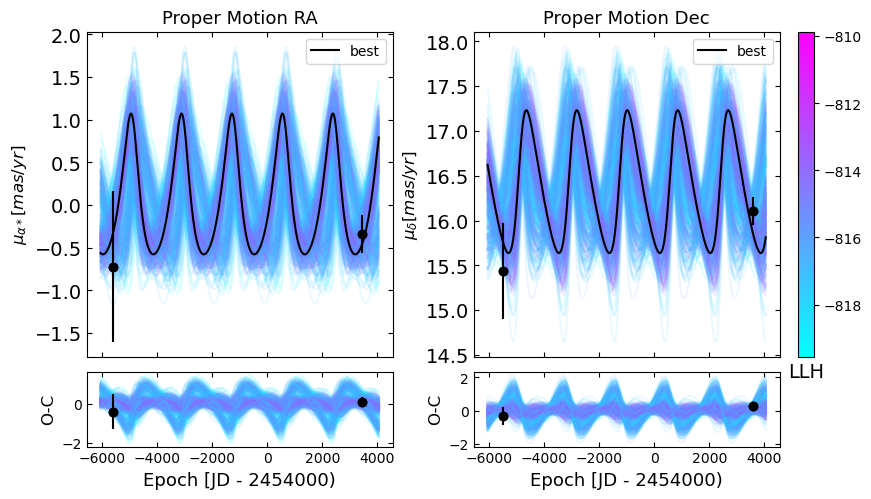}
\includegraphics[width=0.88\textwidth]{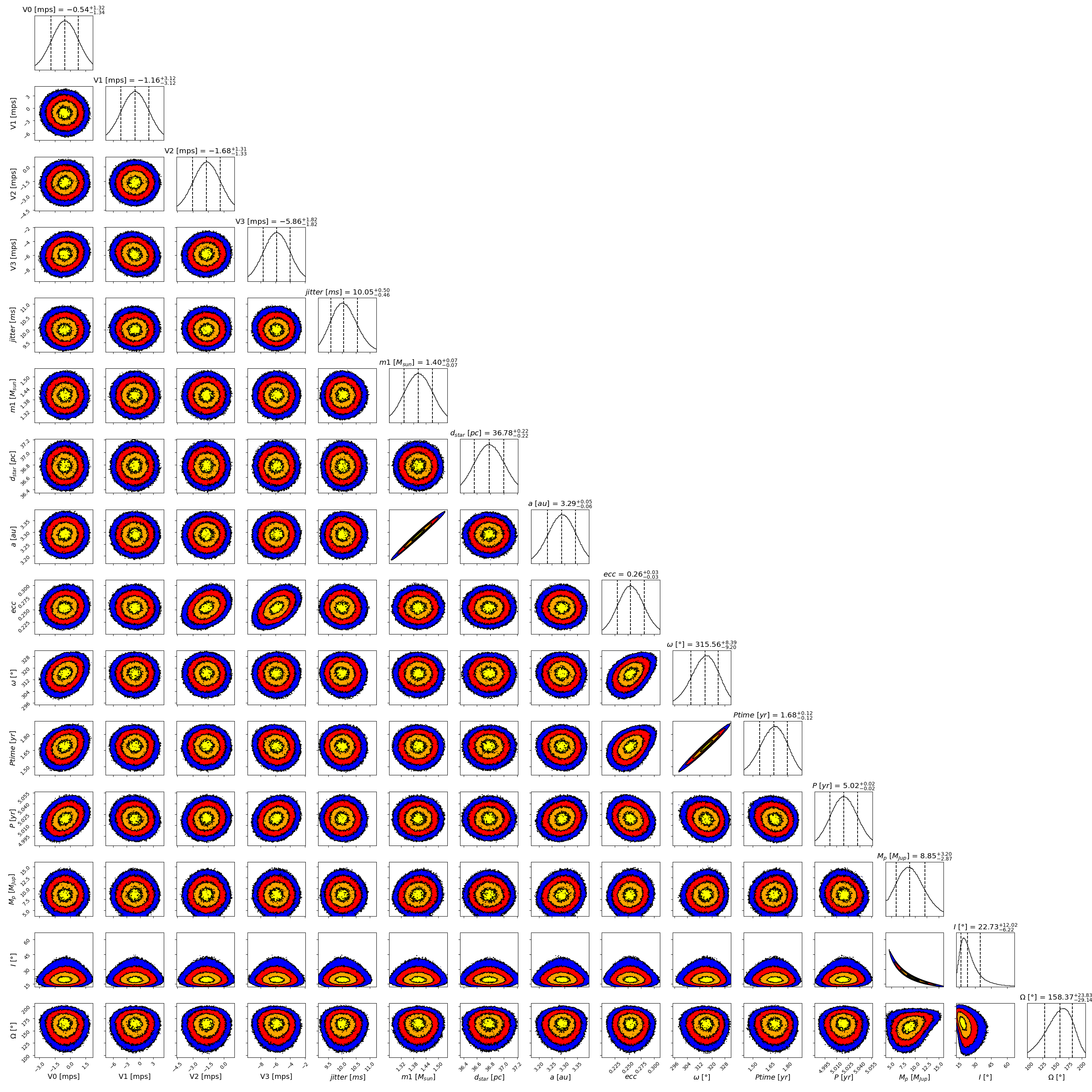}
\caption{Fit of the orbit of HD169830. \textit{Top left}: Fit of the HD169830 RV measurements after subtracted the signal of the inner planet. \textit{Top right}: Fit of the HD169830 astrometric acceleration in right ascension (left) and declination (right). The black points correspond to the Hipparcos and Gaia EDR3 proper motion measurements. In each plot, the black curve corresponds to the best fit. The color bar indicates the log-likelihood corresponding to the different fits plotted. \textit{Bottom}: Corner plot of the fitted parameters of HD169830 c, considering a prior on \textit{I} between 0 and 90°. V0, V1, V2, and V3 correspond to C98, Hir94, Hir04, and H03, respectively.
\label{RV_AA_HD169830}} 
\end{figure}

\begin{figure}[h!]
\centering
\includegraphics[width=1.0\textwidth]{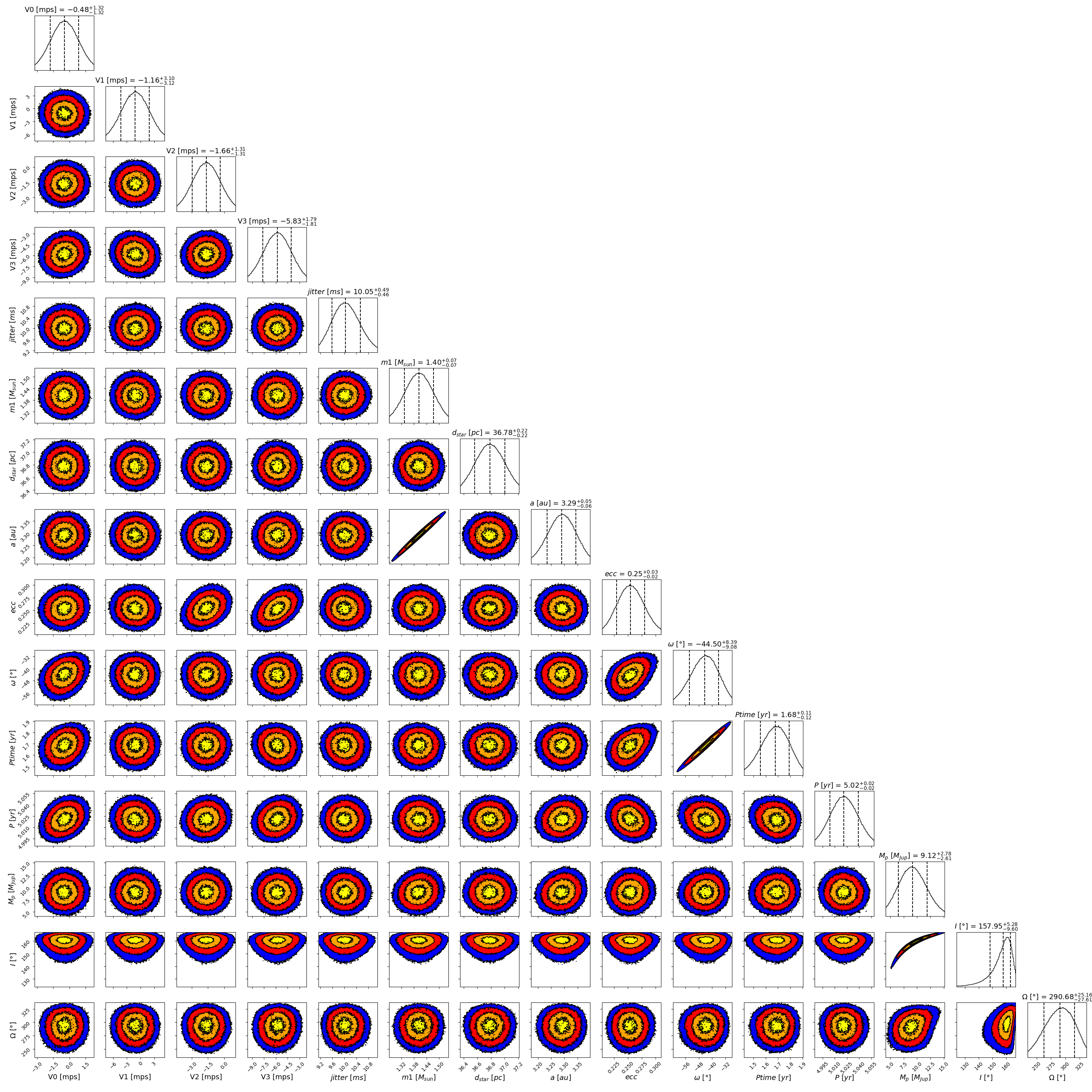}
\caption{Corner plot of the fitted parameters of HD169830 c, considering a prior on \textit{I} between 90 and 180°. V0, V1, V2, and V3 correspond to C98, Hir94, Hir04, and H03, respectively.
\label{RV_AA_HD169830_sup}} 
\end{figure}

 \begin{figure}[h!]
 \centering
 \includegraphics[width=0.33\textwidth]{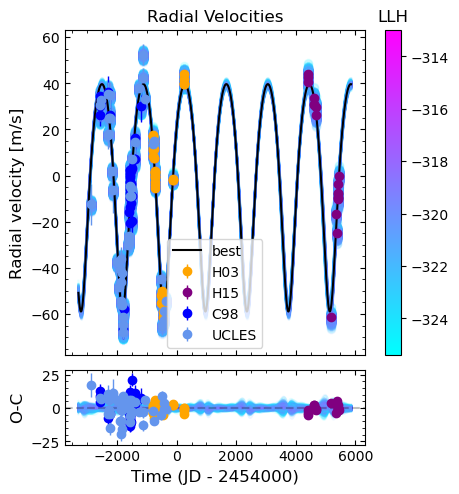}
 \includegraphics[width=0.6\textwidth]{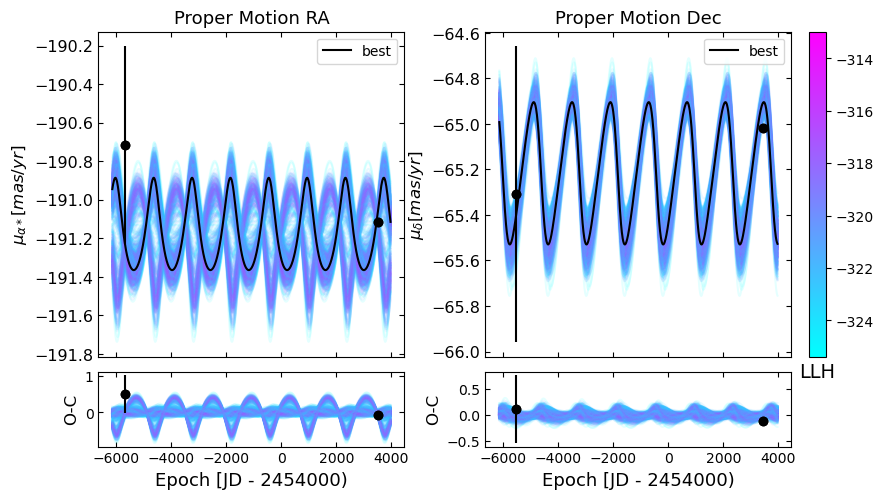}
 \includegraphics[width=0.88\textwidth]{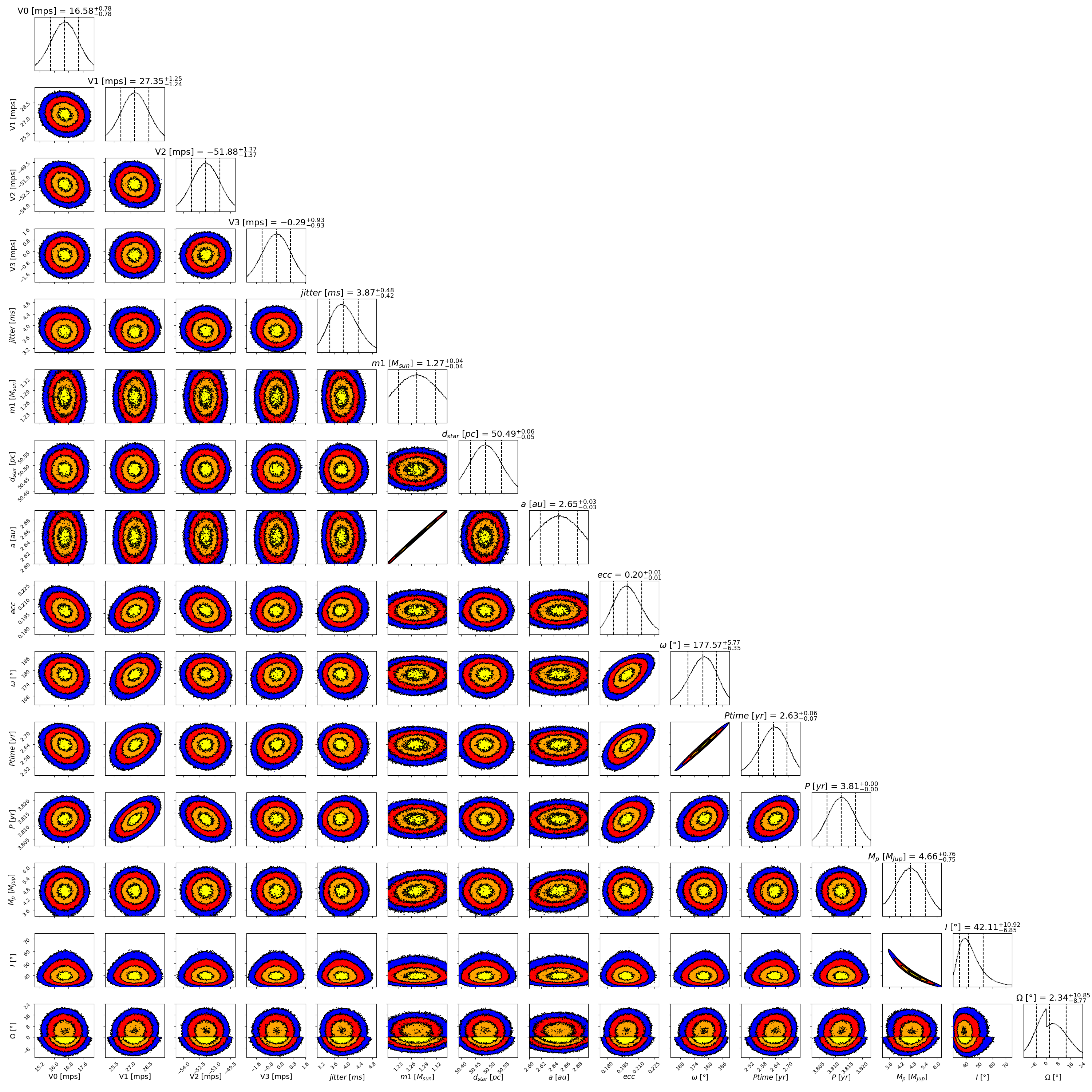}
 \caption{Fit of the orbit of HD196050. \textit{Top left}: Fit of the HD196050 RV measurements. \textit{Top right}: Fit of the HD196050 astrometric acceleration in right ascension (left) and declination (right). The black points correspond to the Hipparcos and Gaia EDR3 proper motion measurements. In each plot, the black curve corresponds to the best fit. The color bar indicates the log-likelihood corresponding to the different fits plotted. \textit{Bottom}: Corner plot of the fitted parameters of HD196050 b, considering a prior on \textit{I} between 0 and 90°. An offset of 61.4 km/s is subtracted to V0 and V1 to improve readability. V0, V1, V2, and V3 correspond to H03, H15, C98, and UCLES, respectively.
 \label{RV_AA_HD196050}} 
 \end{figure}

 \begin{figure}[h!]
 \centering
 \includegraphics[width=1.0\textwidth]{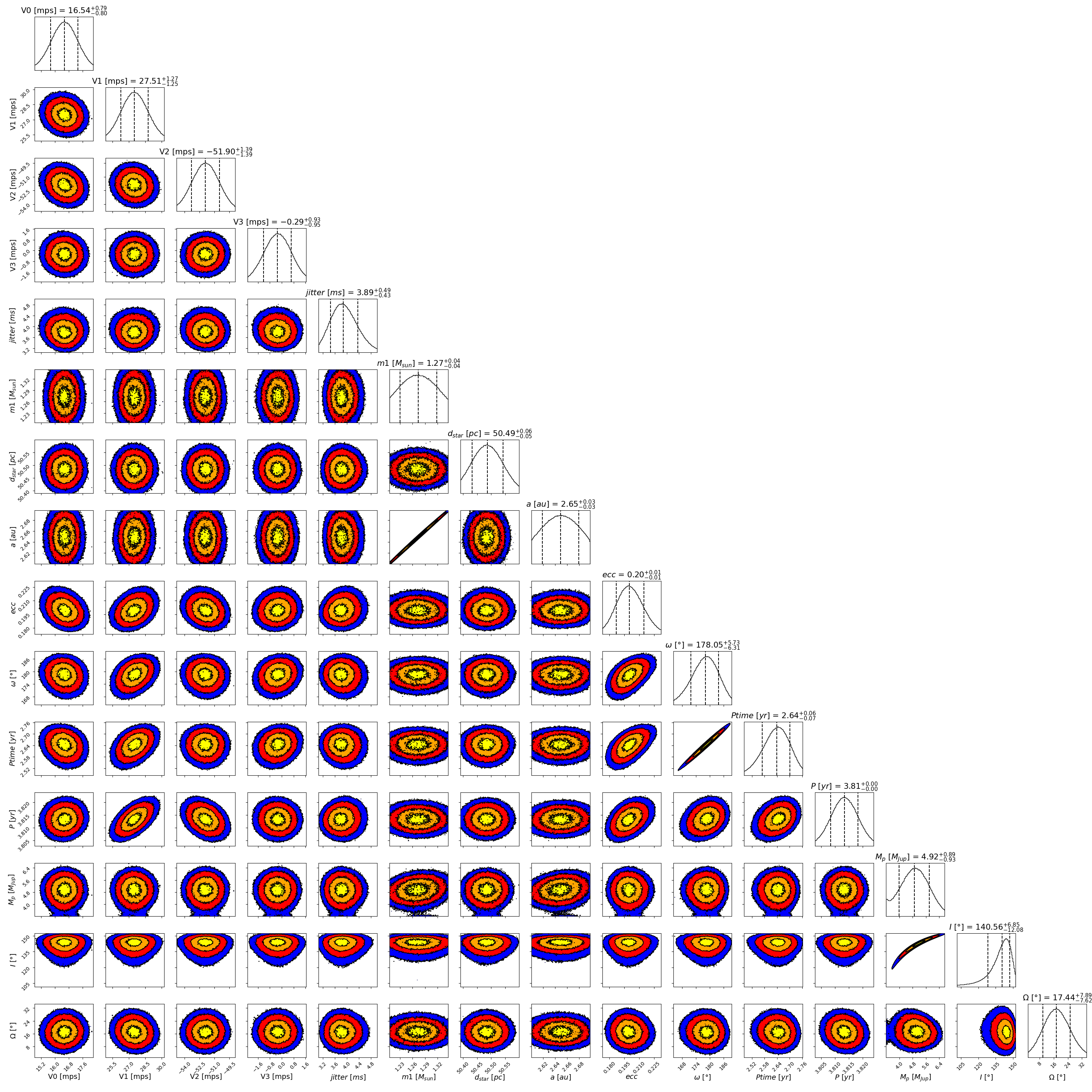}
 \caption{Corner plot of the fitted parameters of HD196050 b, considering a prior on \textit{I} between 90 and 180°. An offset of 61.4 km/s is subtracted to V0 and V1 to improve readability. V0, V1, V2, and V3 correspond to H03, H15, C98, and UCLES, respectively.
 \label{RV_AA_HD196050_sup}} 
 \end{figure}

 \begin{figure}[h!]
 \centering
 \includegraphics[width=0.33\textwidth]{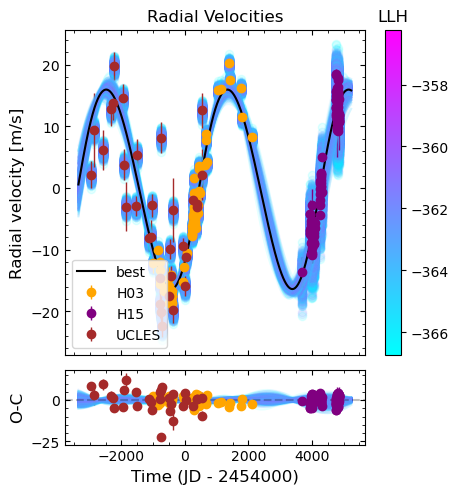}
 \includegraphics[width=0.6\textwidth]{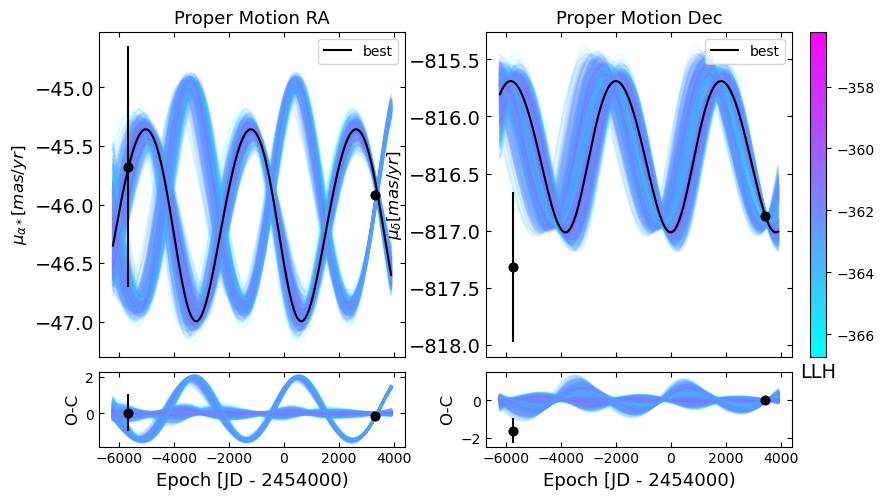}
 \includegraphics[width=0.88\textwidth]{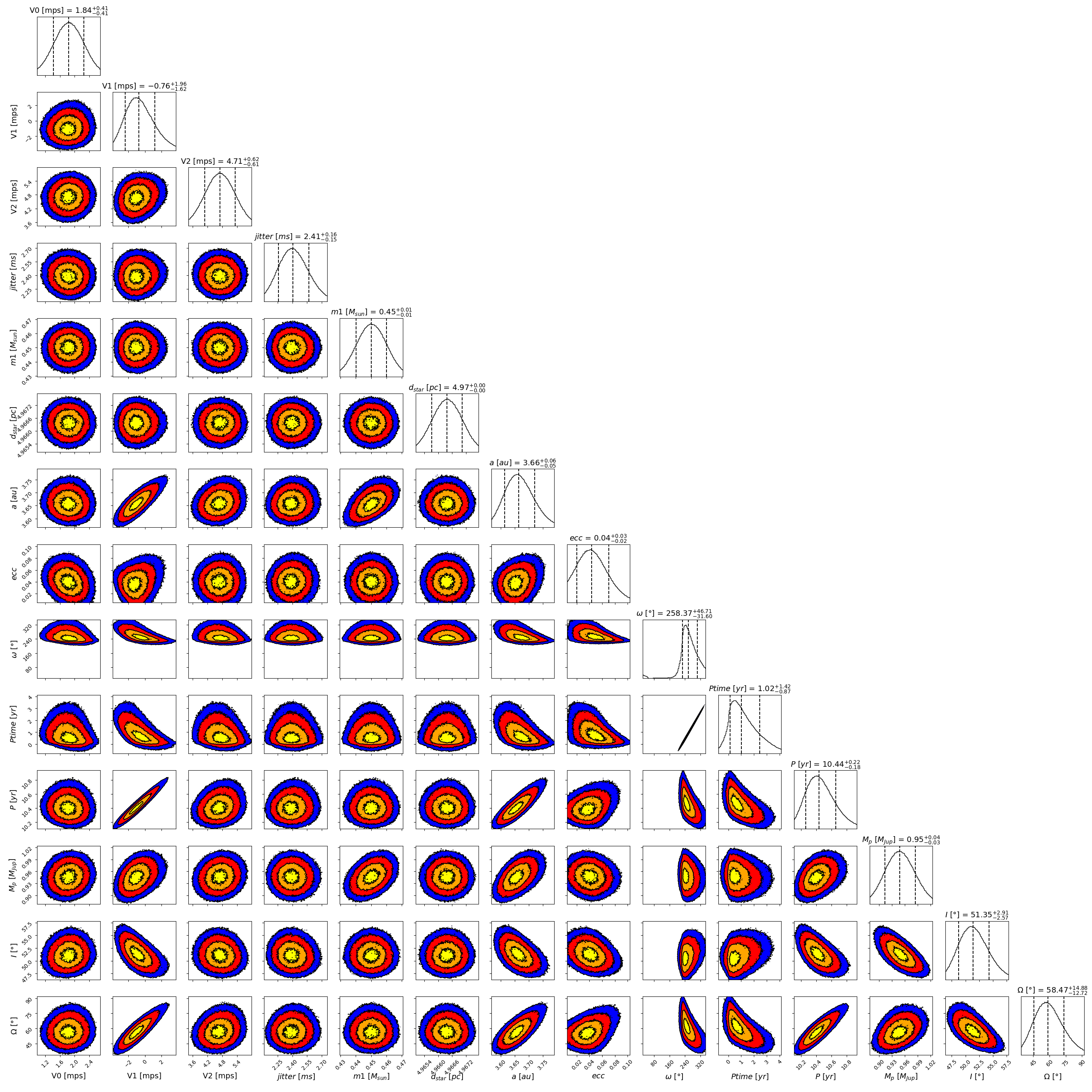}
 \caption{Fit of the orbit of HD204961. \textit{Top left}: Fit of the HD204961 RV measurements. \textit{Top right}: Fit of the HD204961 astrometric acceleration in right ascension (left) and declination (right). The black points correspond to the Hipparcos and Gaia EDR3 proper motion measurements. In each plot, the black curve corresponds to the best fit. The color bar indicates the log-likelihood corresponding to the different fits plotted. \textit{Bottom}: Corner plot of the fitted parameters of HD204961 b. An offset of 13.0 km/s is subtracted to V0 and V1 to improve readability, considering a prior on \textit{I} between 0 and 90°. V0, V1, and V2 correspond to H03, H15, and UCLES, respectively.
 \label{RV_AA_HD204961}} 
 \end{figure}

 \begin{figure}[h!]
 \centering
 \includegraphics[width=1.0\textwidth]{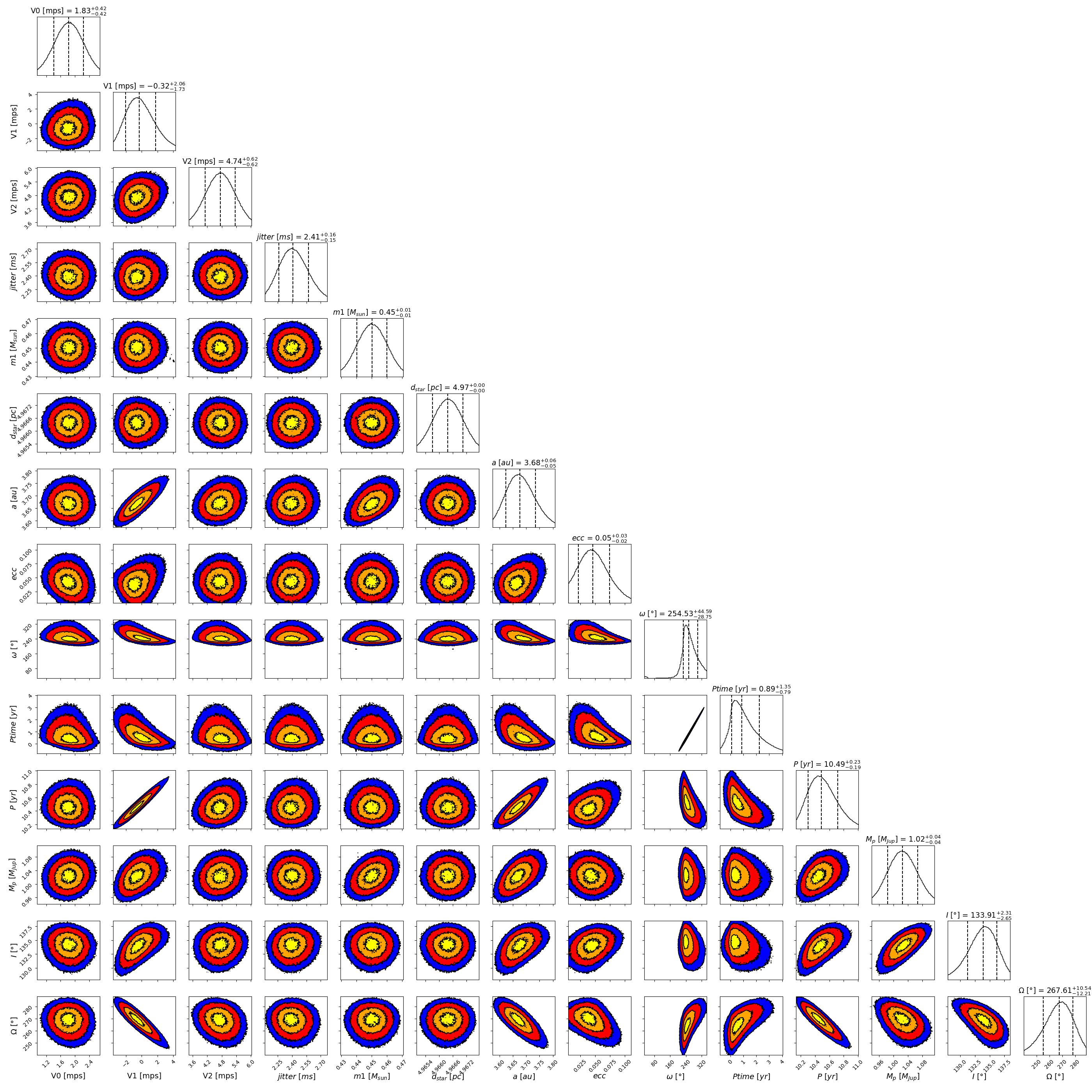}
 \caption{Corner plot of the fitted parameters of HD204961 b, considering a prior on \textit{I} between 90 and 180°. An offset of 13.0 km/s is subtracted to V0 and V1 to improve readability. V0, V1, and V2 correspond to H03, H15, and UCLES, respectively.
 \label{RV_AA_HD204961_sup}} 
 \end{figure}

 \begin{figure}[h!]
 \centering
 \includegraphics[width=0.33\textwidth]{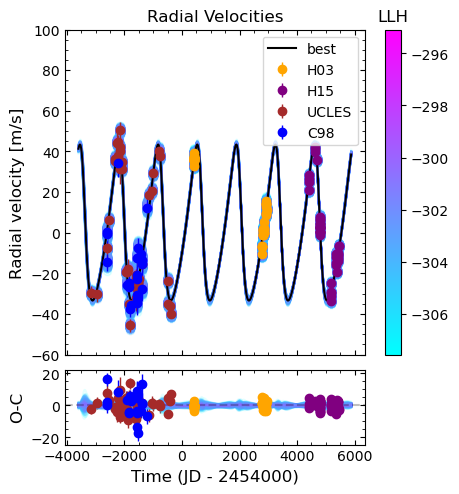}
 \includegraphics[width=0.6\textwidth]{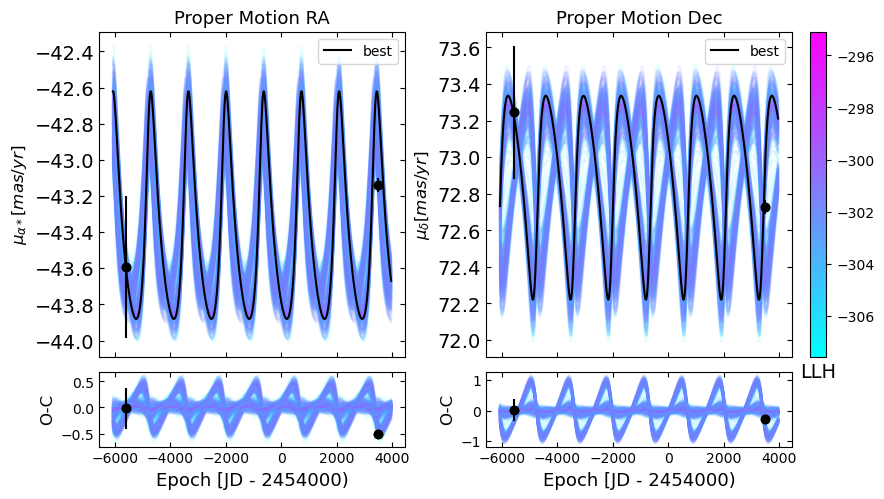}
 \includegraphics[width=0.88\textwidth]{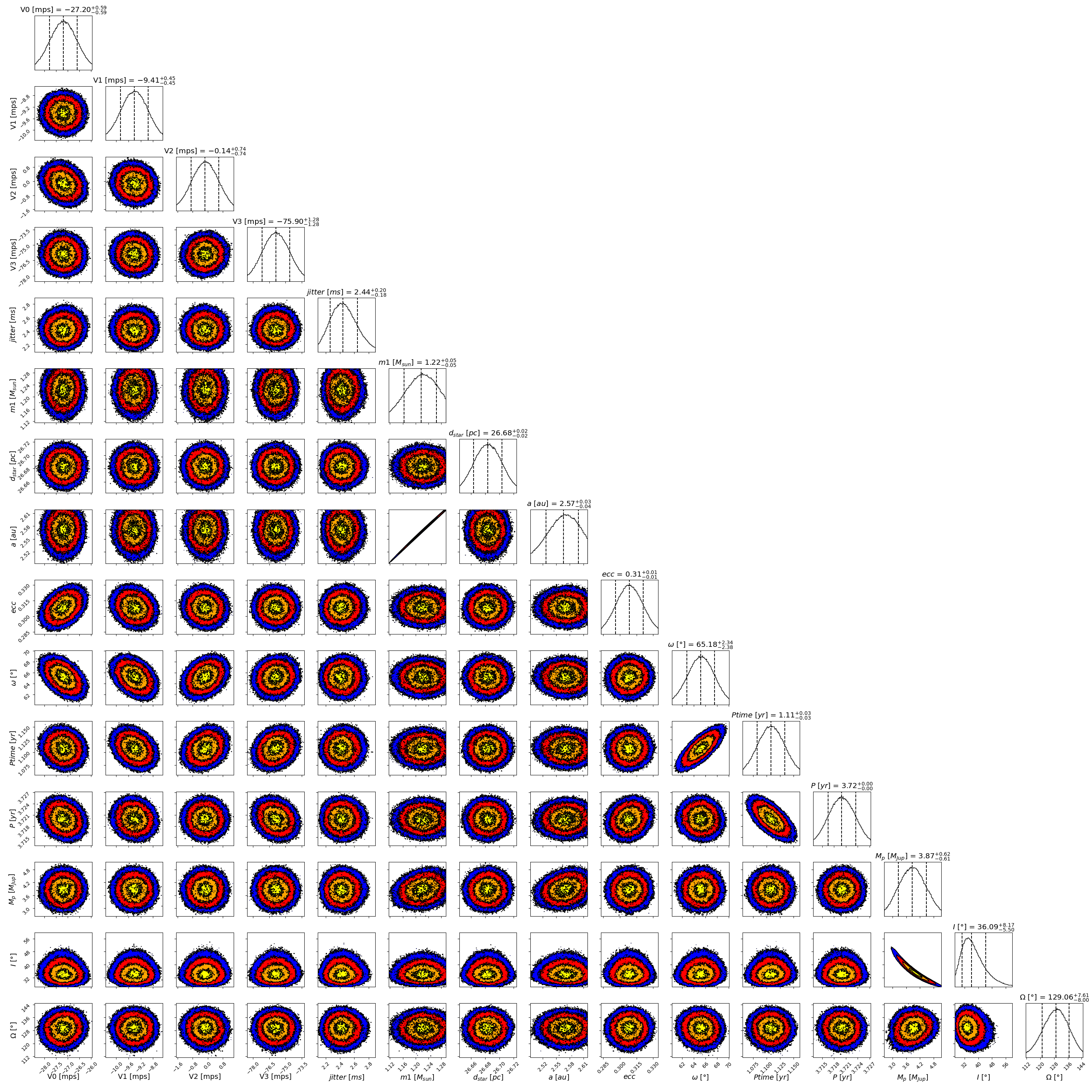}
 \caption{Fit of the orbit of HD216437. \textit{Top left}: Fit of the HD216437 RV measurements. \textit{Top right}: Fit of the HD216437 astrometric acceleration in right ascension (left) and declination (right). The black points correspond to the Hipparcos and Gaia EDR3 proper motion measurements. In each plot, the black curve corresponds to the best fit. The color bar indicates the log-likelihood corresponding to the different fits plotted. \textit{Bottom}: Corner plot of the fitted parameters of HD216437 b, considering a prior on \textit{I} between 0 and 90°. An offset of 2.2 km/s is added to V0 and V1 to improve readability. V0, V1, V2, and V3 correspond to H03, H15, UCLES, and C98, respectively.
 \label{RV_AA_HD216437}} 
 \end{figure}

 \begin{figure}[]
 \centering
 \includegraphics[width=1.0\textwidth]{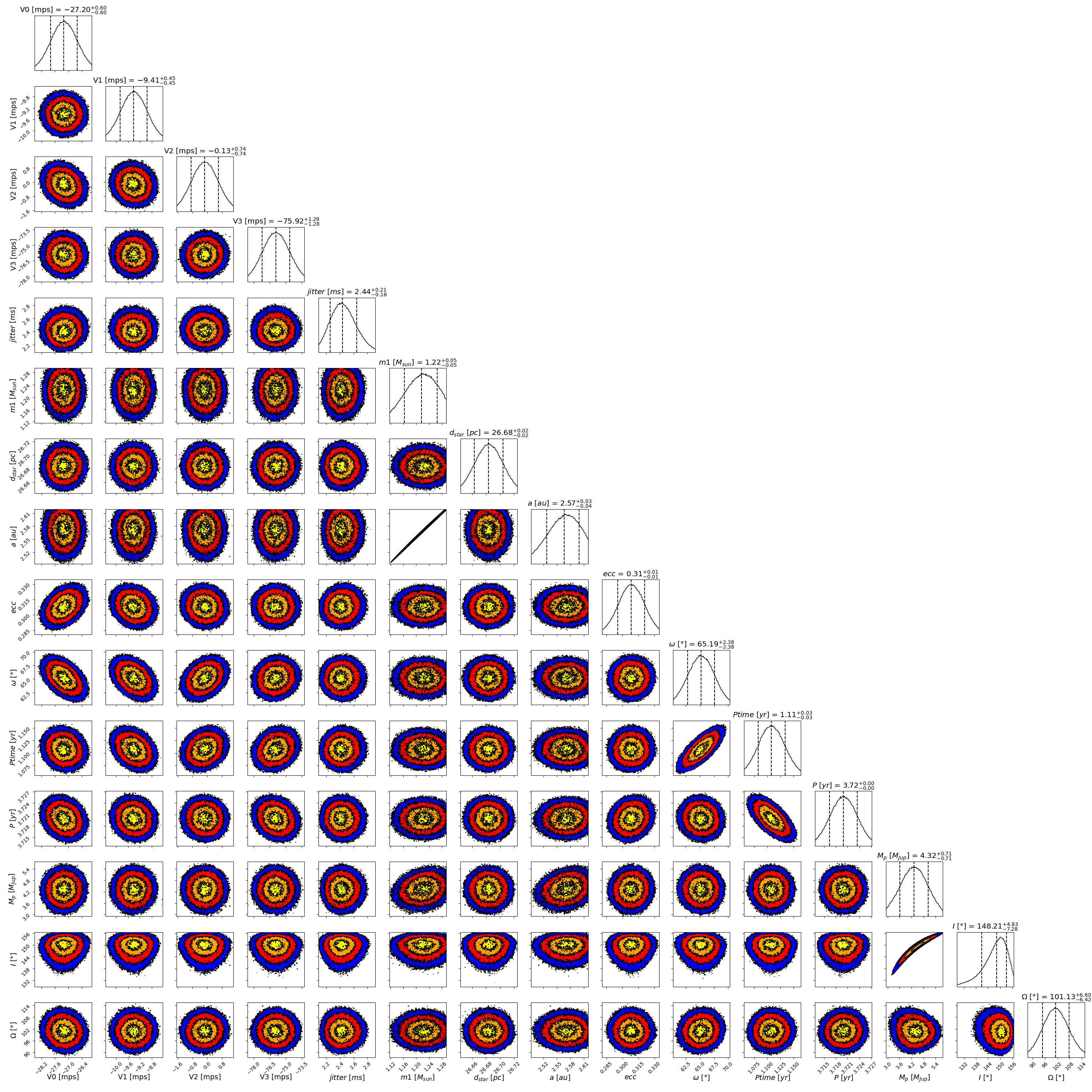}
 \caption{Corner plot of the fitted parameters of HD216437 b, considering a prior on \textit{I} between 90 and 180°. An offset of 2.2 km/s is added to V0 and V1 to improve readability. V0, V1, V2, and V3 correspond to H03, H15, UCLES, and C98, respectively.
 \label{RV_AA_HD216437_sup}} 
 \end{figure}

 \begin{figure}[]
 \centering
 \includegraphics[width=0.33\textwidth]{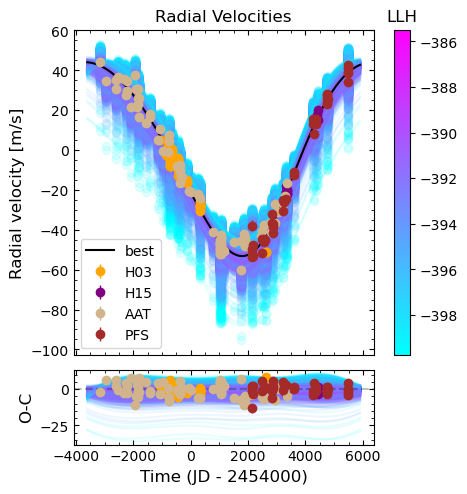}
 \includegraphics[width=0.6\textwidth]{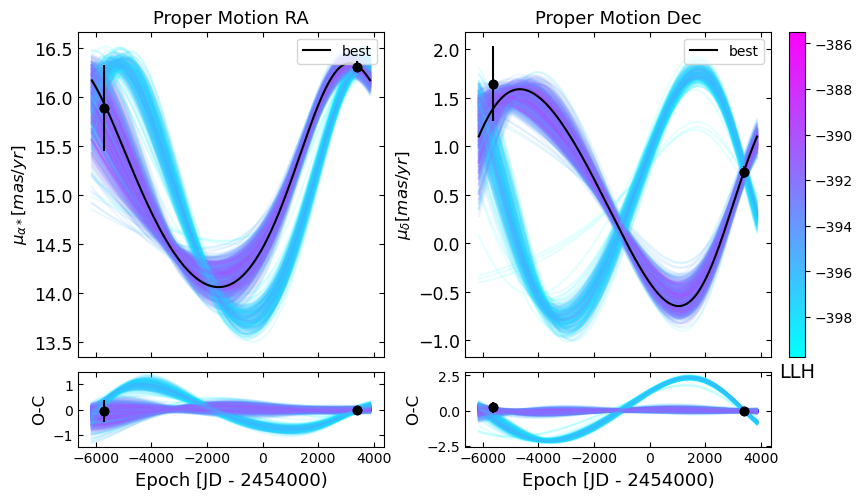}
 \includegraphics[width=0.88\textwidth]{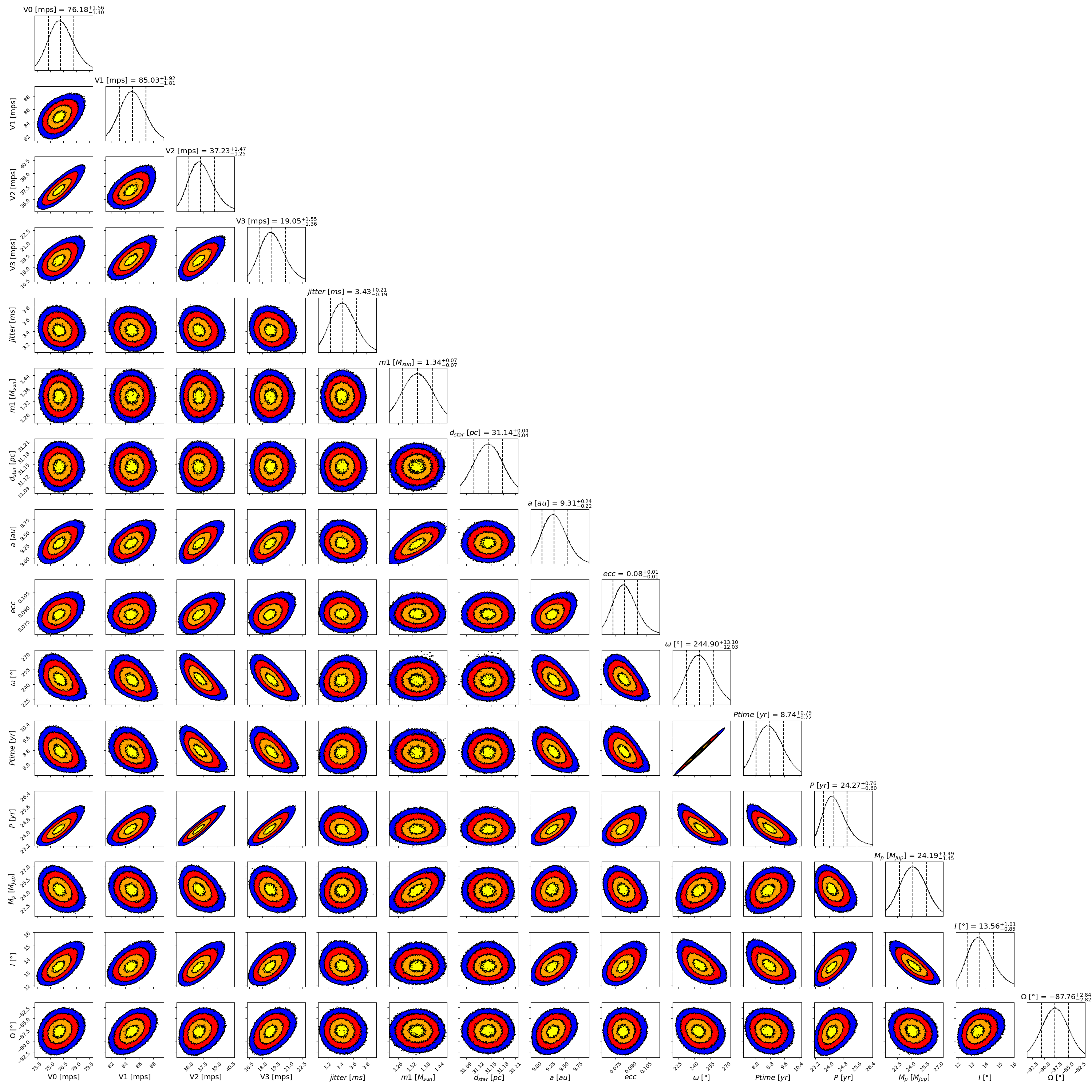}
 \caption{Fit of the orbit of HD221420. \textit{Top left}: Fit of the HD221420 RV measurements. \textit{Top right}: Fit of the HD221420 astrometric acceleration in right ascension (left) and declination (right). The black points correspond to the Hipparcos and Gaia EDR3 proper motion measurements. In each plot, the black curve corresponds to the best fit. The color bar indicates the log-likelihood corresponding to the different fits plotted. \textit{Bottom}: Corner plot of the fitted parameters of HD221420 B, considering a prior on \textit{I} between 0 and 90°. An offset of 26.5 km/s is substracted to V0 and V1 to improve readability. V0, V1, V2, and V3 correspond to H03, H15, AAT, and PFS, respectively.
 \label{RV_AA_HD221420}} 
 \end{figure}

 \begin{figure}[]
 \centering
 \includegraphics[width=1.0\textwidth]{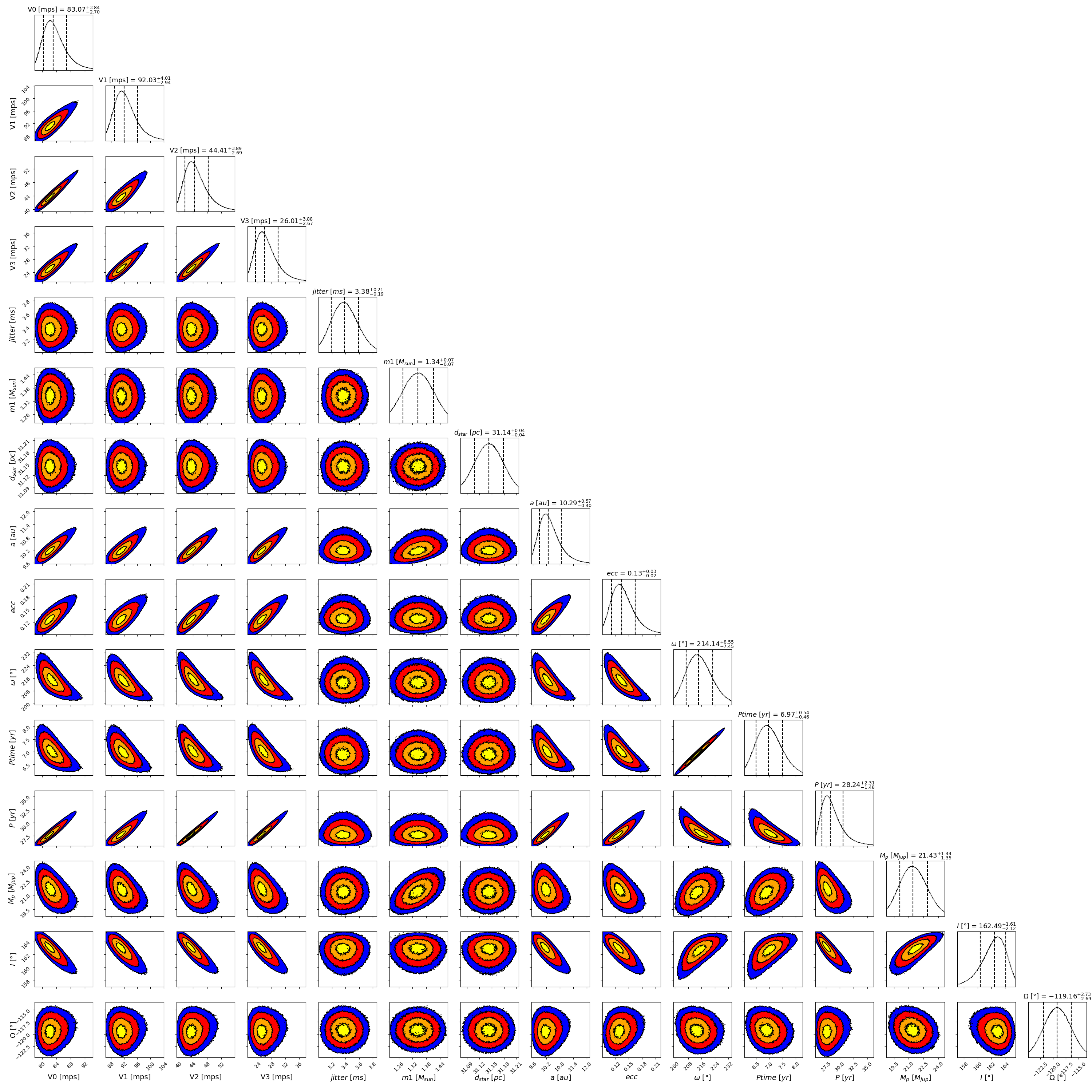}
 \caption{Corner plot of the fitted parameters of HD221420 B, considering a prior on \textit{I} between 90 and 180°. An offset of 26.5 km/s is substracted to V0 and V1 to improve readability. V0, V1, V2, and V3 correspond to H03, H15, AAT, and PFS, respectively.
 \label{RV_AA_HD221420_sup}} 
 \end{figure}

 \begin{figure}[]
 \centering
 \includegraphics[width=0.33\textwidth]{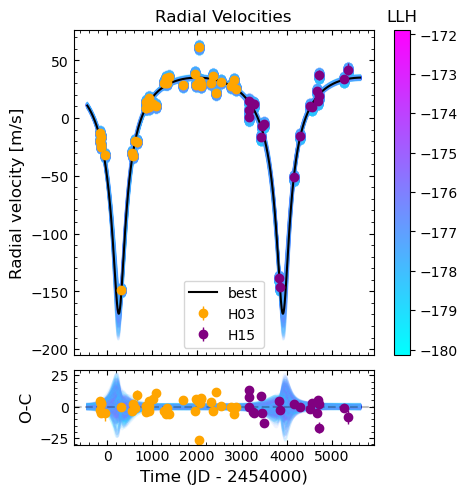}
 \includegraphics[width=0.6\textwidth]{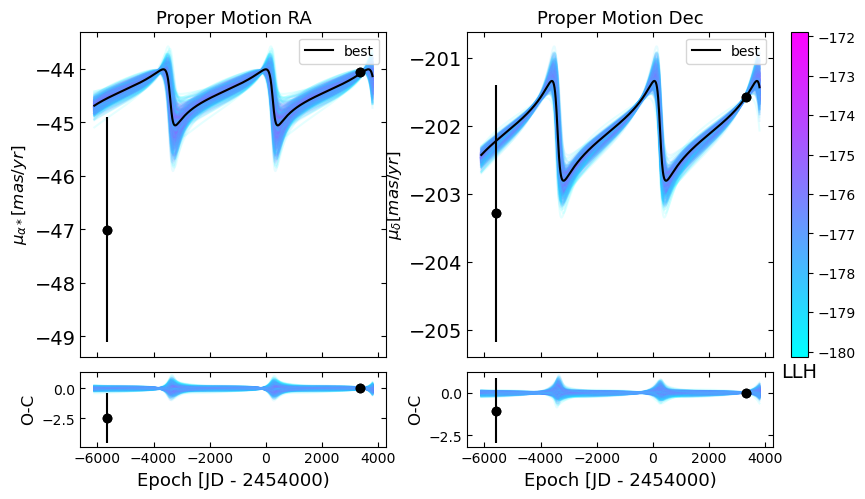}
 \includegraphics[width=0.88\textwidth]{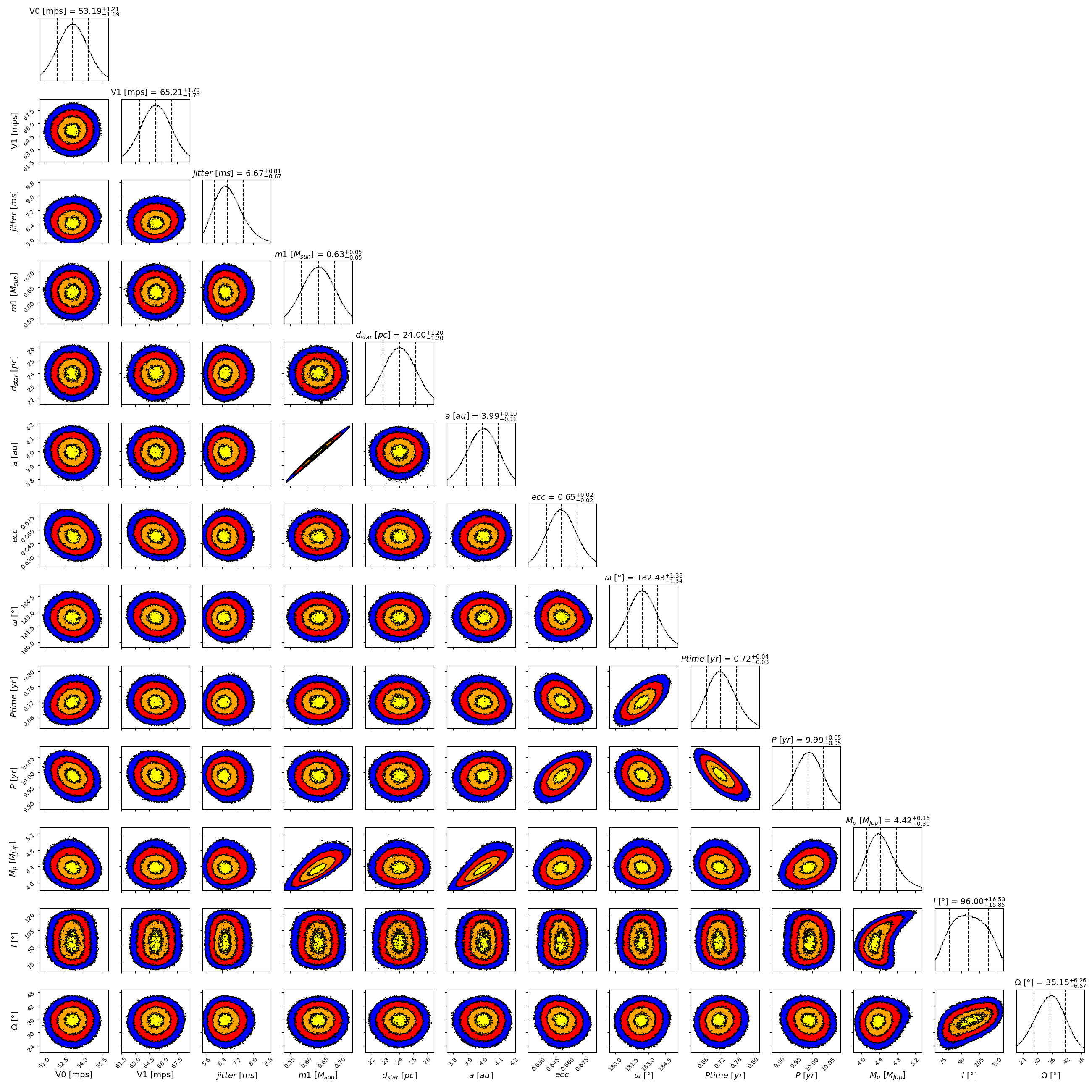}
 \caption{Fit of the orbit of HIP70849. \textit{Top left}: Fit of the HIP70849 RV measurements. \textit{Top right}: Fit of the HIP70849 astrometric acceleration in right ascension (left) and declination (right). The black points correspond to the Hipparcos and Gaia EDR3 proper motion measurements. In each plot, the black curve corresponds to the best fit. The color bar indicates the log-likelihood corresponding to the different fits plotted. \textit{Bottom}: Corner plot of the fitted parameters of HIP70849 b. V0 and V1 correspond to H03 and H15, respectively.
 \label{RV_AA_HIP70849}} 
 \end{figure}

\subsubsection{short-period companions}

\begin{figure}[]
 \centering
\includegraphics[width=0.8\textwidth]{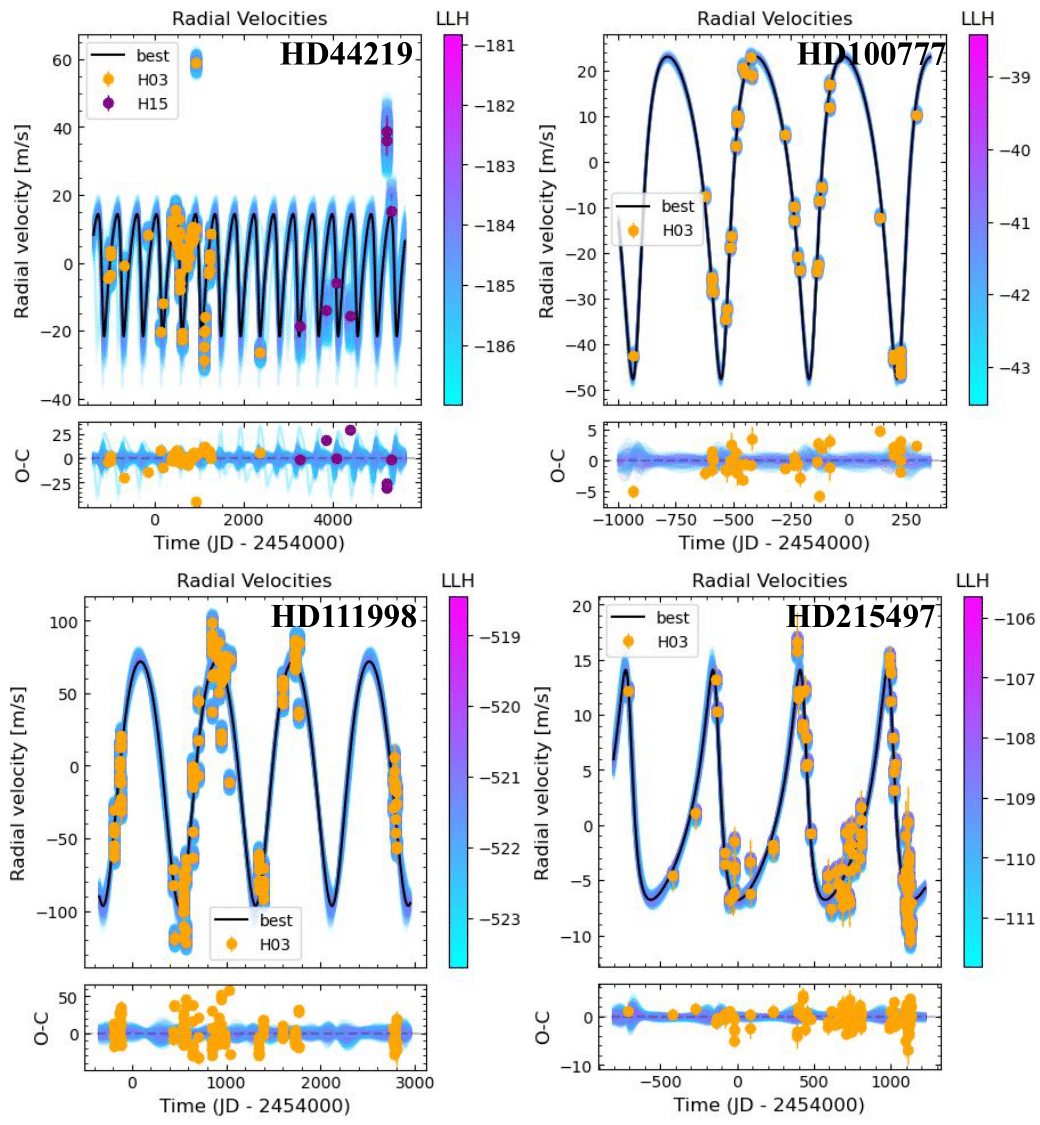}
\caption{Fit of RV data points of the stars hosting short-period planets with MCMC. The stars shown are, from left to right and top to bottom, HD44219, HD100777, HD111998, and HD215497.
\label{RV_short_period}} 
\end{figure}

\end{document}